\documentclass[oneside,12pt,final]{sty/ucthesis-CA2012}
\pdfoutput=1

\usepackage{fancyhdr}
\usepackage{caption}
\usepackage{subcaption}
\usepackage{hyperref}
\hypersetup{hidelinks}
\usepackage{amsmath, amssymb, graphicx}
\graphicspath{{./fig/}{./NCFT/}{./FreezeTwin/}{./LLP/}{./TwinCosmo/}{./TwinCosmo/Figures/}}
\usepackage{xspace}
\usepackage{braket}
\usepackage{color}
\usepackage[dvipsnames]{xcolor}
\usepackage{subfiles}
\usepackage{epigraph}
\usepackage{captdef}
\usepackage[backgroundcolor=green]{todonotes}
\usepackage{chngcntr}
\counterwithout{footnote}{chapter}
\usepackage{wrapfig}
\usepackage{pdfpages}

\usepackage{titlesec}

\newcommand{\be}{\begin{equation}}
\newcommand{\ee}{\end{equation}}

\newcommand{\half}{\frac{1}{2}}
\newcommand{\pl}{\text{pl}}

\usepackage{booktabs}
\usepackage[english]{babel}
\usepackage{amsmath,amssymb,amsbsy,amstext,simplewick}
\usepackage{graphicx}
\usepackage[export]{adjustbox}
\usepackage{upgreek}
\usepackage{cancel}
\usepackage{color}
\usepackage{slashed}
\usepackage[dvipsnames]{xcolor}
\usepackage{tikz-cd}
\usepackage{color}
\usepackage{soul}
\usepackage{dsfont}
\usepackage[force]{feynmp-auto}
\usepackage[utf8]{inputenc}

\usepackage{graphicx}
\usepackage{amsmath} 
\usepackage[T1]{fontenc} 
\usepackage{amssymb}
\usepackage{cleveref}
\usepackage{upgreek}
\usepackage{listings}
\usepackage{slashed}
\usepackage{mathtools}
\usepackage{ctable}
\usepackage{feynmp-auto}

\newcommand{\prn}[1]{ \left(  #1 \right) }

\newcommand{\magn}[1]{\left| #1 \right|}

\newcommand{\al}[1]{\begin{align} #1 \end{align}}

\newcommand{\Neff}{N_{\text{eff}}}
\newcommand{\gamp}{{\gamma '}}
\newcommand{\QCD}{\Lambda_{\text{QCD}}}
\newcommand{\dd}{\partial}
\newcommand{\ep}{{e'}}
\newcommand{\ebp}{{\bar{e}'}}
\newcommand{\oo}{\infty}
\newcommand{\MM}{\mathcal{M}}
\newcommand{\avg}[1]{\left< #1 \right>}

\def\beq{\begin{equation}}
\def\eeq{\end{equation}}
\def\bea{\begin{eqnarray}}
\def\eea{\end{eqnarray}}
\newcommand{\mpl}{M_\text{pl}}
\newcommand{\appropto}{\mathrel{\vcenter{
			\offinterlineskip\halign{\hfil$##$\cr
				\propto\cr\noalign{\kern2pt}\sim\cr\noalign{\kern-2pt}}}}}

\usepackage{multicol}
\usepackage{graphicx}
\usepackage{booktabs}
\usepackage{amssymb,bm,mathrsfs,bbm,amscd}
\usepackage{upgreek,fancyhdr}
\usepackage{multirow}
\usepackage{url}
\usepackage{soul}

\newcommand{\eehz}{e^+e^- \to hZ}

\usepackage{letltxmacro}
\makeatletter
\let\oldr@@t\r@@t
\def\r@@t#1#2{%
	\setbox0=\hbox{$\oldr@@t#1{#2\,}$}\dimen0=\ht0
	\advance\dimen0-0.2\ht0
	\setbox2=\hbox{\vrule height\ht0 depth -\dimen0}%
	{\box0\lower0.4pt\box2}}
\LetLtxMacro{\oldsqrt}{\sqrt}
\renewcommand*{\sqrt}[2][\ ]{\oldsqrt[#1]{#2}}
\makeatother

\begin{document}

\begin{frontmatter}

\title{The Hierarchy Problem: \\ From the Fundamentals to the Frontiers}

\author{Seth Koren}

\report{Dissertation} \degree{Doctor of Philosophy} \degreemonth{September} \degreeyear{2020}
\defensemonth{August} 
\defenseyear{2020}

\chair{Professor Nathaniel Craig}  
\othermemberA{Professor Mark Srednicki} 
\othermemberB{Professor Claudio Campagnari} 

\numberofmembers{3} 

\field{Physics}
\campus{Santa Barbara}
	\maketitle
	\includepdf{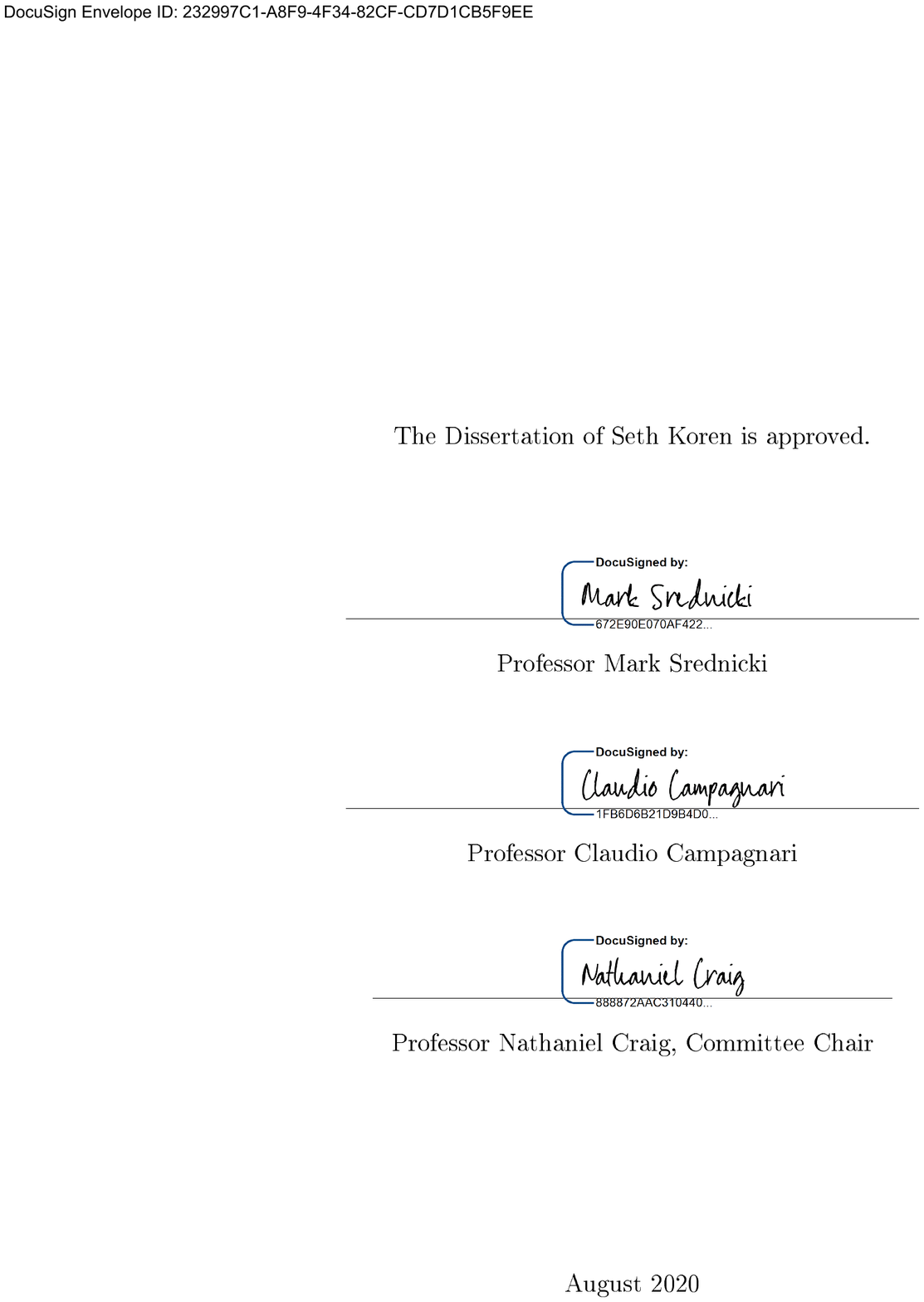}
	\copyrightpage
	\begin{dedication}

\bigskip

${}$ \\

\bigskip

${}$ \\

\bigskip

\begin{center}
\begin{large}
	
``You seem very clever at explaining words, Sir'', said Alice. \\ ``Would you kindly tell me the meaning of the poem `Jabberwocky'?'' \vspace{1cm}

``Let's hear it'', said Humpty Dumpty. \\ ``I can explain all the poems that ever were invented \\ --- and a good many that haven't been invented just yet.'' \vspace{1cm}

Lewis Carroll \\
\textit{Through the Looking Glass} (1871) \cite{carroll_through}

\end{large}
\end{center}

\end{dedication} \newpage
	\section*{Acknowledgements} \addcontentsline{toc}{chapter}{Acknowledgements}

My thanks go first and foremost to Nathaniel Craig for his continual support and encouragement. From Nathaniel I learned not only an enormous amount of exciting physics, but also how to prepare engaging lessons and talks, how to be an effective mentor, how to think intuitively about the natural world and determine the particle physics to describe it, and how to be a caring, welcoming, community-focused academic. It goes without saying that Nathaniel's influence pervades every word written below and the physical understanding behind them.

I would not have made it here were it not for the many senior academics who have charitably given their own time to encourage and support me. As a young undergraduate, Tom Lubensky's patient help during many office hours and his explicit encouragement for me to continue studying physics were vital. Cullen Blake took a risk on me as an undergrad with little to show but excitement and enthusiasm, taught me how to problem-solve like a physicist and a researcher, and truly made a huge difference in my life. And there were many such professors who kindly gave me their time and support---Mirjam Cveti\v{c}, Larry Gladney, Justin Khoury, H. H. ``Brig'' Williams, Ned Wright, and others---and if I listed all the ways they have all supererogatorily supported me and my education this section would be too long.

In graduate school I have also benefited from the kindness of a cadre of senior academics. Dave Sutherland and John Mason spent hours engaging me in many elucidating conversations. Don Marolf generously included me in gravity theory activities and answered my endless questions. Nima Arkani-Hamed has graciously given me his time and made me feel welcome at every turn. And there have been many other particle theorists who have offered me their advice and support over the years---Tim Cohen, Patrick Draper, Matthew McCullough, and Flip Tanedo, among others.


Of course I have not worked alone, and little of this science would have been accomplished were it not for my many collaborators who have helped me learn and problem-solve and provided guidance and been patient when I was overwhelmed with being pulled in too many directions. Let me especially mention those undergrads I have spent significant time mentoring during my time in graduate school, namely Aidan Herderschee, Samuel Alipour-fard, and Umut Can \"{O}ktem. Indeed, they each took a chance on me as well, and working with them has taught me how to be a better teacher and physicist---not to mention all of the great science we worked out together.

My physics knowledge would have also been stunted were it not for the countless hours spent discussing all manner of high energy theory with friends and peers---primarily Matthew Brown, Brianna Grado-White, Alex Kinsella, Robert McGehee, and Gabriel Trevi\~{n}o Verastegui. And my enjoyment of graduate school would have been stunted were it not for the board games, hiking, trivia, art walks, biking, and late-night philosophical discussions with them and with Dillon Cisco, Neelay Fruitwala, Eric Jones, Farzan Vafa, Sicheng Wang, and others. No one has enriched my life here moreso than Nicole Iannaccone---a summary statement of far too few words.

Nor would this have been possible were it not for my `medical support team' consisting primarily of endocrinologists Dr. Mark Wilson and Dr. Ashley Thorsell, student health physician Dr. Miguel Pedroza, and my therapist Dr. Karen Dias, who have helped me immensely in my time here. Even putting aside physiological issues, graduate school can be and has been incredibly mentally taxing. I'm not sure I could not have made it through were it not for the psychological assistance I have received, both psychotherapeutic and pharmacological.

Of course a special `shout-out' goes to Mother Nature. The idyllic weather, scenic ocean, beautiful mountains, and clear night skies of Santa Barbara may have just been too distracting for me to progress through my degree were it not for the Rey fire, the Whittier fire, the Cave fire, the Santa Barbara microburst, the recurring January flooding, the Ridgecrest earthquakes, the month of unbreathable air from the Thomas fire, the threat of the Thomas fire itself, the Montecito mudslides, and of course the COVID-19 pandemic, all of which have kept me indoors thinking about physics.  

Finally, I would like to thank all those who kindly read drafts of (parts of) this thesis and provided invaluable feedback, including Samuel Alipour-fard, Ian Banta, Manuel Buen-Abad, Changha Choi, Nathaniel Craig, David Grabovsky, Adolfo Holguin, Samuel Homiller, Lucas Johns, Soubhik Kumar, Umut Can \"{O}ktem, Robert McGehee, Alex Meiburg, Gabriel Trevi\~{n}o Verastegui, Farzan Vafa, and George Wojcik.

	\begin{vitae}
\addcontentsline{toc}{chapter}{Curriculum Vitae}

\begin{vitaesection}{Education}
\vspace{-0.1cm}
\item [2020]	Ph.D. in Physics, University of California, Santa Barbara.
\item [2017]	M.A. in Physics, University of California, Santa Barbara.
\item [2015]	M.S. in Physics, University of Pennsylvania.
\item [2015]	B.A. in Mathematics and Physics, Astrophysics Concentration, Honors Distinction in Physics, University of Pennsylvania.
\end{vitaesection}

\textbf{Publications}

``Supersoft Stops''\\
T. Cohen, N. Craig, S. Koren, M. McCullough, J. Tooby-Smith\\
Accepted to Phys. Rev. Lett., [arXiv:2002.12630 [hep-ph]] \cite{Cohen:2020ohi}\\

``IR Dynamics from UV Divergences: UV/IR Mixing, NCFT, and the Hierarchy Problem'' \\
N. Craig and S. Koren \\
JHEP 03 (2020) 037, [arXiv:1909.01365 [hep-ph]] \cite{Craig:2019zbn} \\

``Freezing-in Twin Dark Matter'' \\
S. Koren and R. McGehee \\
Phys. Rev. D101 (2020) 055024, [arXiv:1908.03559 [hep-ph]] \cite{Koren:2019iuv} \\

``The Weak Scale from Weak Gravity'' \\
N. Craig, I. Garcia Garcia, S. Koren \\
JHEP 09 (2019) 081, [arXiv:1904.08426 [hep-ph]] \cite{Craig:2019fdy} \\

``Exploring Strong-Field Deviations From General Relativity via Gravitational Waves'' \\
S. Giddings, S. Koren, G. Treviño \\
Phys. Rev. D100 (2019) 044005, [arXiv:1904.04258 [gr-qc]] \cite{Giddings:2019ujs} \\

``Neutrino - DM Scattering and Coincident Detections of UHE Neutrinos with EM Sources'' \\
S. Koren \\
JCAP 09 (2019) 013, [arXiv:1903.05096 [hep-ph]] \cite{Koren:2019wwi} \\

``Constructing N=4 Coulomb Branch Superamplitudes'' \\
A. Herderschee, S. Koren, T. Trott \\
JHEP 08 (2019) 107, [arXiv:1902.07205 [hep-th]] \cite{Herderschee:2019dmc} \\

``Massive On-Shell Supersymmetric Scattering Amplitudes'' \\
A. Herderschee, S. Koren, T. Trott \\
JHEP 10 (2019) 092, [arXiv:1902.07204 [hep-th]] \cite{Herderschee:2019ofc} \\

``The second Higgs at the lifetime frontier'' \\
S. Alipour-fard, N. Craig, S. Gori, S. Koren, D. Redigolo \\
JHEP 07 (2020) 029, [arXiv:1812.09315 [hep-ph]] \cite{Alipour-fard:2018mre} \\

``Discrete Gauge Symmetries and the Weak Gravity Conjecture'' \\
N. Craig, I. Garcia Garcia, S. Koren \\
JHEP 05 (2019) 140, [arXiv:1812.08181 [hep-th]] \cite{Craig:2018yvw} \\

``Long Live the Higgs Factory: Higgs Decays to Long-Lived Particles at Future Lepton Colliders'' \\
S. Alipour-fard, N. Craig, M. Jiang, S. Koren \\
Chin. Phys. C43 (2019) 053101, [arXiv:1812.05588 [hep-ph]] \cite{Alipour-Fard:2018lsf} \\

``Cosmological Signals of a Mirror Twin Higgs'' \\
N. Craig, S. Koren, T. Trott \\
JHEP 05 (2017) 038, [arXiv:1611.07977 [hep-ph]] \cite{Craig:2016lyx} \\

``The Low-Mass Astrometric Binary LSR1610-0040'' \\
S. C. Koren, C. H. Blake, C. C. Dahn, H. C. Harris \\
The Astronomical Journal 151 (2016) 57, [arXiv:1511.02234 [astro-ph.SR]] \cite{2016AJ....151...57K} \\

``Characterizing Asteroids Multiply-Observed at Infrared Wavelengths'' \\
S. C. Koren, E. L. Wright, A. Mainzer \\
Icarus 258 (2015) 82-91, [arXiv:1506.04751 [astro-ph.EP]] \cite{2015Icar..258...82K}


\end{vitae}
	%
%

\begin{abstract}
\addcontentsline{toc}{chapter}{Abstract}

We begin this thesis with an extensive pedagogical introduction aimed at clarifying the foundations of the hierarchy problem. After introducing effective field theory, we discuss renormalization at length from a variety of perspectives. We focus on conceptual understanding and connections between approaches, while providing a plethora of examples for clarity. With that background we can then clearly understand the hierarchy problem, which is reviewed primarily by introducing and refuting common misconceptions thereof. We next discuss some of the beautiful classic frameworks to approach the issue. However, we argue that the LHC data have qualitatively modified the issue into `The Loerarchy Problem'---how to generate an IR scale without accompanying visible structure---and we discuss recent work on this approach. In the second half, we present some of our own work in these directions, beginning with explorations of how the Neutral Naturalness approach motivates novel signatures of electroweak naturalness at a variety of physics frontiers. Finally, we propose a New Trail for Naturalness and suggest that the physical breakdown of EFT, which gravity demands, may be responsible for the violation of our EFT expectations at the LHC.

\end{abstract}

 \newpage
	\section*{Permissions and Attributions}\addcontentsline{toc}{chapter}{Permissions and Attributions}
	\begin{enumerate}
		
		\item The content of Chapter \ref{sec:InTheSky} is the result of collaboration with Nathaniel Craig and Timothy Trott, and separately with Robert McGehee. This work previously appeared in the Journal of High Energy Physics (JHEP \textbf{05} (2017) 038) and Physical Review D (Phys. Rev. \textbf{D101} (2020) 055024), respectively. 
		\item The content of Chapter \ref{sec:InTheGround} is the result of collaboration with Samuel Alipour-fard, Nathaniel Craig, and Minyuan Jiang, and previously appeared in Chinese Physics C (Chin. Phys. \textbf{C43} (2019) 053101). 
		\item The content of Chapter \ref{sec:newtrail} is the result of collaboration with Nathaniel Craig and previously appeared in the Journal of High Energy Physics (JHEP \textbf{03} (2020) 037).
		
	\end{enumerate}\newpage
	\section*{Preface}\addcontentsline{toc}{chapter}{Preface}
	The first four chapters of this thesis are introductory material which has not previously appeared in any public form. My intention has been to write the guide that would have been most useful for me toward the beginning of my graduate school journey as a field theorist interested in the hierarchy problem. My aim has been to make these chapters accessible to beginning graduate students in particle physics and interested parties in related fields---background at the level of a single semester of quantum field theory should be enough for them to be understandable in broad strokes.
	
	Chapter 1 introduces fundamental tools and concepts in quantum field theory which are essential for particle theory, spending especial effort on discussing renormalization from a variety of perspectives. Chapter 2 discusses the hierarchy problem and how to think about it---primarily through the pedagogical device of refuting a variety of common misconceptions and pitfalls. Chapter 3 introduces in brief a variety of classic strategies and solutions to the hierarchy problem which also constitute important frameworks in theoretical particle physics beyond the Standard Model. Chapter 4 discusses more-recent ideas about the hierarchy problem in light of the empirical pressure supplied by the lack of observed new physics at the Large Hadron Collider. Throughout I also make note of interesting research programs which, while they lie too far outside the main narrative for me to explain, are too fascinating not to be mentioned.  
	
	The first half of this thesis is thus mostly an introduction to and review of material I had no hand in inventing. As always, I am `standing on the shoulders of giants', and I have benefited enormously from the pedagogical efforts of those who came before me. When my thinking on a topic has been especially informed by a particular exposition, or when I present an example which was discussed in a particular source, I will endeavor to say so and refer to that presentation. As to the rest, it's somewhere between difficult and impossible to distinguish precisely how and whose ideas I have melded together in my own understanding of the topics---to say nothing of any insight I may have had myself---but I have included copious references to reading material I enjoyed as a guide. Ultimately this is a synthesis of ideas in high energy theory aimed toward the particular purpose of understanding the hierarchy problem, and I have attempted to include the most useful and pedagogical explanations of these topics I could find, if not invent.
	
	I then present some work on the subject by myself and my collaborators. Chapter 5 contains work constructing a viable cosmological history for mirror twin Higgs models, an exemplar of the modern Neutral Naturalness approach to the hierarchy problem. Chapter 6 focuses on searching for long-lived particles produced at particle colliders as a discovery channel for a broad class of such models. Chapter 7 is an initial exploration of a new approach to the hierarchy problem which follows a maximalist interpretation of the lack of new observed TeV scale physics, and so relies on questioning and modifying some core assumptions of conventional particle physics. In Chapter 8 we conclude with some brief parting thoughts.
	
	If you enjoy reading this work, or find it useful, or have questions, or comments, or recommendations for good references, please do let me know---at whatever point in the future you're reading this. As of autumn 2020, I can be reached at sethk@uchicago.edu.
	
	\setcounter{secnumdepth}{2}
	\setcounter{tocdepth}{3}
	\tableofcontents
\end{frontmatter}

\begin{mainmatter}

\pagestyle{fancy}
\renewcommand{\chaptermark}[1]{\markboth{{\sf #1 \hspace*{\fill} Chapter~\thechapter}}{} }
\renewcommand{\sectionmark}[1]{\markright{ {\sf Section~\thesection \hspace*{\fill} #1 }}}
\fancyhf{}

\makeatletter \if@twoside \fancyhead[LO]{\small \rightmark} \fancyhead[RE]{\small\leftmark} \else \fancyhead[LO]{\small\leftmark}
\fancyhead[RE]{\small\rightmark} \fi

\def\cleardoublepage{\clearpage\if@openright \ifodd\c@page\else
  \hbox{}
  \vspace*{\fill}
  \begin{center}
    This page intentionally left blank
  \end{center}
  \vspace{\fill}
  \thispagestyle{plain}
  \newpage
  \fi \fi}

\providecommand*{\input@path}{}
\g@addto@macro\input@path{{./NCFT//}{./sty//}{./tex//}{./fig//}{./FreezeTwin//}{./LLP//}{./TwinCosmo//}}

\makeatother
\fancyfoot[c]{\textrm{\textup{\thepage}}} 
\fancyfoot[C]{\thepage}
\renewcommand{\headrulewidth}{0.4pt}

\fancypagestyle{plain} { \fancyhf{} \fancyfoot[C]{\thepage}
\renewcommand{\headrulewidth}{0pt}
\renewcommand{\footrulewidth}{0pt}}

\chapter{Effective Field Theory}
\label{sec:EffFieldTh}
The formulation and understanding of the hierarchy problem is steeped heavily in the principles and application of effective field theory (EFT) and renormalization, so we begin with an introductory overview to set the stage for our main discussion. As is clear from the table of contents, I have prioritized clarity over brevity---especially when it comes to renormalization. The reader with a strong background in particle physics may find much of this to be review, so may wish to skip ahead directly to Chapter \ref{sec:hierarchy} and circle back to sections of this chapter if and when the subtleties they discuss become relevant. 

We will endeavor to discuss the conceptual points which will be useful later in understanding the hierarchy problem, and more generally to clarify common confusions with ample examples. Of course we will be unable to discuss everything, and will try to provide references to more detailed explanations when we must needs say less than we would like. Some generally useful introductions to effective field theory can be found from Cohen \cite{Cohen:2019wxr} and Georgi \cite{Georgi:1994qn}, and useful, pedagogical perspectives on renormalization are to be found in Srednicki \cite{Srednicki:2007qs}, Peskin \& Schroeder \cite{Peskin:1995ev}, Zee \cite{Zee:2003mt}, Polchinski \cite{Polchinski:1983gv}, and Schwartz \cite{Schwartz:2013pla}, among others.

\section{EFT Basics}
\label{sec:EFT}

Effective field theory is simply the familiar strategy to focus on the important degrees of freedom when understanding a physical system. For a simple example from an introductory Newtonian mechanics course, consider studying the motion of balls on inclined planes in a freshman lab. It is neither necessary nor useful to model the short-distance physics of the atomic composition of the ball, nor the high-energy physics of special relativity. Inversely, it is also unnecessary to account for the long-distance physics of Hubble expansion or the low-energy physics of air currents in the lab. In quantum field theories this intuitive course of action is formalized in decoupling theorems, showing precisely the sense in which field theories are amenable to this sort of analysis: the effects of short-distance degrees of freedom may be taken into account as slight modifications to the interactions of long-distance degrees of freedom, instead of including explicitly those high-energy modes.

Of course when one returns to the mechanics laboratory armed with an atomic clock and a scanning tunneling microscope, one begins to see deviations from the Newtonian predictions. Indeed, the necessary physics for describing a situation depends not only on the dynamics under consideration but also on the precision one is interested in attaining with the description. So it is crucial that one is able to correct the leading-order description by systematically adding in subdominant effects, as organized in a suitable power series in, for example, $(v/c)$, where $v$ is the ball's velocity and $c$ is the speed of light. Of course when the full description of the physics is known it's in principle possible to just use the full theory to compute observables---but I'd still rather not begin with the QED Lagrangian to predict the deformation of a ball rolling down a ramp.

\begin{figure}
	\centering
		\includegraphics[width=0.5\textwidth]{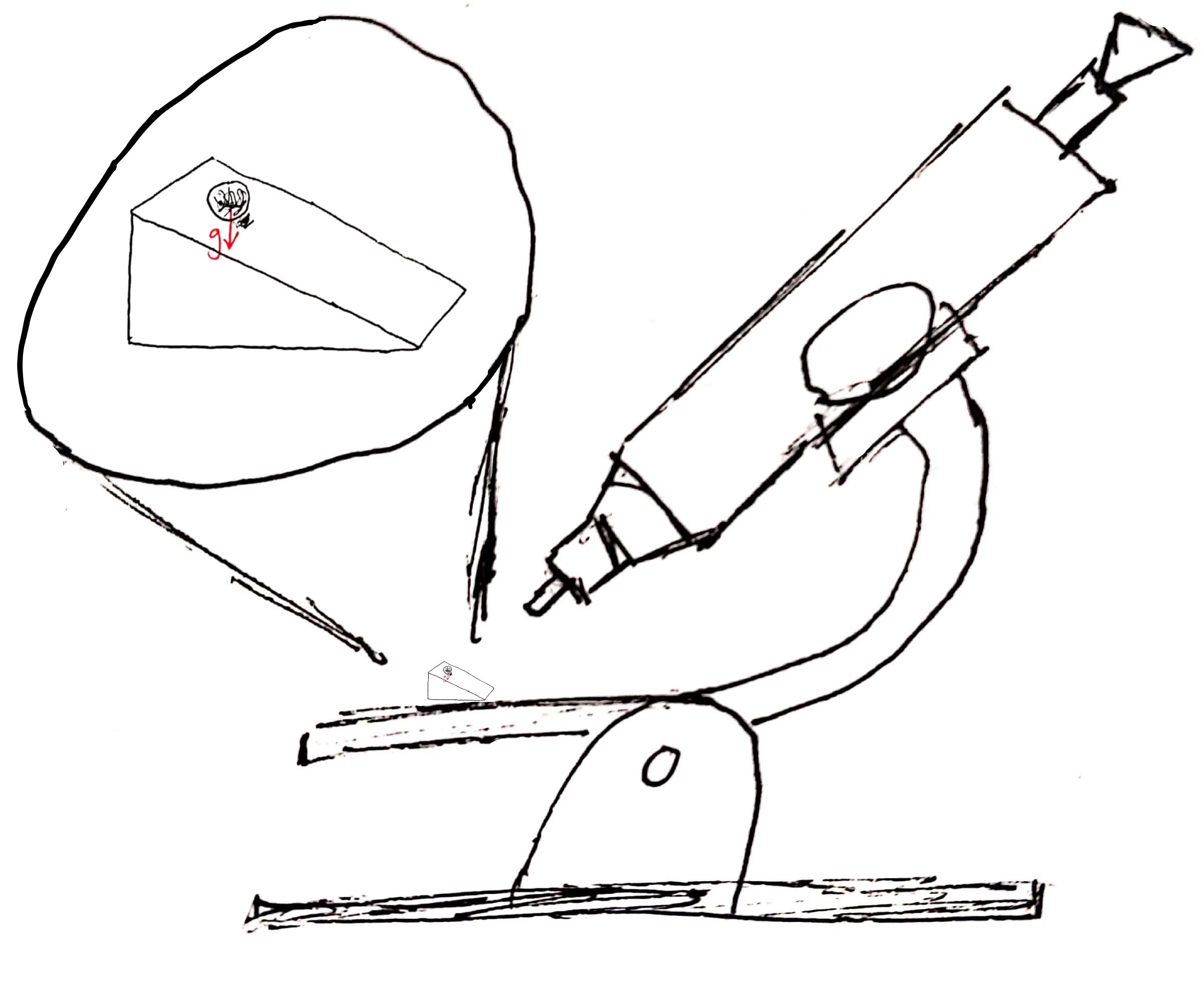}
	\caption{\ssp A cartoon of an experimental setup to look at miniature balls rolling down ramps, for which proper accounting of subleading effects may be necessary.}
	\label{fig:microscope}
\end{figure}

The construction of an appropriate effective description relies on three ingredients. The first is a list of the important degrees of freedom which specify the system under consideration---in particle physics this is often some fields $\left\lbrace\phi_i\right\rbrace$. The second is the set of symmetries which these degrees of freedom enjoy. These constrain the allowed interactions between our fields and so control the dynamics of the theory. Finally we need a notion of power counting, which organizes the effects in terms of importance. This will allow us to compute quantities to the desired precision systematically. Frequently in effective field theories of use in particle physics this role is played by $E/\Lambda$, where $E$ is an energy and $\Lambda$ is a heavy mass scale or cutoff above which we expect to require a new description of the physics.

\subsection{Scale-dependence}

We will often be interested in determining the appropriate description of a system at some scale, so it is necessary to understand which degrees of freedom and which interactions will be important as a function of energy. We can gain insight into when certain modes or couplings are important by studying the behavior of our system under scale transformations. Consider for example a theory of a real scalar field $\phi$, with action 
\begin{equation}\label{eqn:scaling}
S = \int \text{d}^{d}x \left(-\half \partial_\mu \phi \partial^\mu \phi - \half m^2 \phi^2 - \frac{1}{4!} \lambda \phi^4 - \frac{1}{6!} \tau \phi^6 + \dots \right),
\end{equation}
where $d$ is the dimensionality of spacetime, $m, \lambda, \tau$ are couplings of interactions involving different numbers of $\phi$, the $\dots$ denote terms with higher powers of $\phi$, and we've imposed a $\mathbb{Z}_2:\phi \rightarrow -\phi$ symmetry for simplicity. We've set $c = \hbar = 1$ leaving us with solely the mass dimension to speak of. From the fact that regardless of spacetime dimension we have $\left[S\right] = 0$ and $\left[x\right]=-1$, where $\left[\cdot\right]$ denotes the mass dimension, we first calculate from the kinetic term that $\left[ \phi \right] = \frac{d-2}{2}$, and then we can read off $\left[m^2\right]=2$, $\left[\lambda\right] = 4-d$, $\left[\tau\right] = 6-2d$. These are known as the classical dimension of the associated operators and are associated to the behavior of the operators at different energies, for the following reason. 

If we wish to understand how the physics of this theory varies as a function of scale, we can perform a transformation $x^\mu \rightarrow s x'^\mu$ and study the long-distance limit $s \gg 1$ with $x'^\mu$ fixed. The measure transforms as $\text{d}^{d}x \rightarrow s^d \text{d}^{d}x'$ and the derivatives $\partial_\mu \rightarrow s^{-1} \partial_{\mu'}$. Then to restore canonical normalization of the kinetic term such that the one-particle states are properly normalized for the LSZ formula to work, we must perform a field redefinition $\phi(x) = s^{\frac{2-d}{2}} \phi'(x')$, and the action becomes 
\begin{equation}
S = \int \text{d}^{d}x' \left(-\half \partial_{\mu'} \phi' \partial^{\mu'} \phi' - \half m^2 s^2 \phi'^2 - \frac{1}{4!} \lambda s^{4-d} \phi'^4 - \frac{1}{6!} \tau s^{6-2d}\phi'^6 + \dots \right).
\end{equation}
As a reminder, in the real world (at least at distances $\gtrsim 1 \ \mu m$) we have $d=4$. As you look at the theory at longer distances the mass term becomes more important, so is known as a `relevant' operator. One says that the operator $\phi^2$ has classical dimension $\Delta_{\phi^2}=2$. The quartic interaction is classically constant under such a transformation, so is known as `marginal' with $\Delta_{\phi^4}=0$, and interactions with more powers of $\phi$ shrink at low energies and are termed `irrelevant', e.g. $\Delta_{\phi^6}=-2$. We have been careful to specify that these are the \textit{classical} dimension of the operators, also called the `engineering dimension' or `canonical dimension', which has a simple relation to the mass dimension as $\Delta_{\mathcal{O}}=d-\left[\mathcal{O}\right]$ for some operator $\mathcal{O}$. If the theory is not scale-invariant then quantum corrections modify this classical scaling by an `anomalous dimension' $\delta_{\mathcal{O}}(m^2,\lambda,\tau,\dots)$ which is a function of the couplings of the theory, and the full behavior is known as the `scaling dimension'. The terms `marginally (ir)relevant' are used for operators whose classical dimension is zero but whose anomalous dimensions push them to one side.

The connection to the typical EFT power counting in $(E/\Lambda)$ is immediate. In an EFT with UV cutoff $\Lambda$, it's natural to normalize all of our couplings with this scale and rename e.g. $\tau \rightarrow \bar{\tau} \Lambda^{6-2d}$ where $\bar{\tau}$ is now dimensionless. It's easy to see that the long-distance limit is equivalently a low-energy limit by considering the derivatives, which pull down a constant $p'^\mu$ and scale as $s^{-1}$---or by simply invoking the uncertainty principle. Operators with negative scaling dimension contribute subleading effects at low energies precisely because of these extra powers of a large inverse mass scale.

\subsection{Bottom-up or Top-down} \label{sec:bottop}

\setlength{\epigraphwidth}{0.65\textwidth}
\epigraph{Then he made the tank of cast metal, 10 cubits across from brim to brim, completely round; it was 5 cubits high, and it measured 30 cubits in circumference.}{God on the merits of working to finite precision \\ 1 Kings 7:23, Nevi'im \\ New Jewish Publication Society Translation (1985) \cite{jewish2007jewish}}
\setlength{\epigraphwidth}{0.6\textwidth}

The procedure of writing down the most general Lagrangian with the given degrees of freedom and respecting the given symmetries up to some degree of power counting is termed `bottom-up EFT' as we're constructing it entirely generally and will have to fix coefficients by making measurements. A great example is the Standard Model Effective Field Theory (SMEFT), of which the Standard Model itself is described by the SMEFT Lagrangian at zeroth order in the power counting. It is defined by being an $SU(3) \times SU(2) \times U(1)$ gauge theory with three generations of the following representations of left-handed Weyl fermions:
\begin{center}
	\begin{tabular}{ |c|c|c|c| } 
		\hline
		$3x$ Fermions & $SU(3)_C$ & $SU(2)_L$ & $U(1)_Y$ \\ \hline
		$Q$ & 3 & 2 & $\frac{1}{6}$ \\ \hline
		$\bar u$ & $\bar 3$ & - & $-\frac{2}{3}$ \\ \hline
		$\bar d$ & $\bar 3$ & - & $\frac{1}{3}$ \\ \hline
		$L$ & - & 2 & $-\frac{1}{2}$ \\ \hline
		$\bar e$ & - & - & 1 \\ \hline
	\end{tabular}
\end{center}
In addition the Standard Model contains one scalar, the Higgs boson, which is responsible for implementing the Anderson-Brout-Englert-Guralnik-Hagen-Higgs-Kibble-'t Hooft mechanism \cite{Anderson:1963pc,Englert:1964et,Guralnik:1964eu,Higgs:1964ia,Higgs:1964pj,tHooft:1971qjg} to break the electroweak symmetry $SU(2)_L\times U(1)_Y$ down to electromagnetism $U(1)_Q$ at low energies:
\begin{center}
	\begin{tabular}{ |c|c|c|c| } 
		\hline
		\hspace{23pt} $H$ \hspace{23pt}  & \hspace{13pt} - \hspace{13pt} & \hspace{12pt} 2 \hspace{12pt} & \hspace{5pt} $-\half$ \hspace{5pt} \\ \hline
	\end{tabular}
\end{center}
The Standard Model Lagrangian contains all relevant and marginal gauge-invariant operators which can be built out of these fields, and has the following schematic form 
\begin{align}
\mathcal{L}_{\text{kin}} &= - \frac{1}{4} F_{\mu\nu} F^{\mu\nu} - i \bar \psi \slashed{D} \psi - (D_\mu H)^\dagger (D^\mu H) \label{eqn:SMkin}\\
\mathcal{L}_{\text{Higgs}} &= - y H \Psi \psi + \text{ h.c. } + m^2 H^\dagger H - \frac{\lambda}{4} (H^\dagger H)^2, \label{eqn:SMHiggs}
\end{align}
with $F$ a gauge field strength, $\psi$ a fermion, $D$ the gauge covariant derivative in the kinetic term Lagrangian on the first line, and the second line containing the Higgs' Yukawa couplings and self-interactions. If a refresher on the Standard Model would be useful, the introduction to its structure toward the end of Srednicki's textbook \cite{Srednicki:2007qs} will suffice for our purposes, while further discussion from a variety of perspectives can be found in Schwartz \cite{Schwartz:2013pla}, Langacker \cite{Langacker:2010zza}, and Burgess \& Moore \cite{Burgess:2007zi}. 

The SMEFT power-counting is in energies divided by an as-yet-unknown UV scale $\Lambda$, so the dimension-$n$ SMEFT Lagrangian consists of all gauge-invariant combinations of these fields with scaling dimension $n-d$. At dimension five there is solely one operator, $\mathcal{L}^{(5)} = (L H)^2 / \Lambda + \text{ h.c.}$, which contains a Majorana mass for neutrinos. In even-more-schematic form, the dimension six Lagrangian contains operators with the field content
\begin{equation}
- \Lambda^2 \mathcal{L}^{(6)} = F^3 + H^6 + D^2 H^4 + \psi^2 H^3 + F^2 H^2 + \psi^2 F H + \psi^2 D H^2 + \psi^4, \label{eqn:SMEFT6}
\end{equation}
where for aesthetics we have multiplied through by the scale $\Lambda$ and haven't bothered writing down couplings. After understanding the structure of the independent symmetry-preserving operators (see e.g. \cite{Grzadkowski:2010es,Gripaios:2018zrz,deBlas:2017xtg}), the job of the bottom-up effective field theorist is to measure or constrain the coefficients of these higher-dimensional operators \cite{Brivio:2017btx}. Useful data comes from both the energy frontier with searches at colliders for the production or decay of high-energy particles through these higher-dimensional operators and from the precision frontier measuring fundamental processes very well to look for deviations from the Standard Model predictions (e.g. \cite{Falkowski:2017pss,Ellis:2018gqa,deBlas:2017wmn}). For more detail, see the introduction to SMEFT by Brivio \& Trott \cite{Brivio:2017vri}.

Another approach is possible when we already have a theory and just want to focus on some particular degrees of freedom. Then we may construct a `top-down EFT' by taking our theory and `integrating out' the degrees of freedom we don't care about---for example by starting with the Standard Model above and getting rid of the electroweak bosons to focus on processes occurring at lower energies (e.g. Fermi's model of the weak interaction \cite{Fermi:1934hr}). We can't necessarily just ignore those degrees of freedom though; what we need to do is modify the parameters of our EFT such that they reproduce the results of the full theory (to some finite precision) using only the low-energy degrees of freedom. Such a procedure can be illustrated formally by playing with path integrals. Consider the partition function for a theory with some light fields $\phi$ and some heavy fields $\Phi$:
\begin{equation}
Z = \int \mathcal{D}\phi\mathcal{D}\Phi \ e^{i S(\phi,\Phi)}.
\end{equation}
This contains all of the physics in our theory, and so in principle we may use it to compute anything we wish. But if we're interested in low-energy processes involving solely the $\phi$ fields, we could split up our path integral and first do the integral over the $\Phi$ fields. The light $\phi$ fields are the only ones left, so we can then write the partition function as 
\begin{equation}
Z = \int \mathcal{D}\phi \ e^{i S_{\text{eff}}(\phi)},
\end{equation}
where this defines $S_{\text{eff}}$. Thus far this still contains all the same physics, as long as we don't want to know about processes with external heavy fields\footnote{Since we haven't made any approximations and have the same object $Z$, one may be confused as to why we've lost access to the physics of the $\Phi$ fields. In fact I've been a bit sloppy. If we want to compute correlation functions of our fields $\phi$, we must couple our fields to classical sources $J_\phi$ as $\mathcal{L} \supset \phi(x) J_\phi(x)$. Physically, those sources allow us to `turn on' particular fields so that we can then calculate their expectation values. Mathematically, we really need the partition function as a functional of these sources $Z[J_\phi]$, and we take functional derivatives with respect to these sources as a step to calculating correlation functions or scattering amplitudes. In integrating out our heavy field $\Phi$, we no longer have a source we can put in our Lagrangian to turn on that field, as it no longer appears in the action. \label{foot:source}}. But having decided that we are interested in the infrared physics of the $\phi$ fields, we can say that the effects of the heavy $\Phi$ fields will be suppressed by factors of the energies of interest divided by the mass of $\Phi$, and we should expand the Lagrangian $\mathcal{L}_{\text{eff}}$ in an appropriate series:
\begin{equation}\label{eqn:topdown}
Z = \int \mathcal{D}\phi \ \text{exp}\left[i \int \text{d}^dx \left(\mathcal{L}_0(\phi) + \sum_{n=0}^{N} M_{\Phi}^{d-n} \sum_i \lambda_i^{(n)} \mathcal{O}_i^{(n)}(\phi) \right) \right],
\end{equation}
where $\mathcal{L}_0(\phi)$ is the part of the full Lagrangian that had no heavy fields in it,
$\mathcal{O}_i^{(n)}(\phi)$ is an operator of classical dimension $n$, $\lambda_i^{(n)}$ is a dimensionless coupling, and $N\geq d$ defines the precision to which one works in this effective theory. This is the procedure to find a top-down effective field theory in the abstract.

\begin{figure}
	\centering
		\includegraphics[width=0.5\textwidth]{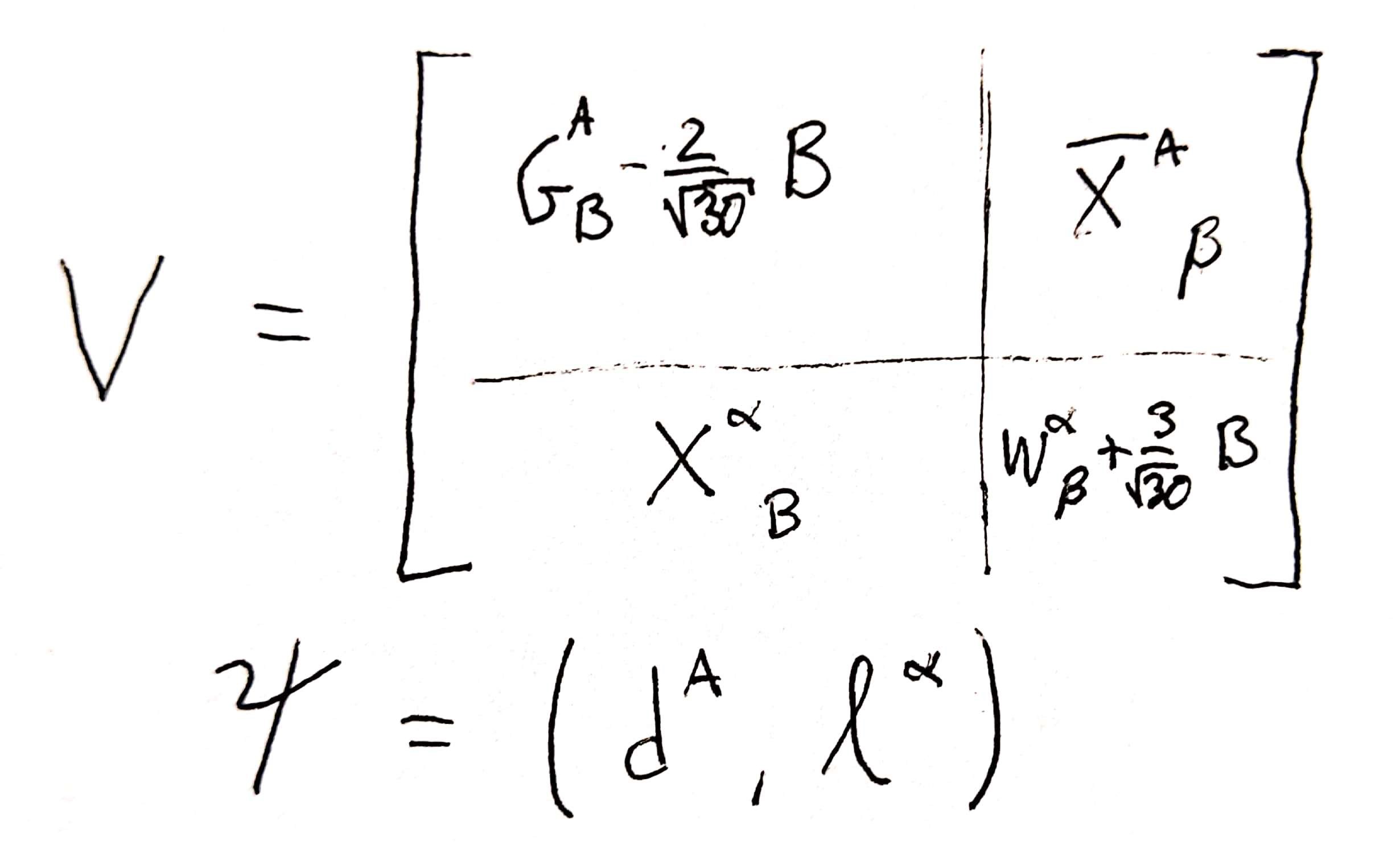}
	\caption{\ssp Schema of how the SM fits into $SU(5)$. The $SU(5)$ vector field $V$ decomposes into the gluons $G^A_B$, the $SU(2)_L$ vectors $W^\alpha_\beta$, hypercharge $B$ along the diagonal, and the new as-yet-unobserved leptoquarks $X^\alpha_\beta$ carrying both color and electroweak charge. The right-handed down-type quarks and the left-handed leptons are unified into an antifundamental. Decomposing the $10$ to find the rest of the SM fermions is left as an exercise. }
	\label{fig:gutreps}
\end{figure}

A great example of a top-down EFT is in studying the Standard Model fields in the context of a Grand Unified Theory (GUT). Broadly, Grand Unification is the hope that there is some simpler, more symmetric theory behind the Standard Model which explains its structure. A GUT is a model in which the gauge groups of the SM are (partially \cite{Pati:1974yy}) unified in the UV. If there is a full unification to a single gauge factor, then this requires `gauge coupling unification' in the UV until the symmetry is broken down to the SM gauge group at a high scale \cite{Georgi:1974sy}. While one's first exposure to this idea today may be in the context of a UV theory like string theory which roughly demands such unification, this was in fact first motivated by the observed infrared SM structure. It is frankly amazing that not only are the values of the SM gauge couplings consistent with this idea, and not only does $SU(3) \times SU(2) \times U(1)$ fit nicely inside $SU(5)$, but the SM fermion representations precisely fit into the $10 \oplus \bar 5$ representations of $SU(5)$ (see Figure \ref{fig:gutreps}). It's hard to imagine a discovery that would have felt much more like one was obviously learning something deep and important about Nature than when Georgi realized how nicely all of this worked out. I'm reminded of Einstein's words on an analogous situation in the early history of electromagnetism---the original unified theory:

\begin{quote}
	The precise formulation of the time-space laws of those fields was the work of Maxwell. Imagine his feelings when the differential equations he had formulated proved to him that electromagnetic fields spread in the form of polarised waves, and with the speed of light! To few men in the world has such an experience been vouchsafed. At that thrilling moment he surely never guessed that the riddling nature of light, apparently so completely solved, would continue to baffle succeeding generations.\\
	--- Albert Einstein \\
	\textit{Considerations concerning the Fundaments of Theoretical Physics}, 1940 \cite{Einstein:1940}
\end{quote}
And just as with Maxwell, the initial deep insight into Nature was not the end of the story. As of yet, Grand Unification remains an unproven ideal, and indeed further empirical data has brought into question the simplest such schemes. But it's hard to imagine all of this beautiful structure is simply coincidental, and I would wager that most high energy theorists still have a GUT in the back of their minds when they think about the UV structure of the universe, so this is an important story to understand. To learn generally about GUTs, I recommend the classic books by Kounnas, Masiero, Nanopoulos, \& Olive \cite{Kounnas:1985cj} and Ross \cite{Ross:1985ai} or the recent book by Raby \cite{Raby:2017ucc} for the more formally-minded. Shorter introductions can be found in Sher's TASI lectures \cite{TASI:2001rya} or in the Particle Data Group's Review of Particle Physics \cite{Tanabashi:2018oca} from Hebecker \& Hisano.

The structure of the simplest $SU(5)$ GUT is that the symmetry group breaks down to the SM at energies $M_{GUT}\sim 10^{16} \text{ GeV}$ via the Higgs mechanism.\footnote{The scale $M_{GUT}$ is determined from low-energy data by computing the scale-dependence of the SM gauge couplings, evolving them up to high energies, and looking for an energy scale at which they meet. Since we have three gauge couplings at low energies, it is quite non-trivial that the curves $g_i(\mu), i = 1,2,3$ meet at a single scale $\mu = M_{GUT}$. The $M_{GUT}$ so computed is approximate not only due to experimental uncertainties on the low-energy values of parameters in the SM, but also because additional particles with SM charges affect slightly how the couplings evolve toward high energies. Indeed, adding supersymmetry makes the intersection of the three curves even more accurate than it is in the SM itself.} More generally, unification may proceed in stages as, for example, $SO(10) \rightarrow SU(4)_c \times SU(2)_L \times SU(2)_R \rightarrow SM$, and the breaking may occur via other mechanisms, as we'll discuss further in Section \ref{sec:classical}. Back to our simple single-breaking example, as is familiar in the SM this means that the gauge bosons corresponding to broken generators get masses of order this GUT-breaking scale. As this is a far higher scale than we are currently able to directly probe, it is neither necessary nor particularly useful to keep these degrees of freedom fully in our description if we're interesting in understanding the effects of GUT-scale fields. Rather than constructing the complete top-down EFT of the SM from a GUT, let's focus on one particularly interesting effect.

One of the best ways to indirectly probe GUTs is by looking for proton decay. The GUT representations unify quarks and leptons, so the extra $SU(5)$ gauge bosons have nonzero baryon and lepton number and fall under the label of `leptoquarks'. It's worth considering in detail why proton decay is a feature of GUTs and not of the SM, as it's a subtler story than is usually discussed. While $U(1)_B$, the baryon number, is an accidental global symmetry of the SM\footnote{`Accidental' here means that imposing this symmetry on the SM Lagrangian does not forbid any operators which would otherwise be allowed. The SM is defined, as above, by the gauge symmetries $SU(3)_C\times SU(2)_L\times U(1)_Y$ and the field content. Writing down the most general dimension-4 Lagrangian \ref{eqn:SMkin},\ref{eqn:SMHiggs} invariant under these symmetries gives a Lagrangian which is \textit{automatically} invariant under $U(1)_B$. This no longer holds at higher order in SMEFT, and indeed the dimension-6 Lagrangian (Equation \ref{eqn:SMEFT6}) does contain baryon-number-violating operators. If one wants to study a baryon-number-conserving version of SMEFT, one needs to explicitly impose that symmetry on the dimension-6 Lagrangian, so $U(1)_B$ is no longer an accidental symmetry of SMEFT.}, it's an anomalous symmetry and so is not a symmetry of the quantum world. The `baryon minus lepton' number, $U(1)_{B-L}$, is non-anomalous, but this is a good symmetry both of the SM and of a GUT and clearly does not prevent e.g. $p^+ \rightarrow \pi^0 e^+$. What's really behind the stability of the proton is that, though $U(1)_B$ and $U(1)_L$ are not good quantum symmetries, the fact that they are good classical symmetries means their only violation is nonperturbatively by instantons. Such configurations yield solely baryon number violation by three units at once, corresponding to the number of generations, and thus the proton with $B=1$ is stable\footnote{Convincing yourself fully that $\Delta B = 3$ is the smallest allowed transition is not straightforward, but let me try to make it believable for anyone with some exposure to anomalies and instantons. The existence of a mixed $U(1)_B SU(2)_L^2$ anomaly---equivalently a nonvanishing triangle diagram with two $SU(2)_L$ gauge legs and a baryon current insertion---means that the baryon current will no longer be divergenceless, $\partial_\mu j_B^\mu \propto \frac{g^2}{32 \pi^2} \text{tr} \tilde{W} W$, where $W_a^{\mu\nu}$ is the field strength and $\tilde{W}$ its Hodge dual. Instantons are field configurations interpolating between vacuum field configurations of different topology, and there are nontrivial instantons in 4d Minkowski space for $SU(2)_L$ but not for $U(1)_Y$ as a result of topological requirements on the gauge group. So while there is also a mixed anomaly with hypercharge, we can ignore this for our purposes. The SM fermions contributing to the $U(1)_B SU(2)_L^2$ anomaly are then only $Q_a$ (the left-handed quark doublet with $B=1/3$, with $a$ a color index) and similarly for the lepton anomaly only the doublet $L$ matters. There are three generations of each, which leads simply to a factor of three in the divergence of the global currents. Thinking about it in terms of triangle diagrams, this is simply because there are thrice as many fermions running in the loop. The extent to which an instanton solution changes the topology is given by the integral of $\frac{g^2}{32 \pi^2} \text{tr} \tilde{W} W$ over spacetime, which as a total derivative localizes to the boundary, and furthermore turns out to be a topological invariant of the gauge field configuration known as the winding number, an integer (technically the change in winding number between the initial and final vacua). Then the anomaly, by way of the nonvanishing divergence, relates the winding number of such a configuration to the change in baryon and lepton number it induces. The factor of the number of generations means that each unit of winding number ends up producing $\Delta B = - \Delta L = n_g = 3$. And \textit{that} is why the proton is stable. Classic, detailed references on anomalies and instantons include Coleman \cite{Coleman:1985rnk}, Bertlmann \cite{bertlmann2000anomalies}, and Rajamaran \cite{rajaraman1982solitons}.}.

\begin{figure}
	\centering
	\includegraphics[width=0.7\linewidth]{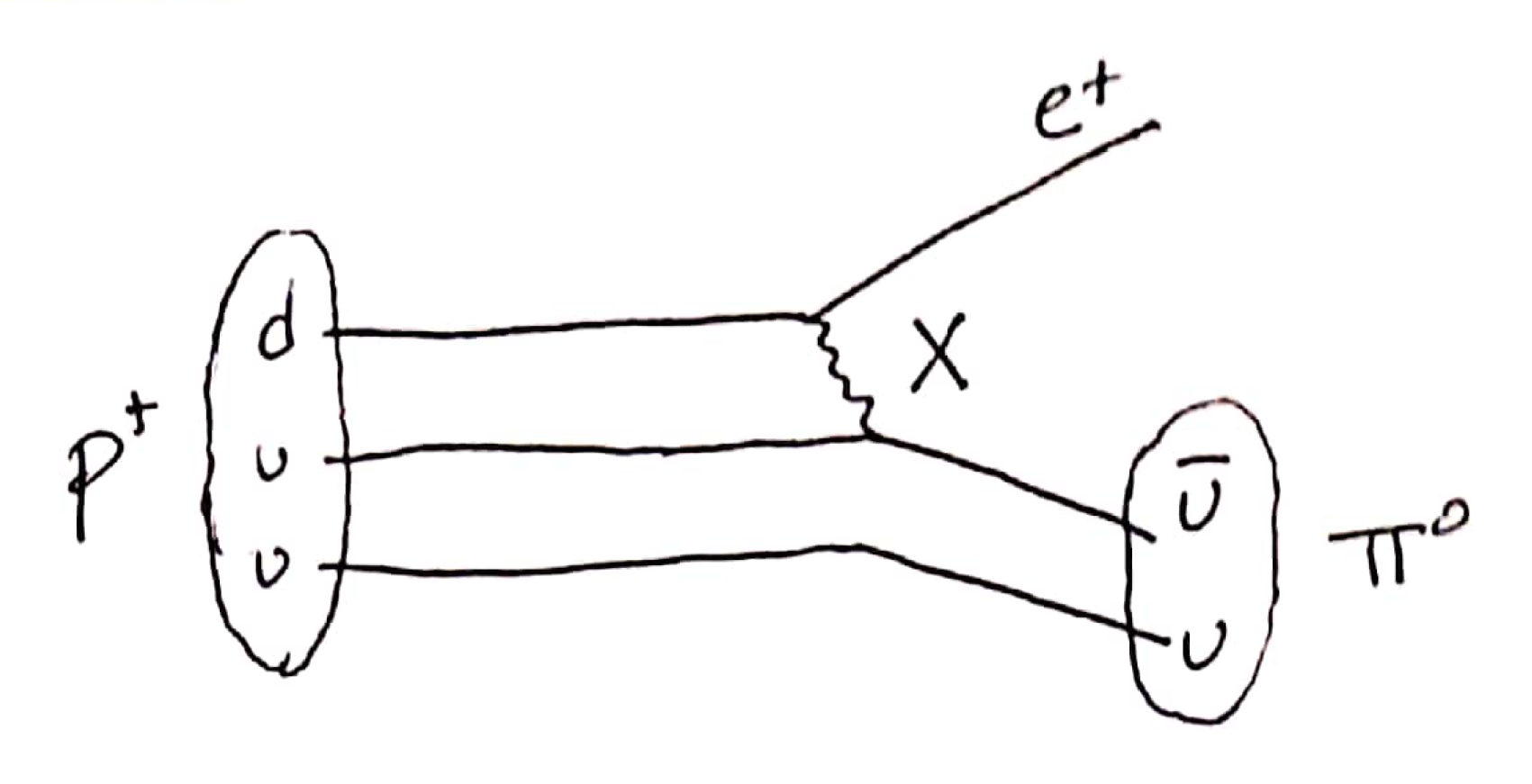}
	\caption{A representative diagram contributing to proton decay due to exchange of a heavy GUT gauge boson X.}
	\label{fig:protondecay}
\end{figure}

But in GUTs, baryon number and lepton number are no longer accidental symmetries, so no such protection is available and the GUT gauge bosons mediate tree-level proton decay processes as in Figure \ref{fig:protondecay}. We can find the leading effect by integrating these out---in particular we'll here look just at a four-fermion baryon-number-violating operator. The tree-level amplitude is simply 
\begin{equation}
i \mathcal{M}(uude) = (ig)^2 v_{u_1} \gamma^\mu u_{u_2} \frac{-i\left(\eta_{\mu\nu} + \frac{p^\mu p^\nu}{M_{GUT}^2}\right)}{p^2 + M_{GUT}^2} v_e \gamma^\nu u_d \approx i g^2 v_{u_1} \gamma^\mu u_{u_2} \frac{\eta_{\mu\nu}}{M_{GUT}^2} v_e \gamma^\nu u_d + \mathcal{O}(\frac{1}{M_{GUT}^2}),
\end{equation}
so integrating out the gauge boson from this diagram gives us one of the contributions to the low-energy operators in Equation \ref{eqn:topdown}
\begin{equation}
\mathcal{O} = \frac{g^2}{2 M_{\text{GUT}}^2} \epsilon_{\alpha\beta\gamma} \left(u_{R\alpha} \gamma_\mu u_{L\beta}\right) \left(e_R \gamma^\mu d_{L\gamma} \right),
\end{equation}
where, in the notation of Equation \ref{eqn:topdown}, $\mathcal{O}_i^{(6)} = \epsilon_{\alpha\beta\gamma} \left(u_{R\alpha} \gamma_\mu u_{L\beta}\right) \left(e_R \gamma^\mu d_{L\gamma} \right), \lambda_i^{(6)} = g^2/2$ and $M_\Phi = M_{GUT}$. The calculation of the proton lifetime from this operator is quite complicated, but the dimensional analysis estimate of $\tau_p \sim M_{GUT}^4 / g^4 m_p^5$ actually works surprisingly well. 

The job of the top-down effective field theorist is to calculate the effects of some particular UV physics on IR observables and by doing so understand how to search for their particular effects. While the effects will, by necessity, be some subset of the operators that the bottom-up effective field theorist has written down, the patterns and correlations present from a particular UV model can suggest or require particular search strategies. In the present context, a GUT may suggest the most promising final states to look for when searching for proton decay. If we wanted to calculate the lifetime and branching ratios more precisely we would have to deal with loop diagrams (among many complications), which of course is a generic feature. So we now turn our attention to the new aspects and challenges of field theory that appear once one goes beyond tree-level.

\section{Renormalization}

Renormalization is a notoriously challenging topic for beginning quantum field theorists to grok, and explanations often get bogged down in the details of one particular perspective or scheme or purpose and `miss the forest for the trees', so to speak.\footnote{Not that I begrudge QFT textbooks or courses for it, to be clear. There is so much technology to introduce and physics to learn in a QFT class that discussion of all of these various perspectives and issues would be prohibitive.} We'll attempt to overcome that issue by discussing a variety of uses for and interpretations of renormalization, as well as how they relate. And, of course, by examining copious examples and pointing out a variety of conceptual pitfalls.

\subsubsection*{Loops are necessary}\addcontentsline{toc}{subsubsection}{\qquad Loops Are Necessary} At the outset the only fact one needs to have in mind is that renormalization is a procedure which lets quantum field theories make physical predictions given some physical measurements. Such a procedure was not necessary for a classical field theory, which is roughly equivalent to a quantum field theory at tree-level. A natural question for beginners to ask then is why we should bother with loops at all: Why don't we just start off with the physical, measured values in the classical Lagrangian and be done with it? That is, if we measure, say, the mass and self-interaction of some scalar field $\phi$, let's just define our theory 
\begin{equation}
\mathcal{L}_{\text{exact}} \stackrel{?}{=} -\half \partial_\mu \phi \partial^\mu \phi - \half m^2_\text{phys} \phi^2 - \frac{1}{4!} \lambda_\text{phys} \phi^4,
\end{equation}
for some definitions of these physical parameters, and compute everything at tree-level. However, this does not constitute a sensible field theory, as the optical theorem tells us this is not consistent. We define $S = 1 + i \mathcal{M}$ as the S-matrix which encodes how states in the theory scatter, where the $1$ is the `trivial' no-interaction part. Quantum mechanics turns the logical necessity that probabilities add up to 1 into the technical demand of `unitarity', $S^\dagger S = 1$, which tells us the nontrivial part must satisfy
\begin{equation} \label{eqn:opticaltheorem}
i \left(\mathcal{M}^\dagger - \mathcal{M}\right) = \mathcal{M}^\dagger \mathcal{M}.
\end{equation}

\begin{figure}
	\centering
	\includegraphics[width=1.0\linewidth]{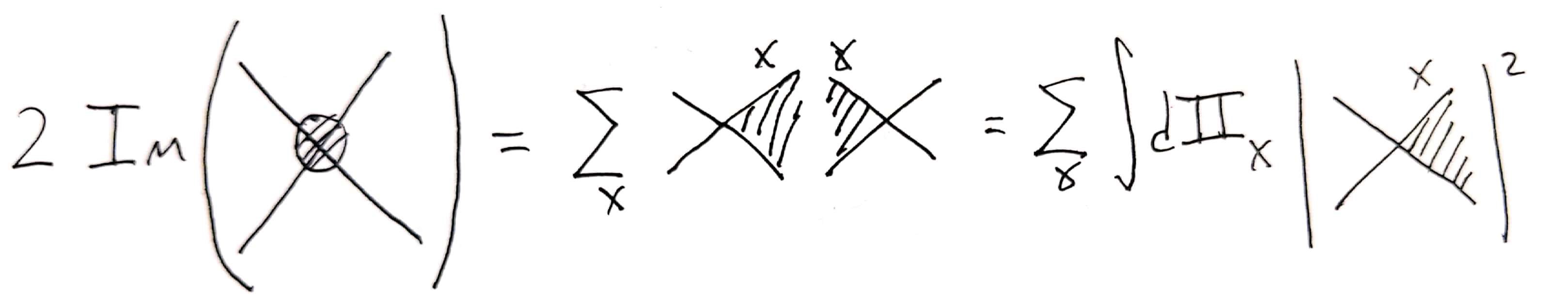}
	\caption{Schematic description of the optical theorem Equation \protect\ref{eqn:opticaltheorem} as applied to a $2 \rightarrow 2$ process. On the right hand side one must sum over all possible intermediate states, including both various state labels $X$ and their phase space $\Pi_X$.}
	\label{fig:opticaltheorem}
\end{figure}

Sandwiching this operator equation between initial and final states, we find that the left hand side is the imaginary part of the amplitude $\mathcal{M}(i\rightarrow f)$, which is nonzero solely due to loops. This is depicted schematically in Figure \ref{fig:opticaltheorem}. We can see why this is by examining a scalar field propagator. Taking the imaginary part one finds
\begin{equation}
\text{Im } \frac{1}{p^2 + m^2 + i \epsilon} = \frac{-\epsilon}{(p^2 + m^2)^2 + \epsilon^2}.
\end{equation}
This vanishes manifestly as $\epsilon \rightarrow 0$ except for when $p^2 = -m^2$, and an integral to find the normalization yields 
\begin{equation}
\text{Im } \frac{1}{p^2 + m^2 + i \epsilon} = -\pi \delta(p^2 + m^2).
\end{equation}
So internal propagators are real except for when the particle is put on-shell. In a tree-level diagram this occurs solely at some measure-zero set of exceptional external momenta, but in a loop-level diagram we integrate over \textit{all} momenta in the loop, so an imaginary part is necessarily present. Now we see the necessity of loops solely from the conservation of probability and the framework of quantum mechanics\footnote{A natural question to ask is whether this structure can be perturbed at all, but in fact it really is quite rigid. After Hawking---motivated by black hole evaporation---proposed that the scattering matrix in a theory of quantum gravity should not necessarily obey unitarity \cite{Hawking:1982dj}, the notion of modifying the S-matrix to a non-unitary `$\$$'-matrix (pronounced `dollar matrix') received heavy scrutiny. This was found to necessarily lead to large violations of energy conservation, among other maladies \cite{Banks:1983by}.}.

The lesson to take away from this is that classical field theories produce correlation functions with some particular momentum dependence, which can be essentially read off from the Lagrangian. But a consistent theory \textit{requires} momentum dependence of a sort that does not appear in such a Lagrangian, which demands that calculations must include loops. In particular it is the analyticity properties of these higher-order contributions that are required by unitarity, and there is an interesting program to understand the set of functions satisfying those properties at each loop order as a way to bootstrap the structure of multi-loop amplitudes (see e.g. \cite{Goncharov:2009,Goncharov:2010jf,Duhr:2011zq,Gaiotto:2011dt,Duhr:2014woa,Arkani-Hamed:2016byb}).

So far from being `merely' a way to deal with seemingly unphysical predictions, renormalization is very closely tied to the physics. We begin in the next section with understanding its use for removing divergences, as this is the most basic application and is often the first introduction students receive to renormalization. We will then move on to discuss other, more physical interpretations of renormalization.

\subsection{To Remove Divergences}\label{sec:removediv}

\subsubsection*{Physical input is required}\addcontentsline{toc}{subsubsection}{\qquad Physical Input is Required}

\begin{figure}
	\centering
	\includegraphics[width=\linewidth]{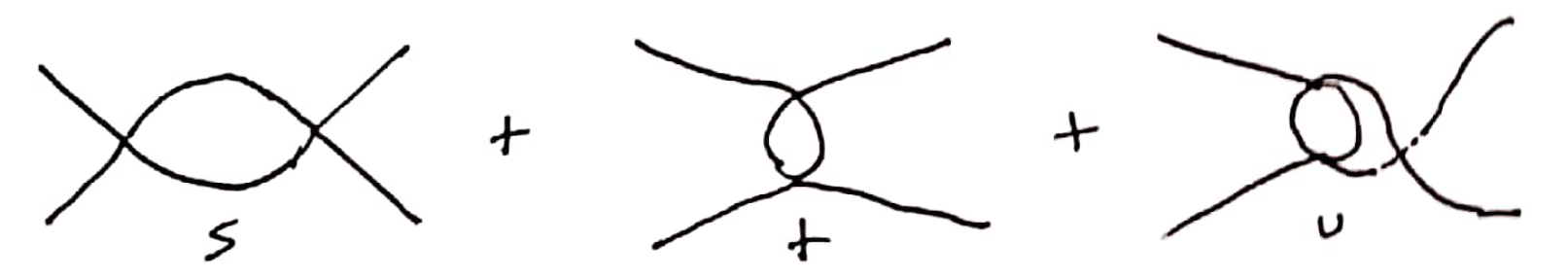}
	\caption{One-loop diagrams contributing to $2 \rightarrow 2$ scattering in the $\phi^4$ theory. Since the external legs are all the same, these three different internal processes must be summed over, corresponding to momentum exchanged in the $s,t,u$ channels, where these are the Mandelstam variables \protect\cite{Mandelstam:1958xc}.}
	\label{fig:phi4scat}
\end{figure}

As a first pass, let's look again at a $\phi^4$ theory 
\begin{equation}\label{eqn:phi4lag}
\mathcal{L} = -\half \partial_\mu \phi \partial^\mu \phi - \half m_0^2 \phi^2 - \frac{1}{4!} \lambda_0 \phi^4,
\end{equation}
and now treat it properly as a quantum field theory. As a simple example, let us consider $2 \rightarrow 2$ scattering in this theory, our discussion of which is particularly influenced by Zee \cite{Zee:2003mt}. At lowest-order this is extremely simple, and the tree-level amplitude is $i \mathcal{M}(\phi\phi\rightarrow \phi\phi) = -i \lambda_0$. But if we're interested in a more precise answer, we go to the next order in perturbation theory and we have the one-loop diagrams of Figure \ref{fig:phi4scat}. \\
Defining $P_s^\mu \equiv p_1^\mu + p_2^\mu$ as the momentum flowing through the loop in the $s$-channel diagram, that diagram is evaluated as
\begin{equation}
i \mathcal{M}(\phi\phi\rightarrow \phi\phi)_{\text{one-loop s-channel}} = (-i \lambda_0)^2 (-i)^2 \half \int \frac{d^4k}{(2\pi)^4} \frac{1}{P_s^2 + m_0^2} \frac{1}{(P_s + k)^2 + m_0^2}.
\end{equation}
Just from power-counting, we can already see that this diagram will be divergent. In the infrared, as $k\rightarrow 0$, the diagram is regularized by the mass of the field, but in the ultraviolet $k \rightarrow \infty$, the integral behaves as $\sim \int d^4k / k^4 \sim \int dk/k$ which is logarithmically divergent.

Though one might be tempted now to give up, we note that this divergence is appearing from an integral over very high energy modes---far larger than whatever energies we've verified our $\phi^4$ model to, so let's try to ignore those modes and see if we can't get a sensible answer. The general term for removing these divergences is `regularization' and we will here regularize (or `regulate') this diagram by imposing a hard momentum cutoff $\Lambda$ in Euclidean momentum space, which is the maximum energy of modes we let propagate in the loop. The loop amplitude may then be calculated with elementary methods detailed in, for example, Srednicki's textbook \cite{Srednicki:2007qs}. First we introduce Feynman parameters to combine the denominators, using $(A B)^{-1} = \int_0^1 dx (x A + (1-x) B)^{-2}$, which here tells us
\begin{align}
\frac{1}{P_s^2 + m_0^2} \frac{1}{(P_s + k)^2 + m_0^2} &= \int_0^1 dx \left[x\left((P_s + k)^2 + m_0^2\right) + (1-x)(P_s^2 + m_0^2)\right]^{-2} \\
&= \int_0^1 dx \left[q^2 + D\right]^{-2},
\end{align}
where we've skipped the algebra letting us rewrite this with $q = k + x P_s$ and $D = x(1-x) P_s^2 + m_0^2$. The change of variables $k \rightarrow q$ has trivial Jacobian, so the next step is to Wick rotate---Euclideanize the integral by defining $q^0 = i \bar q^d$, such that $q^2 \equiv q^\mu \eta_{\mu\nu} q^\nu = \bar q^\mu \delta_{\mu\nu} \bar q^\nu \equiv \bar q^2$. The measure simply picks up a factor of $i$, $d^dq = i d^d\bar q$, and we can then go to polar coordinates via $d^d\bar q = \bar q^{d-1} d\bar q d\Omega_{d}$. Lorentz invariance then means the angular integral gives us the area of the unit sphere in $d$ dimensions, $\Omega_d = 2 \pi^{d/2} /\Gamma(d/2)$, where $\Omega_4 = 2 \pi^2$, and the radial integral becomes
\begin{align}
\int_{0}^\Lambda d\bar q \ \bar q^3 \left[q^2 + D\right]^{-2} &= \left. \half \left[\frac{D}{\bar q^2 + D} + \log \left(\bar q^2 + D\right)\right] \right|_{0}^\Lambda \\
&= -\half \left[\frac{\Lambda^2}{\Lambda^2 + D} + \log \frac{D}{\Lambda^2 + D}\right].
\end{align}
In fact it is possible to do the $x$ integral analytically here, but we'll take $\Lambda^2 \gg \left| P^2 \right| \gg m^2$ to find a simple answer
\begin{equation}
-\half \int_0^1 dx \left[\frac{\Lambda^2}{\Lambda^2 + D} + \log \frac{D}{\Lambda^2 + D}\right] = \half\left(1 - \log \frac{P_s^2}{\Lambda^2}\right) + \dots
\end{equation}
Now putting all that together and including all the diagrams up to one-loop, we get the form 
\begin{equation} \label{eqn:phi4loopcalc}
\mathcal{M}(\phi\phi\rightarrow \phi\phi) = -\lambda_0 + C \lambda_0^2 \left[\log\left(\frac{\Lambda^2}{s}\right)+\log\left(\frac{\Lambda^2}{t}\right)+\log\left(\frac{\Lambda^2}{u}\right)\right] + \text{ subleading},
\end{equation}
where $s,t,u$ are the Mandelstam variables and $C$ is just a numerical coefficient. Now we see explicitly that the divergence has led to dependence of our amplitude on our regulator $\Lambda$. Of course this is problematic because we introduced $\Lambda$ as a non-physical parameter, and it would not be good if our calculation of a physical low-energy observable depended sensitively on how we dealt with modes in the far UV. But let us try to connect this with an observable anyway. We note that the theory defined by the Lagrangian in Equation \ref{eqn:phi4lag} can not yet be connected to an observable because we have not yet given a numerical value for $\lambda_0$. So let's imagine an experimentalist friend of ours prepares some $\phi$s and measures this scattering amplitude at some particular angles and energies corresponding to values of the Mandelstam variables $s_0, t_0, u_0$. They find some value $\lambda_{\text{phys}}$, which is a pure number. If our theory is to describe this measurement accurately, this tells us a relation between our parameters and a physical quantity
\begin{equation}
-\lambda_{\text{phys}} = -\lambda_0 + C \lambda_0^2 \left[\log\left(\frac{\Lambda^2}{s_0}\right)+\log\left(\frac{\Lambda^2}{t_0}\right)+\log\left(\frac{\Lambda^2}{u_0}\right)\right] + \mathcal{O}(\lambda^3).
\end{equation}
This is known as a `renormalization condition' which tells us how to relate our quantum field theories to observations at non-trivial loop order. Since the left hand side is a physical quantity, it may worry us that the right hand side contains a non-physical parameter $\Lambda$. But we still haven't said what $\lambda_0$ is, so perhaps we'll be able to find a sensible answer if we choose $\lambda_0 \equiv \lambda(\Lambda)$ in a correlated way with our regularization scheme. We call this `promoting $\lambda$ to a running coupling' by changing from the `bare coupling' $\lambda_0$ to one which depends on the cutoff. So let's solve for $\lambda$ in terms of $\lambda_{\text{phys}}$ and $\Lambda$. Rearranging we have
\begin{align}
\lambda_0 &= \lambda_{\text{phys}} + C \lambda_0^2 \left[\log\left(\frac{\Lambda^2}{s_0}\right)+\log\left(\frac{\Lambda^2}{t_0}\right)+\log\left(\frac{\Lambda^2}{u_0}\right)\right] + \mathcal{O}(\lambda_0^3) \\
\lambda(\Lambda) &= \lambda_{\text{phys}} + C \lambda_{\text{phys}}^2 \left[\log\left(\frac{\Lambda^2}{s_0}\right)+\log\left(\frac{\Lambda^2}{t_0}\right)+\log\left(\frac{\Lambda^2}{u_0}\right)\right] + \mathcal{O}(\lambda_{\text{phys}}^3) \label{eqn:lambdaoneloop}
\end{align}
where in the second line the replacement $\lambda_0^2 \mapsto \lambda_{\text{phys}}^2$ modifies the right side only at higher-order and so that is absorbed into our $\mathcal{O}(\lambda^3)$ uncertainty. To see what this has done for us, let us plug this back into our one-loop amplitude Equation \ref{eqn:phi4loopcalc}. This will impose our renormalization condition that our theory successfully reproduces our experimentalist friend's result. We find
\begin{align} 
\mathcal{M}(\phi\phi\rightarrow \phi\phi) = -\lambda_{\text{phys}} &- C \lambda_{\text{phys}}^2 \left[\log\left(\frac{\Lambda^2}{s_0}\right)+\log\left(\frac{\Lambda^2}{t_0}\right)+\log\left(\frac{\Lambda^2}{u_0}\right)\right] \\
&+ C \lambda_{\text{phys}}^2 \left[\log\left(\frac{\Lambda^2}{s}\right)+\log\left(\frac{\Lambda^2}{t}\right)+\log\left(\frac{\Lambda^2}{u}\right)\right] + \mathcal{O}(\lambda_{\text{phys}}^3) \nonumber
\end{align}
where again we liberally shunt higher-order corrections into our uncertainty term. Taking advantage of the nice properties of logarithms, we rearrange to get
\begin{equation} \label{eqn:phi4looprenorm}
\mathcal{M}(\phi\phi\rightarrow \phi\phi) = -\lambda_{\text{phys}} - C \lambda_{\text{phys}}^2 \left[\log\left(\frac{s}{s_0}\right)+\log\left(\frac{t}{t_0}\right)+\log\left(\frac{u}{u_0}\right)\right] + \mathcal{O}(\lambda_{\text{phys}}^3).
\end{equation}
We see that our renormalization procedure of relating our theory to a physical observable has enabled us to write the full amplitude in terms of \textit{physical} quantities, and remove the divergence entirely. This one physical input at some particular kinematic configuration has enabled us to fully predict any $2 \rightarrow 2$ scattering in this theory. 

We thus see how the renormalization procedure removes the divergences in a na\"{i}ve formulation of a field theory and allows us to make finite predictions for physical observables. While we did need to introduce a regulator, once we make the replacement $\lambda_0 \rightarrow \lambda(\Lambda)$ as defined in Equation \ref{eqn:lambdaoneloop} (and similar replacements for the coefficients of the other operators), the one-loop divergences are gone. We are guaranteed that any one-loop correlation function we calculate is finite in the $\Lambda \rightarrow \infty$ limit, which removes the regulator. If we wanted to increase our precision and calculate now at two loops, we would first renormalize the theory at two loops analogously to the above, and would find a more precise definition for $\lambda(\Lambda)$ which included terms of order $\mathcal{O}(\lambda_\text{phys}^3)$. At each loop order, replacing the bare couplings with running couplings suffices to entirely rid the theory of divergences.

\subsubsection*{Renormalizability}\addcontentsline{toc}{subsubsection}{\qquad Renormalizability}
An important question is for which quantum field theories do a finite set of physical inputs allow the theory to be fully predictive, in analogy to the example above. Such a theory is called `renormalizable' and means that after some finite number of experimental measurements, we can predict any other physics in terms of those values. Were this not the case, and no finite number of empirical measurements would fix the theory, it would not be of much use. Within the context of perturbation theory, a theory will be renormalizable if its Lagrangian contains solely relevant and marginal operators, and indeed for our $\phi^4$ theory three renormalization conditions are needed---one for each such operator.

\begin{figure}
	\centering
	\begin{subfigure}{.7\textwidth}
		\centering
		\includegraphics[width=\linewidth]{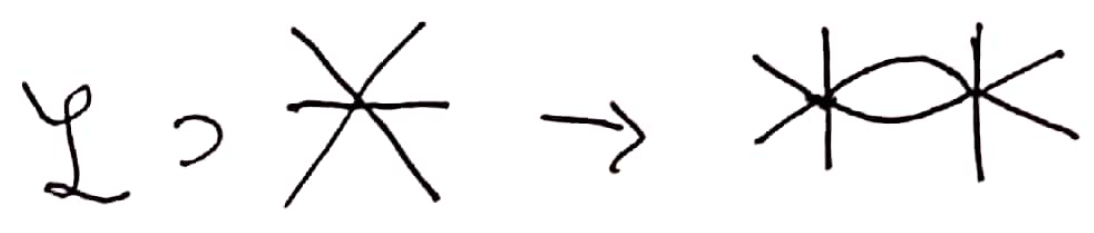}
		\caption{}
		\label{fig:IrrelInta}
	\end{subfigure}%
	\begin{subfigure}{.3\textwidth}
		\centering
		\includegraphics[width=0.7\linewidth]{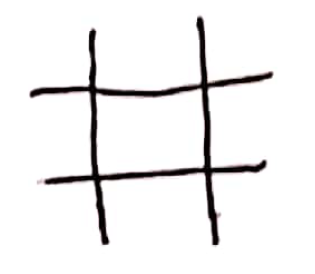}
		\caption{}
		\label{fig:IrrelIntb}
	\end{subfigure}
	\caption{In (a), a pictoral representation of the nonrenormalizability of theories described by Lagrangians with irrelevant operators. A tree-level six-point coupling leads to a new divergence in $4 \rightarrow 4$ scattering, which must be absorbed into a renormalized eight-point coupling, which would then beget a divergent $6 \rightarrow 6$ amplitude \dots. Note that this does not mean there are not one-loop contributions to $4 \rightarrow 4$ scattering. Such a diagram is depicted in (b), but it is finite and does not require additional local operators. }
	\label{fig:IrrelInt}
\end{figure}

The simplest way to understand why we must restrict to relevant and marginal operators is that irrelevant operators inevitably lead to the generation of a tower of more-and-more irrelevant operators. To see this, imagine now including a $\phi^6$ interaction, as depicted in Figure \ref{fig:IrrelInta}. At one loop this leads to a $4 \rightarrow 4$ scattering process with the same sort of divergence we saw in our previous loop diagram. So this loop is probing the UV physics, but we cannot absorb the unphysical divergence into a local interaction in our Lagrangian unless we now include a $\phi^8$ term. But then we can draw a similar one-loop diagram with the $\phi^8$ interaction which will require a $\phi^{10}$ interaction, and so on. Note that in our $\phi^4$ theory we also have $4 \rightarrow 4$ scattering at one loop, seen in Figure \ref{fig:IrrelIntb}, but there it comes from a box diagram which is finite, and so there is no need to include more local operators. \\
However, we emphasized above that the most useful description of a system depends on the precision at which one wishes to measure properties of the system. Thus in the study of effective field theories a broader definition of renormalizability should be used. For a theory with cutoff $\Lambda$, one decides to work to precision $\mathcal{O}(E/\Lambda)^n$ where $E$ is a typical energy scale of a process and $n$ is some integer. There are then a finite number of operators which contribute to processes to that precision---only those up to scaling dimension $n$---and so there is a notion of `effective renormalizability' of the theory. We still require solely a finite number of inputs to set the behavior of the theory to whatever precision we wish, but such a theory nevertheless fails the original criterion, which may be termed `power counting renormalizability' in comparison.

\subsubsection*{Wilsonian renormalization of $\phi^4$}\addcontentsline{toc}{subsubsection}{\qquad Wilsonian Renormalization of $\phi^4$}
Above we characterized our cutoff as an unphysical parametrization of physics at high scales that we do not know and we found that its precise value dropped out of our physically observable amplitude. To some extent this is rather surprising, as it's telling us that the high energy modes in our theory have little effect on physics at long distances---we can compensate for their effects by a small shift in a coupling. We can gain insight into the effects of these high energy modes by taking the cutoff seriously and looking at what happens when the cutoff is lowered. This brilliant approach due to Wilson \cite{Wilson:1974mb} is aimed at providing insight as to the particular effects of these high-energy modes by integrating out `shells' of high-energy Euclidean momenta and looking at the low-energy effects. This discussion is closely inspired by that in Peskin \& Schroeder's chapter 12 \cite{Peskin:1995ev}, as well as Srednicki's chapter 29 \cite{Srednicki:2007qs}.

It is easiest to see how to implement this by considering the path integral formulation. We can equally well integrate over position space paths as over momentum modes:
\begin{equation}
Z = \int \mathcal{D} \phi(x) e^{i \int \mathcal{L}(\phi)} =\left({\displaystyle \prod_{k}} \int \text{d}\phi(k)\right) e^{i \int \mathcal{L}(\phi)},
\end{equation}
and here it is clear that we may integrate over particular momentum modes separately if we so choose. In the condensed matter application in which Wilson originally worked, a cutoff appears naturally due to the lattice spacing $a$ giving an upper bound on momenta $k_{\text{max}} \sim 1/a$. In a general application we can imagine defining the theory with a fundamental cutoff $\Lambda$ by including in the path integral only modes with Euclidean momentum $k^2 \leq \Lambda^2$.\footnote{It is necessary to define this cutoff in Euclidean momentum space for the simple fact that in Lorentzian space a mode with arbitrarily high energy $k^0$ may have tiny magnitude by being close to the light cone $|k^0|\simeq|\vec{k}|$. It is left as an exercise for the reader to determine what deep conclusion should be taken away from the fact that we generally perform all our QFT calculations in Euclidean space.} The theory is defined by the values of the parameters in the theory with that cutoff---our familiar relevant and marginal operators $m^2(\Lambda), \lambda(\Lambda)$ and in principle all of the `Wilson coefficients' of irrelevant operators as well, since the theory is manifestly finite. The idea is to effectively lower the cutoff by explicitly integrating out modes with $b \Lambda \leq k \leq \Lambda$ for some $b < 1$. This will leave us with a path integral over modes with $k^2 \leq b^2 \Lambda^2$---which is a theory of the same fields now with a cutoff $b \Lambda$. By integrating out the high-energy modes we'll be able to track precisely their effects in this low-energy theory.

Peskin \& Schroeder perform this path integral explicitly by splitting the high energy modes into a different field variable and quantizing it, but since we've already introduced the conceptual picture of integrating fields out we'll take the less-involved approach of Srednicki. To repeat what we discussed above, the diagrammatic idea of integrating out a field is to remove it from the spectrum of the theory and reproduce its effects on low-energy physics by modifying the interactions of the light fields. Performing the path integral over some fields does not change the partition function, so the physics of the other fields must stay the same. We want to do the same thing here, but integrate out solely the high energy modes of a field and reproduce the physics in terms of the light modes.

We'll continue playing around with $\phi^4$ theory and define our (finite!) theory with a cutoff of $\Lambda$, which in full generality looks like:
\begin{equation}
\mathcal{L}(\Lambda) = -\half Z(\Lambda) \partial_\mu \phi \partial^\mu \phi - \half m^2(\Lambda) \phi^2 - \frac{1}{4!} \lambda(\Lambda) \phi^4 - \sum\limits_{n=3}^\infty \frac{c_n(\Lambda)}{(2n)!} \phi^{2n}.
\end{equation}
For simplicity we will decree that at our fundamental scale $\Lambda$ we have a canonically normalized field $Z(\Lambda)=1$ and no irrelevant interactions $c_n(\Lambda)=0$, but just some particular $m^2(\Lambda)$ and $\lambda(\Lambda)$. 

Let's look first at the one-loop four-point amplitude, which we must ensure is the same in both the theory with cutoff $\Lambda$ and that with cutoff $b \Lambda$. In the original theory, the amplitude at zero external momentum is
\begin{equation}
i V_4^{\Lambda}(0,0,0,0) = -i \lambda(\Lambda) + \frac{3}{2} (-i\lambda(\Lambda))^2 \int_{|k|<\Lambda} \frac{d^4k}{(2\pi)^4} \frac{(-i)^2}{(k^2 + m^2(\Lambda))^2} + \mathcal{O}(\lambda^3)
\end{equation}
When we evaluate this in the theory with a lowered cutoff $b\Lambda$, the modification is simply to everywhere make the replacement $\Lambda \mapsto b \Lambda$. In order for the physics to remain the same without the high-energy modes, the vertex function must not change. We'll take full advantage of the  perturbativity of the result---that is, $\lambda(\Lambda) - \lambda(b\Lambda) \sim \mathcal{O}(\lambda^2)$, $m^2(\Lambda) - m^2(b\Lambda) \sim \mathcal{O}(\lambda^2)$---to swap out quantities evaluated at $b\Lambda$ in the second-order term for those evaluated at $\Lambda$ at the cost solely of higher-order terms which we ignore. 
\begin{align}
0 &\equiv V_4^{\Lambda}(0,0,0,0) - V_4^{b\Lambda}(0,0,0,0) \\
&= - \lambda(\Lambda) + \frac{3}{2} \lambda(\Lambda)^2 \int_{|\bar k|<\Lambda} \frac{d^4\bar k}{(2\pi)^4} \frac{1}{(\bar k^2 + m^2(\Lambda))^2} \nonumber\\
&+\lambda(b\Lambda) - \frac{3}{2} \lambda(b\Lambda)^2 \int_{|\bar k|<b\Lambda} \frac{d^4\bar k}{(2\pi)^4} \frac{1}{(\bar k^2 + m^2(b\Lambda))^2} \nonumber\\
&= - \lambda(\Lambda) + \lambda(b\Lambda) + \frac{3}{2} \lambda(\Lambda)^2 \int_{|\bar k|=b\Lambda}^{\Lambda} \frac{d^4\bar k}{(2\pi)^4} \frac{1}{(\bar k^2 + m^2(\Lambda))^2} + \mathcal{O}(\lambda^3) \nonumber\\
&= - \lambda(\Lambda) + \lambda(b\Lambda) + \frac{3}{16 \pi^2}\lambda(\Lambda)^2 \log\left(\frac{\Lambda}{b\Lambda}\right) + \mathcal{O}(\lambda^3,\frac{m^2}{\Lambda^2}) \nonumber \\
\Rightarrow \lambda(b\Lambda) &= \lambda(\Lambda) - \frac{3}{16 \pi^2} \lambda(\Lambda)^2 \log\left(\frac{1}{b}\right) + \dots \label{eqn:lambdarun} 
\end{align}
The effect of high energy modes on the four-point vertex function can be simply absorbed into a shift in the coupling constant! This procedure explicitly transfers loop-level physics in the theory defined at $\Lambda$ into tree-level physics at $b \Lambda$.

We can repeat this for the two point function to find the behavior of $Z(\Lambda)$ and $m^2(\Lambda)$.
\begin{align}
0 &\equiv \Sigma^{\Lambda}(p) - \Sigma^{b\Lambda}(p) \\
&= Z(\Lambda)p^2 + m^2(\Lambda) + \lambda(\Lambda) \int_{|\bar k|<\Lambda} \frac{d^4\bar k}{(2\pi)^4} \frac{1}{(Z(\Lambda)\bar k^2 + m^2(\Lambda))} \nonumber\\
&- Z(b\Lambda)p^2 - m^2(b\Lambda) - \lambda(b\Lambda) \int_{|\bar k|<b\Lambda} \frac{d^4\bar k}{(2\pi)^4} \frac{1}{(Z(b\Lambda)\bar k^2 + m^2(b\Lambda))}\nonumber\\
&= \left[Z(\Lambda)-Z(b\Lambda)\right] p^2 + m^2(\Lambda) - m^2(b\Lambda) + \lambda(\Lambda) \int_{|\bar k|=b\Lambda}^{\Lambda} \frac{d^4\bar k}{(2\pi)^4} \frac{1}{(Z(\Lambda)\bar k^2 + m^2(\Lambda))} \nonumber \\
&= \left[Z(\Lambda)-Z(b\Lambda)\right] p^2 + \left[m^2(\Lambda) - m^2(b\Lambda)\right] + \frac{\lambda(\Lambda)}{16 \pi^2} \left[\Lambda^2 - b^2 \Lambda^2\right] + \lambda(\Lambda) \frac{m^2(\Lambda)}{8 \pi^2} \log\left(\frac{1}{b}\right) \nonumber \\
\Rightarrow Z(b\Lambda) &= Z(\Lambda) + \mathcal{O}(\lambda^2) \label{eqn:zrun} \\
\Rightarrow m^2(b\Lambda) &= m^2(\Lambda) + \frac{\lambda(\Lambda)}{16 \pi^2} \Lambda^2 \left[1 - b^2\right] - \lambda(\Lambda)\frac{m^2(\Lambda)}{8 \pi^2} \log\left(\frac{1}{b}\right) + \dots \label{eqn:massrun}
\end{align}
We have again liberally ignored subleading terms. We see that the wavefunction normalization $Z$ does not run at one-loop in this theory, since the only one-loop diagram contributing to the two-point function does not have external momentum routed through the loop. This is merely `accidental' as $Z$ is not symmetry-protected and does run at two-loops. We also see the first hints of a somewhat worrisome situation with scalar masses. The mass $m^2(\Lambda)$ receives large one-loop corrections which tend to raise the mass up to near the cutoff, regardless of whether we originally had $m^2(\Lambda) \ll \Lambda^2$. We will investigate this in great detail later.

Now imagine we want to measure some properties of $\phi$ particles with external momenta far below our fundamental cutoff $p_i \sim \mu \ll \Lambda$. By construction, our procedure of integrating out high-energy momentum modes keeps the physics of these low-energy particles the same. But if we calculate this scattering amplitude using $\mathcal{L}(\Lambda)$, it is not easy to see from the Lagrangian how these low-energy modes will behave, since important effects are hidden in loop diagrams. If we instead first integrate out momentum shells down to some $\mathcal{L}(b \Lambda)$ with $\mu < b \Lambda \ll \Lambda$, then the effects of the high energy modes have been absorbed into the parameters of our Lagrangian, and we can read off much more of how $\phi$ particles will behave at low energies simply by looking at the parameters.

We can see further value in this approach if we consider scattering more low-energy $\phi$. Let's look at the $6$-point vertex function at zero momentum---in the theory at $\Lambda$, we start with $c_6(\Lambda) = 0$ and a one-loop diagram where momenta up to $\Lambda$ run in the loop: 

\begin{align}
V_6^\Lambda = \frac{\lambda^3(\Lambda)}{48} \int_{|k|<\Lambda} \frac{d^4k}{(2\pi)^4} \left(\frac{1}{Z(\Lambda) k^2 + m^2(\Lambda)}\right)^3.
\end{align} 
Now in the theory at $b \Lambda$, the loop only contains momenta up to $b\Lambda$, so we must account for the difference with a contact interaction $c_6(b\Lambda)$:
\begin{align}
V_6^{b\Lambda} = c_6(b\Lambda) + \frac{\lambda^3(b\Lambda)}{48} \int_{|k|<b\Lambda} \frac{d^4k}{(2\pi)^4} \left(\frac{1}{Z(b\Lambda) k^2 + m^2(b\Lambda)}\right)^3.
\end{align} 
Again we should ensure that the physics is the same upon lowering the cutoff:
\begin{align}
0 &\equiv V_6^\Lambda - V_6^{b\Lambda} \\
c_6(b\Lambda) &= \frac{\lambda^3(\Lambda)}{48} \int_{b\Lambda}^{\Lambda} \frac{d^4k}{(2\pi)^4} \left(\frac{1}{Z(\Lambda) k^2 + m^2(\Lambda)}\right)^3 \\
&= \frac{\lambda(\Lambda)^3}{3 \times 256 \pi^2} \left(\frac{1}{(b\Lambda)^2}-\frac{1}{\Lambda^2}\right) + \dots
\end{align} 
So this renormalization procedure is especially useful for understanding the behavior of irrelevant interactions. In our original theory nothing about the six-particle interaction was obvious from the Lagrangian, but in our theory with cutoff $b\Lambda$ we can simply read off the strength of this interaction at lowest order.

Note also the inverse behavior to that of the mass corrections---for the irrelevant interaction, the most significant contributions to the infrared behavior come from the \textit{low}-energy part of the loop integral, and the UV contributions are suppressed relative to this. Similarly, if we had started with a nonzero $c_6(\Lambda)$ which was small in units of the cutoff $c_6(\Lambda) \ll \Lambda^{-2}$ (so perturbative), such a UV contribution will also be subdominant. Then fully generally here, we have 
\begin{equation}\label{eqn:irrelrun}
c_6(b\Lambda) \simeq \frac{\lambda(\Lambda)^3}{16 \pi^2} \frac{1}{b^2 \Lambda^2} + \mathcal{O}(b^0)
\end{equation}
as we evolve to low scales $b \ll 1$.

The Wilsonian approach we have discussed here gives useful intuition for how renormalization works as a coarse-graining procedure wherein one changes the fundamental `resolution' of the theory, but in practice can make calculations cumbersome. Furthermore, the hard momentum-space cutoff we used is not gauge-invariant, which causes difficulties in realistic applications. 

The benefit, however, is that this is a `physical renormalization scheme' in which the renormalization condition relates the bare parameters to physical observables. For this reason, this renormalization scheme satisfies the Appelquist-Carrazone decoupling theorem \cite{Appelquist:1974tg}, which is enormously powerful. This guarantees for us that the effects of massive fields can, at low energies, simply be absorbed into modifications of the parameters in an effective theory containing solely light fields. 

In the next section we'll return to a clarifying example of the meaning of the decoupling theorem, and also discuss a renormalization scheme which does not satisfy the requirements for this theorem to operate, but is far simpler to use for calculations. The winning strategy will be to input decoupling by hand, which will allow us to get sensible physical results without the computational difficulty. Before we get to that, though, we'll take a couple detours.

\subsubsection*{Renormalization and locality}\addcontentsline{toc}{subsubsection}{\qquad Renormalization and Locality}\label{sec:locality}

We have seen that the need to remove divergences in our theory led to the introduction of running couplings which change as a function of scale. In our example above we see that renormalization has the operational effect of transferring loop-level physics into the tree-level parameters. This is an interesting perspective which bears further exploration---if there is hidden loop-level physics that really has the same physical effects as the tree-level bare parameters, perhaps this is a sign that there is a better way to organize our perturbation theory. Indeed, at some very general level renormalization can be thought of as a method for improving the quality of perturbation theory. For useful discussions at this level of abstraction of how renormalization operates, see \cite{Delamotte:2002vw} for its natural appearance whenever infinities are encountered in na\"{i}ve perturbative calculations, and \cite{neumaier_2015} for its usefulness even when infinities are not present. We'll discuss this perspective on renormalization further in the next section.

However it's clear that loops also give rise to physics that is starkly different from the lowest-order result (e.g. non-trivial analytic structure), so how do we know what higher-order physics we \textit{can} stuff into tree-level? In a continuum quantum field theory, a Lagrangian is a \textit{local} object $\mathcal{L}(x)$---that is, it contains operators like $\phi(x)^3$ which give an interaction between three $\phi$ modes at a single spacetime point $x$. Such effects are known as `contact interactions', but even at tree-level a local Lagrangian can clearly produce non-local (that is, long-range) physics effects. For example, consider the amplitude for $2 \rightarrow 2$ scattering in a $\phi^3$ theory at second order in the coupling.\\
\begin{figure}
	\centering
	\includegraphics[width=0.4\linewidth]{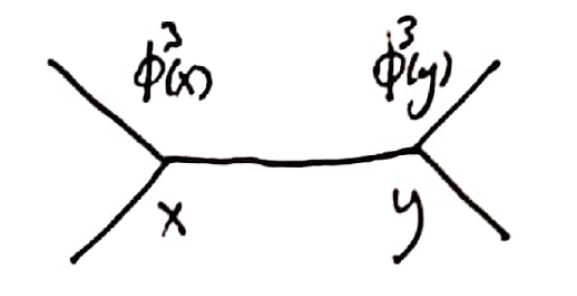}
	\caption{A position space Feynman diagram with vertices labeled.}
	\label{fig:position_space}
\end{figure}

In position space the non-locality here is obvious, as in Figure \ref{fig:position_space}: A simple tree-level diagram corresponds to a particle at point $x$ and a particle at point $y$ exchanging a $\phi$ quantum, but one may forget this important fact when working in momentum space. There the result is
\begin{equation} \label{eqn:treeexchange}
i\mathcal{M}(\phi\phi\rightarrow \phi\phi) = \frac{i g^2}{p^2 + m^2},
\end{equation}
and indeed, Fourier transforming the cross-section for this process yields a Yukawa scattering potential for our $\phi$ particles, showing that they mediate a long-range force over distances $r \sim 1/m$. We obviously cannot redefine the Lagrangian to put this effect into the lowest order of perturbation theory since this is not a local effect. 

But if we do have a continuum quantum field theory, then because of locality it describes fluctuations on \textit{all} scales. When we go to loop-level, we must integrate over all possible internal states, which includes integrating over arbitrarily large momenta or equivalently fluctuations on arbitrarily small scales. Heuristically, when the loop integral is sensitive to the ultraviolet of the theory, it is computing effects that operate on all scales---that is, it gives a contribution to \textit{local} physics. This tells us that the pieces of this higher-order contribution which we can reshuffle into our Lagrangian are connected with ultraviolet sensitivity, leading to a close connection of renormalization with divergences. 

A couple notes are warranted about the notion of locality we rely on here. Firstly, it's clear that this criterion of local effects appears because we began with a local Lagrangian. If we postulated that our fundamental theory contained nonlocal interactions, say $\mathcal{L}(x) \supset g \phi(x) \left( \int dy \phi(y)\right) \left( \int dz \phi(z)\right)$, then we could clearly absorb further nonlocalities with the same structure into this coupling as well. However, this sort of nonlocality is different from the nonlocality we saw appearing out of the local theory at tree-level. In particular it would break the standard connection between locality and the analytic structure of amplitudes---see e.g. Schwartz \cite{Schwartz:2013pla} or Weinberg \cite{Weinberg:1995mt} on `polology' and locality.

Secondly, our notion of locality should be modified in a low-energy theory with an energy-momentum cutoff $\Lambda$, as can be seen in hindsight in our Wilsonian discussion above. As $\Lambda$ defines a maximum energy scale we can probe, there is equivalently a minimum time and length scale we can probe due to the uncertainty principle, heuristically $\Delta x^\mu \gtrsim 1/\Delta p^\mu \gtrsim 1/\Lambda$. As a result, any fluctuations on shorter length scales are effectively local from the perspective of the low-energy theory. An exchange of a massive field with $M > \Lambda$ or of a light field with high frequency $\omega > \Lambda$ appears instantaneous to low-energy observers. This explains how it's sensible to use renormalization techniques in, for example, condensed matter applications, where systems are fundamentally discrete.

We can see this concretely by imagining the light $\phi$s in the tree-level example above instead exchanged a heavier scalar $\Phi$ with mass $M\gg m$. While the amplitude $\mathcal{M} = g^2/(p^2 + M^2)$ is still nonlocal in the continuum theory, if we're only interested in physics at energies $E \ll M$ we may Taylor expand the result $\mathcal{M} = g^2/M^2 + g^2 p^2/2 M^4 + \dots$. We may then absorb the leading effects of this heavy scalar into an effectively local interaction $g^2 \phi^4/M^2$ among the light fields---as long as we work at energies below $M$.

In the next section we'll explore concretely how these insights enable us to transfer loop-level physics to tree-level physics, and so improve our calculations.

\subsection{To Repair Perturbation Theory}\label{sec:repairpert}

\subsubsection*{Renormalization group equations}\addcontentsline{toc}{subsubsection}{\qquad Renormalization Group Equations}
The astute reader may notice a potential issue with our one-loop results in the $\phi^4$ theory, Equations \ref{eqn:lambdarun},\ref{eqn:zrun},\ref{eqn:massrun}. We've derived this behavior as the first subleading terms in a series expansion in the number of loops. In relation to the tree-level results, the one-loop contributions are suppressed by a factor $\sim \lambda / 16\pi^2 \log(\Lambda_0/\Lambda)$, where I've switched notation to now have $\Lambda_0$ as a high scale and $\Lambda$ as the lower scale we integrate down to. Higher $n$-loop contributions will be further suppressed by $n$ loop factors. But what if we wanted to study physics at a scale far lower than $\Lambda_0$? Eventually this factor becomes large enough that we will need to compute many loops to obtain high precision, and then large enough that we have reason to question the convergence of the series\footnote{Please excuse my slang. Perturbative series in QFT are quite generally \textit{not} convergent but we can trust the answers anyway to order $n \sim \exp 1/ \text{expansion parameter}$ because they are \textit{asymptotic} series. So really when this parameter becomes large enough we worry that our series is not even asymptotic. Thinking deeply about this leads to many interesting topics in field theory, from accounting for nonpertubative instanton effects which are (partially) behind the lack of convergence; to the program of `resurgence', the idea that there are secret relations between the perturbative and nonperturbative pieces. This is all far outside my purview here, but some introductions aimed at a variety of audiences can be found in \cite{Schafer:1996wv,Vandoren:2008xg,Dunne:2012ae,Dunne:2014bca,Dorigoni:2014hea,Marino:2015yie}.}. Keep in mind that we are in the era of precision measurements of the Standard Model, so these one-loop expressions are very restrictive. 

For a concrete example, say we wanted to check the SM prediction for a measurement of a coupling $\lambda(\Lambda)$ with $\lambda(\Lambda_0)=1$ whose experimental uncertainty was $1\%$. Let's define a theoretical uncertainty on a perturbative calculation to $n^{\text{th}}$ order as 
\begin{equation}
\epsilon_n \equiv \frac{n^{\text{th}} \text{ order result} - \text{ estimated size of } (n+1)^{\text{th}} \text{ order result} }{n^{\text{th}} \text{ order result}},
\end{equation}
where our Wilsonian calculation in Section \ref{sec:removediv} gave at first order, as a reminder (and with modified notation)
\begin{equation}
\lambda(\Lambda) = \lambda(\Lambda_0) - \frac{3}{16 \pi^2} \lambda(\Lambda_0)^2 \log \frac{\Lambda_0}{\Lambda} + \mathcal{O}(\lambda^3),
\end{equation}
and our heuristic estimate for the size of the second order correction is $(\frac{3}{16 \pi^2})^2 \lambda(\Lambda_0)^3 \log^2 \frac{\Lambda_0}{\Lambda}$. When the result is simply a series, the uncertainty is very simple to calculate, as the numerator is then our estimate of the $(n+1)^{\text{th}}$ order correction, which is roughly the square of the first order correction in this case for $n=1$. Here we have $\epsilon_1 = \frac{3}{16 \pi^2} \lambda(\Lambda_0) \log \frac{\Lambda_0}{\Lambda}$.
	
Then in order for our theoretical uncertainty to be subdominant to the experimental precision, $\epsilon_1 < 0.01$, we must go past the one-loop result if we wish to look at energies below $\Lambda \sim \exp(-16 \pi^2 \times 0.01/3) \Lambda_0 \sim \Lambda_0 / 2$, the two-loop result is only sufficient until $\Lambda \sim \Lambda_0 / 400$, and if we can manage to calculate the three-loop corrections that only gets us down to something like $\Lambda \sim \Lambda_0 / 8\times 10^4$. If we're interested in taking the predictions of a Grand Unified Theory defined at $\Lambda_0 \sim 10^{16} \text{ GeV}$ and comparing them to predictions at SM energies, how in the world are we to do so?

Fortunately, we can do better by applying our one-loop results more cleverly. It is clear by looking at Equations \ref{eqn:lambdarun},\ref{eqn:zrun},\ref{eqn:massrun} that the results have the same form no matter the values of $\Lambda_0,\Lambda$. So if we take $\Lambda$ to be only slightly smaller than $\Lambda_0$ (corresponding to $1 - b \ll 1$ in our previous notation) the expansion parameter becomes very small and the one-loop result becomes very trustworthy. What we would like is some sort of iterative procedure to gradually lower the cutoff, which we could then use to find the one-loop result for energies far lower than the range of our perturbative series. This is in fact precisely the sort of problem that a differential equation solves, and we can derive such an equation by differentiating both sides by $\ln \Lambda$ and then taking $\Lambda$ infinitesimally close to $\Lambda_0$. That exercise yields
\begin{align}
\frac{\text{d} Z(\Lambda)}{\text{d}\log \Lambda} &= 0 \\
\frac{\text{d} m^2(\Lambda)}{\text{d}\log \Lambda} &= \frac{\lambda(\Lambda)}{8 \pi^2} \left( m^2(\Lambda)-\Lambda^2 \right) \\
\frac{\text{d} \lambda(\Lambda)}{\text{d}\log \Lambda} &= \frac{3}{16 \pi^2} \lambda(\Lambda)^2
\end{align}
These are known as `renormalization group equations' and they indeed allow us to evolve the coupling down to low energies---one says we use them to `resum the logarithm'. Then to study physics at a very low scale we can bring these couplings down to a lower scale and do our loop expansion using those couplings, which is known as  `renormalization group improved perturbation theory', and which we will discuss in more detail soon. Explicitly solving with our boundary condition at $\Lambda_0$ yields
\begin{equation}
\lambda(\Lambda) = \frac{\lambda(\Lambda_0)}{1 + \frac{3 \lambda(\Lambda_0)}{16 \pi^2} \log \frac{\Lambda_0}{\Lambda}}
\end{equation}
Turning back to our effective field theory language, we see that quantum corrections have generated an anomalous dimension for $\lambda$, $\delta_{\phi^4}= 3 \lambda^2/(16 \pi^2)$, correcting the leading order scaling behavior. Since $\delta_{\phi^4}>0$, we've determined that the quartic interaction is marginally irrelevant, which we will return to later.

We can now look at the theoretical uncertainty in this one-loop resummed calculation by including an estimate of the next order correction to the running of the quartic $\frac{\text{d} \lambda(\Lambda)}{\text{d}\log \Lambda} = \frac{3}{16 \pi^2} \lambda(\Lambda)^2 + \left(\frac{3}{16 \pi^2}\right)^2 \lambda(\Lambda)^3$ and resumming that expression. This can no longer be done analytically, but numerical evaluation easily reveals that the theoretical uncertainty here stays below $1\%$ for many, many orders of magnitude below $\Lambda_0$. Resumming the logarithmic corrections allows us to use our loop results to far greater effect.

\subsubsection*{Decoupling}\addcontentsline{toc}{subsubsection}{\qquad Decoupling}

The physical meaning and technical statement of the decoupling theorem commonly confuse even prominent practitioners of effective field theory,
so it's worth going clearly through an example to refine our understanding. Indeed, one may be confused just at zeroth order about how decoupling is sensible against the background of the hierarchy problem---which is an issue of sensitivity of a scalar mass to heavy mass scales. How can we claim QFT obeys a decoupling theorem and then go on to worry at length about quantum corrections $\delta m^2 \propto M^2$?

The correct way to think about the decoupling theorem is \textit{not} whether a top-down calculation could yield a result that depends on heavy mass scales, but whether a bottom-up effective field theorist and low-energy observer could gain information about the heavy mass scales through low-energy measurements. We can clarify this important difference by looking at a one-loop mass correction to a light scalar $\phi$ of mass $m$ from a heavy scalar $\Phi$ of mass $M$ through the interaction $\lambda \phi^2 \Phi^2$. We again take a Wilsonian perspective and begin at a scale $\Lambda_0 > M$. In close analogy to what we had before, we now find
\begin{align}
0 &\equiv \Sigma^{\Lambda_0}(0) - \Sigma^{\Lambda}(0) \\
&= m^2(\Lambda_0) + \lambda(\Lambda_0) \int_{|k|<\Lambda_0} \frac{d^4k}{(2\pi)^4} \frac{1}{(k^2 + M^2(\Lambda_0))} \nonumber\\
& - m^2(\Lambda) - \lambda(\Lambda) \int_{|k|<\Lambda} \frac{d^4k}{(2\pi)^4} \frac{1}{(k^2 + M^2(\Lambda))}\nonumber\\
&= m^2(\Lambda_0) - m^2(\Lambda) + \lambda(\Lambda_0) \int_{|k|=\Lambda}^{\Lambda_0} \frac{d^4k}{(2\pi)^4} \frac{1}{(k^2 +M^2(\Lambda_0))} \nonumber \\
&= \left[m^2(\Lambda_0) - m^2(\Lambda)\right] + \frac{\lambda(\Lambda_0)}{16 \pi^2} \left[\Lambda_0^2 - \Lambda^2\right] - \lambda(\Lambda_0) \frac{M^2(\Lambda_0)}{16 \pi^2} \log\left(\frac{\Lambda_0^2 + M^2(\Lambda_0)}{\Lambda^2 + M^2(\Lambda)}\right) \nonumber \\
\Rightarrow m^2(\Lambda) &= m^2(\Lambda_0) + \frac{\lambda(\Lambda_0)}{16 \pi^2}\left(\left[\Lambda_0^2 - \Lambda^2\right] + M^2(\Lambda_0) \log \frac{\Lambda_0^2 + M^2(\Lambda_0)}{\Lambda^2 + M^2(\Lambda)} \right) + \mathcal{O}(\lambda^2).
\end{align}
And we may already exhibit the confusion. If we use this to calculate the mass at a low scale $\Lambda \ll M$, we see that $m^2(\Lambda)$ does depend on the heavy mass scale, and gets a contribution which goes like $m^2(\Lambda) \sim M^2(\Lambda_0) \log \left(1 + \Lambda_0^2/M^2(\Lambda_0)\right)$.

However, the effect on the light scalar is an additive shift of the mass. If we go out and measure the mass at a single scale $m^2(\Lambda)$ we can't tell empirically which `parts' of that came from $m^2(\Lambda_0)$ and which came from $M^2(\Lambda_0)$ or whatever else is in there, so we have no idea of how this low-energy measurement depends on heavy scales. To get information about the various contributions to the light scalar mass, we can measure it at different scales and look at how it changes. Of course this information is contained in the renormalization group equation for $m^2(\Lambda)$. At $\mathcal{O}(\lambda)$, we can find this by differentiating the above, and we find
\begin{equation}
\frac{\text{d}m^2(\Lambda)}{\text{d}\log \Lambda} = \frac{\lambda(\Lambda)}{8 \pi^2} \Lambda^2 \left[\frac{M^2(\Lambda)}{\Lambda^2 + M^2(\Lambda)}-1\right].
\end{equation}
Now we can see the difference. If we perform low-energy observations where we can take the cutoff below the mass of the heavy scalar $\Lambda \ll M$, then the physics of the heavy scalar decouples from the \textit{running} of the light scalar mass. It is only by studying this running at low energies that we can gain information about the ultraviolet, and we see that this information is contained solely in small corrections scaling as $\Lambda^2/M^2$. At low energies, to learn about short-distance physics we must make very precise measurements of the low-energy physics. \textit{This} is the sense in which heavy mass scales decouple from the theory in the infrared.

\subsubsection*{Renormalized perturbation theory}\addcontentsline{toc}{subsubsection}{\qquad Renormalized Perturbation Theory}
Now let us study another, slightly more complex theory and apply renormalization techniques to simplify our calculations. We avoid the complication of gauge symmetries and focus instead on a Yukawa theory of a Dirac fermion interacting with a parity-odd scalar.

Our first improvement to perturbation theory will be to switch from `bare' to `renormalized perturbation theory'. Let's first recap our procedure in Section \ref{sec:removediv}. We began with a Lagrangian with bare parameters $m_0, \lambda_0, \dots$, introduced a regulator, computed the physical parameters $m_\text{phys},\lambda_{\text{phys}}$ in terms of the bare ones, inverted those relationships, and then plugged in for the bare parameters in terms of the renormalized ones, after which we were left with an amplitude which remains finite as we remove the regulator. This procedure works to remove the divergences in any renormalizable theory, but is obviously rather cumbersome. 

Furthermore one may question the validity of performing a perturbative expansion in a bare parameter which we later discover is formally infinite in the continuum theory $\lambda_0 \sim \log(\Lambda_0^2) \rightarrow \infty$. It is both conceptually and computationally easier to instead start off by performing perturbation theory in terms of the renormalized parameters which we know to be finite by definition. Fortunately we can improve our accounting simply by reshuffling the Lagrangian as follows.

In terms of the bare parameters and fields, the Lagrangian reads
\begin{align}
\mathcal{L}_0 &= \bar \psi_0 \left( i \slashed{\partial} -M_0\right) \psi_0 + \half \phi_0 \left(\Box - m^2_0\right) \phi_0 \\
\mathcal{L}_1 &= i g_0 \phi_0 \bar \psi_0 \gamma_5 \psi_0 - \frac{1}{24} \lambda_0 \phi_0^4 
\end{align}
where we've split up the free and interaction parts. Just as in our earlier example, when we compute at one-loop these parameters will get corrections such that the bare parameters are no longer the physical parameters we measure. Anticipating that fact, let us rewrite the Lagrangian to explicitly account for those corrections from the outset. 

Although it was not a feature of our simple example above, in general there will be `wavefunction renormalization' which changes the normalization of our field operators, so we define $\phi_0 = Z_\phi^{1/2} \phi, \psi_0 = Z_\psi^{1/2} \psi$ where $\psi, \phi$ are now renormalized fields, We do the same to define renormalized masses related to the bare masses as $M_0 = Z_M Z_\psi^{-1} M, m_0^2 = Z_\phi^{-1} Z_m m^2$, and for the couplings $g_0 = Z_\phi^{-1/2} Z_\psi^{-1}  Z_g g, \lambda_0 = Z_\phi^{-2} Z_\lambda \lambda$. Next we use the brilliant strategy of adding zero to split these $Z$-factors into a piece with the same form we started with and a `counterterm' proportional to $(Z-1)$. Since at tree-level there's no renormalization needed, we know $Z = 1 + \mathcal{O}(\text{couplings})$. At nontrivial loop level, we must choose the $Z$-factors to implement our chosen renormalization scheme.

The Lagrangian now takes the form
\begin{align}
\mathcal{L}_0 &= \bar \psi \left( i \slashed{\partial} -M\right) \psi + \half \phi \left(\Box - m^2\right) \phi \\
\mathcal{L}_1 &= i Z_g g \phi \bar \psi \gamma_5 \psi - \frac{1}{24} Z_\lambda \lambda \phi^4 + \mathcal{L}_{\text{ct}} \\
\mathcal{L}_{\text{ct}} &= i (Z_\psi - 1) \bar \psi \slashed{\partial} \psi - (Z_M - 1) M \bar \psi \psi - \half (Z_\phi - 1) \partial^\mu \phi \partial_\mu \phi - \half (Z_m - 1) m^2 \phi^2
\end{align}
where we've split off the counterterms into $\mathcal{L}_{\text{ct}}$. We can now treat the terms in $\mathcal{L}_{\text{ct}}$ simply as additional lines and vertices contributing to our Feynman diagrams. We'll see how useful this is once we begin renormalizing the theory. This is done in full in Srednicki's chapters 51-52 \cite{Srednicki:2007qs}, so we will not go through every detail.

\subsubsection*{Continuum renormalization}\addcontentsline{toc}{subsubsection}{\qquad Continuum Renormalization}
We'll regulate this theory using `dimensional regularization' (dim reg) which analytically continues the theory to general dimension $d = 4 - \epsilon$. That this will regulate our theory is not obvious, but I recommend Georgi \cite{Georgi:1994qn} to convince yourself of this and Collins \cite{Collins:1984xc} for a full construction of dim reg; we'll content ourselves with seeing it in action. Our renormalization scheme will be `modified minimal subtraction' and denoted $\overline{\text{MS}}$, where `minimal subtraction' means we'll choose our counterterms solely to cancel off the divergent pieces (rather than to enforce some relation to physical observables, as we did previously) and `modified' means that actually it's a bit nicer if we cancel off a couple annoying constants as well. Since we're using $\overline{\text{MS}}$, the mass parameters $m, M$ will not quite be the physical masses, which are always the locations of the poles in the propagators, and the fields will not be normalized to have unity residue on those poles. So we'll have to relate these parameters to the physical ones later.

We'll briefly go through renormalizing the scalar two-point function at one loop to evince dim reg and $\overline{\text{MS}}$. In our one-loop diagrams we use propagators given by $\mathcal{L}_0$, since we know the counterterms begin at higher order. The full details of the one-loop renormalization of this theory can be found in Srednicki's Chapter 51. 

\begin{figure}
	\centering
	\includegraphics[width=0.7\linewidth]{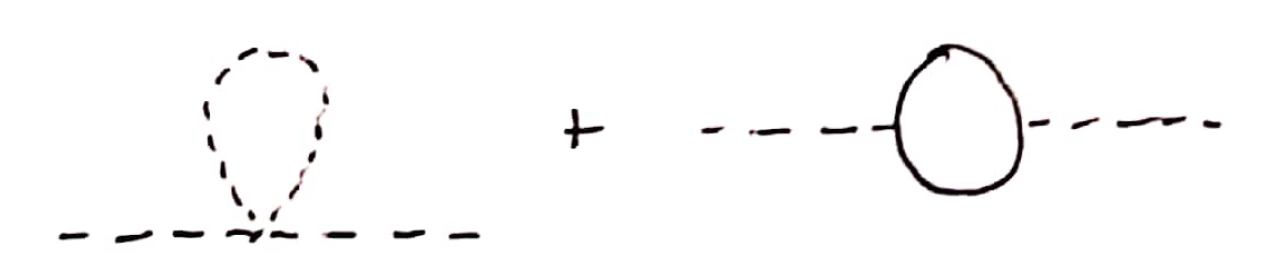}
	\caption{Diagrams giving the one-loop correction to the scalar propagator in Yukawa theory.}
	\label{fig:yukscalarprop}
\end{figure}

At one-loop, the scalar two-point function gets corrections due to both interactions, as seen in Figure \ref{fig:yukscalarprop}. There is the diagram we had in the $\phi^4$ theory above, but we must recompute this in dim reg
\begin{align}
i \Sigma_{\phi \text{ loop}}(p^2) &= -i\half\lambda \int \frac{d^4k}{(2\pi)^4}\frac{-i}{k^2 + m^2} \\ 
&= -i\half\lambda \tilde{\mu}^\epsilon \int \frac{d^d\bar k}{(2\pi)^d}\frac{1}{\bar k^2 + m^2} \\ 
&= -i\frac{\lambda}{2}  \Gamma\left(-1+\frac{\epsilon}{2}\right) \frac{m^2}{(4\pi)^2} \left(\frac{4 \pi \tilde{\mu}^2}{m^2}\right)^{\epsilon/2} \\ 
&\simeq i\frac{\lambda}{2(4\pi)^2} m^2 \left(\frac{2}{\epsilon}+1-\gamma_E\right)\left(1 + \frac{\epsilon}{2} \log \frac{4 \pi \tilde{\mu}^2}{m^2} \right) \\
&\simeq i\frac{\lambda}{2(4\pi)^2} m^2 \left(\frac{2}{\epsilon}+1-\gamma_E-\log \frac{4 \pi \tilde{\mu}^2}{m^2}\right) \\
&= i\frac{\lambda}{(4\pi)^2} m^2 \left(\half + \frac{1}{\epsilon}+\half\log \frac{\mu^2}{m^2}\right) + \mathcal{O}(\epsilon),
\end{align}
where we have analytically continued to $d=4-\epsilon$ dimensions including replacing $\lambda \mapsto \lambda \tilde{\mu}^\epsilon$, with $\tilde{\mu}$ a mass scale, to keep $\lambda$ dimensionless; performed the integral in general dimension; expanded for $\epsilon \simeq 0$; and defined $\mu^2 \equiv 4 \pi \tilde{\mu}^2 e^{-\gamma_E}$, where $\gamma_E$ is the Euler-Mascheroni constant, to simplify the expression. Details on these calculational steps are laid out in Srednicki's Chapter 14, but there's an important and na\"{i}vely surprising feature of the above that we should discuss.

In previous sections our regularization scheme explicitly introduced a mass scale $\Lambda$ which we could think of as having a physical interpretation as some sort of short-distance cutoff, if we wished. We then saw that by cleverly studying how the parameters in the theory are modified as we change $\Lambda$ and demand the physics stays the same, we could make a variety of things easier to calculate and make the physical content of the theory more transparent. Note that in doing so, we're stretching the meaning of the scale $\Lambda$ away slightly from that physical picture---we don't care what the `real' cutoff of our system is, or if there really is any sort of cutoff; we simply know that allowing such scale-dependence in our couplings and studying the theory at different values of $\Lambda$ makes our lives easier, so we imagine varying it.

Now the way this new regularization scheme works is somewhat opaque, but it still necessitates the introduction of a new scale. In this case, the unphysical scale $\mu$ is required to ensure that our couplings remain dimensionless away from $d=4$. This scheme thus invites us to further broaden our notion of varying a scale to study the theory at different energies---this time the scale explicitly never had a physical interpretation. We can view $\mu$ as labeling a one-parameter family of calculational schemes. We've ensured by construction that the physics is the same no matter what $\mu$ we choose, but by cleverly using the scale-dependence we can make our lives far easier. The intuition should be the same as in the previous case, and lowering $\mu$ does likewise transfer loop-level physics to tree-level physics and can be used to improve the convergence of perturbative calculations. The connection is now slightly more opaque, which is why we began by discussing a cutoff in Euclidean momentum space, but the calculations become far simpler.

There's another diagram with a $\psi$ loop
\begin{align}
i \Sigma_{\psi \text{ loop}}(p^2) &= (-1)(ig)^2\int \frac{d^4k}{(2\pi)^4}(-i)^2\frac{\text{Tr}\left[(-\slashed{k}+M)\gamma_5(-\slashed{k}-\slashed{p}+M)\gamma_5\right]}{(k^2 + M^2)((k+p)^2 + M^2)}\\
&=- \frac{g^2}{4 \pi^2} \left[\frac{1}{\epsilon}(k^2 + 2 M^2) + \frac{1}{6} k^2 + M^2 - \int_0^1 dx (3x(1-x)k^2+M^2)\ln\frac{D}{\mu^2}\right],
\end{align}
with $D = x(1-x) k^2 + m^2$, whose evaluation follows similar steps but we skip for brevity. Adding these together, $\overline{\text{MS}}$ tells us the $\phi$ counterterms must take the values 
\begin{align} 
Z_\phi &= 1 - \frac{g^2}{4\pi^2} \frac{1}{\epsilon} \\
Z_m &= 1 + \left( \frac{\lambda}{16 \pi^2} - \frac{g^2}{2 \pi^2} \frac{M^2}{m^2}\right) \frac{1}{\epsilon}.
\end{align}
For the fermion, evaluating the one-loop diagrams gives us the counterterms
\begin{align}
Z_\psi &= 1 - \frac{g^2}{16 \pi^2} \frac{1}{\epsilon} \\
Z_M &= 1 - \frac{g^2}{8 \pi^2}\frac{1}{\epsilon}.
\end{align}
Since we didn't choose the counterterm to keep the location of the pole in the propagator fixed, $m$ is no longer the physical scalar mass. But we can find the physical, `pole' mass precisely from that condition:
\begin{align}
0 &\equiv \Delta^{-1}(k^2 = -m_{\text{phys}}^2) \\
&= \left. k^2 + m^2 + \Sigma(k^2)\right|_{k^2 = -m_{\text{phys}}^2}\\
&= -m_{\text{phys}}^2 + m^2 - \Sigma(-m_{\text{phys}}^2)\\
\Rightarrow m_{\text{phys}}^2 &= m^2 -\Sigma(-m^2) + \mathcal{O}(\lambda^2, g^4,\lambda g^2),
\end{align}
where we have used our favorite trick to replace $m_{\text{phys}}^2$ with $m^2$ in the one-loop correction, since it is already higher order in couplings.

\begin{figure}
	\centering
	\begin{subfigure}{.45\textwidth}
		\centering
		\includegraphics[width=0.7\linewidth]{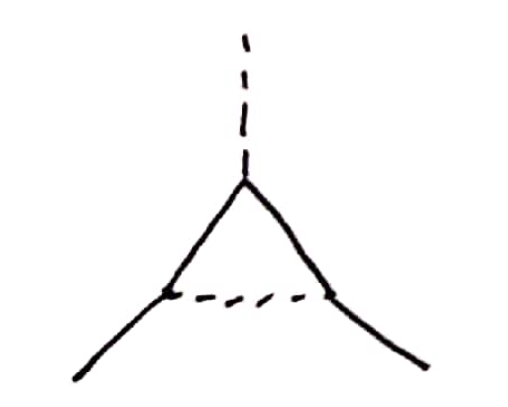}
		\caption{}
		\label{fig:YukInta}
	\end{subfigure}%
	\begin{subfigure}{.45\textwidth}
		\centering
		\includegraphics[width=0.7\linewidth]{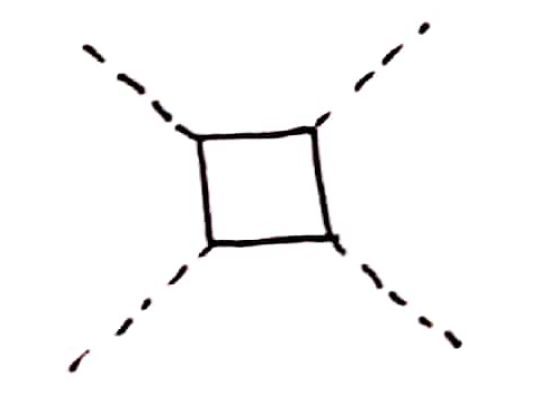}
		\caption{}
		\label{fig:YukIntb}
	\end{subfigure}
	\caption{Some one-loop diagrams in Yukawa theory correcting the interactions. In (a), a triangle diagram correcting the Yukawa interaction. In (b), a box diagram correcting the scalar quartic interaction.}
\end{figure}

As for the interactions, we have a triangle diagram for the Yukawa coupling and a new contribution to the quartic with a fermion running in the loop, as depicted in Figures \ref{fig:YukInta} and \ref{fig:YukIntb} respectively. These lead to the counterterms
\begin{align}
Z_g &= 1 + \frac{g^2}{8 \pi^2} \frac{1}{\epsilon} \\
Z_\lambda &= 1 + \left(\frac{3\lambda}{16\pi^2}-\frac{3 g^4}{\pi^2 \lambda}\right) \frac{1}{\epsilon},
\end{align}
from which we'll be able to understand how the strength of the interactions changes as a function of the energy at which the theory is probed. 

\subsubsection*{Renormalization group improvement}\addcontentsline{toc}{subsubsection}{\qquad Renormalization Group Improvement}
Now the second improvement to perturbation theory is the RG-improved perturbation theory we mentioned above. This takes on an even more useful role in our continuum renormalization scheme here. In the Wilsonian picture, $\Lambda$ was a high cutoff and we ensured the physics was invariant under evolution of $\Lambda$, but this scale still needed to stay far above the scales of interest in the problem $\Lambda \gg m, M, -k^2,\dots$. Now the scale $\mu$ is entirely unphysical and we are free to bring it all the way down to the scales of kinematic interest---in fact doing so will vastly simplify calculations. As a result we are able to make even more use of the RG-improvement than we could above.

We find the running couplings by again using the fact that the bare parameters are independent of the unphysical renormalization scale $\mu$. Having utilized a mass-independent regulator, a Wilsonian interpretation of couplings running with the value of the regulator is nonsensical here and so renormalization group improvement is the way to extract predictions from this theory. We already have the relations between the bare and renormalized quantities, e.g.
\begin{equation}
g_0 = Z_\phi^{-1/2} Z_\psi^{-1} Z_g g \tilde{\mu}^{\epsilon/2}.
\end{equation}
And since we know that the bare parameters are independent of $\mu$ by definition, we have
\begin{align}
\ln g_0 &= \ln\left(1 + \frac{g^2}{8\pi^2} \frac{1}{\epsilon}\right) + \ln\left(1 + \frac{g^2}{16\pi^2} \frac{1}{\epsilon}\right) + \ln\left(1 + \frac{g^2}{8\pi^2} \frac{1}{\epsilon}\right) + \ln g + \half \epsilon \ln \tilde{\mu}\\
&= \frac{5 g^2}{16\pi^2} \frac{1}{\epsilon} + \ln g + \half \epsilon \ln \tilde{\mu} + \mathcal{O}(g^4)\\
\frac{\text{d}\ln g_0}{\text{d}\ln \mu} &= 0\\
&= \frac{10 g}{16\pi^2} \frac{1}{\epsilon} \frac{\text{d}g}{\text{d}\ln \mu} + \frac{\text{d}\ln g}{\text{d}\ln \mu} + \half \epsilon \\
&= \frac{\text{d}g}{\text{d}\ln \mu} \left(1 + \frac{5 g^2}{8\pi^2} \frac{1}{\epsilon}\right) + \epsilon g.
\end{align}
If we expand $\frac{\text{d}g}{\text{d}\ln \mu} = a_1 \epsilon + a_0 + \dots$ order by order in $\epsilon$, then matching the $\mathcal{O}(\epsilon)$ terms gives $a_1 = -g/2$ and matching the $\mathcal{O}(\epsilon^0)$ terms tells us that, in the $\epsilon \rightarrow 0$ limit, we have
\begin{equation}
\frac{\text{d}g}{\text{d}\ln \mu} = \frac{5}{16 \pi^2} g^3 + \mathcal{O}(g^4).
\end{equation}
This $\epsilon$-independent piece is known as the `beta function' for the coupling, $\beta(g)= \frac{5}{16 \pi^2} g^3$. Of course there are higher-order terms in $\frac{\text{d}g}{\text{d}\ln \mu}$ which are needed to match the $\mathcal{O}(\epsilon^{-n})$ terms, and which one can solve for. But these vanish in the $\epsilon\rightarrow 0$ limit, so will not contribute to the running of $g(\mu)$.

We can now resum this logarithm to find the evolution of this coupling with renormalization scale
\begin{equation}
g(\mu) = \frac{\bar g}{\sqrt{1 - \frac{5 \bar g^2}{8 \pi^2} \log \frac{\mu}{\bar \mu}}},
\end{equation}
where we've used the boundary condition $g(\bar \mu) = \bar g$. As before, the resummed version will allow us to maintain precision to far lower scale than we could with simply its leading order approximation.

It's useful to keep in mind the Wilsonian picture as a clearer example because our regulator had a physical interpretation. The point is that the logarithms are really what's encoding how couplings change as a function of scale; in the Wilsonian calculation it was obvious that the logarithmic contribution $\log \frac{1}{b}$ is present no matter the initial cutoff. One says that couplings which receive logarithmic quantum corrections `get contributions from all scales'. Then it's clear why this RG-improvement is sensible---though we may start at some particular $\Lambda$ or $\mu$, a one-loop calculation offers information on the lowest-order logarithmic running over \textit{all} momenta, and we may sum up those modifications to improve our perturbation theory.

\subsection{To Relate Theories}\label{sec:relatetheories}

\subsubsection*{Mass-independent schemes and matching}\addcontentsline{toc}{subsubsection}{\qquad Mass-Independent Schemes and Matching}
We've seen already the necessity of renormalization when a theory produces na\"{i}vely divergent results, and its enormous use in improving the precision of perturbative calculations in a given theory. The last facet we'll discuss is its use in connecting theories. This is necessary to use the computationally-simple scheme of dim reg with $\overline{\text{MS}}$ in theories with different mass scales, and is very closely related to the effective field theory philosophy we discussed in Section \ref{sec:EFT}. Cohen's monograph \cite{Cohen:2019wxr} goes into far more depth than I will be able to, and is a fantastic introduction to these ideas and their application. This perspective on renormalization has also been of enormous use in condensed matter to understand behaviors that appear in many distinct systems in the long-distance limit, and has applications in formal field theory to understand better the properties of QFT itself.

In the previous section we derived the beta function for Yukawa theory in the $\overline{\text{MS}}$ scheme. As promised by our terming of this as a `mass-independent' scheme, the beta functions indeed have no reliance on the masses. But this should seem remarkably peculiar, as it suggests that there is no decoupling at all. Were that the case, by measuring the low-energy beta functions of QED we could tell how many charged particles existed up to arbitrarily large mass scales! What has gone wrong is that $\overline{\text{MS}}$ does not meet the criterion of a physical scheme which is necessary for the Appelquist-Carrazone theorem to operate. In $\overline{\text{MS}}$ the renormalization condition has nothing to do with physical values of the parameters so, while it makes calculations far simpler, $\overline{\text{MS}}$ has broken decoupling.

To restore decoupling and allow us to properly use a mass-independent scheme, we must implement the mass-dependence ourselves by `matching' the Yukawa theory at energies above the fermion mass to a theory of solely light scalars at energies below the spinor mass. To `match', we consider some process which exists in both theories---for example, $\phi^4$ scattering---and ensure that at the matching scale $M$ both theories agree on the physics.

In the high-energy Yukawa theory we can run the RG scale all the way down from a high scale $\bar \mu$ to $\mu = M_\text{phys}$, the physical, pole mass of the fermion. To get simple closed-form expressions, we'll take the couplings small enough that working to lowest order gives a good approximation. We'll denote all of the UV values with bars, e.g. $M(\mu=\bar \mu)=\bar M$. Firstly, we use the counterterms to find the anomalous dimension of the fermion mass
\begin{align}
\frac{d \log M}{d \log \mu} &= - \frac{d}{d \log \mu}\left(Z_\psi^{-1} Z_M\right) \\
 &= - \frac{g^2}{16 \pi^2} \\
\Rightarrow M(\mu) &\simeq \bar M \left(\frac{\bar \mu}{\mu} \right)^{\frac{\bar g^2}{16\pi^2}} \\
&\simeq \bar M \left( 1 + \frac{\bar g^2}{16\pi^2} \log \frac{\bar \mu}{\mu}\right)  + \dots
\end{align}
We then find the fermion pole mass as
\begin{align}
0 &= - M_\text{phys} + M(\mu=M_\text{phys}) - \Sigma(-M_\text{phys}) \\
M_\text{phys} &= \bar M \left( 1 + \frac{\bar g^2}{16\pi^2} \log \frac{\bar \mu}{\bar M} \right) - \Sigma(-\bar M) + \text{ subleading}\\
&= \bar M \left( 1 + \frac{\bar g^2}{16\pi^2} \log \frac{\bar \mu}{\bar M} \right) - \frac{\bar g^2}{16 \pi^2} \bar M \int_0^1 dx x \log\left(\frac{x^2 \bar M^2 + (1-x)\bar m^2}{\bar M}\right) + \dots \\
&= \bar M \left[1 + \frac{\bar g^2}{16\pi^2}\left(\half + \log \frac{\bar \mu}{\bar M}\right)\right] + \dots
\end{align}
Now we need the value of the other parameters at that mass threshold
\begin{align}
\frac{1}{m}\frac{d m}{d \log \mu} &= \frac{1}{4 \pi^2} \left(\frac{1}{8} \lambda - g^2\frac{M^2}{m^2}+\half g^2\right) \\
\Rightarrow m(M_\text{phys}) &= \bar m \left(1 - \frac{1}{4 \pi^2}\left[\frac{1}{8} \bar \lambda - \bar g^2\frac{\bar M^2}{\bar m^2}+\half \bar g^2\right]\log\frac{\bar \mu}{\bar M} \right) + \dots \\
\frac{dg}{d\log \mu} &= \frac{5}{16 \pi^2} g^3 \\
\Rightarrow g(M_\text{phys}) &= \bar g \left(1 - \frac{5 \bar g ^2}{16 \pi^2} \log \frac{\bar \mu}{\bar M} \right) + \dots \\
\frac{d\lambda}{d\log \mu} &= \frac{1}{16\pi^2} \left(3\lambda^2 + 8 \lambda g^2 - 48 g^4\right) \\
\Rightarrow  \lambda(M_\text{phys}) &= \bar \lambda \left(1 - \frac{1}{16 \pi^2}\left[3\bar \lambda^2 + 8 \bar \lambda \bar g^2 - 48 \bar g^4\right]\log \frac{\bar \mu}{\bar M}\right) + \dots
\end{align}
Now we are ready to proceed to even lower energies. We enforce decoupling by matching to the low-energy theory of just a self-interacting scalar. We have 
\begin{align}\label{eqn:yukctsplit}
\mathcal{L}_0 &= \half \phi\left(\Box - m^2\right) \phi \\
\mathcal{L}_1 &= - \frac{1}{24} Z_\lambda \lambda \phi^4 + \mathcal{L}_{\text{ct}} \\
\mathcal{L}_{\text{ct}} &= - \half (Z_\phi - 1) \partial^\mu \phi \partial_\mu \phi - \half (Z_m - 1) m^2 \phi^2
\end{align}
The counterterms and beta functions in this theory can be conveniently found by truncating those found above. Of course, by construction, we find that the heavy fermion $\psi$ no longer contributes to the running of parameters at low-energies. To make sure we're getting the physics correct, we must impose the boundary condition that the predictions match at $\mu = M_\text{phys}$, which here is quite simple---we just use the values at $M_\text{phys}$ in the UV theory as literal boundary conditions for our running in the IR theory. In the IR theory, for $\mu \leq M_\text{phys}$, we have
\begin{align}
\frac{d\lambda}{d\log \mu} &= \frac{3}{16\pi^2} \lambda^2 \\
\Rightarrow  \lambda(\mu) &= \frac{\lambda(M_\text{phys})}{1 + \frac{3}{16 \pi^2} \lambda(M_\text{phys}) \log \frac{M_\text{phys}}{\mu}} \\
&\simeq \bar \lambda \left(1 - \frac{1}{16 \pi^2}\left[3\bar \lambda^2 + 8 \bar \lambda \bar g^2 - 48 \bar g^4\right]\log \frac{\bar \mu}{\bar M}\right) \left(1 - \frac{3}{16 \pi^2} \bar \lambda \log \frac{\bar M}{\mu}\right) + \dots\\
&\simeq \bar \lambda \left(1 - \frac{1}{16 \pi^2}\left[3\bar \lambda^2 + 8 \bar \lambda \bar g^2 - 48 \bar g^4\right]\log \frac{\bar \mu}{\bar M} - \frac{3}{16 \pi^2} \bar \lambda \log \frac{\bar M}{\mu}\right)+ \dots
\end{align}
The benefit is now clear. While the RGEs in the UV theory were very complicated, the running of $\lambda$ in the low energy theory is simple. Our mass-independent scheme allows us to explicitly factorize these and contain all the UV physics in the boundary condition, which lets us study the low-energy theory in a simple manner.

\begin{wrapfigure}{r}{0.3\textwidth}
	\vspace{-20pt}
	\begin{center}
		\includegraphics[width=0.3\textwidth]{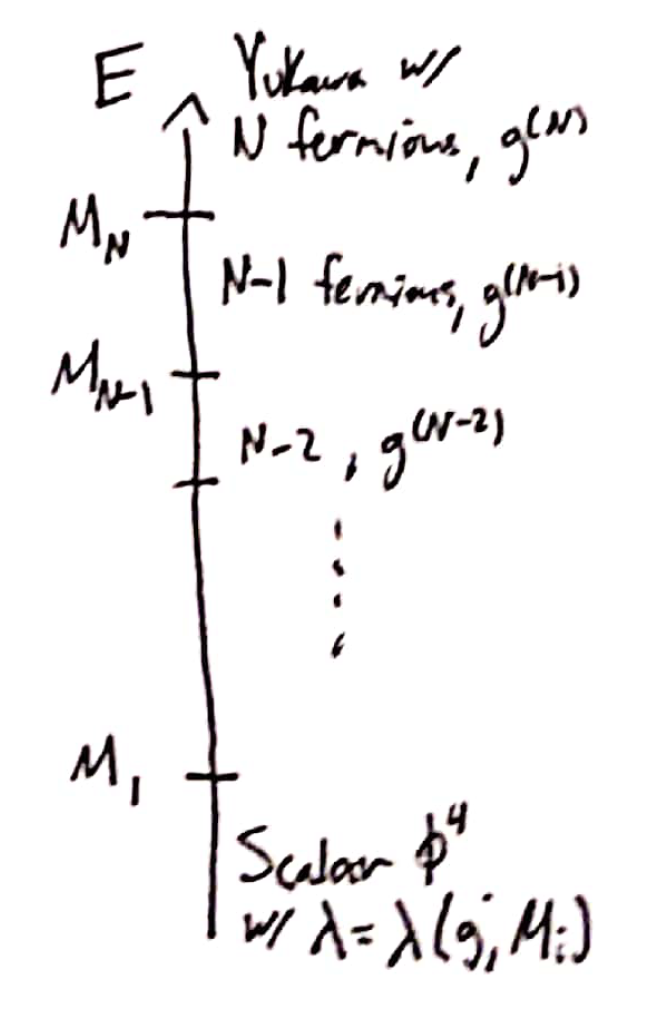}
	\end{center}
	\vspace{-20pt}
	\caption{\ssp A schematic depiction of the cascade of effective field theories as one transitions from a Yukawa theory of $N$ fermions at high energies through integrating these out sequentially until one finds an effective $\phi^4$ theory at low energies.}
	\vspace{-10pt}
	\label{fig:yuk_eft}
\end{wrapfigure}

The general procedure of renormalization group evolution in mass-independent schemes is called `running and matching'. The parameters in the Lagrangian run as you evolve down in energies, but at a mass threshold $M$ we must match the UV theory at $\mu = M$ from above to a theory without this field at $\mu = M$ from below. When we match we ensure that the physics of the low-energy fields stays constant as we cross that threshold and remove that particle from the spectrum of our theory. This becomes less trivial when we have multiple mass scales, so consider now upgrading our Yukawa theory with additional fermions.
\begin{equation}
\mathcal{L} \supset \sum\limits_{i=1}^N M_i \bar \psi_i \psi_i + g^{(N)} \phi \sum\limits_{i=1}^N \bar \psi_i \gamma_5 \psi_i
\end{equation}
Now if we are given the theory at very high energies $\mu \gg M_i$ and we want to understand what it looks like at very low energies, there is a cascade of EFTs we evolve through. As depicted in Figure \ref{fig:yuk_eft}, we run the parameters down to the largest fermion mass, match to a theory with one less fermion, run down again until the next mass scale, and so on.

As an example which is closer to the real world, consider QED with our three generations of leptons (and ignoring the strong sector for simplicity). At low energies we measure the asymptotic value $\alpha(\mu \simeq 0) \simeq 1/137$, and in colliders we measure the value of the gauge coupling at the $Z$ mass $m_Z \simeq 91 \text{ GeV}$. To compare these, we must run the high-energy value all the way down into the infrared. Above $m_\tau \simeq 1.7 \text{ GeV}$ we have a theory where all of $e, \mu, \tau$ run in loops and giving an $\overline{\text{MS}}$ beta function $\beta_\alpha = 2 \alpha^2 / \pi$. But at $m_\tau$ we should remove the $\tau$ from our theory, such that from $m_\tau$ down to $m_\mu \simeq 105 \text{ MeV}$ we have $\beta_\alpha = 4 \alpha^2 / 3 \pi$. Below the $\mu$ we solely have the electron and recover the textbook $\beta_\alpha = 2 \alpha^2 / 3 \pi$, and finally as we cross the electron mass threshold $m_e \simeq 511 \text{ keV}$ we remove the electron from the spectrum and find that the gauge coupling stops running $\beta_\alpha \equiv 0$. Physically this corresponds to the fact that pure QED is scale-invariant, meaning that the coupling will not evolve at all in a theory with no charged particles. This is the regime in which classical electrodynamics holds precisely (up to the presence of additional interactions suppressed by powers of $m_e$, that is). 

A possible confusion is to conflate the mass-independence of the regularization scheme with that of the renormalization scheme, and conclude that dimensional regularization cannot be used if one wants decoupling without having to integrate out and match. So lest one confuse the roles let's quickly look at an example of using dimensional regularization with a renormalization scheme which \textit{does} satisfy decoupling, known as `off-shell momentum subtraction'. For simplicity, we'll look at the anomalous dimension of our Yukawa scalar $\phi$, and we'll perform wavefunction renormalization by subtracting the value of the graphs at the off-shell momentum scale $k^2 = \mu^2_E$. In symbols this amount to the prescription
\begin{equation}
Z_\phi = 1 + \Sigma_{\text{loop}}(k^2 = \mu_E^2), \quad \text{ where } \Pi_{\text{loop}}(k^2)=k^2 \Sigma_{\text{loop}}(k^2) + \text{ mass renorm pieces}.
\end{equation}
Since we're still using the same regularization scheme, we have the same result for $\Pi_{\text{loop}}(k^2)$ as above. We can then simply calculate the anomalous scaling dimension as defined by $\gamma_\phi \equiv \half \frac{\text{d}\log Z_\phi}{\text{d}\log \mu_E}$,
\begin{align}
\gamma_\phi &= \half \frac{\text{d} \Sigma_{\text{loop}}(\mu_E^2)}{\text{d} \log \mu_E} \\
&= \frac{3g^2}{4 \pi^2} \int\limits_0^1 \text{d}x \frac{x^2 (1-x)^2 \mu_E^2}{M^2 + (1-x) x \mu_E^2}.
\end{align}
This integral can be performed analytically, but the full expression is unilluminating. However, it is useful to look at the limits
\begin{align*} 
\gamma_\phi = \begin{cases}
\frac{g^2}{8 \pi^2}, &\mu_E^2 \gg M^2 \\
\frac{g^2}{40 \pi^2} \frac{\mu_E^2}{M^2}, &\mu_E^2 \ll M^2
\end{cases}
\end{align*}
to check if they agree with our expectations. At energies far above the fermion mass its contribution to the scalar anomalous dimension cannot know about that scale, and at energies far below its mass we expect inverse dependence on the mass for decoupling to occur. This is precisely what we find, so the lesson is that even if we didn't want to go through the trouble of integrating out the fermion and matching, we could still make use of the magical regularization scheme that is dim reg.

\subsubsection*{Flowing in theory space}\addcontentsline{toc}{subsubsection}{\qquad Flowing in Theory Space}

Our interpretation of the renormalization group thus far has been as a way of understanding what a particular theory looks like at different energies. But there is another way of looking at it that is also useful, for which we shall follow an example of Peskin \& Schroeder, though I recommend Skinner's lecture notes \cite{skinner} for clear explanation of these concepts which goes farther than we have time to. Let's return to the idea of the Wilsonian path integral and successively integrating out Euclidean momentum shells. In the previous section we began with a scalar field theory
\begin{equation}
Z =\left(\int {\displaystyle \prod_{k=0}^{\Lambda}} \text{d}\phi(k)\right) \exp\left[- \int \text{d}^dx \left(\half (\partial_\mu \phi)^2 + \half m^2 \phi^2 + \frac{1}{4!} \lambda \phi^4\right)\right].
\end{equation}
We then integrated over momentum shells from $\Lambda$ down to $b \Lambda$ with $0 < b < \Lambda$, and found we could express our result as (schematically; see \ref{eqn:lambdarun},\ref{eqn:zrun},\ref{eqn:massrun},\ref{eqn:irrelrun})
\begin{align}
Z &=\left(\int {\displaystyle \prod_{k=0}^{b\Lambda}} \text{d}\phi(k)\right) \exp\left[- \int \text{d}^dx \mathcal{L}_{\text{eff}} \right] \\
\mathcal{L}_{\text{eff}} &=\left(\half (1 + \Delta Z)(\partial_\mu \phi)^2 + \half (m^2+\Delta m^2) \phi^2 + \frac{1}{4!} (\lambda+\Delta \lambda) \phi^4 + \frac{1}{6!} \Delta c_6 \phi^6 + \dots\right).
\end{align}
Above we interpreted this in terms of looking at the same theory at lower energies, having coarse-grained over the largest momentum modes, which is a useful way of comparing the two path integrals. Another useful way to compare is to get them to a form where they look similar, so let's now define a change of variables $k' = k/b, x' = x b$, in terms of which the path integral now looks like 
\begin{align}
Z &=\left(\int {\displaystyle \prod_{k'=0}^{\Lambda}} \text{d}\phi(k')\right) \exp\left[- \int \text{d}^dx' \mathcal{L}_{\text{eff}} \right] \\
\mathcal{L}_{\text{eff}} &= b^{-d} \left(\half (1 + \Delta Z)b^2 (\partial_{\mu'} \phi)^2 + \half (m^2+\Delta m^2) \phi^2 + \frac{1}{4!} (\lambda+\Delta \Lambda) \phi^4 + \frac{1}{6!} \Delta c_6 \phi^6 + \dots\right).
\end{align}
We can transform the kinetic terms back to the canonical form with the field redefinition $\phi'(x') = b^{(2-d)/2} (1 + \Delta Z)^{1/2} \phi(x')$, after which we can write the effective Lagrangian as 
\begin{equation}
\mathcal{L}_{\text{eff}} = \half (\partial_{\mu'} \phi')^2 + \half m'^2 \phi'^2 + \frac{1}{4!} \lambda' \phi'^4 + \frac{1}{6!} c'_6 \phi'^6 + \dots.
\end{equation}
with $m'^2 = (m^2 + \Delta m^2)(1 + \Delta Z)^{-1} b^2, \ \lambda' = (\lambda+\Delta \lambda)(1 + \Delta Z)^{-2} b^{d-4}, \ c_6' = \Delta c_6 (1 + \Delta Z)^{-3} b^{2d-6}, \dots$, and it's clear that we could write such an effective action regardless of what sort of coefficients we began with before integrating out this momentum shell.

Now our series of transformations has effected the change (up to normalization)
\begin{equation}
Z =\left(\int {\displaystyle \prod_{k=0}^{\Lambda}} \text{d}\phi(k)\right) \exp\left[- \int \text{d}^dx \mathcal{L} \right] \rightarrow Z =\left(\int {\displaystyle \prod_{k'=0}^{\Lambda}} \text{d}\phi'(k')\right) \exp\left[- \int \text{d}^dx' \mathcal{L}' \right].
\end{equation}
Since all of our dynamical variables are integrated over in calculating the partition function, we can view this as a transition in the space of Lagrangians, $\mathcal{L} \rightarrow \mathcal{L}'$. So this gives us an interpretation of the renormalization group as a flow in `theory space'.

\begin{figure}
	\centering
	\includegraphics[width=0.5\linewidth]{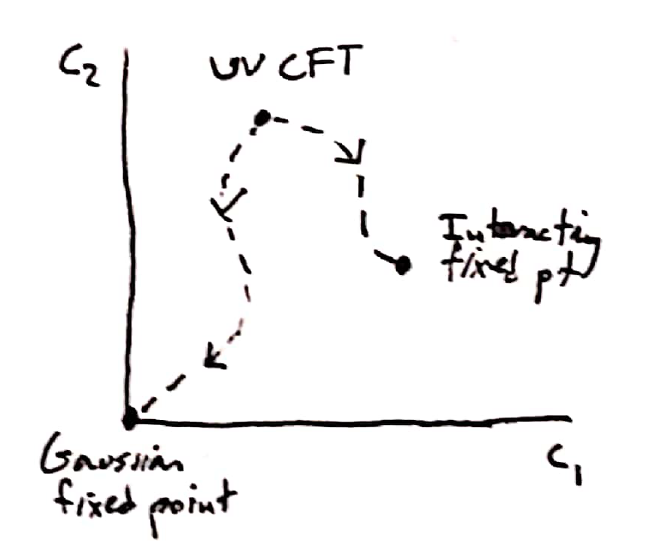}
	\caption{Schematic two-dimensional projection of some RG flow trajectories from an interacting UV fixed point which has been perturbed by one or another relevant operator, and which flow to IR fixed points that are either interacting or free. Perturbing the UV CFT by an irrelevant operator would lead to a flow directly back to the same CFT.}
	\label{fig:rgflow}
\end{figure}

This interpretation invites us to conceptualize renormalization group flow as a path through theory space between two conformal field theories (CFTs), as depicted in Figure \ref{fig:rgflow}. CFTs are quantum field theories with an enlarged spacetime symmetry group\footnote{This statement may appear confusing if you have come across the Coleman-Mandula theorem \cite{Coleman:1967ad}, which roughly says that the most amount of symmetry you can have in a QFT is the direct product of the Poincar\'{e} spacetime symmetry and whatever internal symmetries you have. However, that beautiful result relies on properties of the S-matrix, and CFTs do not have S-matrices because they do not have mass gaps, meaning this enlarged spacetime symmetry group does not violate the theorem. We'll see another, even more interesting loophole in this theorem exploited in Section \ref{sec:SUSY}, which comes from enlarging our notion of what sorts of symmetries an S-matrix could possess.}, consisting essentially of a scaling symmetry. CFTs are fixed points of RG flows---since they possess scaling symmetry they look the same at all energy scales, so if an RG flow is to have an endpoint it clearly must be a CFT\footnote{I've elided a subtlety here, which is that it is not entirely known whether scale invariance in fact implies conformality in four-dimensional QFTs, the latter of which includes also invariance roughly under inversion of spacetime through a point. No counterexamples have been found, despite much effort. Polchinski's early paper on the topic is a classic \cite{Polchinski:1987dy}, and a recent review can be found from Nakayama \cite{Nakayama:2013is}.}.

There's a terminological confusion here, which is that the `renormalization group' isn't actually a group at all, since the operation of integrating out a momentum shell is irreversible. This came up already above when we saw that integrating out heavy fields means we can no longer compute processes which have them as external fields (see Footnote \ref{foot:source}). Flowing to lower energies, or toward the IR, is really a coarse-graining operation which does lose information about small scales, in precise analogy to decreasing the resolution of an image. This means that RG evolution is a directed flow, so there is a difference between fixed points in the UV and in the IR.

Quantum field theories can have different sorts of fixed points. As a familiar example, if the theory has a `mass gap'---no zero-energy excitations in the infrared, which may be because one began solely with massive fields or through dynamical mechanisms like Higgsing and confinement---then one finds a `trivial' fixed point. In the far infrared, everything has been integrated out and there is not enough energy to excite any modes. We know phenomenologically that this happens in QCD. In the other familiar case one can have a `Gaussian' or `free' fixed point if the theory contains massless fields which don't interact, such as in QED. At energies far below the mass of the lightest charged particle this is a theory of free electromagnetism, though one can still excite photons of arbitrarily-long wavelength. 

Such a Gaussian fixed point occurs in the UV for QCD---the celebrated result of `asymptotic freedom' \cite{Gross:1973id,Politzer:1973fx}---because the strong coupling flows toward zero, giving a free theory. This famously cannot occur for $U(1)$ gauge theories whose couplings necessarily grow with increasing energies, leading them to herald their own breakdown with the prediction of a `Landau pole' \cite{Landau}, a finite UV energy where the perturbative theory predicts the coupling becomes infinite. From one perspective this is an inverse to the prediction of a confinement scale in QCD, where the perturbative prediction is a blowup of the coupling at low energies, as we'll discuss further in Section \ref{sec:dimtrans}. In either case the theory cannot make predictions for energies above the Landau pole or below the confinement scale, respectively, and so can be called inconsistent.

It's clear that the divergence of a coupling either in the IR or the UV is problematic for a complete, consistent interpretation of a QFT. This is precisely why the formal perspective on a well-defined QFT is that it describes an RG flow between two CFTs, such that in neither direction does a coupling grow uncontrollably. In fact this provides an incredibly important perspective on renormalizability, which we'll get to momentarily.

First, let us introduce somewhat of a generalization of the `relevant/irrelevant' terminology which we introduced in Section \ref{sec:EFT}. We implicitly had in mind that we were studying a theory in the vicinity of the Gaussian fixed point which we perturbed with various field operators---indeed, this is precisely how we normally carry out perturbative calculations---and our terminology depended on that. Generally one wishes to imagine perturbing a CFT by a particular operator, flowing down in energy, and seeing whether the operator grows in importance---a relevant operator---or shrinks---an irrelevant operator. Perturbing a CFT by an irrelevant operator does not induce an RG flow (to a different CFT), so interesting dynamical RG flows come from CFTs perturbed by relevant operators. This clearly agrees with our power-counting notion of relevance when we're near the Gaussian fixed point, but works also if one is near a strongly-coupled, interacting fixed point where one may not know how to do such power-counting and anomalous dimensions of operators may grow to overpower their classical dimensions. 

The importance of this language was realized in particular by Polchinski in his pioneering article \cite{Polchinski:1983gv}. He showed that in fact the intuitive notion of `power-counting renormalizability' that the field had been building---that for a theory near the Gaussian fixed point we could see whether it was renormalizable merely by checking whether it has any coefficients of negative mass dimensions---in fact maps on to a very general statement. This is enormously powerful, as prior arguments for renormalizability were made on a case by case basis and were complicated and messy and graph-theoretic. His derivation of this fact is brilliant but requires much work, so we'll merely try to get a sense for why it should be true by building on our intuition from our Wilsonian renormalization of $\phi^4$ theory in Section \ref{sec:removediv}.

So long as your theory has a finite number of fields, all of which have mass dimension $\left[\phi_i\right] >0$ when canonically normalized, then there are a finite number of relevant or marginal operators. As we flow down in energy through theory space, we saw above the sense in which those coefficients are UV-sensitive---the IR coefficients of those operators are determined primarily by their UV values and IR corrections are subdominant. Contrariwise, the coefficients of the infinite number of possible irrelevant operators are UV-insensitive, being determined primarily by the IR physics of the relevant and marginal operators. That is, the RG flows are attracted to a finite-dimensional submanifold of the space of Lagrangians. 

Consider a Wilsonian RG flow where we start off, as above, by specifying a cutoff $\Lambda_0$ and the values of coefficients $\lambda_0^i$ of all the $n$ marginal and relevant operators at that scale, as well as the coefficients $c_0^i$ of however many irrelevant operators we wish to turn on. We can then flow downwards in energy as normal, and at an energy scale $\Lambda_R \ll \Lambda_0$ let's say we measure the relevant and marginal coefficients $\lambda_R^i$ at that scale. The coefficients of the irrelevant operators are then dominated by the infrared $\lambda_R^i$ and $\Lambda_R$ up to precision $(\Lambda_R/\Lambda_0)^{\Delta_i}$ from subleading corrections, with $\Delta_i > 0$ the scaling dimension of the irrelevant operator. So indeed, the RG flow is attracted to the $n$-dimensional surface described by $c_i = c_R^i(\lambda_R^i; \Lambda_R)$ and separate trajectories through theory space as a function of scale $\Lambda$ which reach the same $\lambda_R^i$ will differ only by positive powers of $(\Lambda_R/\Lambda_0)$. For less abstract discussion, Polchinski goes through a simple example which may provide further insight, and Schwartz discusses the same example in Chapter 23 of his textbook \cite{Schwartz:2013pla}.

To see why this implies renormalizability, recall that the program of Wilsonian renormalization is to define renormalized, `running' couplings as a function of scale to keep infrared physics at $\Lambda_R$ fixed while `removing the cutoff'. A bit more formally, we want a family of Lagrangians $\mathcal{L}(\Lambda_0;\lambda_0^i, c_0^i)$ with coefficients chosen as a function of $\Lambda_0$ such that each Lagrangian yields the same low energy physics $\lambda_R^i$, in terms of which all the IR observables can be calculated up to subleading corrections in $\Lambda_R/\Lambda_0$. When the cutoff is removed by taking $\Lambda_0 \rightarrow \infty$, one thus recovers precisely the correct physics, specified by those chosen values of the $\lambda_R^i$, which are the renormalization conditions. If we can find such a family of Lagrangians, then we say this theory is power counting renormalizable. Polchinski's argument shows that this can be done so long as one wishes solely to fix the IR values of relevant and marginal couplings. 

\subsubsection*{Trivialities}\addcontentsline{toc}{subsubsection}{\qquad Trivialities}

The attentive reader may at this point notice an inconsistency due to imprecise language. We've seen now that the criterion for renormalizability, which Polchinski provided a  robust basis for, is power-counting of the operators near the IR fixed point. This would suggest that the $\lambda \phi^4$ theory we've studied by means of an example is renormalizable. However, recall the result of resumming its renormalization group equation:
\begin{equation}
\lambda(\Lambda) = \frac{\lambda(\Lambda_0)}{1 + \frac{3 \lambda(\Lambda_0)}{16 \pi^2} \log \frac{\Lambda_0}{\Lambda}},
\end{equation}
which has a Landau pole for $\Lambda = \Lambda_0 \exp(\frac{16 \pi^2}{3 \lambda(\Lambda_0)})$, preventing us from taking the limit we required above. A mathematical physicist would say $\lambda \phi^4$ theory is `trivial' or `quantum trivial', as if we demand the existence of a continuum limit, that sets $\lambda(\Lambda) = \lambda(\Lambda_0) = 0$.

The issue is that the tree-level, classical scaling dimension captures only the scaling of the operators infinitesimally close to the infrared Gaussian fixed point. If we move a finite distance upward in energy scale, we've seen above that the $\phi^4$ operator gets an anomalous dimension $\delta_{\phi^4}>0$ and so is marginally irrelevant. So it's clear that Polchinski's picture of renormalization is only getting at a perturbative sense of renormalizability, and cannot tell us whether there truly exists an RG flow from a UV fixed point down to the IR theory we want to study.

So what are we to make of $\lambda \phi^4$ theory---or for that matter of QED, which has the same problem? Of course we know empirically that QED works fantastically well and we can absolutely make finite, accurate predictions after a finite number of inputs. To understand this, we must appeal the language of effective field theory, which we've already discussed. In fact the feature we're really relying on is effective renormalizability, which tells us we require solely a finite number of inputs to set the behavior of the theory to a given precision. It's clear that in this sense QED itself is an effective field theory whose validity breaks down somewhere below its Landau pole.

Finally, let me mention another reason not to be too worried that our most beloved quantum field theories don't exist in the continuum limit: Our universe is not described by a QFT at its smallest scales! It's indeed true that only RG flows between a UV CFT and an IR CFT can hope to define fully consistent and mathematically well-defined QFTs. But the existence of gravity---and the very strong evidence that a quantum field theory of gravity is inconsistent---means that at some energy scale effects not present in quantum field theory must become relevant. And since gravity couples universally to everything \cite{vonMeyenn1990, Weinberg:1964ew}, we have no strict empirical need for a UV complete, interacting quantum field theory that does not include gravity. It is entirely consistent, and overwhelmingly likely, that a quantum-field-theoretic description of the world works only approximately and some inherently quantum gravitational theory provides a sensible UV complete theory.

\subsection{To Reiterate}

Before moving on, let us reiterate what we've discussed about renormalization. As we've seen, renormalization is \textit{so} important and \textit{so} useful and fulfills \textit{so many} purposes that an entirely general statement risks becoming vague. But if a single sentence summary is demanded: Renormalization reveals for us the scale-dependence of a quantum mechanical field theory.

The effects of this seemingly innocuous statement, however, are powerful and manifold, including:
\begin{itemize}
	\item Correctly accounting for this scale-dependence is necessary to have well-defined quantum field theories, which otherwise appear nonpredictive.
	\item Bringing the non-scale-invariance of the quantum mechanical theory into clear scale-dependence of the couplings makes it simple to read off the qualitative behavior at different scales from the renormalized Lagrangian.
	\item Including this scale-dependence in the couplings allows us to reorganize our perturbative series such that we can efficiently calculate the behavior of the theory over a far wider range of energies than a na\"{i}ve treatment allows.
	\item Properly accounting for the scale dependence allows us to harness the full power of effective field theory, as we can study a theory of low-energy fields which correctly accounts for corrections from the high-energy physics.
	\item Understanding the perspective of single quantum field theories as flows through theory space as a function of the scale allowed us to develop a nonperturbative definition of a fully UV-complete quantum field theory and how it behaves.
\end{itemize}

All of these various perspectives will be of use in the following chapters as we apply this technology to understanding what the hierarchy problem is and how we can solve it.

\section{Naturalness} \label{sec:Naturalness}

Naturalness is the notion that we have the right to ask about the origins of the dimensionless numbers in our theories---past solely fitting them to the data. It was Dirac who first introduced such a notion to particle physics in 1938 \cite{Dirac:1938mt}. In modern language, what is referred to as `Dirac naturalness' consists of the idea that dimensionless parameters in a fundamental physical theory should take values of order 1. In the language of EFT, in a theory with a cutoff $\Lambda$ and an operator $O$ with scaling dimension $\Delta$, we expect its coefficient to take a value $c_O \sim \mathcal{O}(1) \Lambda^{d-\Delta}$. As stated this is essentially dimensional analysis, as $\Lambda$ is the only scale we've introduced, but we will discuss below that quantum mechanics gives additional credence to this expectation---indeed we've already seen this feature in our one-loop examples above.

't Hooft pointed out a refinement of this principle \cite{tHooft:1979rat}, which has come to be called `technical naturalness'. If the operator $O$ breaks a symmetry which is respected by the action in the limit $c_O \rightarrow 0$, then one says it is `technically natural' for $c_O$ to take on a small value. The reasoning here is simple---as we saw above, in a quantum field theory defined at a high scale, one finds corrections $\delta c_O$ to such coupling constants as they run to low energies. If there is an enhanced symmetry of the theory in the limit that $c_O \rightarrow 0$, then such quantum mechanical corrections cannot generate that operator and break the symmetry, so we know that $\delta c_O \propto c_O$. The low-energy effective field theorist says of such couplings that one can `set it and forget it': if one has $c_O \ll 1$ at the cutoff $\Lambda$, that coupling will remain small as one flows to lower energies. 

Conversely, we can emphasize the connection to Dirac naturalness by looking at this picture in reverse. We know of mechanisms to generate small technically natural couplings at low energies from Dirac natural ones, as we will discuss in detail below. Imagine one measures a small coupling $c_O$ at long distances in the low-energy theory with cutoff $\Lambda$ that does not have a Dirac natural explanation. If that parameter is technically natural, it remains small up to the cutoff, and so the next generation of physicists can explain its small size at $\Lambda$ in the UV theory, as emphasized nicely by Zhou \cite{zhou}. If $c_O$ is not technically natural, then its RG evolution up to the cutoff yields a value to which the low-energy physics is very sensitive, and we must explain why it has a very specific value such that the correct physics emerges at long distances.

\subsection{Technical Naturalness and Fine-Tuning}
\epigraph{A model is fine-tuned if a plot of the allowed parameter space makes you wanna puke.}{David E. Kaplan (2007)}

It's useful to make this less abstract by looking at a simple example. Consider a $d=6$ dimensional effective field theory of a real scalar field $\phi$ of mass $m$ which is odd under a $\mathbb{Z}_2$ symmetry, which we expect is a good description of our system up to a cutoff $\Lambda$ with  $m \ll \Lambda$. If we add a small explicit breaking $\sigma \phi^3$ with $\sigma \ll 1$ at low energies, $\sigma$ is technically natural and stays small up to the cutoff, so we can easily write down a UV completion which generates this small value Dirac naturally. 

However, if we add another $\mathbb{Z}_2$-odd real scalar $\Phi$ and give it a large $\mathbb{Z}_2$-breaking interaction with $\phi$, then $\sigma$ is no longer technically natural. Its low-energy value becomes extremely sensitive to the values of the parameters at the cutoff. It then becomes difficult to understand an ultraviolet reason for why those values take the precise values they need to realize small $\sigma$ in the far IR. Consider the bare action
\begin{equation}
S = \int \text{d}^6x \left[-\half (\partial \phi_0)^2 - \half m_0^2 \phi_0^2 - \half (\partial \Phi_0)^2- \half m_0^2 \Phi_0^2 - \frac{\sigma_0}{3!} \phi_0^3  - \frac{y_0}{2} \phi_0 \Phi_0^2\right],
\end{equation}
where we've given the two fields the same mass for simplicity. This is not stable under radiative corrections, but that's a higher-order effect which will not come into our one-loop calculation of the RG evolution of the cubic couplings.

We will renormalize this theory at one loop using dim reg with $\overline{\text{MS}}$. As discussed above, we will compute the one-loop 1PI diagrams and add counterterms to cancel solely the $\frac{1}{\epsilon}$ pieces of the results. With counterterms, the action is 
\begin{equation}
S = \int \text{d}^dx \left[-\half Z_\phi (\partial \phi)^2 - \half Z_{m\phi} m^2 \phi^2 - \half Z_\Phi (\partial \Phi)^2- \half Z_{m\Phi} m^2 \Phi^2 + \frac{\sigma}{3!} Z_\sigma \phi^3  + \frac{y}{2} Z_y \phi \Phi^2\right]
\end{equation}
where these parameters and fields are the renormalized parameters, and for compactness we have not written down the split of these terms as we did above in Equation \ref{eqn:yukctsplit}. At tree level the relation to the bare quantities is trivial and so $Z = 1 + \dots$. To get an accurate picture of how the strength of the interactions vary as they're probed at different energy scales, we must fully renormalize the theory. Since our focus is on the interactions, we simply state the results for the quadratic part of the action, where we have
\begin{align}
Z_\phi &= 1 - \frac{1}{6(4\pi)^3}\left(\sigma^2 + y^2\right)\frac{1}{\epsilon} + \dots \\
Z_\Phi &= 1 - \frac{1}{3(4\pi)^3}\left(\sigma^2 + y^2\right)\frac{1}{\epsilon} + \dots \\
Z_{m\phi} &= 1 - \frac{1}{(4\pi)^3}\left(\sigma^2 + y^2\right)\frac{1}{\epsilon} + \dots \\
Z_{m\Phi} &= 1 - \frac{2}{(4\pi)^3} y^2\frac{1}{\epsilon} + \dots 
\end{align}
These tell us that the physical mass of the fields and the normalization of their one-particle states has changed. The relation to these can be found using the quantum-corrected propagator $\Delta(k^2)^{-1}$ as $\Delta(-m_\text{phys}^2)^{-1}\equiv 0$ to define the mass and $R^{-1} = \left.\frac{d}{dk^2}\left[\Delta(k^2)^{-1}\right]\right|_{k^2 = - m_\text{phys}^2}$ to define the normalization $R$, but solving for these relations explicitly will not be necessary for our purposes.

\begin{figure}
	\centering
	\begin{subfigure}{.4\textwidth}
		\centering
		\includegraphics[width=\linewidth]{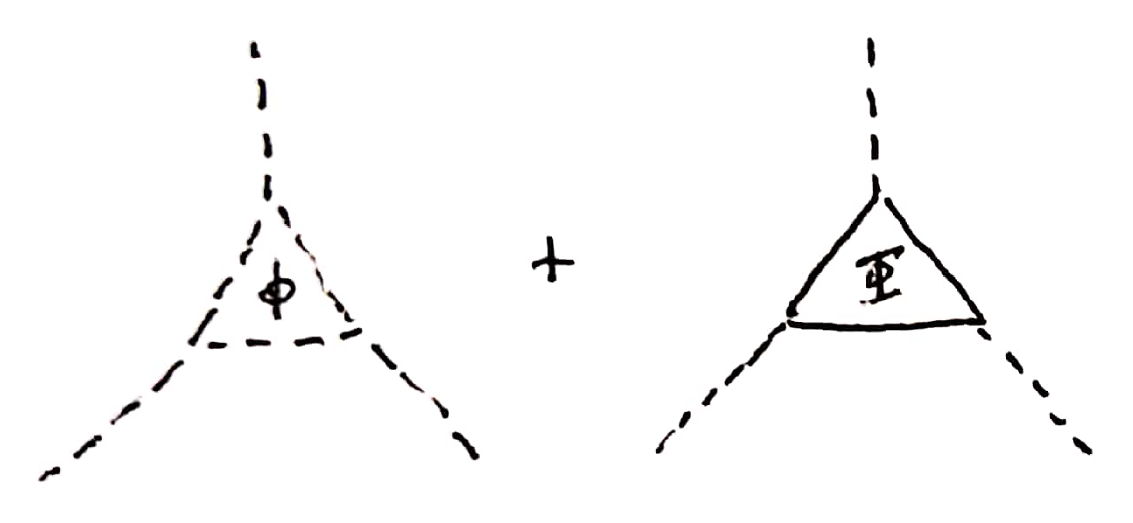}
		\caption{}
		\label{fig:technatcubic1}
	\end{subfigure}%
	\begin{subfigure}{.4\textwidth}
		\centering
		\includegraphics[width=\linewidth]{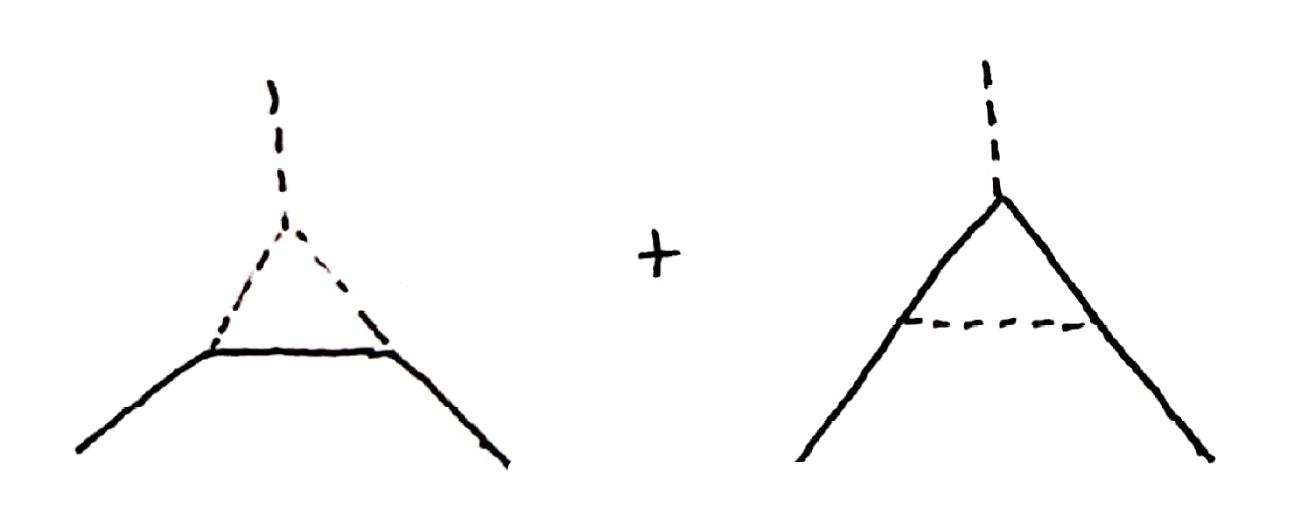}
		\caption{}
		\label{fig:technatcubic2}
	\end{subfigure}
	\caption{Diagrams contributing to the one-loop corrections to the three-point functions in our $6d$ scalars with cubic interactions. Dashed lines denote $\phi$ and full lines denote $\Phi$. In (a), the diagrams renormalizing $\sigma$, and in (b), the diagrams renormalizing $y$.}
	\label{fig:technatcubic}
\end{figure}

The one-loop three-point functions each have two diagrams, whose evaluation only differs in the coupling constants. For the correction to $\sigma$, we have triangle diagrams with either $\phi$ or $\Phi$ running in the loop. We can evaluate them in $d=6-\epsilon$ dimensions as
\begin{align}
i \frac{1}{\sigma} V_\sigma &= \frac{1}{\sigma}\left[(i\sigma)^3 + (i y)^3\right]\int \frac{d^6q}{(2\pi)^6} \frac{(-i)^3}{(q^2 + m^2)^3} \\
&= \frac{i}{\sigma}\left[\sigma^3 + y^3\right]\tilde{\mu}^\epsilon \int \frac{d^d\bar q}{(2\pi)^d} \frac{1}{(\bar q^2 + m^2)^3} \\
&= \frac{i}{\sigma}\left[\sigma^3 + y^3\right]\tilde{\mu}^\epsilon \frac{\Gamma\left(\frac{\epsilon}{2}\right)}{2(4\pi)^3}\left(\frac{m^2}{4\pi}\right)^{-\frac{\epsilon}{2}} \\
&= \frac{i}{2(4\pi)^3 \sigma}\left[\sigma^3 + y^3\right] \left(\frac{2}{\epsilon}-\gamma_E\right) \left(1 + \frac{\epsilon}{2} \log \frac{4 \pi \tilde{\mu}^2}{m^2} \right) \\
&= \frac{i}{2(4\pi)^3 \sigma}\left[\sigma^3 + y^3\right] \left(\frac{2}{\epsilon}+ \log \frac{\mu^2}{m^2}\right).
\end{align}
The counterterm vertex contributes to this as $-i(Z_\sigma - 1)$, meaning that $\overline{\text{MS}}$ prescribes we set
\begin{equation}
Z_\sigma = 1 + \frac{1}{(4\pi)^3}\left(\sigma^2 + \frac{y^3}{\sigma}\right) \frac{1}{\epsilon}
\end{equation}
For the other vertex correction there are diagrams with either one or two of each internal line, which give
\begin{align}
i \frac{1}{y} V_y &= \frac{1}{y}\left[(iy)^3 + (i\sigma)(iy)^2\right]\int \frac{d^6q}{(2\pi)^6} \frac{(-i)^3}{(q^2 + m^2)^3} \\
&= \frac{i}{2(4\pi)^3}\left[y^2 + \sigma y\right] \left(\frac{2}{\epsilon} + \log \frac{\mu^2}{m^2}\right), 
\end{align}
leading to the counterterm
\begin{equation}
Z_y = 1 + \frac{1}{(4\pi)^3}\left(y^2 + \sigma y \right) \frac{1}{\epsilon}.
\end{equation}
This gives us the beta functions 
\begin{align}
\beta_\sigma &= \frac{1}{4(4\pi)^3} \left(-3 \sigma^3 - 4 y^3 + y^2 \sigma\right) + \dots \\
\beta_y &= \frac{1}{12(4\pi)^3} \left(\sigma^2 y - 12 \sigma y^2 - 7 y^3\right) + \dots
\end{align}
Now we are finally in a place to mathematically evince our physics point about technical naturalness. Without $\Phi$, the coupling $\sigma$ is the only one which violates the $\mathbb{Z}_2$ and so the beta function is necessarily proportional to $\sigma$. \footnote{This is trivial in our case as $\sigma$ would be the only interaction, but you'll find the feature persists if you add other symmetry-respecting interactions, for example a $\mathbb{Z}_2$-even scalar $\psi$ with an interaction $\phi^2 \psi$. The general argument for this fact is given in the next section.} Let's say we recruit an experimentalist friend of ours to measure the $2 \rightarrow 2$ scattering cross-section of $\phi$s and we find that at $\mu = m$, the theory fits the data for $\sigma(m) = \sigma_0$ with $\sigma_0 \ll 1$. Solving the beta function, we find that to lowest order
\begin{equation}
\sigma(\mu) \simeq \sigma_0 \left(1 - \frac{3}{4 (4\pi)^3} \sigma_0^2 \log\left[\frac{\mu}{m}\right]\right).
\end{equation}
So $\sigma$ is indeed radiatively stable. If $\sigma(m) = \sigma_0$ is small, then it takes until the enormous scale $\mu \simeq m \exp\left((4\pi)^3/\sigma_0^2\right)$ for $\sigma$ to change by an order one fraction. So running $\sigma(\mu)$ up to wherever the cutoff $\Lambda$ of our theory is, $\sigma(\Lambda)$ will still be small. If by $\Lambda$ we haven't discovered any explanation for the size of $\sigma(m)$, we can ask the theory above $\Lambda$ to produce this small value of $\sigma(\Lambda)$ from Dirac natural parameters at yet higher energies. Perhaps the high-energy Dirac natural value of $\sigma$ has been relaxed toward zero by an axion-type mechanism, or perhaps $\sigma$ is the vev of another $\mathbb{Z}_2$-odd field which spontaneously broke the $\mathbb{Z}_2$ via confinement. We don't need to know a particular mechanism; the fact that $\sigma$ is technically natural means that if we don't find an explanation for its size there is hope yet that our academic descendants will. One says that here $\sigma$ is UV-insensitive as its low-energy value does not depend strongly on the physics at high energies. 

On the other hand, if $\sigma$ is \textit{not} technically natural, we have a much more difficult issue. If we now have the theory with both $\phi$ and $\Phi$, and our experimentalist measures $\sigma(m) = \sigma_0 \ll 1$ and $y(m) = y_0 =\mathcal{O}(1)$, then to lowest order the RG evolution of $\sigma$ will be $\beta_\sigma \simeq - y^3/(4\pi)^3$ leading to 
\begin{equation}
\sigma(\mu) = \sigma_0 - \frac{1}{(4\pi)^3} y_0^3 \log\left[\frac{\mu}{m}\right].
\end{equation}
And we can see the issue, as we are no longer guaranteed that a small $\sigma(m)$ is related to a small $\sigma(\Lambda)$. For concreteness, if $\sigma_0 = 10^{-3}$, $y_0 = -5$ and $\Lambda = 10^5 m$, then (using the full one-loop RG), we have $\sigma(\Lambda) \simeq 0.55$ and $y(\Lambda) \simeq 4.16$. How are we to ensure \textit{these} values at $\Lambda$? We know how to produce small numbers, but not incredibly specific ones.

To see that we do need to produce these values very precisely, let's switch directions and consider the RG evolution down in energy from $\Lambda$ to $m$. In the theory with solely $\phi$, the coupling $\sigma$ runs incredibly slowly, so an $\mathcal{O}(1)$ change in $\sigma(\Lambda)$, evolved down to the scale $m$, results in an $\mathcal{O}(1)$ change to $\sigma(m)$. But in the theory with two sources of breaking, $\sigma(m)$ is \textit{enormously} sensitive to the values of the couplings at $\Lambda$. With the same cutoff $\Lambda = 10^5 m$, if we very slightly change the input value to $\sigma(\Lambda) = 0.56$ and leave $y(\Lambda)$ as above, evolving down now gives us $\sigma(m) \simeq 10^{-2}$---a $<2\%$ change in input parameters has resulted in a $1000\%$ change in our low energy observable! It's even more sensitive to the input value of $y$; a $\sim 1\%$ modification solely to $y(\Lambda) = -4.20$ trickles down to give $\sigma(m) \simeq 2\times 10^{-{2}}$, a $2000\%$ change. In this theory $\sigma$ is now a UV-sensitive parameter, whose low-energy behavior depends strongly on the high-energy physics. To say the least, it seems difficult to find a natural way to achieve the precise values needed to reproduce the observed low-energy physics in this theory. We'll return to this issue at length in Section \ref{sec:philosophy}. 

\subsubsection*{Technical naturalness and masses}\addcontentsline{toc}{subsubsection}{\qquad Technical Naturalness and Masses}
Our understanding of technical naturalness allows us to already see another warning sign of the hierarchy problem. An elementary spin-1 field comes along with a gauge symmetry $A^\mu \rightarrow A^\mu - \partial^\mu \alpha(x)$ which is broken by a mass term $m_A^2 A^\mu A_\mu$. So a mass for a gauge boson is technically natural and one necessarily finds $\delta m_A^2 \propto m_A^2$. Similarly, a massive Dirac fermion $\Psi = (\psi, \chi^\dagger)$ has a $U(1)$ global symmetry under which $\psi\rightarrow e^{i \alpha} \psi, \chi\rightarrow e^{-i \alpha} \chi$. In the $m \rightarrow 0$ limit, the symmetry is enhanced to $U(1)^2$ as arbitrary rephasings of the two Weyl fermions become symmetries, so again $\delta m_\Psi \propto m_\Psi$. But an elementary scalar does not automatically come with any such protective symmetry, and we've already seen in all our examples above that scalar mass corrections indeed get contributions not proportional to the mass itself.  

In fact for discussing the technical naturalness of masses there is an even simpler argument: A massless spin-1 boson has two degrees of freedom and a massive one has three. Quantum corrections cannot generate a degree of freedom \textit{ex nihilo}, so a massless gauge boson must be protected. Similarly a massless chiral fermion has two degrees of freedom, but a massive Dirac fermion has four. So for charged spinors and for vectors, it is simply the representation theory of the Lorentz group that is responsible for the stability of their masses. In either of these cases mass must arise from interactions of the field with a scalar (very broadly defined) as in the Higgs mechanism, which can pair up chiral spinors together and lend vectors another degree of freedom. But a massless scalar and a massive scalar have the same number of degrees of freedom. If we want a scalar mass to be technically natural, it must come from some symmetry past simply the Lorentz group. We'll see some examples of how to arrange this in Chapter \ref{sec:classical}.

\subsection{Spurion Analysis}

An important tool for understanding symmetries and their violation is known as `spurion analysis'. The basic idea is simple: for a theory which respects a symmetry except for the coupling $c$, this coupling parametrizes the breaking of the symmetry and any effects which violate the symmetry are proportional to $c$. More concretely, one assigns such couplings spurious transformation properties under the symmetry such that the action becomes invariant under the symmetry. Physically one can imagine that the observed values of such couplings come from the vacuum expectation values of some heavy fields which are above the cutoff of the theory. This can be viewed as imagining a UV completion where the explicit symmetry breaking in the low-energy effective theory comes microscopically from some spontaneous symmetry breaking, but the value of spurions is not dependent upon a particular realization of the UV completion.

We can quickly see the utility of this by looking at a simple example of a complex scalar field $\phi$ with an interaction which explicitly breaks the $U(1)$ global symmetry $\phi \rightarrow \phi e^{i\alpha}$.
\begin{equation}
S = \int \text{d}^{4}x \left(-\partial_\mu \phi^\dagger \partial^\mu \phi - m^2 \phi^\dag \phi - \frac{1}{3!} \lambda \phi^3 + \ h.c.\right),
\end{equation}
where ``$+ \ h.c."$ denotes the addition of the Hermitian conjugate of the non-Hermitian interaction term. A na\"{i}ve effective field theorist would say that our Lagrangian has no symmetries, and so we have no control and should expect that quantum corrections give us any polynomials in $\phi, \phi^\dag$ at low energies. 

However, we may note that if we assign $\lambda$ a charge of $-3$ such that $\lambda \rightarrow \lambda e^{-3i\alpha}$, then the theory \textit{is} invariant under that $U(1)$ global symmetry. So quantum corrections cannot violate our spurious symmetry, and as a result we know that we can only generate terms like $\lambda^2 \phi^6$, and there are no $\phi^4$ or $\phi^5$ interactions generated at any order in perturbation theory.

We can also usefully apply this to the example of technical naturalness studied in the preceding section. If $\epsilon$ is given the spurious transformation $\mathbb{Z}_2: \epsilon \rightarrow -\epsilon$ then the $\epsilon \phi^3$ term is invariant. Then it's simple to see that no matter what other sorts of interactions we add, so long as they respect the $\mathbb{Z}_2$ symmetry we must have $\delta \epsilon \propto \epsilon$. But having added $y \phi \Phi^2$ we see that this term can also be made invariant with $y \rightarrow -y$, and this allows $\delta \epsilon \propto y$ as well.

An important real-world example where spurion analysis is useful is in understanding the flavor structure of the SM. With all masses turned off, the SM has a large global symmetry group $U(3)^5 = U(3)_Q\times U(3)_u\times U(3)_d\times U(3)_L\times U(3)_e = SU(3)^5 \times U(1)^5$, corresponding to arbitrary unitary reshufflings of the three generations of each fermion representation. These symmetries are explicitly broken by the Yukawa matrices which generate hierarchically different masses for the three generations. 
\begin{equation}
\mathcal{L} \supset - Y_d^{ij}Q_iHd_j - Y_u^{ij} Q_i H^\dagger u_j - Y_e^{ij} L_i H e_j
\end{equation}
where $i,j=1,2,3$ are generation indices, and the Yukawa couplings are matrices in this generation space.

We don't understand why these hierarchies are present, but we can carry out a spurion analysis to see how worried we should be. We see that our theory will be invariant under the full flavor group if we assign the Yukawa matrices the following transformation properties under the various $SU(3)$ symmetry groups
\begin{align}
Y_d &\sim (3, \bar 3, 1) \text{ under } SU(3)_Q \times SU(3)_d \times SU(3)_u \\
Y_u &\sim (3, 1, \bar 3) \text{ under } SU(3)_Q \times SU(3)_d \times SU(3)_u \\
Y_e &\sim (3, \bar 3) \text{ under } SU(3)_L \times SU(3)_e 
\end{align}
Since these are the \textit{only} flavor-violating couplings in the SM and they are all in distinct spurious flavor representations, this tells us the quantum corrections to these matrices must be proportional to the matrices themselves e.g. $\delta Y_e \propto Y_e$. Thus this pattern of Yukawa couplings is stable under RG evolution to higher scales, and we are justified in thinking that the generation of this pattern may take place at large, currently-inaccessible scales. 

This eases our minds about when we need to discover the origin of these flavor hierarchies, but this holds true only as long as these remain the only flavor-violating couplings. Fantastic work in precision flavor measurements and theory has provided lower bounds on the scale at which additional flavor violation can occur. Searches for flavor-violating processes have constrained these violations to take place at scales enormously higher than scales we are able to directly probe at colliders, which poses a puzzle. If there is new physics near the TeV scale, how is it arranged to respect the flavor structure of the SM? A phenomenological approach known as Minimal Flavor Violation \cite{DAmbrosio:2002vsn} demands that \textit{all} flavor violation is proportional to these Yukawa couplings, but no fundamental explanation for this is known. For  recent introductions to flavor in the Standard Model, see e.g. \cite{Grossman:2017thq,Zupan:2019uoi,Gori:2019ybw}.

\subsection{Dimensional Transmutation} \label{sec:dimtrans}

Perhaps the most important example of a Dirac natural field theory generating a small number is that of `dimensional transmutation'. In particular, in quantum chromodynamics (QCD) the theory is `asymptotically free'---meaning that the interaction strength vanishes in the far UV---and the gauge coupling $g$ grows as one goes to lower energies. We skip the Nobel-worthy calculation (see e.g. Srednicki's chapter 73 \cite{Srednicki:2007qs})
and simply quote the results for the beta function of QCD (here parametrized via $\alpha_s = g^2/4\pi$), which dictates the dependence of the gauge coupling on energy. In $\overline{\text{MS}}$, the calculation finds 
\begin{equation}
\beta(\alpha_s) \equiv \frac{\text{d}\alpha_s}{\text{d}\ln\mu} = - \frac{\alpha_s^2}{2\pi} (11 - \frac{2}{3} n_f) + \mathcal{O}(\alpha_s^3)
\end{equation}
where $\mu$ is the energy scale of interest and $n_f$ is the number of quarks with masses below $\mu$, which at high energies is $n_f = 6$. Then if we know the gauge coupling at a fundamental scale like $M_{pl}$, we can follow the procedure discussed in Section \ref{sec:relatetheories} of running and matching to sequentially evolve the coupling down to low energies. We end up with a result like
\begin{equation}
\frac{1}{\alpha(M_{pl})} - \frac{1}{\alpha(\mu)} = \frac{b}{2\pi} \ln \frac{M_{pl}}{\mu}
\end{equation}
where $b$ is somewhere between the $11-6\times 2/3=7$ value it takes above the top quark mass and the $11-3\times 2/3=8$ value it has below the charm quark mass. This tells us that eventually the QCD coupling blows up in the infrared, and the theory becomes strongly coupled---we expect our perturbative understanding of the theory to break down. While there is no proof of the precise effects of this, there is strong evidence that this is responsible for the observed phenomenon of `color confinement'--- at low energies colored particles form bound states which are color-neutral. The intuition being that the gluon interaction is so strong that trying to pull quarks in a color-singlet apart from each other requires so much energy that it is energetically favorable for a quark-antiquark pair to be created out of the vacuum and to end up with two color singlets. We may define a new scale $\Lambda_{QCD}$ as being the energy at which $\alpha$ diverges 
\begin{equation}
\Lambda_{QCD} \equiv M_{pl} e^{-\frac{2\pi}{b} \frac{1}{\alpha(M_{pl})}}
\end{equation}
So for some reasonable fundamental coupling $\alpha(M_{pl})$ at high energies, the theory generates a new scale which is exponentially far removed from the fundamental physics. Since the mass of the proton is mainly from QCD binding energy, $m_p \sim \Lambda_{QCD}$ this explains the huge hierarchy $m_p \lll M_{pl}$. This is an extremely important mechanism and historically one of the first suggestions for how to generate the electroweak scale was by copying this strategy, as we'll discuss in Section \ref{sec:composite}.

\chapter{The Hierarchy Problem}
\label{sec:hierarchy}

\section{The Higgs in the Standard Model}

The physical question of the hierarchy problem is how to get an infrared scale $v$ out of a microscopic theory whose degrees of freedom live at the much higher scale $\Lambda$, with $v/\Lambda \ll 1$. The tools introduced in Section \ref{sec:EffFieldTh} have already provided a window into why this can be difficult in a quantum field theory. Our aim in this section is to expand on that notion for the generation of the electroweak symmetry breaking scale in the Standard Model, where this scale is provided by the Higgs mass. In the Standard Model the Higgs mass is not technically natural, so the example discussed in Section \ref{sec:Naturalness} suggests the issue that may appear.

It is well-known that the Higgs plays a central role in the Standard Model, but the tagline `it provides mass' doesn't go far enough in underscoring its importance. The Higgs is needed because the fermions of the Standard Model have a \textit{chiral} spectrum: There are no representations with opposite charges under the full SM gauge group. This means that there are no gauge-invariant fermion bilinears, so no fermions can be paired up to form mass terms. If the Standard Model were \textit{not} chiral, we could write mass terms directly in the Lagrangian. In such a case there's no reason to expect small masses for the fermions, and indeed in the familiar case of right-handed neutrinos---which are SM gauge-neutral themselves, so can have Majorana masses---we generally expect them to be very heavy. In some sense the natural expectation for such `vector-like' (non-chiral) fermions would be to have Planck-scale masses, as in the absence of other particle physics, this is the only scale. 

So macroscopic structure in the universe is solely made possible by the chiral nature of the Standard Model.\footnote{We note that it's also true that the existence of a macroscopic \textit{universe} relies on the smallness of the cosmological constant, which is the other pressing fine-tuning issue present in the Standard Model. This way of viewing the naturalness problems of the Standard Model has been beautifully articulated by Nima Arkani-Hamed in \cite{Arkani-Hamed:2012ekl} and in many seminars.} There may well be other vector-like sectors which indeed contain Planck-scale masses. The fact that the particles which comprise us have a weakly-coupled description where quantum gravitational effects are suppressed---and so notions like locality and Riemannian geometry work well---would not hold in such a vector-like sector.  

Thus the fact that the Standard Model is chiral and so requires the Higgs mechanism to provide masses answers a deep and important question about our place in the universe. But it doesn't provide a full answer, as the scale at which the Standard Model sits still needs to be generated somehow. And in the absence of a mechanism to make it light, we must worry once more about losing our macroscopic existence with a mass scale which is again naturally of order the only other mass scale, the Planck scale. 

So, far more than the hierarchy problem being a small detail to clean up after having empirically verified the structure of the Standard Model, the question of why $m_H \ll M_\pl$ has serious physical importance.\footnote{It's worth noting that in the complete absence of the Higgs, electroweak symmetry is broken by the QCD chiral condensate \cite{Samuel:1999am,Quigg:2009xr}. It's interesting to ponder why Nature did not choose to let QCD confinement solely fill the role, but we know empirically that there is EWSB at $\sim 100 \text{ GeV}$ scales, so we need to understand the generation of that separate scale.} If we want an answer to why we live in a world with macroscopic structure, we must grapple with the hierarchy problem.

This section is devoted to understanding the technical statement of this physical question in the framework of effective field theory. We pursue this by introducing, discussing, and refuting some common confusions about the hierarchy problem. 

\section{\textit{Non}solutions to the Hierarchy Problem}


In this section I will introduce a few common confusions and misconceptions about the hierarchy problem. Discussion and refutation of these arguments provides a natural backdrop for introducing how the hierarchy problem should be properly understood and why it is important.

\subsection{An End to Reductionism} \label{sec:input}

A first point of confusion is that the Higgs mass is a free input parameter in the Standard Model, so a natural objection is that we should just set $m_H = 125 \text{ GeV}$ and call it a day. Indeed, this hits on a basic and important point: \textit{There is no hierarchy problem in the Standard Model.} The hierarchy problem exists for a more-fundamental theory which \textit{predicts} the Higgs mass---that is, one in which the Higgs mass is an \textit{output} parameter.

We can evince this in a simple toy model of a scalar $\phi$ interacting with other general fields $\psi_i$ where a tree-level, `bare' mass term is allowed by the symmetries. This is our toy version of the Standard Model, in which the scalar mass is likewise an input.
\begin{equation}\label{eqn:SMtoy}
S = \int d^dx \left[-\half (\partial_\mu \phi)^2 - \half m_0^2 \phi^2 - V(\phi) - \phi g\mathcal{O}(\psi_i \psi_j) \right]
\end{equation} 
Of course, as is familiar, $m_0$ itself is not measurable. When one calculates the two-point function of $\phi$ perturbatively in couplings\footnote{Perturbative calculations are an expansion in couplings, \textit{not} $\hbar$ \cite{Holstein:2004dn}, though this subtlety is commonly elided.} one finds quantum corrections $\Gamma^{(2)}(p) = p^2 + m_0^2 + g^2 \left(m_1^2 + \dots\right) + \mathcal{O}(g^4)$, where these are generically large because the mass is not technically natural---there's no symmetry protecting it. When one measures the mass of $\phi$ with e.g. some scattering experiment, it is $m^2_\text{phys} = m_0^2 + g^2 m_1^2 + \dots$ which one measures. And luckily so, because $m_1^2$ may well be formally infinite in a continuum quantum field theory, and a similarly infinite bare mass term is necessary to end up with the correct finite physical mass. Indeed, we are justified in this theory in choosing $m_0$ such that $m_\text{phys}$ matches the measured value. This is just the familiar procedure of renormalization, stretching back many decades and first understood in the context of quantum electrodynamics. To define QED one needs to input some definitions of the electron mass and the electromagnetic coupling based on experimental data, and these two inputs then determine all other predictions of the theory---e.g. the differential cross section of Coulomb scattering, the lifetime of positronium, and anything else you could hope to measure.

However, consider now a theory which has a global $SU(2)$ symmetry which is spontaneously broken at a high scale $M$. We want to understand how to get a light scalar degree of freedom out of this theory---that is, we've measured $m_\text{phys} \ll M$. This is a toy model of a Grand Unified Theory, where the microscopic physics exists at a high scale $M_\text{GUT}$, and indeed it was in this guise that the hierarchy problem was first recognized. Let's say our light scalar degree of freedom $\phi$ originated from a doublet $\Phi^a = (\varphi, \phi)^\intercal$. Our microscopic theory now does not have a bare mass term for $\phi$ but rather solely for $\Phi$ as a whole, since it must respect the symmetry. A difference in the masses of $\phi$ and $\varphi$ can only come from the spontaneous breaking of the symmetry---let's say when another fundamental scalar $\Sigma$ gets a vev $\nu = M$. This vev is a physical, measurable parameter related to the mass of $\Sigma$ and its self-interactions. Our action is controlled by the symmetries, as ever.
\begin{equation}\label{eqn:GUTtoy}
S = \int d^dx \left[-\half (\partial_\mu \Phi)^\dagger (\partial^\mu \Phi) - \half M_0^2 \Phi^\dagger \Phi - \lambda_0 \Phi^\dagger \Sigma \Sigma^\dagger \Phi - V(\Phi) \right]
\end{equation}
where $M_0$ is the bare mass of the $\Phi$ doublet and $\lambda_0$ is its bare interaction strength. In the absence of any other scales we generally expect $M_0 \sim M$ and $\lambda \sim \mathcal{O}(1)$. In this theory there is no reason for the values of these parameters to be connected to each other in any way.

When $\Sigma$ gets a vev, $\langle \Sigma \rangle = (0, \nu)^\intercal$, it breaks the $SU(2)$ symmetry and gives a mass splitting between the two degrees of freedom in $\Phi$, since $\langle \Sigma \rangle^\dagger \Phi = \nu \phi$. We then have the masses $m_\varphi^2 = M_0^2, \ m_\phi^2 = M_0^2 + \lambda_0 \nu^2$. In this theory our tree-level inputs are $M_0$ and $\lambda_0$ (and the interactions controlling the value of $\nu$, which for simplicity we don't write down) and the scalar mass $m_\phi$ is an \textit{output}. In fact here the hierarchy problem occurs at tree-level, simply as a result of wishing to produce a small mass via splitting a multiplet. If we wish to have, say, $M \simeq 10^{16} \text{ GeV}$ and $m_\phi \simeq 100 \text{ GeV}$---the values of the GUT scale and the electroweak scale in the real world---we need to fine-tune $\lambda$ enormously so it takes a value like $-1.0000000000000000000000000001\times \frac{M_0^2}{\nu^2}$. 

Of course when we look at our theory at loop level there will again be quantum corrections to our tree-level parameters $m_0$ and $\lambda_0$, and again it will be their corrected values which are physical and measurable $M^2_\text{phys} = M_0^2 + g M_1^2 + \dots$, $\lambda_\text{phys} = \lambda_0 + g \lambda_1 + \dots$. But if our theory is renormalizable, we know that quantum corrections will merely change the values of these parameters, and not the operators we have. The point is that the quantum corrections are $SU(2)$ invariant, so the masses of both $\phi$ and $\varphi$ will receive the same loop contributions. We will then still predict the mass of $\phi$ as $m_\phi^2 = M^2_\text{phys} + \lambda_\text{phys} \nu^2$. Now at the level of inventing the theory we may still tune these parameters to get a small $m_\phi$, but we're tuning \textit{physical, observable} parameters.

In the theory described by Equation \ref{eqn:SMtoy}, one might have also said that we needed to fine-tune $m_1$ against $m_0$ in order to get a small $m_\phi$, especially if we calculated that the quantum correction $m_1$ was large. But there the fine-tuning was of a different sort, since $m_0$ and $m_1$ were only ever observable in the combination $m_\text{phys}$. Here the fine-tuning has a much sharper meaning. This tuning translates into a physical demand on our theory that at high energies the strength of the interaction between our two scalars $\Phi, \Sigma$ for some reason has a value \textit{extremely} close to $-M^2_\text{phys}/\nu^2$, despite having nothing to do with either of these parameters. The tuning is now a physical feature of our theory and demands explanation.\footnote{Let me mention parenthetically a confusion one may encounter if one reads older literature on the hierarchy problem in GUTs. It was common to speak of having to `re-tune the parameters at every order in perturbation theory', as if imagining an algorithmic process where one first computed a tree-level prediction, tuned that to be correct, and then computed the one-loop corrections, re-tuned those parameters to get it right again, etc. This is framed as being `worse' than just requiring `one' set of tunings. This is moronic, for the simple physical reason that Nature does not compute via perturbation theory. There is a physical problem, which is how to get the electroweak scale out of other physical parameters in the theory. Whether you compute the predictions in perturbation theory, or on a lattice, or whilst standing on your head is immaterial.}

So we see explicitly that the hierarchy problem is present when the light scalar mass is an \textit{output} of the theory, rather than an input. If one is so inclined, one can say the words that the Higgs mass is simply an input, but this possibility spells the end of scientific reductionism. It is indeed \textit{conceivable} that this is how the universe works. However, we know there is physics beyond the Standard Model at smaller length scales, and our best ideas for what those could be involve theories where the inputs are defined in the ultraviolet and the Higgs mass comes out. Whether the Higgs ultimately originates as a component of a larger multiplet, or a bound state of fermions, or an excitation of a string, we expect that the Higgs mass is a parameter that comes out in the low-energy theory. 

\subsection{Waiter, there's Philosophy in my Physics} \label{sec:philosophy}

\epigraph{You can always use the history of physics to illustrate any polemic point you want to make.}{Nima Arkani-Hamed}

Now having exhibited that getting a light scalar truly does involve some fine-tuning of physical parameters, one may still say `\textit{So?}'. In the real world, we've observed (the analogue of) $m_\phi$, but the physics we've discussed at the heavy scale $M$ is new physics, and we don't have experimental measurements of $\lambda$ telling us the value is \textit{not} that perfect value to get our light scalar. One might thus object that there's no fundamental inconsistency, and as long as our theory can fit the data anything else is \textit{just philosophy}.

But the criterion of it being \textit{not literally impossible} for a theory to fit the data is an incredibly low bar, and scientists always use additional criteria to select theories. While there is ultimately a degree of subjectivity in any notion of `naturalness', this is really the same subjectivity that one constantly uses in science to decide which of two explanations for some data to accept.

\begin{figure}
	\centering
	\includegraphics[width=\linewidth]{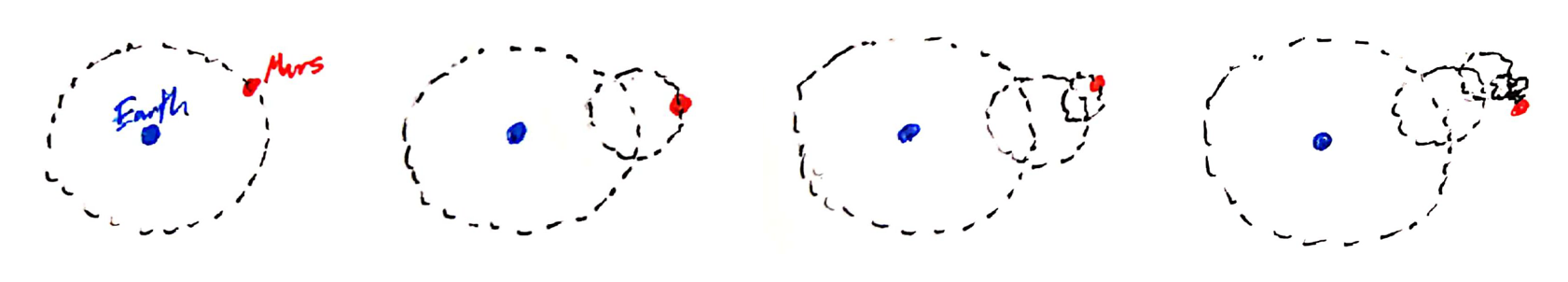}
	\caption{Schematic progression of epicycles needed in the Ptolemaic model of Mars orbiting the Earth as astronomical measurements increased in precision.}
	\label{fig:epicycles}
\end{figure}

A simple (approximately) historical example of this can be seen in epicyclic theories of the motion of the planets. The Ptolemaic, geocentric model of the universe predicted at first that the heavenly bodies orbited the Earth in circles, but eventually astronomical data was accurate enough to show that the motion of the planets and sun around the earth was not circular. In the Ptolemaic model, this was dealt with by adding an \textit{epicycle}, the suggestion that the heavenly bodies moved on smaller circular orbits about their circular orbit around the Earth. As astronomical observations became more and more detailed over the ensuing centuries, multiple layers of epicyclic motion were needed to explain the data---circles on circles on circles, as depicted in Figure \ref{fig:epicycles}. The description never stopped working though; if you'd like to, you may describe the orbit of any solar system object via $r(\theta) = \sum\limits_{n=1}^N r_n \sin\frac{\theta}{n}$, and for some large but finite $N$ you'll be able to fit any orbit to within observational precision. 

So why did sixteenth century physicists favor Kepler's laws and heliocentrism? The discriminating factor is manifestly \textit{not} which model better-fitted the data. Rather, the choice comes down to Occam's razor, to explanatory power, to simplicity and to fine-tuning of parameters. Physicists favor theories which do more with less---theories which explain more about the world while requiring fewer inputs. This is a subjective bias about how we think the universe should work, and it's possible that this philosophy will ultimately fail---but it's been working well thus far.\footnote{As a semi-autobiographical aside, I had the honor and pleasure of being in the inaugural cohort of the Integrated Studies Program for Benjamin Franklin Scholars in the School of Arts \& Sciences at the University of Pennsylvania. The program, founded and spearheaded by the classicist Prof. Peter Struck, offered a dedicated interdisciplinary experience wherein, each semester, three diverse fields gave courses offering perspectives around a central topic, which were concurrently collectively compared and contrasted. In my year we studied biology, anthropology, classics, political science, physics, and literature, all taught by preeminent professors in their respective fields. After noticing a pattern, the group kept track of (among other things) how many times each professor mentioned, discussed, or appealed to `beauty'. The winner in this regard was Prof. Vijay Balasubramanian---lecturing on the way the universe works---by a country mile.}

While keeping the above intuition firmly in the back of our minds, it can be useful to introduce a mathematical classification of this fine-tuning, with the understanding that no such measure is god-given and so what to do with such a measure is up to us. We'll discuss a couple such schemes, the first being a mathematical formalization of the dependence of an output of interest on the values of the inputs. This has the benefit that it is intuitive and simple to compute, so it is widely used in the particle physics literature. However, it lacks independence under how variables are parametrized so can lead to misleading conclusions if used without care. 

Furthermore, it will assign a measure of fine-tuning to individual points in the parameter space of a model, whereas we'd like to characterize the naturalness of a model as a whole---if a model only produces predictions that match the real world in a small region of its parameter space, that's another important element of fine-tuning \cite{Anderson:1994dz}. For example, if new data removes all but a small fraction of the viable parameter space in a given model, we want to regard that model as being less natural afterward. An approach based on Bayesian statistics allows us to incorporate these issues and gives unambiguous comparisons of the relative naturalness of models upon the collection of new data \cite{Fichet:2012sn}, but loses out on simplicity. 

A simple and often-used measure was introduced by Giudice and Barbieri \cite{Barbieri:1987fn}, who suggested the definition
\begin{equation}
\Delta_X \equiv \left| \frac{\text{d} \ln m^2}{\text{d} \ln X} \right| = \left| \frac{X}{m^2}\frac{\text{d} m^2}{\text{d} X} \right|
\end{equation}
which may be called a measure of the fine-tuning of the input parameter $X$ necessary to get out the correct output parameter $m^2$. The logarithmic dependence naturally gives a measure of relative sensitivity and removes dependence on overall scale or choice of units. If $\Delta_X$ is large, this denotes a large sensitivity of $m^2$ to the value of $X$, and implies that one must choose the value of $X$ very carefully to get out the right physics. In the example in Equation \ref{eqn:GUTtoy} above we have $\Delta_\lambda = \frac{\lambda}{m_\phi^2} \nu^2 \simeq \frac{M^2}{m_\phi^2} \simeq 10^{28}$, indeed signaling enormous fine-tuning, and likewise $\Delta_{\nu^2} = \lambda \frac{\nu^2}{m_\phi^2}$. Contrast this with the familiar case of a seesaw mechanism where the light neutrino mass is given by a formula like $m_\nu^2= m^4/M^2$, with $M$ a heavy mass scale and $m$ a weak scale mass. We can check whether this mechanism requires fine-tuning with $\Delta_{M^2} = \frac{M^2}{m^2_\nu} \frac{m^4}{M^4} = \frac{1}{m_\nu^2} \frac{m^4}{M^2} = 1$, and we find that seesaw mechanisms are natural. So $\Delta_X$ matches our intuition here, and can be quite useful.

\begin{figure}
	\centering
	\includegraphics[width=0.8\linewidth]{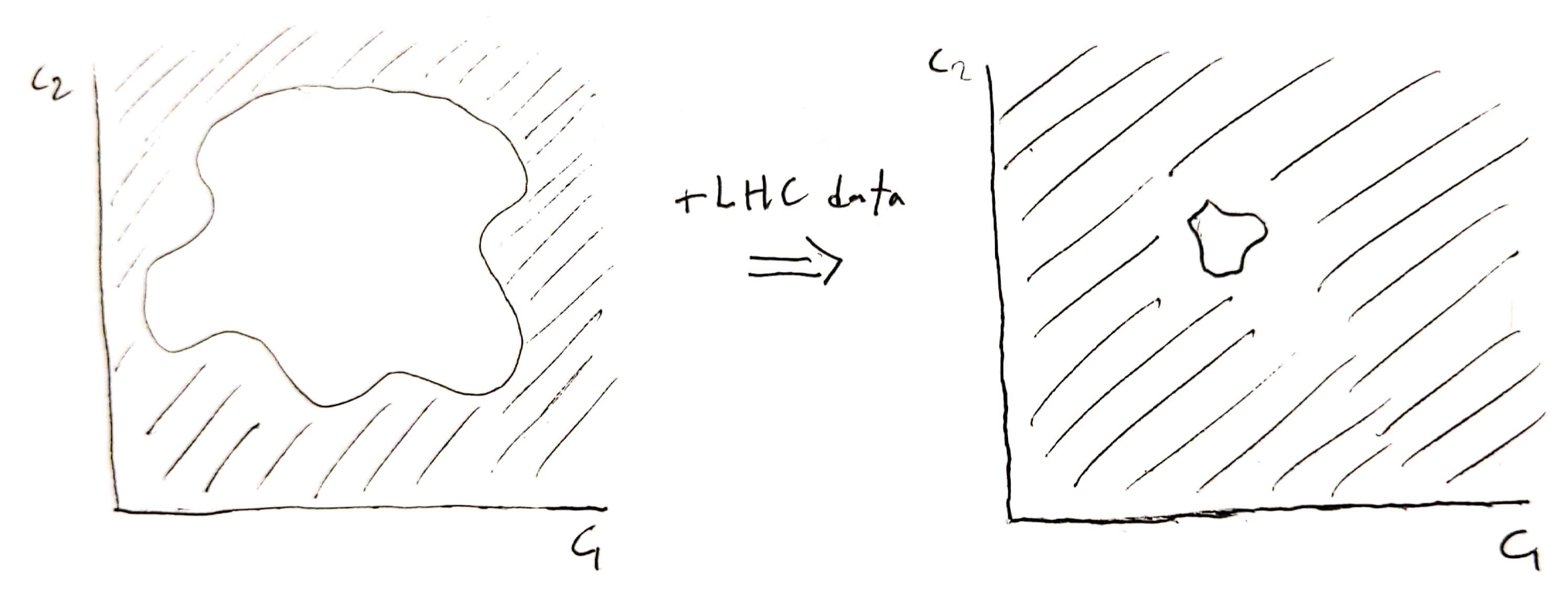}
	\caption{A schematic representation of the LHC ruling out previously-viable parameter space in some model. Surely afterwards we should view this model as more fine-tuned compared to a model on which the LHC results had no impact.}
	\label{fig:paramspace}
\end{figure}

However, for a model with free parameters a notion of fine-tuning at a single point in parameter space does not capture the full picture, and we should incorporate a notion of the volume of viable parameter space into our naturalness criterion \cite{Athron:2007ry}, as diagrammed in Figure \ref{fig:paramspace}. This necessity should be intuitive in the context of constraining models of new physics. Models which achieve their aim throughout parameter space are viewed more favorably than models which only work in some small region of their parameter space. For a relevant example, there are still corners of the MSSM parameter space that are natural under the Giudice-Barbieri measure. But the fact that the LHC has ruled out large swathes of this parameter space means that we should surely view weak-scale supersymmetry as less natural now than we did a decade ago. A pointwise measure of fine-tuning misses this.

An approach based on Bayesian statistics can be used to better match what we want from a measure of naturalness, as is discussed well in \cite{zhou}. The definition of a model in a Bayesian framework requires priors on its free parameters $\lbrace\theta_i\rbrace$, denoted $p(\theta_i|\mathcal{M})$, and different choices of $\lbrace p(\theta_i|\mathcal{M})\rbrace$ should be considered different models. The probabilist's notation $p(A|B)$ may be interpreted as `the probability of $A$ given that $B$ is true'. We also require a prior probability $p(\mathcal{M})$ that the model $\mathcal{M}$ is true as a whole.
After we receive data $d$, Bayes' theorem gives us the posterior probability for the model as 
\begin{equation}
p(\mathcal{M}|d) = \frac{p(d|\mathcal{M})p(\mathcal{M})}{p(d)}
\end{equation}
where $p(d|\mathcal{M})$ is known as the likelihood. Both $p(\mathcal{M})$ and $p(d)$ are explicitly subjective, but if we take the ratio of the posterior probabilities for two models $\mathcal{M}_1$ and $\mathcal{M}_2$ we find
\begin{equation}
\frac{p(\mathcal{M}_1|d)}{p(\mathcal{M}_2|d)} = \frac{p(d|\mathcal{M}_1)}{p(d|\mathcal{M}_2)} \frac{p(\mathcal{M}_1)}{p(\mathcal{M}_2)}
\end{equation}
This expresses how the ratio of the likelihood of these two models changes after receiving new data. So while different physicists may disagree on the prior and posterior probabilities, the `Bayes update factor' $B \equiv p(d|\mathcal{M}_1)/p(d|\mathcal{M}_2)$ is unambiguous and shows that the physicists agree on how the relative naturalness of the two models is affected by the new data. 

The likelihood in a model with free parameters $\lbrace\theta_i\rbrace$ is calculated as
\begin{equation}
p(d|\mathcal{M}) = \int \left(\prod_i d\theta_i\right) p(d|\mathcal{M},\theta_i) p(\theta_i|\mathcal{M})
\end{equation}
This is an integration over parameter space of the likelihood of producing the data in this model, weighted by the prior probability distribution we've placed on our parameters. This balances the competing effects of how well parameter points fit the data with the principle of parsimony---models with large regions of parameter space which don't fit the data are penalized. The need to compare models to define naturalness in a Bayesian formalism is easily seen by the fact that the likelihood for any new physics model decreases monotonically as more data is collected and previously-viable regions of parameter space are ruled out (in the absence of a discovery, of course). 

An interesting playground for these ideas is the strong CP problem. In brief, this is the smallness of the so-called `theta angle' $\theta$ in QCD, which controls the amount of CP breaking in the strong sector. While $\theta \in\left[0,2\pi\right)$, empirical measurements now constrain $\theta \lesssim 10^{-10}$. Although $\theta$ is not technically natural, in the Standard Model it runs very slowly---since the other source of CP breaking is from the CKM matrix---such that if one sets it tiny in the UV, it stays small down to the IR. The Guidice-Barbieri measure would thus produce $\Delta_{\theta_{\text{UV}}} \simeq 1$, as the measured value is insensitive to small changes in the UV value. And yet, the small theta angle is regarded as a naturalness problem, which can be justified in a Bayesian approach. The necessity of a prior does help quantify one's surprise at the small value of $\theta$ in the context of the Standard Model, but to really think about naturalness we need to have a comparison. We know of simple theories which produce vanishing $\theta$ starting from generic values of the parameters---for example, an axion naturally produces $\theta = 0$. Then if we have new data which pushes down the upper bound on $\theta$, axion models receive large Bayesian update factors in relation to the Standard Model. When we know of a simple model which automatically explains some data, it's puzzling if our current model requires precise choices for free parameters in order to explain the data. 

We can see these notions play out for the hierarchy problem by comparing our toy model of a GUT to one with weak-scale supersymmetry. Our toy model in Section \ref{sec:input} produced a scalar through a cancellation of GUT-scale $\sim 10^{16} \text{ GeV}$ contributions. A toy model with supersymmetry (to be discussed in Section \ref{sec:SUSY}) would remove sensitivity to the ultraviolet, and replace the scale of GUT-breaking with the effective scale of supersymmetry-breaking in the SM sector $\tilde{m}$. And while the GUT scale is (more or less) fixed by the running of SM couplings, the SUSY-breaking scale in the SM can be far lower. We use $\tilde{m}$ here as a one-parameter avatar of the scale of superpartners. The general prediction for supersymmetry before the LHC was $\tilde{m} \sim \mathcal{O}(100 \text{ GeV})$, with multiple species of superpartners appearing below the TeV scale. Such a model produces an improvement in Giudice-Barbieri tuning of $\delta \Delta \sim 10^{28}$ over our model without SUSY. 

Let's take $p(\tilde{m}|\text{weak scale SUSY})$ to be a logarithmic prior from $m_\text{low}$ to $m_\text{high}$, where an upper limit $m_\text{high} \sim 500 \text{ GeV} - 1 \text{ TeV}$ is justified by the requirement that the model give small Higgs mass corrections and a weak-scale dark matter candidate. If we collect collider data for a decade and find that much of the parameter space is ruled out, say with a limit $\tilde{m} \geq \mathfrak{m}$ we should update our thoughts on the naturalness of the model. This data has no effect on our non-SUSY GUT model, as it predicts no new light particles, but there has been a large effect on our SUSY model as its most favored parameter space has been ruled out. So we have a large Bayesian update factor. \vspace{-0.3cm}
\begin{align}
B  &\equiv \frac{p(\text{LHC data}|\text{non-SUSY GUT})}{p(\text{LHC data}|\text{weak scale SUSY GUT})} \\
&= \frac{1}{\left(\log\frac{m_\text{high}}{m_\text{low}}\right)^{-1}\int_{\mathfrak{m}}^{m_\text{high}} \frac{\text{d} \tilde{m}}{\tilde{m}}} \nonumber \\
&= \log\frac{m_\text{high}}{m_\text{low}} / \log\frac{m_\text{high}}{\mathfrak{m}} \nonumber
\end{align} 
Of course supersymmetric extensions of the SM have a large number of parameters, and how to translate to an upper bound $\mathfrak{m}$ on our one-parameter version isn't well-defined, but certainly much of the previously-favored parameter space has been ruled out. The general lesson is that models which don't predict new visible states near the weak scale have received large Bayesian update factors from the LHC. This doesn't give a strict mandate for our overall relative belief, as one may argue there are good reasons to take a large prior for supersymmetry and expect always to find superpartners right around the corner.
After all, even if $\tilde{m} \sim 1000 \text{ TeV}$, it would still have $\delta \Delta \sim 10^{20}$ Giudice-Barbieri tuning better than the GUT without SUSY. But it does motivate further investigation of models which don't succumb to this issue, be they neutral naturalness modules which push the scale of new visible states up by a loop factor or more radical ideas about the origin of the electroweak scale.

Furthermore, we can directly input physics into making a sensible choice of the prior one places on a model, and there has been much discussion of justifying simple choices. The dictum of Dirac naturalness mandates priors peaked at $\mathcal{O}(1)$ values. But with a model which makes some parameter technically natural one, a prior that allows small values is justified, such as a logarithmic prior. An explicit example of this modification of priors by physics can be seen in the application of the Weak Gravity Conjecture to the hierarchy problem \cite{Cheung:2014vva, Ooguri:2016pdq, Ibanez:2017kvh, Ibanez:2017oqr, Hamada:2017yji, Lust:2017wrl, Gonzalo:2018tpb, Gonzalo:2018dxi, Craig:2018yvw, Craig:2019fdy,March-Russell:2020lkq}, which will be discussed in more detail in Section \ref{sec:graveft}. This mechanism explicitly modifies the prior one should have on UV theories by linking the notion of the Swampland---that some string vacua don't admit universes like ours---to the allowed range of Higgs masses. This addresses the hierarchy problem by constructing a model where the priors are forced by UV physics to favor a light Higgs. 

While these measures of naturalness are useful to help us clarify our expectations, we emphasize again that they must be used sensibly. But it's clear that some notion of naturalness appears solely from the axioms of probability, and is indeed baked-in to the practice of science inquiry. That's not to say that we should be epistemologically committed to the naturalness of the universe, but we can still see it as a useful guide toward new physics which has worked well in the past.

\subsection{The Lonely Higgs}\label{sec:lonely}

\setlength{\epigraphwidth}{0.45\textwidth}
\epigraph{All models are wrong, but some are useful.}{George E. P. Box \\ \textit{Robustness in the Strategy of \\ Scientific Model Building} (1979) \cite{BOX1979201}}
\setlength{\epigraphwidth}{0.6\textwidth}

There is another obvious suggestion that is useful to discuss: perhaps the Higgs does not interact with any physics at higher mass scales, such that despite in principle worries about the hierarchy problem there is no hierarchy problem \textit{in practice}. The physics which can destabilize the Higgs mass must, as seen in the above example, both be heavier than the Higgs and interact with it in order to generate a contribution. Since the Higgs mass in the SM is not technically natural (that is, it is UV sensitive), large mass corrections from such particles are generic, as we saw in Section \ref{sec:Naturalness}. One can explore the idea that perhaps there are no such particles. This faces a number of challenges.

The first difficulty is that we \textit{know} there must be new physics. The Standard Model does not explain neutrino masses nor dark matter, though it is possible that both of these be resolved without introducing new heavy particles. But a deeper issue is that we know the SM cannot be a fundamental theory because the Landau pole in hypercharge makes it inconsistent. This demands that \textit{something} must happen to rid the theory of this pole, and it will interact with the Higgs because the Higgs is hypercharged. This is one motivation for thinking that something like Grand Unification must take place, and of course its breaking introduces a heavy fundamental scale. One can try to get around this by appealing to quantum gravity coming in at scales below that of the Landau pole. But the Higgs certainly interacts gravitationally, so for this program to succeed one needs a quantum gravitational theory which does not introduce any scales. This is interesting to explore, but does not seem to be the way the universe works, though we leave detailed criticism of this idea to the gravity theorists.

\begin{figure}
	\centering
	\begin{subfigure}{.45\textwidth}
		\centering
		\includegraphics[width=0.8\linewidth]{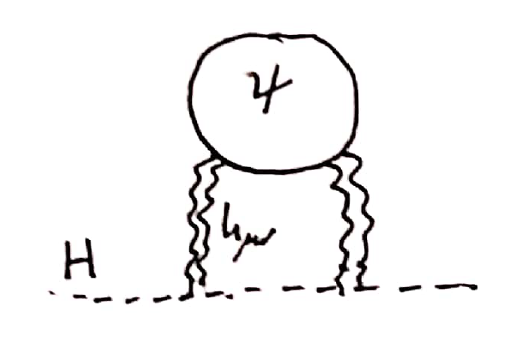}
		\caption{}
		\label{fig:grav2loop}
	\end{subfigure}%
	\begin{subfigure}{.45\textwidth}
		\centering
		\includegraphics[width=0.8\linewidth]{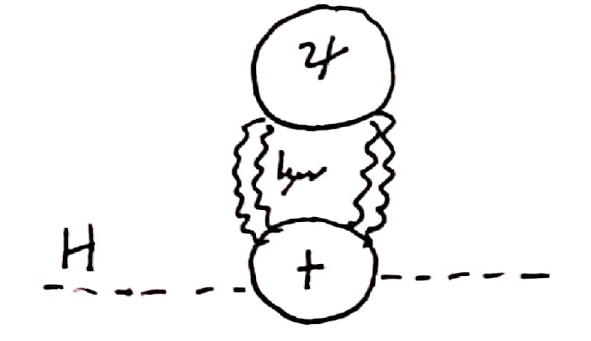}
		\caption{}
		\label{fig:grav3loop}
	\end{subfigure}
	\caption{In (a), a representative two-loop diagram giving gravitational corrections to the Higgs mass from a new dark fermion $\psi$. These diagrams do not destabilize the Higgs mass. In (b), a representative three-loop diagram in which the gravitons couple to an off-shell top quark, which is no longer proportional to the Higgs mass. }
	\label{fig:gravloop}
\end{figure}

Furthermore, even if somehow all heavy particles are neutral under the SM gauge groups, there is no way for them to escape gravitational interactions with the Higgs. This leads to irreducible three-loop corrections to the Higgs mass \cite{deGouvea:2014xba}, as depicted in Figure \ref{fig:gravloop}. Consider a fermion $\psi$ with mass $M_\psi$. The obvious two-loop diagram where the graviton couples to the Higgs directly gives a correction proportional to $m_H^2$, as the graviton coupling to a massless on-shell particle vanishes at zero momentum. However, we can draw a diagram where a $\psi$ loop talks gravitationally to an off-shell top loop contribution to the Higgs two-point function, and integrating this particle out yields a correction 
\begin{equation}
\delta m_H^2 \simeq \frac{y_t^2}{(16 \pi^2)^3} \frac{M_\psi^4}{M_{pl}^4} M_\psi^2 + \text{ subleading}.
\end{equation} 
The powers of $M_\psi$ appear because the only other possible mass scale for the numerator is $m_t = y_t v$, and the top loop power-law correction to the Higgs mass does not vanish in the $v \rightarrow 0$ limit.  The sensitivity of the Higgs mass is softened by three loop factors as well as by the Planck mass from the gravitational couplings, but insisting that $\delta m_H^2 \lesssim (1 \text{ TeV})^2$ places an upper `naturalness' limit on such fermions of $\sim 10^{14} \text{ GeV}$. While far better than the $\sim 1 \text{ TeV}$ limit for SM-charged particles, this is still well below the Planck scale and amounts to an enourmous constraint on UV physics. In fact the problem is a bit worse than this estimate, as we should sum over all SM loops that couple to the Higgs, but this suffices already to see the issue. 

So asking for the Higgs to be lonely enough to cure the hierarchy problem is a humongous requirement on the ultraviolet of the universe, and this approach faces a number of important hurdles. We mention that there is work on interesting theories which touch on some of these points, 
but as a whole Nature seems not to have taken this approach.

\subsection{Mass-Independent Regulators}

One may hear the statement that the hierarchy problem disappears if you use a mass-independent regularization scheme, for example dimensional regularization. Unlike the prior nonsolutions we've considered, this one is definitively incorrect. The mass scale one introduces in EFT is a stand-in for genuine physical effects of any sort which appear at shorter distance scales. So the cutoff regularization is useful for seeing an avatar of the hierarchy problem even when one does not know the ultraviolet theory. With a mass-independent scheme, one must instead put in specific short-distance physics to see the problem, but we can easily see the general issue.

\begin{figure}
	\centering
	\includegraphics[width=0.4\linewidth]{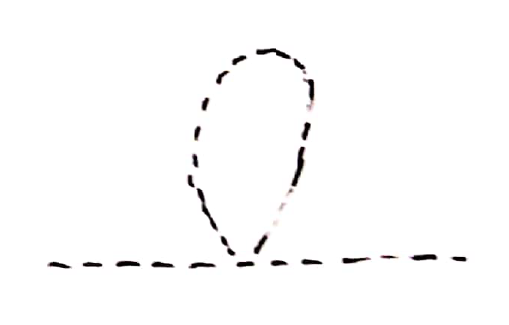}
	\caption{The one-loop diagram for a scalar quartic interaction contributing to a scalar mass.}
	\label{fig:scalarloop}
\end{figure}

As a simple example, take a theory with two real scalars - our light $\phi$ and a heavier $\varphi$. If we impose a $\mathbb{Z}_2$ symmetry for simplicity, the action is
\begin{equation}
S = \int d^4x \left[-\half (\partial_\mu \phi)^2 - \half m_0^2 \phi^2 - \half M_0^2 \varphi^2 - \frac{\lambda}{4} \phi^2 \varphi^2 \right]
\end{equation} 
Doing continuum effective field theory with dimensional regularization and the $\overline{\text{MS}}$ renormalization scheme, we must upgrade the masses and couplings to running parameters which depend on the renormalization scale $\mu$, as usual (introduced in Section \ref{sec:repairpert}). Since our renormalization scheme is mass-independent, if we want to study physics at an energy scale $\mu \sim m_\phi \ll M_\varphi$, we should implement the decoupling theorem by hand. We integrate out the heavy degree of freedom at the scale $\mu \sim M_\varphi$ and match to a low-energy effective field theory which only contains the low-energy degree of freedom $\phi$ (introduced in Section \ref{sec:relatetheories}). So let us go ahead and integrate out the heavy scalar $\varphi$. The effect on the $\phi$ mass comes from the simple diagram of Figure \ref{fig:scalarloop}.

We go to general dimension $d=4-\epsilon$ and replace $\lambda \rightarrow \lambda \tilde\mu^\epsilon$, where $\tilde\mu$ is an arbitrary scale which soaks up the mass dimension of $\lambda$ away from $d=4$. The resources needed to compute the integrals for general dimension and to take the limit $\epsilon \rightarrow 0$ can be found in Srednicki's textbook \cite{Srednicki:2007qs}.
\begin{align}
-i \delta m^2 &= \frac{-i \lambda \tilde\mu^\epsilon}{2} \int \frac{d^dk}{(2\pi)^d} \frac{-i}{(k^2 + M_\varphi^2)} \nonumber\\
&=  \frac{-i \lambda \mu^\epsilon}{2} \int \frac{d^d\bar k}{(2\pi)^d} \frac{1}{(\bar k^2 + M_\varphi^2)} \nonumber\\
&= -i \frac{\lambda \tilde{\mu}^\epsilon}{2 (4 \pi)^2} \Gamma(-1+\frac{\epsilon}{2}) (4\pi)^{\frac{\epsilon}{2}} (M_\varphi^2)^{1-\frac{\epsilon}{2}} \nonumber\\
&= i \frac{\lambda}{2(4 \pi)^2} M_\varphi^2 \left[\frac{2}{\epsilon} - \gamma_E + 1 + \ln 4\pi - \ln M_\varphi^2 + 2 \ln \tilde{\mu} + \mathcal{O}(\epsilon)\right] \nonumber\\
&= i \frac{\lambda}{(4 \pi)^2} M_\varphi^2 \left[\frac{1}{\epsilon} + \half + \ln \frac{\mu^2}{M_\varphi^2} + \mathcal{O}(\epsilon)\right] 
\end{align}
Where $\gamma_E$ is the Euler-Mascheroni constant, and we have Wick rotated $k^0 \rightarrow i \bar k^d$, integrated in general dimension, expanded in the limit $\epsilon \rightarrow 0$ and defined $\mu^2 \equiv 4 \pi \tilde{\mu}^2 e^{-\gamma_E}$ to soak up the annoying constants. Indeed, we see that there's no quadratic divergence, which ultimately is due to the fact that scaleless integrals vanish in dimensional regularization $\int \frac{d^dk}{k^n} = 0$, as must be true simply by dimensional analysis.

Now we follow the $\overline{\text{MS}}$ renormalization scheme by adding a counterterm which cancels off the divergent piece and we match at $\mu = M_\varphi$ to ensure that our low-energy EFT produces the same predictions as the UV theory, as discussed in Section \ref{sec:relatetheories}. This gives us
\begin{equation}
S = \int d^dx \left[-\half (\partial_\mu \phi)^2 - \half \left(m_0^2 + \frac{\lambda}{(4 \pi)^2} M_\varphi^2\right) \phi^2 - \dots \right]
\end{equation} 
We see that the high-energy degree of freedom $\varphi$ contributes a \textit{threshold correction} when we flow to lower energies and remove it from the spectrum. While there was never a quadratic divergence, we still found a large quadratic correction to the mass of the scalar $\phi$ which is proportional to the scale of new physics.

This underscores the importance of not getting confused by unphysical features of renormalization. There is a physical issue, which is the sensitivity of the physical low-energy scalar mass to the physics in the ultraviolet. Indeed if we tame the UV in different ways, we find different avatars of this sensitivity. It's true that a Wilsonian cutoff acts as a stand-in for arbitrary scaleful physics, which is why it's more direct to see the issue in that picture, but the same \textit{physical} problem appears regardless of the regularization.

\section{The Hierarchy Problem}

After discussing those misconceptions, we may give a one-sentence description of the hierarchy problem: \\

\noindent\fbox{%
	\parbox{\textwidth}{%
In a theory of physics beyond the Standard Model where the Higgs mass is an output, physical parameters must be finely-tuned in order to produce a mass which is far below the scale of new physics, in tension with the principle of parsimony.
	}%
}\\

With that in hand, we are prepared to delve in to how the hierarchy problem may be solved in the next section. Our discussion below will not take place within a UV complete extension of the Standard Model, so one might worry that we are attacking a problem without knowing its source. While true, the point from Section \ref{sec:lonely} is that the sensitivity to UV physics is so general that we expect to need a mechanism which stabilizes the Higgs mass to \textit{whatever} new heavy physics is out there. Our toy calculation of the relative naturalness of SUSY already evinces this point---SUSY tamps UV sensitivity no matter what it is, so the details of the UV completion are immaterial. As a result of this idea, we will mostly worry just about finding a way to produce a light scalar and assume that it can be embedded into whatever physics exists at high scales, rather than committing to a particular framework of grand unification or what have you. Of course it's possible that interesting mechanisms to produce an IR scale do rely on particular properties of the UV, and we'll discuss this important idea in Section \ref{sec:EFTbar}. But even there, our initial goal is simply to produce a light scalar which is compatible with the Standard Model, rather than to write down a full theory of the universe on all scales. If we can first solve the problem in a toy model which shares some features of our universe, then we can hope to abstract what we learn from that to solve the real problem.

\chapter{The Classic Strategies}
\label{sec:classical}

\setlength{\epigraphwidth}{0.4\textwidth}
\epigraph{Without going out of my door \\I can know all things on earth\\Without looking out of my window\\I could know the ways of heaven}{George Harrison satirizing the \\ past decades of particle theory \\ in light of LHC data \\\textit{The Inner Light} (1968) \cite{harrison_1968}}
\setlength{\epigraphwidth}{0.6\textwidth}

\section*{Fantastic Symmetries and How To Break Them}

In large part the story of particle physics over the past decades is the story of attempts to solve the hierarchy problem. Much theoretical effort has been put into understanding interesting symmetries and mechanisms for breaking them, and more generally ways that small numbers can pop out of physical theories; and much experimental effort has focused on locating empirical hints of these ideas. However, the past few years have seen many practitioners turn their attention toward topics like dark matter, cosmology, and astrophysics. And for good reason---on the experimental side, this is largely where the new data is and will be for the foreseeable future, and at the purely theoretical level the hierarchy problem has become a lot more challenging, as we will argue below. But this has lead to a new generation of particle theorists who are largely unfamiliar with the fantastic and brilliant ideas which drove the field in the prior couple decades. 

Despite the fact that we will argue below that these ideas largely appear to not be the way the world works at the weak scale, understanding this prior work can be enormously helpful for inventing novel ideas in the future. It is with this in mind that we introduce below the basics of a variety of interesting ideas and methods in particle physics against the backdrop of their relevance to the hierarchy problem. These are ideas that have not yet made their way into standard textbooks on field theory, but are nevertheless essential topics for students of particle theory to absorb. We will endeavor to explicate the core of these ideas in the simplest models possible, and will largely avoid discussing phenomenological considerations past producing a light scalar. The discussion will not be at the level of depth required for research in the field, but will hopefully be a nice overview of interesting topics for which references to serious introductions and reviews will be provided as well.

So how does one solve the hierarchy problem? The classical solutions may be conceptually divided into two steps. First one introduces some structure above the electroweak scale which protects the Higgs mass from large contributions due to UV physics. This could be something like a new symmetry which forbids a scalar mass term, or a modification to spacetime on small length scales, or the dissolution of a non-fundamental Higgs into component fields.

However, the Higgs is not exactly massless, which is due to the fact that whatever structure we add is not a feature of the low-energy Standard Model. There must thus be some IR dynamics that break that UV structure at the electroweak scale to ensure that we end up with the Standard Model at low energies. Depending on the UV structure this may be something like spontaneous symmetry breaking or moduli stabilization or dimensional transmutation. 

There are two big categories of classical solutions. One is to find a field-theoretic mechanism which prevents contributions to the Higgs mass in the UV. Supersymmetry is the prime example here. The other is to bring the fundamental cutoff of the theory down to the infrared, such that in the UV there's no Higgs to talk about. This is exemplified by composite Higgs theories or theories where the cutoff of quantum gravity is lowered to the weak scale. 

To evince these strategies, we'll go through a couple examples of ways to forbid scalar masses and to break those structures. Our aim here is not to construct realistic theories of the Higgs but rather to understand these general principles, so we'll study simple toy models which allow us to appreciate the essential points.

\section{Supersymmetry} \label{sec:SUSY}
\setlength{\epigraphwidth}{0.4\textwidth}
\epigraph{Superpartners aren't essential \\ But would have been consequential \\ Such wasted superpotential \\ Super once,  super twice \\ Super chicken soup with rice}{Maurice Sendak on his disappointment with the LHC data \\ Lost Stanza of \textit{Chicken Soup with Rice} (1962) \cite{sendak1962chicken}}
\setlength{\epigraphwidth}{0.6\textwidth}

Supersymmetry exploits a loophole in the classic Coleman-Mandula theorem \cite{Coleman:1967ad} by introducing \textit{fermionic} symmetry generators, which in layman's terms turn bosons into fermions and vice-versa. By the Haag–Łopuszański–Sohnius theorem \cite{Haag:1974qh}, this is the unique extension to the Poincar\'{e} algebra. Since we know that symmetries tend to make physics easier, it is not surprising that supersymmetry is an indispensable tool in high energy theory, regardless of how or whether it is realized in the real world. Some useful general introductions to supersymmetry in $d=4$ and its application to the real world are Terning's book \cite{Terning:2006bq}, Martin's periodically-updated lecture notes \cite{Martin:1997ns}, and Shih's video lectures \cite{shih_2014}, in roughly increasing order of friendliness to neophytes.

In a supersymmetric theory fields come in multiplets which include particles of different spins (so called `supermultiplets') all having the same mass and quantum numbers. We add fermionic generators $Q_\alpha$ and $Q^\dagger_{\dot{\alpha}}$, called supercharges, with the defining (anti)commutation relations
\begin{gather}\label{eqn:susyalg}
\left\lbrace Q_\alpha, Q^\dagger_{\dot{\alpha}}\right\rbrace = 2 \sigma^\mu_{\alpha \dot{\alpha}} P_\mu \quad \left\lbrace Q_\alpha, Q_\beta\right\rbrace = 0 = \left\lbrace Q^\dagger_{\dot{\beta}}, Q^\dagger_{\dot{\alpha}}\right\rbrace \\
\left[ Q_\alpha, P_\mu \right] = 0 = \left[P_\mu, Q^\dagger_{\dot{\alpha}}\right]
\end{gather}
where $P_\mu$ is the generator of spacetime translations and $\sigma^\mu = (1,\vec{\sigma^i})$ with $\sigma^i$ the Pauli matrices. These may be determined simply by writing down all objects with the correct index structure. For later use, recall that spinor indices are raised/lowered with the invariant antisymmetric symbols $\epsilon_{\alpha \beta}, \epsilon_{\dot{\alpha} \dot{\beta}}$, as used for example in defining the conjugate invariants $(\overline{\sigma}^\mu)^{\dot{\alpha} \alpha} \equiv \sigma^\mu_{\beta \dot{\beta}} \epsilon^{\alpha \beta} \epsilon^{\dot{\alpha} \dot{\beta}}$.

We want to find irreducible representations of the supersymmetry algebra, called supermultiplets. Since $P_\mu P^\mu = m^2$ commutes with the generators $Q_\alpha$, $Q^\dagger_{\dot{\alpha}}$, the different particles in a supermultiplet will have the same mass. As an example of how to generate supermultiplet states, consider a massive particle. We can go to its rest frame, where it has momentum $P_\mu = (m,0,0,0)$ with $m$ its mass. Then the supersymmetry algebra greatly simplifies to $\left\lbrace Q_\alpha, Q^\dagger_{\dot{\alpha}}\right\rbrace = 2 m \mathds{1}_{\alpha \dot{\alpha}}$, and we see that this is just a Clifford algebra of raising and lowering operators. We define a lowest weight state, or Clifford vacuum $\ket{\Omega_s}$ such that it is annihilated by the undotted generators 
\begin{gather}
\ket{\Omega_s} = Q_1 Q_2 \ket{m,s,s_z} \\
\Rightarrow Q_1 \ket{\Omega_s} = 0 = Q_2 \ket{\Omega_s}
\end{gather}
Now we can use the dotted generators as raising operators to generate the entire multiplet.
\begin{gather}
\ket{\Omega_s} \\ 
Q^\dagger_{\dot{1}}\ket{\Omega_s}, Q^\dagger_{\dot{2}}\ket{\Omega_s} \\
Q^\dagger_{\dot{1}}Q^\dagger_{\dot{2}}\ket{\Omega_s}
\end{gather}
A single fermionic supersymmetry generator must change the spin of a state by $\frac{1}{2}$. Starting at the top with a spin $j$ particle gives us states of spin $j - \frac{1}{2}$ and $j + \frac{1}{2}$ on the middle line, and another state of spin $j$ on the bottom line. In $d=4$ the supermultiplet formed from a vacuum state of spin $\frac{1}{2}$ is called a `vector multiplet' and contains four states with spins $(0, \frac{1}{2}, \frac{1}{2}, 1)$. That formed from spin 0 is called a `chiral multiplet', and has states with spins $(0,0,\frac{1}{2})$. Note that this is fewer degrees of freedom, since negative spins are not allowed. Beginning with spins higher than $\frac{1}{2}$ leads to states with spins greater than $1$, which will take us into supergravity and will not be necessary for our purposes.

We could repeat this exercise for massless supermultiplets, labelling states by their energy and helicity $\ket{E,\lambda}$. We would find a Clifford algebra with only one set of raising/lowering operators, and find supermultiplets with helicities $\lambda$ and $\lambda + \frac{1}{2}$ for some starting $\lambda$. Then CPT invariance would force us to add states of helicity $-\lambda$ and $-\lambda - \frac{1}{2}$.

Merely from the definition of the symmetry group there are already a few interesting immediate results. For a start, we show that physical states have nonnegative energy in a supersymmetric theory, and the vacuum energy is an order parameter for supersymmetry breaking. First, let's give a simple expression for the Hamiltonian operator of supersymmetry. We act on our anticommutation relation with $(\overline{\sigma}^\nu)^{\dot{\alpha}\alpha}$ and recall various identities to note that  $\sigma^\mu_{\alpha \dot{\alpha}} (\overline{\sigma}^\nu)^{\dot{\alpha}\alpha} = 2 \eta^{\mu\nu}$, which gives us
\begin{align*}
4 P^\nu = (\overline{\sigma}^\nu)^{\dot{\alpha}\alpha} \left\lbrace Q_\alpha, Q^\dagger_{\dot{\alpha}}\right\rbrace \\
\Rightarrow 4P^0 = 4H = \mathds{1}^{\dot{\alpha}\alpha} \left\lbrace Q_\alpha, Q^\dagger_{\dot{\alpha}}\right\rbrace \\
4 H = Q_1 Q^\dagger_{\dot{1}} + Q^\dagger_{\dot{1}} Q_1 + Q_2 Q^\dagger_{\dot{2}} + Q^\dagger_{\dot{2}} Q_2,
\end{align*}
where we have used the fact that the zeroth component of the generator of spacetime translations is the generator of time translations, which is the Hamiltonian operator. Then we can write the energy of some state state $S$ as
\begin{equation}
\bra{S}H\ket{S} = \frac{1}{4} \left( \vert\vert Q_1 \ket{S} \vert\vert^2 + \vert\vert Q^\dagger_{\dot{1}} \ket{S} \vert\vert^2 + \vert\vert Q_2 \ket{S} \vert\vert^2 + \vert\vert Q^\dagger_{\dot{2}} \ket{S} \vert\vert^2\right) \geq 0,
\end{equation}
so the energy of $S$ is non-negative. Furthermore, consider a vacuum state $\ket{0}$ of our theory. In a standard QFT, the vacuum energy $\bra{0} H \ket{0} = E_\text{vac}$ is non-physical---we can just shift the Hamiltonian arbitrarily to remove it. But here, the supersymmetry algebra gives a preferred frame. If a vacuum state $\ket{0}$ is supersymmetric then it is annihilated by the supercharges $Q_\alpha \ket{0} = 0, Q^\dagger_{\dot{\alpha}} \ket{0} = 0$, otherwise the vacuum would not be invariant under supersymmetry transforms. This implies that it will have vanishing total energy $\bra{0} H \ket{0} = 0$. Conversely, if the vacuum state is non-supersymmetric, then its energy is strictly positive. We say supersymmetry is broken in such a state. Thus the vacuum energy acts as an order parameter for SUSY breaking. 

Connected to that fact is that each supermultiplet contains the same number of fermionic and bosonic degrees of freedom. We can see this by defining an operator $F$ which counts the fermion number of a state, so that bosonic states have eigenvalue $1$ under $(-1)^F$, and fermionic states have eigenvalue $-1$. Since the SUSY generators interchange bosonic and fermionic states, they must anticommute with $(-1)^F$. 

Now, for a given supermultiplet consider the states $\ket{a}$ with the same given four-momentum $p_\mu$, $p_0 = E \neq 0$. Since the supercharges commute with $P^\mu$, we know that these must form a complete set of states in this subspace $\sum\limits_{a} \ket{a}\bra{a} = 1$. Now consider the trace of the weighted energy operator $(-1)^F H/4$.
\begin{align}
\sum\limits_{a} \bra{a} (-1)^F H \ket{a} &= \sum\limits_{a} \bra{a} (-1)^F QQ^\dagger \ket{a} + \sum\limits_{a} \bra{a} (-1)^F Q^\dagger Q \ket{a} \nonumber \\
&= \sum\limits_{a} \bra{a} (-1)^F QQ^\dagger \ket{a} + \sum\limits_{a} \sum\limits_{b} \bra{a}  (-1)^F Q^\dagger\ket{b}\bra{b}Q \ket{a} \nonumber \\
&= \sum\limits_{a} \bra{a} (-1)^F QQ^\dagger \ket{a} + \sum\limits_{b} \bra{b} Q(-1)^FQ^\dagger\ket{b} \nonumber \\
&= \sum\limits_{a} \bra{a} (-1)^F QQ^\dagger \ket{a} - \sum\limits_{b} \bra{b} (-1)^F QQ^\dagger\ket{b} \nonumber \\
&= 0
\end{align}
where we have suppressed the contracted spinorial indices. This implies that the number of bosonic degrees of freedom is the same as the number of fermionic degrees of freedom in our supermultiplet, which we found to be true in the example we considered above.

There is a beautiful formalism of `superspace' which can be used to make supersymmetric theories far more transparent, but introducing this would be too large of a digression for our purposes.\footnote{Martin's notes \cite{Martin:1997ns} serve as a good introduction to traditional `off-shell' superspace for $\mathcal{N}=1, d=4$ theories, and Thaler's TASI lecture notes \cite{Bertolini:2013via} are also a fantastic resource. There is a related but distinct formalism of `on-shell' superspace, which falls under the heading of the amplitudes/on-shell/S-matrix program. This was first introduced very early on by Nair \cite{Nair:1988bq} and was used to great effect by Arkani-Hamed, Cachazo, \& Kaplan \cite{ArkaniHamed:2008gz} much later. A pedagogical introduction to on-shell techniques including superspace can be found in the textbook by Elvang \& Huang \cite{Elvang:2015rqa}. Until recently, the on-shell program was mostly restricted to massless particles. As it so happens, after Arkani-Hamed, Huang, \& Huang \cite{Arkani-Hamed:2017jhn} introduced a beautiful extension of the formalism to include massive particles, it was Timothy Trott, my undergrad mentee Aidan Herderschee, and myself who formulated an extension of the on-shell superspace formalism for massive particles \cite{Herderschee:2019ofc}. The on-shell program is another fascinating line of work that I suggest any aspiring particle or field theorist learn about.} We simply want to see the effects of supersymmetry on (in)sensitivity of low-energy physics to the ultraviolet, for which studying a simple theory of chiral superfields will do. The Wess-Zumino model is the simplest such example which is not free, consisting of a single self-interacting chiral supermultiplet, and was historically the first non-trivial four-dimensional theory proved to be supersymmetric.

We may write down the Wess-Zumino Lagrangian as 
\begin{equation}\label{eqn:WZLag}
\mathcal{L} = \int \text{d}^4x \left(-\partial_\mu\phi^\star\partial^\mu\phi - m^2 \phi^\star \phi - i \psi^\dagger \bar{\sigma}^\mu \partial_\mu \psi - \half m \psi \psi - \half y \phi \psi \psi - \half y m \phi^2 \phi^\star - \frac{1}{4} |y^2| \phi \phi \phi^\star \phi^\star \right),
\end{equation}
where for compactness we've left off the Hermitian conjugate terms.	To avoid introducing certain technical complications we eschew the proof that this is indeed invariant under a supersymmetric transformation and instead evince its UV insensitivity. For fun we'll compute without assuming the masses are the same $m_\phi \neq m_\psi$, as can happen in the presence of soft breaking of the symmetry.

Let's look at the vacuum energy, which we'll calculate generally but schematically. A quantum harmonic oscillator has ground state energy $\pm \half \hbar \omega$ for bosonic and fermionic states respectively, with the sign being familiar from the Casimir effect. If we consider a box of side length $V^{1/D}$, the energy of the fields inside it is
\begin{equation}
E_0 = \sum\limits_{\vec{k}}^{\text{bosons}} \half \hbar \omega_{\vec{k}} - \sum\limits_{\vec{k}}^{\text{fermions}} \half \hbar \omega_{\vec{k}} ,
\end{equation}
where $\vec{k} = (k_1, k_2, \dots, k_D) (2 \pi /V^{1/D})$, $k_i \in \mathbb{Z}$. Now in QFT each mode has energy $\omega_{\vec{k}} = \sqrt{k^2 + m^2}$, and as we make the box bigger $V^{1/D}\rightarrow \infty$, the sum turns into an integral
\begin{equation}
E_0 = V\int\limits \frac{d^Dk}{(2\pi)^D} \left( \half \sqrt{k^2 + m_B^2} - \half \sqrt{k^2 + m_F^2} \right),
\end{equation}
with $m_B$ a boson mass and $m_F$ a fermion mass, where the sum over species is implicit. We also recognize $E_0/V$ as the vacuum energy density, denoted $\Lambda$, which we can write as
\begin{equation}
\Lambda \simeq \frac{1}{2 (2\pi)^D} \int\limits d^Dk \left(\sqrt{k^2 + m_B^2} - \sqrt{k^2 + m_F^2}\right).
\end{equation}
Now if we specialize to $D=4$ and introduce a cutoff $k_{max}$ up to which we're confident that our description of particle physics holds, the schematic form is simply
\begin{equation}
\Lambda \sim k_{max}^4 \left(\sum\limits^{\text{bosons}} 1 - \sum\limits^{\text{fermions}} 1\right) + k_{max}^2 \left(\sum\limits^{\text{bosons}} m_B^2 - \sum\limits^{\text{fermions}} m_F^2 \right) + \dots
\end{equation}
Now we see quite generally and explicitly that in a supersymmetric state the vacuum energy vanishes, since there are equal numbers of bosonic and fermionic fields with degenerate masses. Furthermore, spontaneous breaking of supersymmetry breaks the degeneracy but does not change the numbers of fields, so softly broken supersymmetry retains protection from the largest contribution. 

\begin{figure}
	\centering
	\includegraphics[width=1\linewidth]{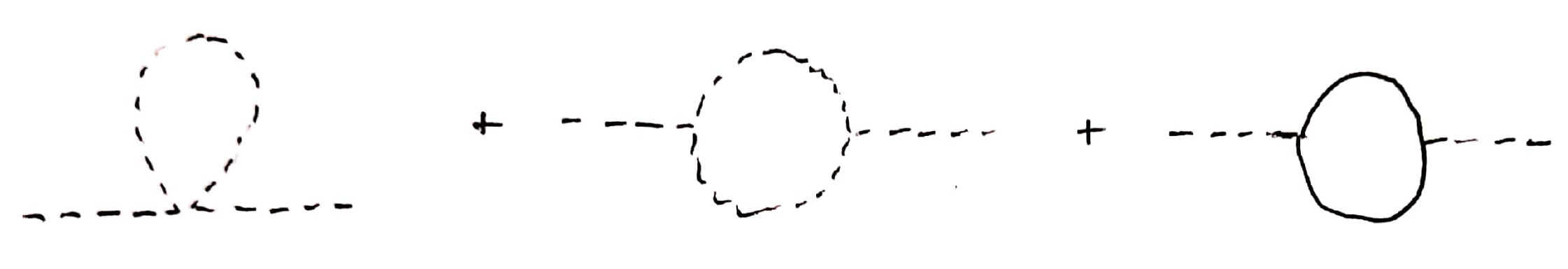}
	\caption{The one-loop diagrams contributing to the scalar mass correction in the Wess-Zumino model.}
	\label{fig:wzdiags}
\end{figure}

Let's look now more sharply at the one-loop contributions to the scalar mass in the Wess-Zumino model, regularized with a hard cutoff $\Lambda$. Take care that we've written the Lagrangian in terms of two-component spinors, an exhaustive guide to which can be found in \cite{Dreiner:2008tw}. The three diagrams are shown in Figure \ref{fig:wzdiags}, and their evaluation proceeds as
\begin{align}
-i \delta m^2 &= (-i m_\psi y)^2 \int \frac{d^4k}{(2\pi)^4} \frac{(-i)^2}{(k^2 + m_\phi^2)^2} + (-i y^2) \int \frac{d^4k}{(2\pi)^4} \frac{(-i)}{(k^2 + m_\phi^2)} \nonumber \\&+ (-1) \half (-i y)^2 \int \frac{d^4k}{(2\pi)^4} \frac{(-i)^2\text{Tr}\left[\sigma^\mu k_\mu \bar \sigma^\nu k_\nu \right]}{(k^2 + m_\psi^2)^2} \\
&= i |y|^2 \int \frac{d^4\bar k}{(2\pi)^4} \left[\frac{m_\psi^2}{(\bar k^2 + m_\phi^2)^2} - \frac{1}{(\bar k^2 + m_\phi^2)} +\frac{\bar k^2}{(\bar k^2 + m_\psi^2)^2}\right] \\
&= i \frac{|y|^2}{16 \pi^2} \int_0^\Lambda d\bar k \ \bar k^3 \left[\frac{m_\psi^2-m_\phi^2 - \bar k^2}{(\bar k^2 + m_\phi^2)^2} +\frac{\bar k^2}{(\bar k^2 + m_\psi^2)^2}\right] \\
&= i \frac{|y|^2}{16 \pi^2} (m_\phi^2 - m_\psi^2) \log \Lambda + \text{ finite}.
\end{align}
We see that the UV sensitivity of the scalar mass in this theory has disappeared, even if the two fields have different masses. In the limit of unbroken supersymmetry, the contribution vanishes identically.

This fact of removing the UV sensitivity of the mass of a scalar persists generally, no matter which other superfields are added, so long as supersymmetry is at most softly broken. The connection to the hierarchy problem is clear: If, in the UV, all fields come in supermultiplets, then the Higgs mass is protected from UV contributions. 

Of course we do not observe mass-degenerate superpartners, so this soft supersymmetry breaking is a necessary feature of any implementation of supersymmetry to the real world. The Minimal Supersymmetric Standard Model \cite{Dimopoulos:1981zb} embeds each of our fermions in a chiral supermultiplet and each gauge boson in a vector multiplet. The Higgs sector must be enlarged to two chiral multiplets containing the up and down Yukawas respectively, as is necessary for anomaly cancellation\footnote{Perhaps the more urgent reason for needing two Higgs multiplets is that the interactions in supersymmetric theories are highly constrained by `holomorphy', a full explanation of which here would require too much machinery but which leads to the conclusion that the same multiplet cannot have Yukawa interactions with both the up- and down-type quarks. However see \cite{Davies:2011mp} for the interesting possibility that at high energies only the up-type Yukawa interactions exist, and the down-type and charged lepton masses are induced by supersymmetry-breaking.}.

The question of supersymmetry-breaking is a very non-trivial one. At the level of a phenomenological accounting of possible soft breaking terms in the MSSM, there are 105 physical parameters \cite{Chung:2003fi}. However, constraints on flavor-violating couplings and on CP violation tell us empirically that the soft terms that appear must be very non-generic. In fact, looking at the MSSM in detail it turns out there are no places for supersymmetry-breaking to enter directly, and indeed there are general arguments that such breaking must take place in another, hidden sector and be indirectly communicated to the MSSM fields (see e.g. Martin's Section 7.4 \cite{Martin:1997ns}).

The origins of supersymmetry-breaking being a separate sector does force us to expand our model of particle physics, but on the upshot this sequestering means we can explore interesting phenomenology in sectors which are unconstrained. One can write down models where supersymmetry breaking is mediated by supergravity effects \cite{Chamseddine:1982jx,Barbieri:1982eh,Ibanez:1982ee,Hall:1983iz,Ohta:1982wn,Ellis:1982wr,AlvarezGaume:1983gj}, communicated to the SM fields by our gauge bosons from a sector with new, massive SM-charged particles \cite{Dine:1981gu,Nappi:1982hm,AlvarezGaume:1981wy,Dine:1993yw,Dine:1994vc,Dine:1995ag}, or takes place at a physically separate location in an extra dimension \cite{Mirabelli:1997aj,Kaplan:1999ac,Chacko:1999mi,Schmaltz:2000ei,Schmaltz:2000gy,Csaki:2001em,Cheng:2001an,Randall:1998uk,Giudice:1998ck,Bagger:1999rd}, for a few examples.
A full discussion of the mechanisms and strategies for models of supersymmetry-breaking is beyond our scope, but we highly recommend Intriligator \& Seiberg's lecture notes \cite{Intriligator:2007cp} as a general reference along with Martin's notes \cite{Martin:1997ns}.

Of course we would like this phase transition to originate as spontaneous symmetry breaking, rather than explicitly putting it in by hand, since we want the far UV to be supersymmetric. Such spontaneous breaking requires the generation of a scale, and so it would be great if such a scale were generated \textit{dynamically}, as in the dimensional transmutation we saw in QCD in Section \ref{sec:dimtrans}. This would then be a natural mechanism for SUSY breaking. This phenomenological prospect lead to and benefited from a fantastic body of work understanding the details of supersymmetric gauge theories e.g. \cite{Witten:1981nf,Affleck:1983mk,Affleck:1983rr,Affleck:1984uz,Affleck:1984xz,Seiberg:1994bz,Intriligator:1995au,Shifman:1995ua,Peskin:1997qi,Poppitz:1998vd,Intriligator:2006dd}, which can be found reviewed in textbooks by Terning \cite{Terning:2006bq} and Shifman \cite{Shifman:2012zz} and in TASI notes by Strassler \cite{Strassler:2003qg}.

We stated above that fields in a given supermultiplet share all the same quantum numbers, but there is in fact one exception. The supersymmetry algebra in Equation \ref{eqn:susyalg} is invariant under opposite rephasings of the supercharges $Q_\alpha \rightarrow Q_\alpha e^{-i\alpha}$, $Q^\dagger_{\dot{\alpha}} \rightarrow Q^\dagger_{\dot{\alpha}} e^{i\alpha}$. So there is a generator of a global internal symmetry that we may add that has nontrivial commutation relations with the supercharges: 
\begin{equation}
\left[R,Q\right] = -Q, \qquad \left[R,Q^\dagger\right] = Q^\dagger.
\end{equation}
This generator is known as an $R$-symmetry generator, and if the theory is invariant under an $R$-symmetry that means that each supermultiplet $\Phi$ can be assigned an $R$-charge $r_\Phi$ and the theory is invariant under transformations of each multiplet $\Phi \rightarrow \Phi e^{i r_\Phi \alpha}$, schematically, where $\Phi$ is the collection of fields in that multiplet. Since $R$ does not commute with the supercharges, the different fields in the supermultiplet have different $R$-charges. For example if $\Phi$ is a chiral superfield consisting of $(\phi, \psi)$ then under a global rotation by $\alpha$ they transform as
\begin{equation}
\phi \rightarrow \phi e^{i r_\Phi \alpha}, \qquad  \psi \rightarrow \psi e^{i (r_\Phi-1) \alpha},
\end{equation} 
which is simply because $\ket{\psi} \sim Q \ket{\phi}$. 	

It's important to emphasize that this $R$-symmetry is not part of the supersymmetry algebra, so one may have supersymmetric theories which do or do not implement $R$-symmetry. Nelson \& Seiberg showed a fascinating connection between $R$-symmetry and supersymmetry-breaking \cite{Nelson:1993nf}. They show roughly that for a low-energy Wess-Zumino model (possibly after having integrated out confined strong dynamics) as long as one has a `generic' potential (in the sense that a generic set of $n$ equations in $n$ unknowns has a solution), then a vacuum spontaneously breaks supersymmetry if and only if it spontaneously breaks $R$-symmetry. The reasoning is simply that such a symmetry imposes an additional constraint on the potential minimization equations, leading to a solution no longer being generically present.

For later use we mention the possibility of `extended supersymmetry', where additional supercharges are added
\begin{equation}
\left\lbrace Q^A_\alpha, Q^{\dagger B}_{\dot{\alpha}}\right\rbrace = 2 \sigma^\mu_{\alpha \dot{\alpha}} P_\mu, \qquad \left\lbrace Q^A_\alpha, Q^B_\beta\right\rbrace = 0 = \left\lbrace Q^{\dagger A}_{\dot{\beta}}, Q^{\dagger B}_{\dot{\alpha}}\right\rbrace
\end{equation}
where $A,B = 1..\mathcal{N}$ index the supercharges. The construction of supermultiplets proceeds as before, but there are now more nonvanishing combinations of supercharges to act on the Clifford vacuum, so supermultiplets are enlarged. In four dimensions, the most supersymmetry we can have without gravity is $\mathcal{N}=4$, which contains enough supercharges to relate the helicity $-1$ vector all the way to the helicity $+1$ vector; any more supercharges would necessarily yield particles of spins $3/2, 2$. If we are willing to include these degrees of freedom we can only go up to $\mathcal{N}=8$ `supergravity' (SUGRA), as more charges would lead to a theory with fundamental particles of spins $s>2$ which is pathological\footnote{There's an important exception here, which is a theory which includes particles of \textit{all} spins. This is necessary for string theory to operate, but has also led to the formulation of novel field theories commonly called `Vasiliev gravity' \cite{Vasiliev:1990en}.}. The classic references for supergravity are Wess \& Bagger \cite{Wess:1992cp} and Freedman \& Van Proeyen \cite{Freedman:2012zz}. Ultimately if supersymmetry is a field-theoretic feature of the ultraviolet of our universe we must have SUGRA as well, as it simply results from applying the supercharges to the graviton field, but we won't discuss SUGRA any further.

When we enlarge our superalgebra we also enlarge the (potential) R-symmetry group---the group of symmetries which do not commute with the supercharges---since we can now shuffle around the supercharges in addition to rephasing them. We'll revisit this in Section \ref{sec:Orbifold} in the context of utilizing non-trivial R-symmetry representations to break supersymmetry.

\section{Extra Dimensions}\label{sec:extradims}

\setlength{\epigraphwidth}{0.5\textwidth}
\epigraph{I exist in the hope that these memoirs, in some manner, I know not how, may find their way to the minds of humanity in Some Dimension, and may stir up a race of rebels who shall refuse to be confined to limited Dimensionality.}{Edwin Abbott Abbott, \\ \textit{Flatland}, 1884 \cite{abbott1885flatland}}
\setlength{\epigraphwidth}{0.6\textwidth}

One of the most important ideas in theoretical physics developed in the latter part of the $20^\text{th}$ century is that there may be additional spatial dimensions past the familiar three of our everyday experiences. Theories in which additional spatial dimensions are present were first studied in the context of unifying gravity and electromagnetism, first by Nordstr\"{o}m \cite{Nordstrom:1988fi} (\textit{before} General Relativity!) and then by Kaluza \cite{Kaluza:1921tu} and Klein \cite{Klein:1926tv}. These ideas saw a resurgence of interest some half-century later with the advent of string theory, and the vision of all features of the universe being fundamentally geometrized. Against that backdrop, it is clearly prudent to consider the interplay of such theories with the puzzle of the hierarchy problem. As we shall see, extra dimensional theories can produce terribly interesting physics, and the possibilities are \textit{manifold}.

\subsection{Technology: Kaluza-Klein Reduction}

To consider the possibility that there are additional microscopic dimensions, we develop a picture of the effects of fundamentally $D$-dimensional fields where $D = 4 + d$. On our manifold $M = \mathcal{M}_4 \times K$, where $K$ is compact, we write down an action
\begin{equation}
S = \int_{\mathcal{M}_4}\int_K \sqrt{- \mathbf{g_{(4+d)}}}\mathcal{L}(\boldsymbol{\phi},\mathbf{A}^M,\mathbf{g}^{MN})
\end{equation}
where our fields are in irreducible representations of the $D$-dimensional Poincar\'{e} algebra, and the Lagrangian manifestly obeys $D$-dimensional Lorentz invariance. We'll use boldface for $D$-dimensional fields and Latin letters for $D$-dimensional Lorentz indices. However, since $K$ is compact, it places constraints on the mode expansions of our fields. Then when we want to study the effective four-dimensional theory we first need to decompose our fields into irreducible representations of the $4$-dimensional Poincar\'{e} algebra. The $D$-dimensional vectors and tensors will become multiplets of $4$-dimensional fields. Then we can explicitly integrate over the compact manifold $K$, which will produce a tower of states. 

As an easy, explicit example, take the compact manifold to be the circle $S^1$ of length $2\pi R$, and consider a single complex scalar. We start with 
\begin{equation}
S_5 = \int \text{d}^4x \int_{0}^{2\pi R} \text{d}y \left[-\half \partial_M \boldsymbol{\phi}^* \partial^M \boldsymbol{\phi} - \half m^2 \boldsymbol{\phi}^* \boldsymbol{\phi}\right]
\end{equation}
From introductory quantum mechanics, we know that the boundary conditions of being single-valued around the circle constrains the mode expansion of $\boldsymbol{\phi}$. We can write
\begin{equation}
\boldsymbol{\phi}(x^\mu,y) = \frac{1}{\sqrt{2 \pi R}} \sum\limits_{n=-\infty}^{+\infty} \phi_n(x^\mu) e^{i n y/R}
\end{equation}
where the normalization will produce canonically normalized kinetic terms in the $4$-dimensional action, and should generally be $\sqrt{\text{Vol}(K)}$. Plugging this into the action gives 
\begin{equation}
S_5 = \frac{1}{2 \pi R} \int \text{d}^4x \sum\limits_{mn} \int_{0}^{2\pi R} \left[-\half \partial^\mu \phi^*_m \partial_\mu \phi_n  - \half \left(\frac{-im}{R}\right)\left(\frac{in}{R}\right) \phi^*_m \phi_n - \half m^2 \phi^*_m \phi_n \right] e^{i(n-m)y/R} 
\end{equation}
where the first and second terms come from either the $\mathcal{M}_4$ derivatives or the $K$ derivatives acting on the mode expansion. The integral over the compact direction now gives, using orthogonality $\int_{0}^{2\pi R} \text{d}y e^{i(n-m)y/R} = 2 \pi R \delta_{nm}$, a $4$-dimensional action

\begin{figure}
	\centering
	\includegraphics[width=0.5\linewidth]{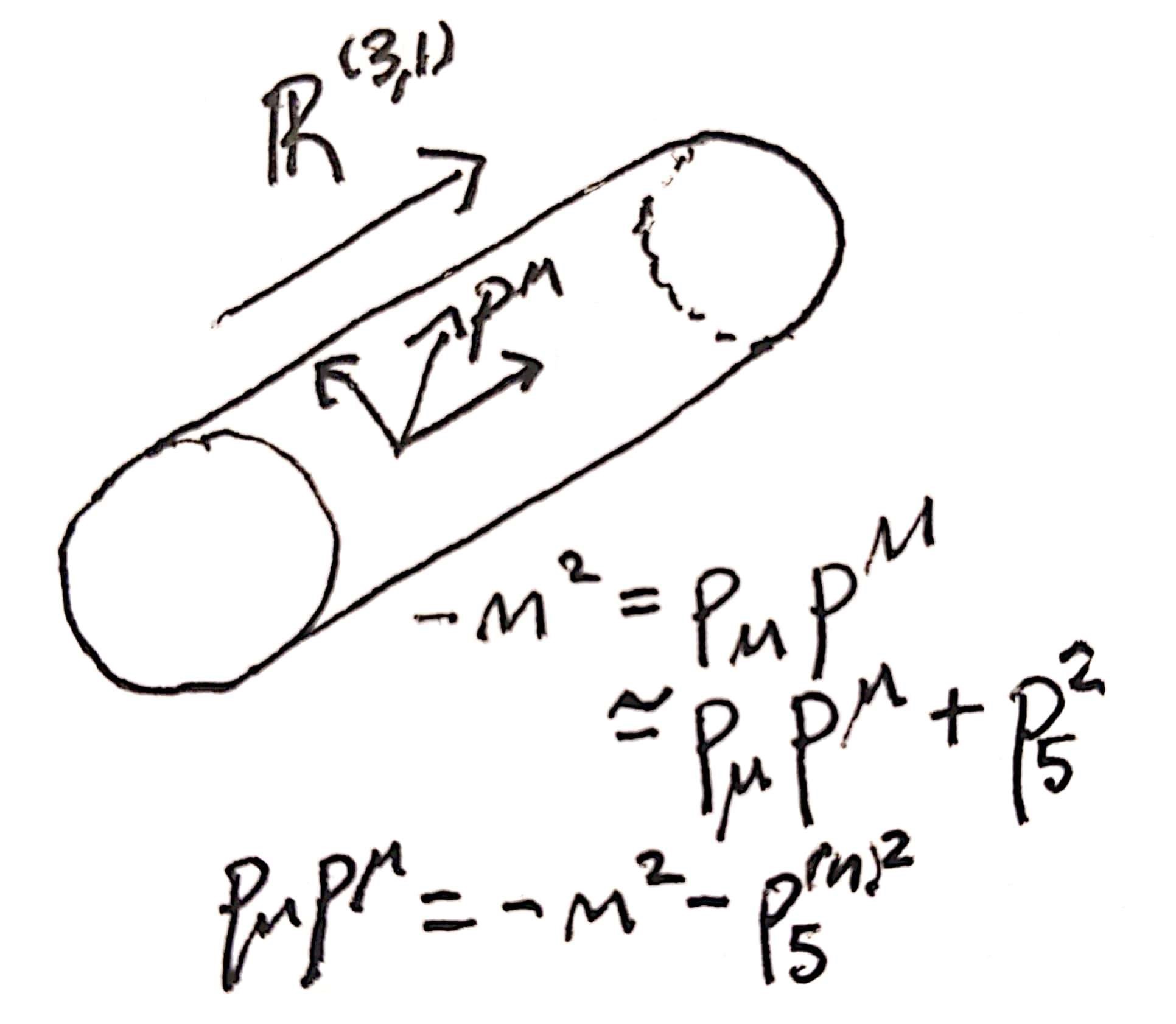}
	\caption{A schematic Kaluza-Klein spacetime for intuition on the appearance of a tower of four-dimensional `Kaluza-Klein modes'. The compact extra dimension demands quantized momentum along the fifth dimension. For an observer on large scales, the extra component of momentum appears as a four-dimensional mass term.}
	\label{fig:kaluzaklein}
\end{figure}

\begin{equation}
S_4 = \int \text{d}^4x \sum\limits_{n} \left[-\half \partial^\mu \phi^*_n \partial_\mu \phi_n - \half \left(m^2 + \frac{n^2}{R^2}\right) \phi^*_n \phi_n  \right] 
\end{equation}
We see that we now have a \emph{tower} of $4$-dimensional states that have arisen from the one $5$-dimensional scalar, as shown schematically in Figure \ref{fig:kaluzaklein}. There is a `zero-mode' which has a mass given by the $5$-dimensional mass term, and then there are states with larger masses. If $R$ is small, these will be heavy, and so we would only see them at, say, a high energy particle collider. Note that all of these higher levels are doubly-degenerate, since $n$ ranges over all the integers.

If we had included the graviton in this compactification procedure, we would see the $5$-dimensional graviton split up into 
\[\mathbf{g}^{MN} = \left(\begin{array}{@{}cccc|c@{}}
g^{\mu\nu} &  & \cdots &  & g^{\mu 4} \\
&  &  & &  \\
\vdots &  & \ddots & & \vdots \\
&  &  &  &  \\\hline
g^{4\nu} &  & \cdots &  & g^{44} 
\end{array}\right)\]
so we see the emergence of a $4$-dimensional scalar, vector, and two-index symmetric tensor. And it was in this context that compactification was originally studied as a unification of general relativity and electromagnetism.

There's another very important effect of the compactification. Let's consider the Einstein Hilbert action
\begin{equation}
S = \frac{1}{16 \pi G_N^{(4+d)}} \int_{\mathcal{M}_4\times K} R \sqrt{-\mathbf{g^{(4+d)}}}  \text{d}^dx
\end{equation}
Noting that the Ricci scalar $R \sim \partial^2$ always has mass dimension $\left[R\right]=2$, we see that the $(4+d)$-dimensional Newton's constant must have $\left[G_N^{(4+d)}\right] = 2-D = -2-d$, so to rewrite the action in natural units we must define a $(4+d)$-dimensional Planck mass as $1/G_N^{(4+d)} = (M_{pl}^{(4+d)})^{2+d}$. If we take the metric to be independent of the $K$ coordinates, then the integration over $K$ just gives a factor of the volume $V_K$ of the compact space
\begin{equation}
S = \frac{(M_{pl}^{(4+d)})^{2+d}}{16 \pi} V_K \int_{\mathcal{M}_4} R \sqrt{-g^{(4)}} \text{d}^4x
\end{equation}
Putting this into the form of the four-dimensional Einstein-Hilbert action, we find $M_\pl^{2} = (M_{pl}^{(4+d)})^{2+d} V_K$. Taking $V_K \simeq R^d$ for some radius $R$, we see that if the compact dimensions are not Planck-sized $R \sim 1/M_\pl$ but larger for whatever reason, then the effective Planck mass at long distances can be much larger than the fundamental Planck mass.

Now we don't see zero-mode, different spin partners of our particles which fill out representations of the higher-dimensional symmetry, so this simple compactification cannot be the real way the world is. If we have compact dimensions, they must be such that the symmetry of zero-modes is somehow broken for the SM fields. This is important---we don't just want to have different $5d$ fields with zero modes or not, we need different four-dimensional components of them to have or not have zero modes. We'll solve this problem in Section \ref{sec:Orbifold}.

\subsection{Quantum Gravity at the TeV Scale}\label{sec:led}

In the previous section we noticed that extra dimensions dilute the fundamental Planck mass in the higher-dimensional theory to produce a weaker effective four-dimensional Planck mass. So perhaps we can fix the hierarchy between the electroweak scale and Planck scale by lowering the fundamental scale of quantum gravity. Arkani-Hamed, Dimopoulos, and Dvali proposed that the fundamental Planck scale can be $M_{pl}^{(4+d)} \sim \text{ TeV}$ with the weakness of four-dimensional gravity resulting from the dilution of gravitational flux into the extra $d$ dimensions \cite{ArkaniHamed:1998rs}. Using 
\begin{equation}
\left(1 \text{ TeV}\right)^{(2+d)/2} \simeq (M_{pl}^{(4+d)})^{(2+d)/2} \simeq M_{pl}^{(4)}/R^{d/2}
\end{equation}
where $R$ is the radius of the extra dimensions, we find that a parsimonious $d=3$ spherical dimensions of radius $R \simeq 1 \text{ nm}$ suffice to remove the hierarchy problem and accord with E\"{o}t-Wash constraints on the behavior of gravity at scales down to $\sim \mathcal{O}(\mu\text{m})$ \cite{Hoyle:2000cv, Adelberger:2009zz, Lee:2020zjt}. 

Now a nanometer is tiny compared to human scales, but the associated energy scale is $1/R \simeq 100 \text{ eV}$, which is a scale we have quite a bit of information about. In particular, if the Standard Model fields were to propagate in all $4+d$ dimensions then it would be easy to excite the `winding modes' with momentum in the extra dimensions, and we should have observed many finely-space Kaluza-Klein resonances with the same quantum numbers. This is obviously not how the universe works, so to dilute gravity with large extra dimensions one must trap Standard Model fields to the four-dimensional manifold we know and love.

This can be done by imagining we live on a $(3 + 1)$-dimensional topological defect which is embedded in the larger $(4+d)$-dimensional space. `Topological defect' sounds exotic, but these are just (semi-)familiar non-perturbative objects such as the branes of string theory \cite{Polchinski:1996fm,Polchinski:1996na,Johnson:2000ch}, or the cosmic strings or domain walls that can appear in Higgsed gauge theories and which are introduced well in Shifman's textbook \cite{Shifman:2012zz}. The original proposal suggests a weak-scale vortex in which zero-modes of our familiar fields are trapped, which is super cool.

Note that the new physics appearing at the TeV scale in this scenario is about as violent as you could imagine: quantum gravity appears at a TeV! This leads to a variety of fascinating signatures and constraints, even in the absence of concrete model of quantum gravity, though embeddings into string theory have also been found \cite{Antoniadis:1998ig,Shiu:1998pa,Ibanez:2001nd}. Very generically there are corrections to the Newtonian gravitational laws \cite{ArkaniHamed:1998nn}, all sorts of effects on precision observables \cite{Rizzo:1999br}, a loss of flux of high energy particles into the ambient space either astrophysically \cite{Cullen:1999hc} or at a collider \cite{Giudice:1998ck},  violations of the global symmetries of the SM since quantum gravity does not respect them \cite{Dienes:1998vg,Dienes:1998vh,ArkaniHamed:1999dc}, and production of Kaluza-Klein gravitons \cite{Hewett:1998sn} and black holes at TeV-scale colliders \cite{Argyres:1998qn,Emparan:2000rs,Giddings:2001bu,Eardley:2002re}. This is a fascinating field which is well worth studying in detail, but unfortunately we do not have the space to do it justice. For more detail we refer to the reviews by Rubakov \cite{Rubakov:2001kp} and Maartens \& Koyama \cite{Maartens:2010ar}, and the more introductory notes from Cs\'{a}ki \cite{Csaki:2004ay}, Kribs \cite{Kribs:2006mq}, P\'{e}rez-Lorenzana \cite{PerezLorenzana:2005iv}, Cheng \cite{Cheng:2010pt}, and Cs\'{a}ki, Hubisz, and Meade \cite{Csaki:2005vy}.

However, shrewd readers will be eager to point out that we \textit{haven't} actually solved the hierarchy problem; we've merely traded the $m_H \ll M_\pl$ hierarchy for a $1/R^{(2+d)/2} \ll M_\pl^{(4+d)}$ hierarchy. And indeed, it can be difficult to stabilize the size of the extra dimensions in this scheme. But the conceptual leap of considering geometric solutions to the hierarchy problem is incredibly important and leads to many further interesting directions. The next ingredient we need is to control the existence of opposite-spin partners.

\subsection{Technology: Orbifold Reduction}\label{sec:Orbifold}

Let us consider a spacetime manifold $M = \mathcal{M}_4 \times K_G$ with $\mathcal{M}_4$ four-dimensional Minkowski space and $K_G$ a $d$-dimensional compact `orbifold'. An orbifold is constructed by `modding out' a manifold $K$ by a discrete symmetry group $G$. A manifold is a space that looks locally like Euclidean space; an orbifold is one which locally looks like the quotient space of Euclidean space quotiented by a finite group. In layman's terms, quotienting or modding out is just identifying points which are transformed into each other under $G$---our space becomes the space of equivalence classes of $K$ under the action of elements of $G$. 

As an illustrative example, consider the line $\mathbb{R}$ and the action of the discrete symmetry $\mathbb{Z}_2: x \mapsto -x$. When we form the quotient space and identify points under this $\mathbb{Z}_2$ action, we find the half line $\mathbb{R}_{\geq 0}$. You'll notice that this space has a boundary at $x = 0$, which was a fixed point of the symmetry group. Mathematicians would say that the $\mathbb{Z}_2$ acts freely except at this point. This is a generic feature, and in fact what makes orbifolds interesting for our purposes. We could have also imagined `orbifolding' $\mathbb{R}$ by the translation $T(2\pi R)$, but this would have produced a compact manifold without boundary, $S^1$, because this symmetry has no fixed points. Part of the power of orbifolds comes because the quotient group structure gives us some information about what happens at fixed points - you could imagine just studying the half-line, but it seems natural to consider smooth structures on the manifold and then look at the effects under the identification. This should be evinced later in our examples.

\begin{figure}
	\centering
	\includegraphics[width=1.0\linewidth]{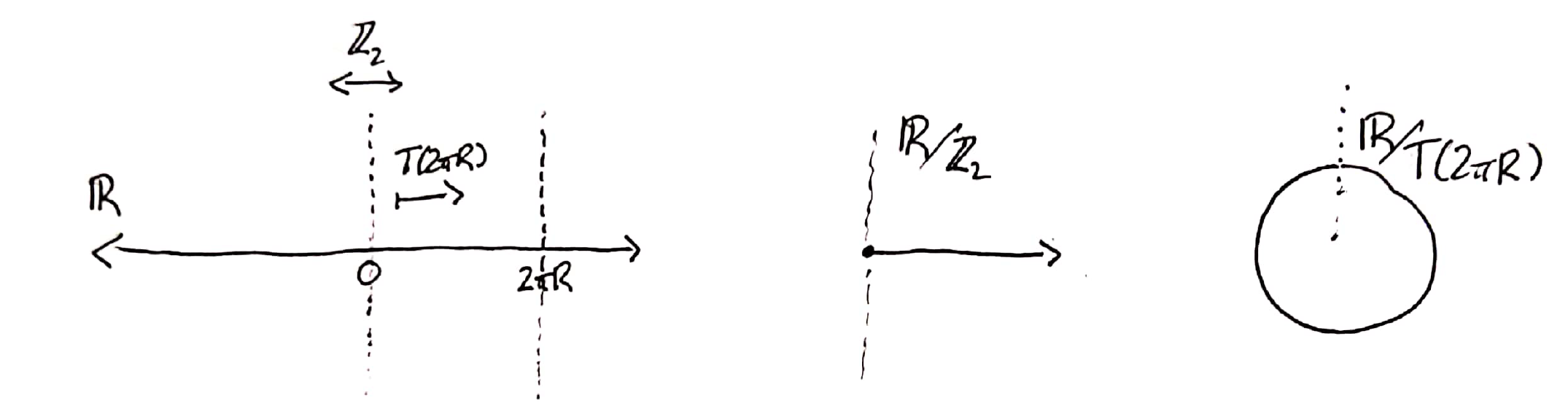}
	\caption{Schematic of the construction of an orbifold from the real line. The real line has a $\mathbb{Z}_2(0): x \mapsto -x$ symmetry and a translational $T(2 \pi R): x \mapsto x + 2 \pi R$ symmetry. Modding out by the former yields an orbifold because it has a fixed point, while modding out by the latter returns a regular manifold.}
	\label{fig:orbifoldbasic}
\end{figure}

\begin{figure}
	\centering
	\includegraphics[width=1.0\linewidth]{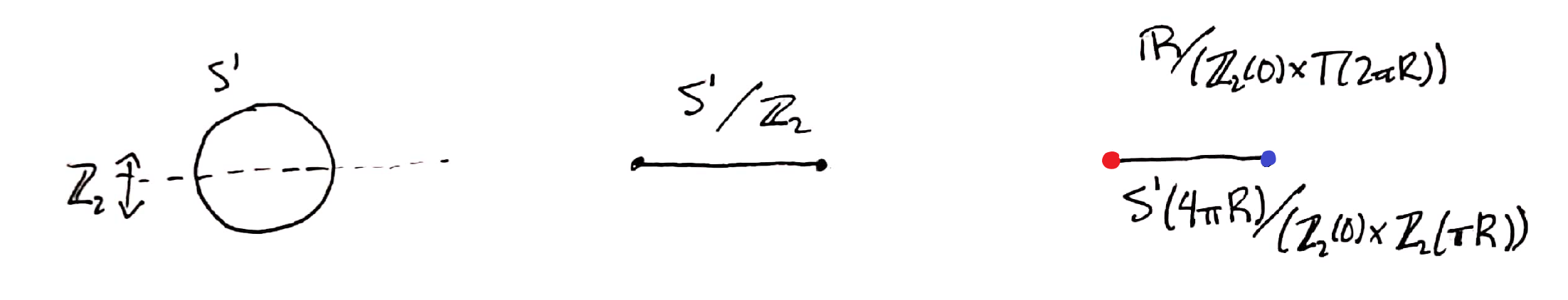}
	\caption{Schematic of the construction of an orbifold from a circle. Quotienting by $\mathbb{Z}_2$ corresponds to folding the circle over on itself. We can produce independent fixed points by uplifting first to the real line, or equivalently by folding the circle into quarters.}
	\label{fig:orbifoldcircle}
\end{figure}

Compactifying on orbifolds also gives us a way to cure our missing partner ills. Consider quotienting the circle by a $\mathbb{Z}_2$ which folds the circle over onto itself, $y \simeq 2 \pi R - y$. This produces a line segment with boundaries at both ends $y=0,\pi R$. We can alternatively think of this as modding out the real line by both translation and mirroring, producing $\mathbb{R}/T(2\pi R)\times \mathbb{Z}_2(0)$. There are then \emph{two} different sorts of fixed points---$y=0$ is fixed by $\mathbb{Z}_2(0)$, and $y=\pi R$ is fixed by $T(2\pi R)\mathbb{Z}_2(0)\sim \mathbb{Z}_2(\pi R)$. You can then envision this as the circle of length $4 \pi R$ with these two discrete identifications, that is $S^1(4 \pi R)/\mathbb{Z}_2(0)\times \mathbb{Z}_2(\pi R)$, as depicted in Figure \ref{fig:orbifoldcircle}.

How does this affect the resulting compactification? As a first step should take our results and rewrite them in terms of eigenfunctions of our $\mathbb{Z}_2$ symmetry. We write
\begin{equation}
\boldsymbol{\phi}(x^\mu,y) = \frac{1}{\sqrt{\pi R}} \left[\frac{1}{\sqrt{2}} \phi_0^{(+)}(x^\mu) +  \sum\limits_{n=1}^{+\infty} \phi^{(+)}_n(x^\mu) \cos n y/R + \sum\limits_{n=1}^{+\infty} \phi^{(-)}_n(x^\mu) \sin n y/R \right] 
\end{equation}
where all we have done is rearrange things by defining
\begin{equation}
\phi^{(+)}_0 \equiv \phi_0, \qquad \phi^{(+)}_{n>0}\equiv \frac{1}{\sqrt{2}}\left(\phi_n + \phi_{-n}\right), \qquad \phi^{(-)}_{n>0}\equiv \frac{i}{\sqrt{2}}\left(\phi_n - \phi_{-n}\right),
\end{equation}
where the superscript denotes their eigenvalue under the $\mathbb{Z}_2$. Now the way to change our $S^1$ compactification to an $S^1/\mathbb{Z}_2$ compactification is to impose in our $5$-dimensional action that $\boldsymbol{\phi}$ transform with definite parity, which must be the case if the $\mathbb{Z}_2$ is a good symmetry\footnote{Note that since we're getting our eigenmodes through this orbifold reduction our reduced action will still need to be produced by integrating from $y=0$ to $y=2 \pi R$ because it's this domain over which our eigenfunctions are orthonormal.}. You'll notice our free action does not demand a particular choice of parity for $\boldsymbol{\phi}$, so we are free to choose $\boldsymbol{\phi}$ either even or odd. But in either case we are forced to get rid of half of our states. If $\boldsymbol{\phi}$ is even we must set $\phi^{(-)}_n = 0$, and if $\boldsymbol{\phi}$ is odd we must set $\phi^{(+)}_n = 0$---including, importantly, getting rid of the zero mode. If we wanted to include certain interactions in the higher-dimensional theory, that could dictate the transformation of $\boldsymbol{\phi}$---for example, the interaction term  $\boldsymbol{\phi}^3$ would necessitate an even $\boldsymbol{\phi}$. This is just the same as we're familiar with in four dimensions.

More interestingly, let us consider the effect of the orbifold on larger Lorentz representations. Imagine a $5$-dimensional gauge field with free action 
\begin{equation}
S_5 = -\frac{1}{4} \int_M \mathbf{F}^{MN} \mathbf{F}_{MN}
\end{equation}
where $\mathbf{F}^{MN} = \partial^M \mathbf{A}^N - \partial^N \mathbf{A}^M$ is the field strength. The mixed terms here read $\mathbf{F}^{\mu 4} = \partial^\mu \mathbf{A}^4 - \partial^4 \mathbf{A}^\mu$, and this must transform coherently under the symmetry in order for the kinetic term to be invariant. But notice that $\mathbb{Z}_2: \partial_4 \mapsto - \partial_4$. So $\mathbf{A}^4$ and $\mathbf{A}^\mu$ are forced to have opposite transformations under the reflection symmetry! Thus in this example one and only one of the four-dimensional vector and scalar multiplet has a massless zero mode; so orbifold compactifications generally enable us to have light fields without partners.

For another example of the usefulness of orbifolding, consider a five dimensional theory with the minimal amount of supersymmetry and we want to ensure our four-dimensional theory has only $\mathcal{N}=1$ supersymmetry. First let's recall why we cannot have more than $\mathcal{N}=1$ in four dimensions. As emphasized above, the SM is a chiral theory, wherein the different chiral components of its Dirac fields are in different gauge symmetry representations. This means that $\mathcal{N} = 2$ supersymmetry in four dimensions is too much for us, since the irreducible representations of super-Poincar\'{e} in that case don't allow for chiral matter. In particular, the $\mathcal{N} = 2$ vector multiplet must transform in the adjoint, and the $\mathcal{N} = 2$ hypermultiplet must transform in a real representation in order for it to be CPT self-conjugate, which means the two Weyl fermions must transform in conjugate representations. Thus if we want to imagine that the world came from a higher-dimensional supersymmetric theory, orbifold compactification is necessary.

The same problem appears just for five dimensional spinor fields, since there is no such thing as chirality in odd spacetime dimensions. So under dimensional reduction, one five-dimensional spinor breaks into two conjugate four-dimensional spinors, and you cannot get chiral matter. This is really the same problem as the above, since the possible supersymmety comes exactly from the possible spinor representations.

Compactifying a theory on an orbifold, rather than a manifold, will allow us to solve both of these problems---obtaining chiral matter, and reducing the amount of supersymmetry we have. First note that we can build $\mathcal{N}=2$ supermultiplets out of two $\mathcal{N}=1$ supermultiplets, which is really just thinking about a particular ordering for the construction of the supermultiplet by acting with supercharges. So for each $\mathcal{N}=1$ superfield we want to have, we need to arrange for it to be even under the parity, and so have a zero-mode, and its partner superfield to be odd under the parity, and so only have $n > 1$ Kaluza-Klein modes. In this way we get a set of zero modes which are chiral and $\mathcal{N}=1$ supersymmetric, while the towers reflect the full $\mathcal{N}=2$ supersymmetry and are non-chiral. 

As an explicit example, consider an $\mathcal{N}=2$ vector multiplet, which consists of a real scalar $\Sigma$, two fermions $\psi^a$ with $a=1,2$, and a vector $A_M$. Under the $SU(2)_R$ symmetry, the scalar and vector are singlets and the fermions form a doublet. As in the example above, the $\mathcal{N}=2$ Lagrangian for this multiplet dictates that the two $\mathcal{N}=1$ multiplets living inside it transform differently under the $\mathbb{Z}_2$. So on the four-dimensional boundary one gets a zero-mode for either $(A_\mu, \psi^1)$ or $(\psi^2, A_4 + i \Sigma)$, corresponding to a vector multiplet or a chiral multiplet respectively. In slightly more group-theoretic language we can say that we've embedded the $\mathbb{Z}_2$ in the $SU(2)_R$, and have been forced by the physics to put the doublet in a two-dimensional representation
\begin{equation}
\mathbb{Z}_2: \psi^a \rightarrow \sigma^{a}_{z \ b} \psi^b, \qquad \sigma_z = \begin{pmatrix}
1 & 0 \\ 0 & -1
\end{pmatrix}
\end{equation}
and to pair each of these fermions with two bosonic degrees of freedom, $\mathbb{Z}_2: A_\mu \rightarrow \pm A_\mu, A_4  \rightarrow \mp A_4, \Sigma \rightarrow \mp \Sigma$. This gives us either an $\mathcal{N}=1$ chiral multiplet or an $\mathcal{N}=1$ vector multiplet on the boundary. More discussion and details can be found in Quir\'{o}s' TASI notes \cite{Quiros:2003gg}, and in e.g. \cite{Delgado:1998qr,Barbieri:2001dm,Hall:2001pg}.

In fact orbifolding can do even more symmetry-breaking for us. As mentioned above, we can alternatively consider compactification on an interval as the orbifold $S^1(4\pi R)/(\mathbb{Z}_2(0) \times \mathbb{Z}_2(\pi R))$, which then means the two boundaries are fixed points of \textit{independent} $\mathbb{Z}_2$ symmetries. In particular this means we can choose different embeddings of the $\mathbb{Z}_2$ in the full symmetry on the two ends $y=0,\pi R$ \cite{Scherk:1978ta,Scherk:1979zr}. Now let's apply this technology to the $\mathcal{N}=2$ case we considered above. We saw we had two different choices for nontrivial embeddings of the $\mathcal{N}=2$ into the R-symmetry to break the supersymmetry down to $\mathcal{N}=1$ at the boundary. These correspond to choosing which half of the fields are even under the $\mathbb{Z}_2$ and so get zero-modes. Now that we have independent $\mathbb{Z}_2$s on either end, we can choose these to leave different $\mathcal{N}=1$ symmetries unbroken.

In a microscopic picture, what we end up with is a theory on $\mathbb{R}^{3,1}\times I[0,\pi R]$ where the bulk is ($\mathcal{N}=2$)-supersymmetric and the boundaries respect different $\mathcal{N}=1$ supersymmetries. When we look at the effective four-dimensional theory at distances much larger than $R$, we have fully broken supersymmetry with the breaking being a nonlocal effect---one must be sensitive to physics on both boundaries in order to see the full breaking. As a result of this nonlocality, supersymmetry-breaking is guaranteed to be `soft' and the effects cannot depend on positive powers of UV scales, as we will discuss further momentarily. This leads to fantastically predictive models of BSM physics from around the turn of the millennium which really do read like they have it all figured out (see e.g. \cite{Antoniadis:1993jp,Delgado:1998qr,Antoniadis:1998sd,Pomarol:1998sd,Barbieri:2000vh,Barbieri:2001dm,Barbieri:2001yz,Hall:2001pg}). It's worth understanding these in some detail simply because of how beautiful they are, but they uniformly lead to lots of (thus far) unseen structure near the TeV scale as KK partners become excited.

\subsection{Nonlocal Symmetry-Breaking}

Consider a single extra circular dimension of radius $R$ with gravity and some gauge field. Upon restriction to the four-dimensional Lorentz group, the five-dimensional graviton breaks up into a four-dimensional graviton, vector, and scalar, while the five-dimensional vector breaks up into a four-dimensional vector and scalar. Each of these must be massless by five-dimensional gauge-invariance above the scale $1/R$, but below that they can pick up mass corrections up to that cutoff. Either of these possibilities then amounts to a mechanism for UV protection of a scalar mass. In implementing this in the SM, the first possibility is known as the Higgs being a radion---the scalar which controls fluctuations of the size of the fifth dimension, $h - \delta h = \langle h \rangle = 1/R$, and the latter is denoted `gauge-Higgs unification' for obvious reasons. A particular motivation for gauge-Higgs unification is as an extension of the strategy of grand unification. As mentioned in Section \ref{sec:EffFieldTh}, the SM gauge bosons and fermions beautifully fit into representations of larger gauge groups, but in 4d GUTs the Higgs is left out in the cold as an extra puzzle piece. But `grand gauge-Higgs unification' in higher dimensions may allow even further frugality of ingredients \cite{Hall:2001zb,Haba:2004qf,Lim:2007jv,Furui:2016owe}.

A very interesting feature of this sort of construction is that the symmetry-breaking which is responsible for producing the light scalar is \textit{nonlocal}---one must traverse around the fifth dimension to see the effects of the breaking. This should be intuitively clear, as at distances small compared to $R$ the theory looks like five-dimensional Minkowski space. If it isn't obvious, I recommend musing by analogy on how we could tell whether or not the universe is a sphere with radius far, far larger than the Hubble scale $R \gg 1/H_0$.

As a result of this nonlocal symmetry-breaking, the scalar mass must be finite and calculable in the low-energy theory below $1/R$. There cannot be any sensitivity to ultraviolet energy scales $\Lambda_\text{UV}$, as this corresponds to a `local counterterm', but in the theory above $1/R$ we know that the mass vanishes identically by gauge invariance, so such a counterterm cannot occur (recall our discussion in Section \ref{sec:locality}). This is powerful because symmetry-breaking generally leaves residual logarithmic dependence on large scales even when quadratic dependence has been eliminated, as we saw in the example of supersymmetry above. 

Of course the theory can generate finite corrections to the mass of such a scalar at and below the scale $1/R$ . If we look only at the low-energy effective theory then these look divergent, but we know they must get cut off at $1/R$. Since we know the high-energy theory we can ask how the scalar gets a mass at all, which seems to be impossible from gauge invariance. In fact the scalar mass comes from the Wilson loop wrapping the non-trivial cycle around the fifth dimension \cite{Manton:1979kb,Forgacs:1979zs,Fairlie:1979at,Fairlie:1979zy,Hosotani:1983vn,Hosotani:1983xw,Hosotani:1988bm}. In any gauge theory there is a gauge-invariant operator called a `Wilson loop',
\begin{equation}
W_C = \mathcal{P} \exp i g \oint_C \mathbf{A}_M \text{d}x^M,
\end{equation}
where $g$ is the gauge coupling, $C$ is some closed path through spacetime, and $\mathcal{P}$ denotes `path ordering' of the operators along $C$ in similarity to the time ordering in the definition of the Feynman propagator. In an Abelian theory, the gauge field transforms as $A_\mu \rightarrow A_\mu - \partial_\mu \Gamma(x)$ and we see that $\oint_C A_\mu \text{d}x^\mu \rightarrow \oint_C A_\mu \text{d}x^\mu - \oint_C \partial_\mu \Gamma \text{d}x^\mu$. Integration by parts leaves us only with gauge transformations at the endpoints of $C$, of which there are none if $C$ is a loop (equivalently if we started more generally looking at a `Wilson line', we can say that for a loop the endpoints are connected and so their gauge transformations cancel each other). This remains true in non-Abelian gauge theories, though we do not go through the proof.

In our case we can consider a path which goes solely around the fifth dimension, in which case the Wilson loop contains only our 4d scalar $A_M \text{d}x^M \rightarrow A_5 \text{d}x^5$ and yet is fully gauge-invariant. Quantum corrections can thus generate the operator 
\begin{equation}
\mathcal{L} \supset \sigma^4 \text{Tr } \mathcal{P} \exp i g_5 \oint \mathbf{A}_5(x_\mu,x_5) \text{d}x^5 + \text{ h.c.}
\end{equation}
We note that on a circle the four- and five-dimensional gauge couplings are related as $g_5^2 = R g_4^2$. Proceeding na\"{i}vely and expanding $\mathbf{A}_5(x_\mu,x_5) = \frac{1}{\sqrt{2\pi R}}\sum\limits_{n=-\infty}^{+\infty} \phi^{(n)}(x_\mu) e^{inx_5/R}$, we see that this operator includes a four-dimensional mass for the zero-mode, 
\begin{equation}
\mathcal{L} \supset \sigma^4 \text{Tr} \mathcal{P} \exp i g_5 \oint \phi^{(0)}(x_\mu) \text{d}x^5 + \text{ h.c.} \simeq \sigma^4 \cos\left[g_5 \sqrt{2\pi R} \phi^{(0)}\right]\supset \sigma^4 g_5^2 R^2 \left(\phi^{(0)}\right)^2,
\end{equation}
where we note that the natural size for the Wilson coefficient $\sigma \simeq 1/R$ yields a scalar mass expectation which is $m^2_\phi \simeq g_4^2/R^2$, unsurprisingly as $R$ is the only scale in the problem, and the dependence on the gauge coupling reveals the scalar's five-dimensional origins. 

For a successful such model of gauge-Higgs unification, we need not only to get a light scalar but also to endow that scalar with dynamics pushing it to break electroweak symmetry \cite{Antoniadis:1990ew,Antoniadis:1993jp,Hatanaka:1998yp,Hall:2001tn,Burdman:2002se,Csaki:2002ur,Scrucca:2003ra,Choi:2003kq,Hosotani:2004ka,Hosotani:2004wv}. There is far too much rich physics involved here to mention, but discussions of the calculation of the one-loop effective potential in these models can be found in \cite{Antoniadis:2001cv,Hasegawa:2004zz,Cacciapaglia:2005da}. Gauge symmetry-breaking by a higher-dimensional gauge field component getting a vev is sometimes referred to as the `Hosotani mechanism' \cite{Hosotani:1983vn,Hosotani:1983xw,Hosotani:1988bm}.

As opposed to the Large Extra Dimensional model of the previous section, in these models the extra dimensions are `universal'---all of the SM fields can propagate around the small dimension, not just gravitons. This means that all of our familiar fields are the zero modes of KK towers with spacing $\sim 1/R \sim m_H$. These KK partners have the same gauge charges, and so should be produced copiously in interactions with $\sqrt{s} \gg 1/R$, yet none have been observed at the LHC (see e.g. \cite{Han:2010rf,Nishiwaki:2011gk,Belanger:2012mc,Belyaev:2012ai} for some constraints). As a result, the lower bound on the KK scale is now far above the Higgs mass, which makes all of these sorts of models increasingly less attractive as solutions to the hierarchy problem.

\section{Compositeness}\label{sec:composite}

We have saved for last what is, in some sense, the most obvious strategy to pursue. As discussed in Section \ref{sec:Naturalness}, the Standard Model already breaks a symmetry and generates a mass scale in a natural manner with the chiral condensate of QCD. Perhaps Nature has only this one trick, and repeats the same mechanism to generate the weak scale. After all, we mentioned above that the QCD condensate does in fact break the electroweak symmetry, just not at the right scale.

\subsection{Technicolor}

The idea is to introduce a new gauge sector which is asymptotically free and so becomes strong and confines at the electroweak scale. If the condensate has electroweak quantum numbers, then it breaks electroweak symmetry just as a Higgs field would, but now without any Higgs. This strategy is known as `technicolor', and was proposed in its simplest form by Weinberg \cite{Weinberg:1975gm} and Susskind \cite{Susskind:1978ms}. A modern pedagogical introduction can be found in TASI lectures from Chivukula \cite{Chivukula:2000mb} or Contino \cite{Contino:2010rs}, which has heavily influenced this discussion, and a more detailed classic review is from Hill \& Simmons \cite{Hill:2002ap}. We introduce a technicolor sector which is an $SU(N_{TC})$ gauge group with $N_D$ technicolor fundamentals which are electroweak doublets and their singlet partners, together enjoying a global $SU(2)_L \times SU(2)_R$ symmetry which is broken in the infrared to $SU(2)_V$ by confinement in the $SU(N_{TC})$ sector at $\Lambda_{TC} \sim v$. This structure exactly matches that of the SM QCD sector (save for the values of $N_{TC}, N_D$), so we expect all the same phenomenology, for example with composite technipions appearing close to the electroweak scale. But in this case since the technicolor chiral condensate is the leading breaking of the gauged $SU(2)_L$, the technipions will be predominantly `eaten' and appear as the longitudinal polarizations of our $W,Z$ bosons. Technicolor can `Higgs' the electroweak gauge symmetry just like a fundamental Higgs field would, but without having any scale at which it looks like there is a scalar field breaking electroweak symmetry.

However, the SM Higgs not only breaks electroweak symmetry but also gives mass to the quarks, and thus far we haven't introduced any coupling between the quarks and the technicolor sector. To get the appropriate couplings we can embed both of these gauge groups in a larger `extended technicolor' group, $SU(N_{ETC})\supset SU(3)_c \times SU(N_{TC})$. After this extended group undergoes spontaneous symmetry breaking at $\Lambda_{ETC}$, the broken gauge bosons generate the appropriate four-Fermi interactions
\begin{equation}
\mathcal{L} \supset \frac{g^2_{ETC}}{\Lambda^2_{ETC}} (\bar q q) (\bar \psi_{TC} \psi_{TC}) + \mathcal{O}\left(\frac{\partial^2}{\Lambda^4_{ETC}}\right),
\end{equation}
and then when the technicolor group confines at a scale $\Lambda_{TC}$, we see the emergence of quark masses
\begin{equation}
m_q \simeq \frac{g^2_{ETC}}{\Lambda^2_{ETC}} \left\langle \bar \psi_{TC} \psi_{TC} \right\rangle \sim \Lambda_{TC}\left(\frac{\Lambda_{TC}^2}{\Lambda^2_{ETC}}\right).
\end{equation}
But this clearly suffices solely to generate a single quark mass scale, since the Yukawa coupling is originating from a single gauge coupling, so it was quickly realized that accounting for flavor physics required far more structure and significantly larger gauge groups \cite{Eichten:1979ah,Dimopoulos:1979es,Dimopoulos:1980fj,Dine:1981za}. To generate the variety of quark mass scales with this mechanism would require a cascade of breakings from $SU(N_{ETC})$ down to $SU(3)_c \times SU(N_{TC})$. Furthermore the same massive gauge boson exchanges which generate those needed four-Fermi operators also generate four-quark interactions $(\bar q q) (\bar q q)$ which can lead to large flavor violation.

While there were insights on how various aspects of this could be tackled, the death-knell for technicolor came with Peskin \& Takeuchi's parametrization of oblique corrections to the two-point functions of the electroweak vector bosons from BSM physics \cite{Peskin:1990zt}. These efficiently parametrize deviations from the tree-level form of the vector boson propagators, and can be simply connected with experiment. Ensuing estimates for the sizes of these parameters in strongly-interacting models were very far from empirical measurements \cite{Peskin:1991sw}. The program of technicolor lives on with `walking technicolor', the idea that the confining dynamics may be due to a strongly-coupled gauge theory which behaves very differently from QCD \cite{Holdom:1981rm,Holdom:1984sk,Yamawaki:1985zg,Akiba:1985rr,Appelquist:1986an,Appelquist:1986tr,Appelquist:1987fc}. This is too large a digression for us to introduce, but we mention that there is interesting recent work relating the existence of `walking' dynamics to proximity of the theory to a fixed point at complex value of the coupling \cite{Gorbenko:2018dtm,Gorbenko:2018ncu,Benini:2019dfy,Faedo:2019nxw,Antipin:2020rdw}.

\subsection{A Composite Goldstone Higgs}

However, there is another way that compositeness can be useful for us in securing a light electroweak-symmetry-breaking scalar: It will allow us to realize the dream of a pseudo-Nambu-Goldstone Higgs. Recall that whenever a continuous global symmetry is spontaneously broken, there appear scalar Goldstone bosons $\pi^i(x)$ which parametrize excitations about the vacuum state in the direction of the broken generators. Since the theory had a global symmetry, the potential along these directions is flat, and thus the Goldstones are massless. More strongly, they contain a shift symmetry: 
\begin{equation}
\mathcal{L}(\pi^i(x)) = \mathcal{L}(\pi^i(x) + \xi) 
\end{equation}
where $\xi$ is independent of $x$. This means they may only be `derivatively coupled'; they may appear in the Lagrangian solely as $\partial_\mu \pi^i(x)$. Thus, a non-zero mass for such a scalar is technically natural---if such an operator is present, $\pi^i$ is said to be a \textit{pseudo}-Goldstone and corresponds to the breaking of an approximate symmetry. This is a familiar story in the context of the QCD condensate breaking the approximate chiral symmetry of the quarks, leading to light but not massless pions.

In the case of technicolor, the phenomenon of confinement of a strongly interacting sector is directly responsible for breaking the gauged electroweak symmetry, leading to no separation between the two scales $\Lambda_{TC}$ and $v$. This means the technipions are immediately eaten, and there is no scale at which it looks like a scalar field is responsible for symmetry-breaking, which leads to large electroweak precision constraints. In a general composite Higgs scenario, we'll attempt to arrange for separation between the scale of confinement and the scale of electroweak symmetry breaking by having confinement break an enhanced \textit{global} symmetry, leading to pseudo-Nambu Goldstone bosons whose masses are protected. We then want to radiatively generate a potential for these pNGBs, leading to them getting a vev and breaking electroweak symmetry at a lower scale. A degree of separation between confinement at the scale $f$ and EWSB at $v$ will reduce the difficulties with electroweak precision constraints, as this scenario returns to the SM elementary Higgs sector in the limit $v/f \rightarrow 0$. More pressingly, now that we have gained experimental access to energies close to $v$ it's even more clear that scale separation is needed---confining dynamics lead generically to \textit{lots} of resonances at the scale $f$, which would be seen in all sorts of ways.

\begin{figure}
	\centering
	\begin{subfigure}{.3\textwidth}
		\centering
		\includegraphics[width=\linewidth]{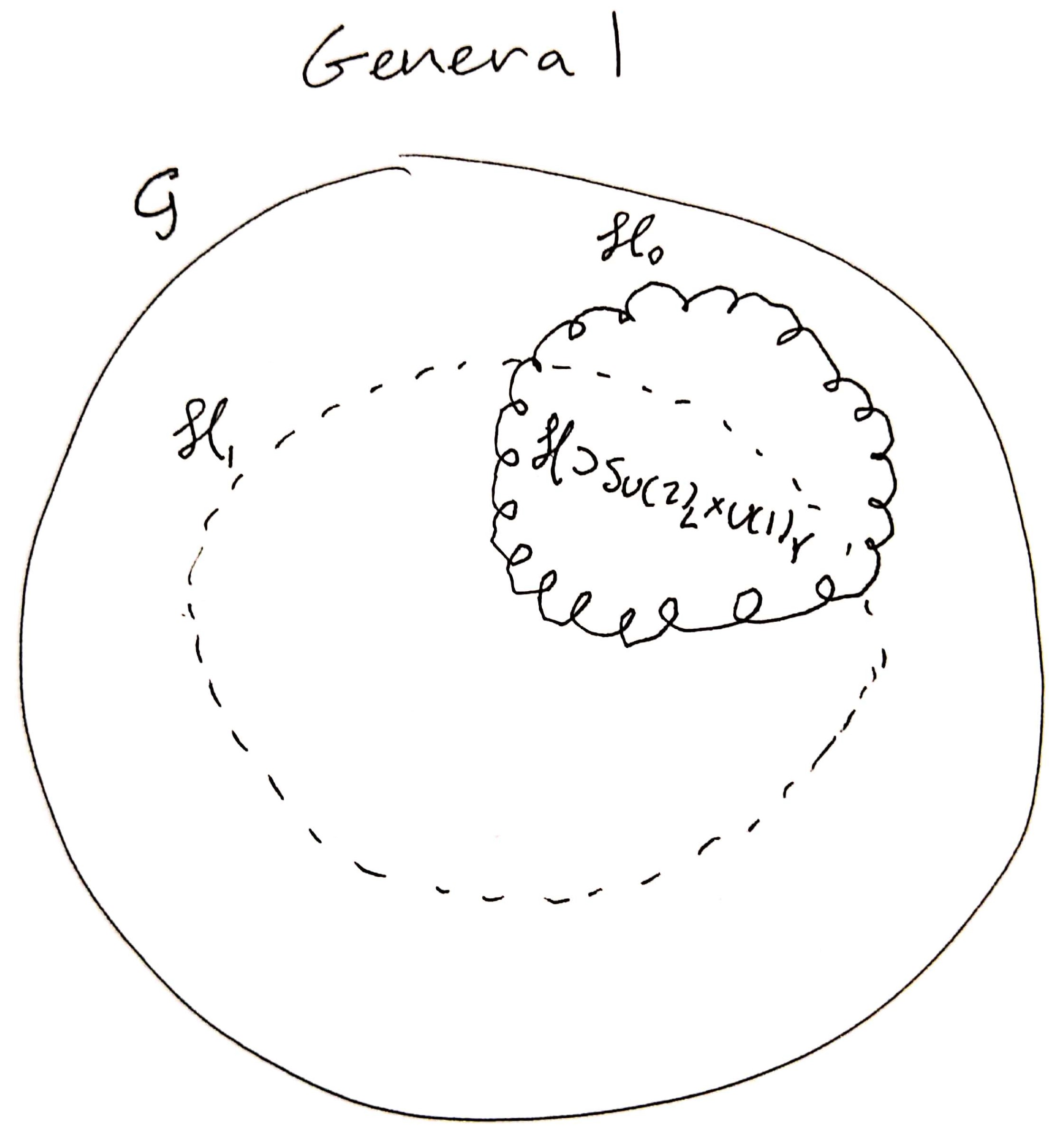}
		\caption{}
		\label{fig:compGen}
	\end{subfigure}%
	\begin{subfigure}{.3\textwidth}
		\centering
		\includegraphics[width=\linewidth]{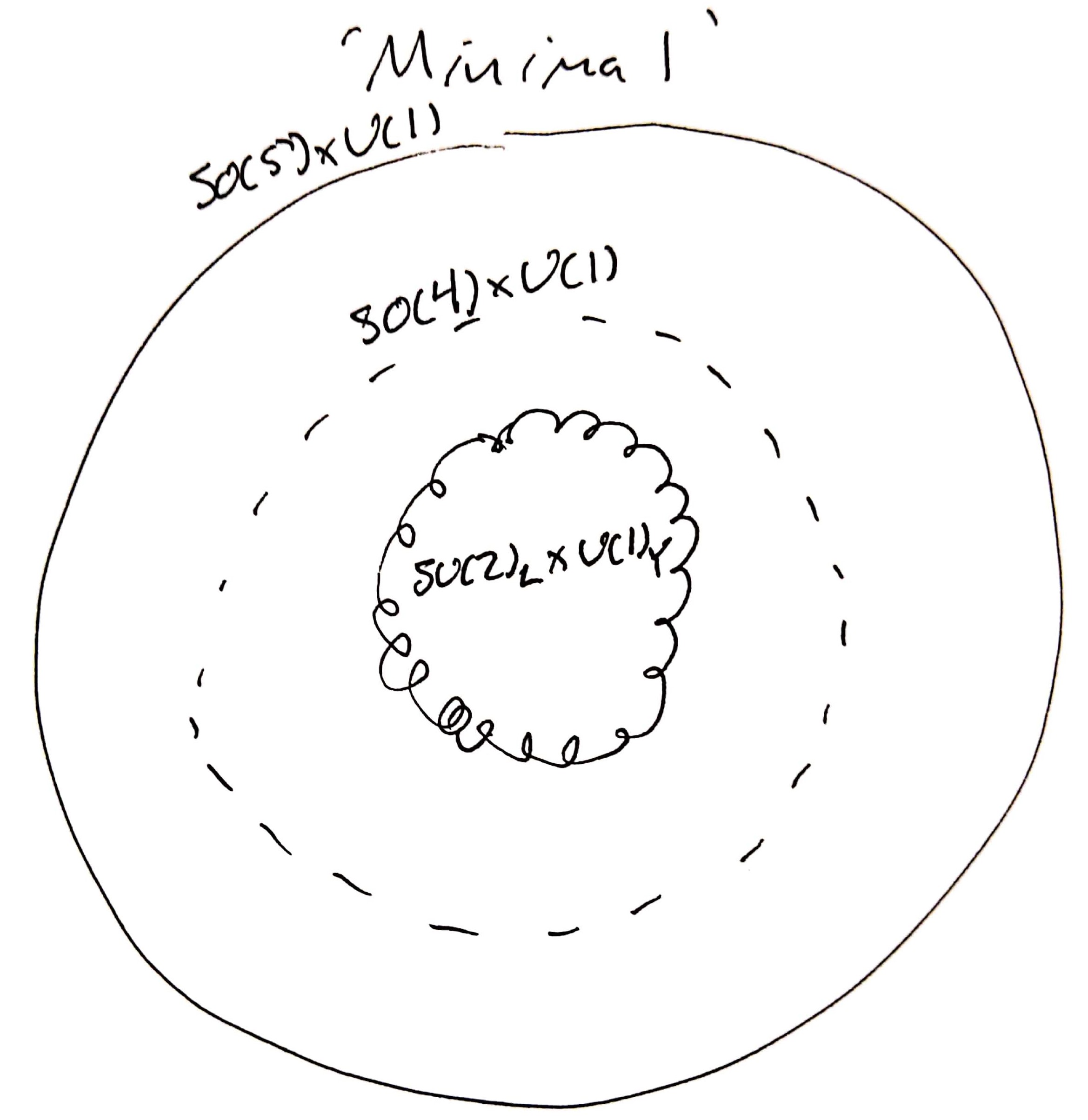}
		\caption{}
		\label{fig:compMin}
	\end{subfigure}
	\begin{subfigure}{.3\textwidth}
		\centering
		\includegraphics[width=\linewidth]{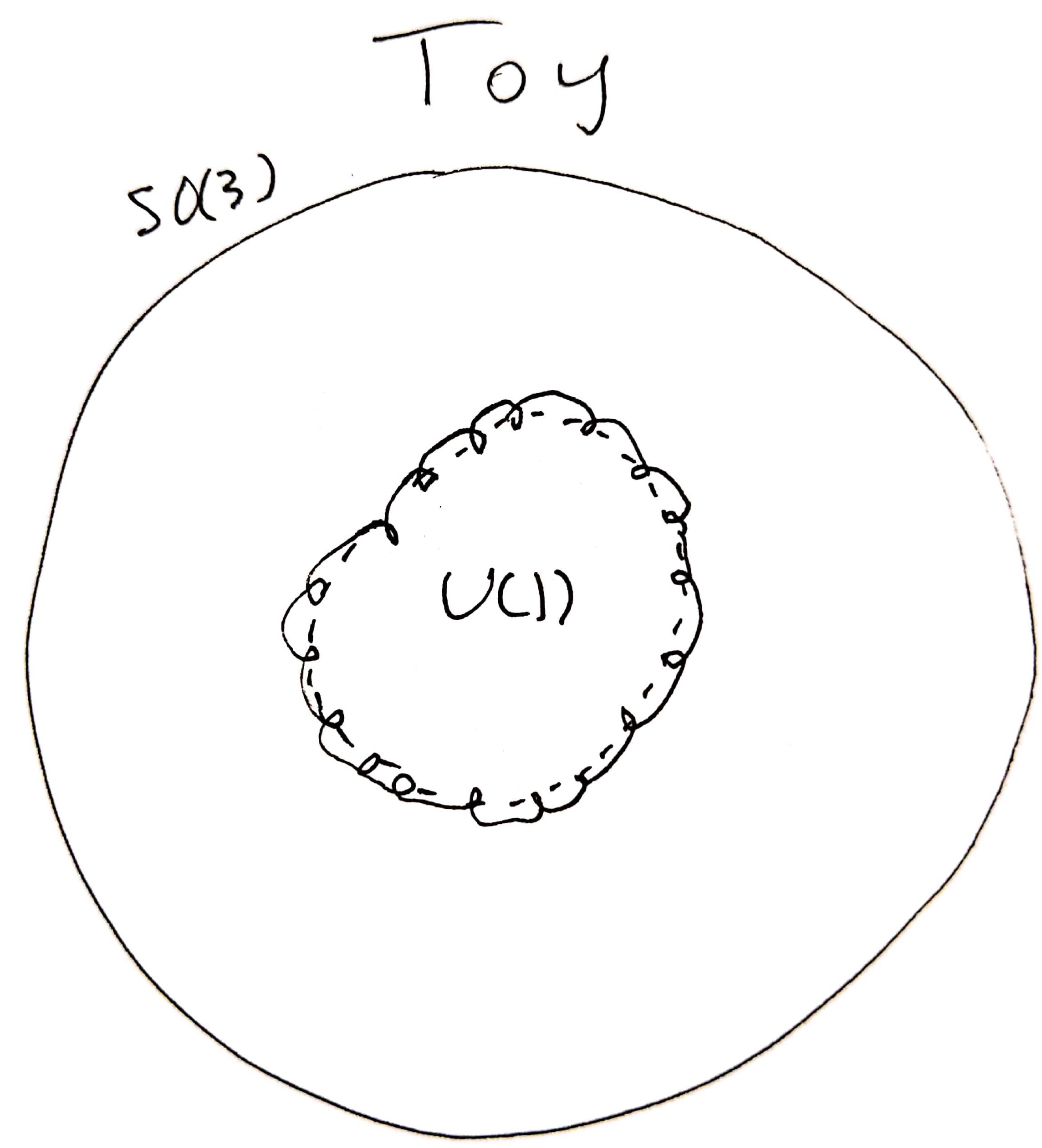}
		\caption{}
		\label{fig:compToy}
	\end{subfigure}
	\caption{In (a), the general symmetry setup for a composite Higgs model. Solid lines are the UV symmetries, with the global symmetry enclosed in the smooth circle and the gauge symmetry in the loopy circle. The symmetry group after confinement is dashed. In (b), the same structure applied to the `minimal' model of \protect\cite{Agashe:2004rs}. In (c), the composite Abelian Higgs toy model discussed in \protect\cite{Panico:2015jxa}, where the symmetry group after confinement coincides with the gauge group.}
	\label{fig:comp}
\end{figure}

The general setup is a group $\mathcal{G}$ of (approximate) global symmetries, of which a subgroup $\mathcal{H}_0$ is gauged. We will have strong dynamics at the scale $f$ break the global symmetry to $\mathcal{H}_1$, and in full generality allow for some part of the gauge symmetries to be broken, such that the unbroken gauge symmetry is $\mathcal{H} = \mathcal{H}_0 \cap \mathcal{H}_1$. This leads to $\left[\mathcal{G}\right]-\left[\mathcal{H}_1\right]$ Goldstone bosons (where $\left[\cdot\right]$ is the dimension of a group) of which $\left[\mathcal{H}_0\right] - \left[\mathcal{H}\right]$ are eaten by the broken gauge bosons. The uneaten, light pNGBs transform non-trivially under the remaining gauge symmetry $\mathcal{H}\supset SU(2)_L \times U(1)_Y$, so we can hope to arrange for them to break this symmetry. In a realistic minimal model, we can have the strong dynamics not break any gauge symmetry, $\mathcal{H}_0 \subset \mathcal{H}_1 \Rightarrow \mathcal{H} = \mathcal{H}_0$. Such a minimal model may be constructed with $\mathcal{G}=SO(5)\times U(1) \rightarrow \mathcal{H}_1 = SO(4) \times U(1) \supset \mathcal{H}_0 = SU(2)_L \times U(1)_Y$ \cite{Agashe:2004rs}. We diagram the general structure in Figure \ref{fig:compGen} and the minimal model in \ref{fig:compMin}.

To evince the ideas in the simplest scenario possible, we'll discuss an even simpler model for a composite pNGB which then breaks a $U(1)$ gauge symmetry. This is just a toy model to understand the features, which has already been kindly worked out in the extensive review from Panico \& Wulzer \cite{Panico:2015jxa}, and which we'll call a `composite Abelian Higgs' model. To find the most minimal model we'll ask for a composite pNGB which breaks the smallest continuous gauge symmetry, $U(1)$, for which we simply need two uneaten degrees of freedom (since a charged scalar is necessarily complex). We can make the even more minimal choice $\mathcal{H}_0 = \mathcal{H}_1=\mathcal{H}$, as we're not worried about having additional unbroken global symmetries. Since $\left[U(1)\right]=1$, we then just need to choose a group $\mathcal{G}$ with at least $\left[\mathcal{G}\right]=3$ to get the right number of pNGBs. We'll study the breaking $SO(3) \rightarrow SO(2) \simeq U(1)$, which is especially nice because we have geometric intuition for the Lie algebras of these groups.

We'll study this using a `linear sigma model', of which the `chiral Lagrangian' describing the QCD pions is the most familiar example. Much of the general technology was developed by Callan, Coleman, Wess \& Zumino \cite{Coleman:1969sm, Callan:1969sn}, and some modern introductions to this technology can be found in Schwartz' textbook \cite{Schwartz:2013pla}, in a pedagogical review of Little Higgs models by Schmaltz \& Tucker-Smith \cite{Schmaltz:2005ky}, and in exhaustive detail in the review by Scherer \cite{Scherer:2002tk}. The big idea is one of bottom-up effective field theory: Given knowledge of the symmetry-breaking structure, we can cleverly parametrize our fields to easily see the structure of the Lagrangian which is demanded both before and after symmetry-breaking.

In our case we must start with an $SO(3)$-invariant Lagrangian of a fundamental field $\vec{\Phi}$, which is a familiar $3d$ vector.
\begin{equation}
-\mathcal{L} = \half \partial_\mu \vec{\Phi}^\intercal \partial^\mu \vec{\Phi} + \frac{g_\star^2}{8} \left(\vec{\Phi}^\intercal \vec{\Phi} - f^2\right)^2,
\end{equation}
where $SO(3)$ rotations act as $\vec{\Phi} \rightarrow g \cdot \vec{\Phi}$ with $g = \exp{i \alpha_A T^A}$, where $T^A$, $A=1..3$ are the generators of $SO(3)$. The potential of $\vec{\Phi}$ is minimized for $\left\langle \left|\left| \vec{\Phi} \right|\right| \right \rangle = f^2$, so $\vec{\Phi}$ gets a nonzero vev and the symmetry is broken down to rotations keeping $\left\langle \vec{\Phi} \right\rangle$ fixed, which is simply $U(1)$. There are a two-sphere worth of vacua corresponding to possible angles for $\left\langle \vec{\Phi} \right\rangle$, which parametrize the Goldstone directions. We can make the split between broken and unbroken generators explicit by parametrizing our field as 
\begin{equation}
\vec{\Phi} = \exp{\left(i \frac{\sqrt{2}}{f} \Pi_i(x) \hat{T}^i\right)} \begin{bmatrix}
0 \\ 0 \\ f + \sigma(x)
\end{bmatrix} = (f + \sigma)\begin{bmatrix}
\sin\left(\frac{\Pi}{f}\right) \frac{\vec{\Pi}}{\Pi} \\ \cos\left(\frac{\Pi}{f}\right)
\end{bmatrix}
\end{equation}
where in the first equality $\hat{T}^i$ are the two broken generators, $\Pi_i(x)$ are the massless Goldstones corresponding to fluctuations along the vacuum manifold, and $\sigma(x)$ is the massive `radial mode' giving fluctuations about the vev. We eschew writing down the generators explicitly and assert that in this case one finds the compact latter expression, with $\Pi = \sqrt{\vec{\Pi}^\intercal \vec{\Pi}}$. We can find an explicit expression for the interaction of the Goldstones and the radial modes by simply plugging this parametrization into the Lagrangian above. We indeed find a mass for $\sigma$ of $m_\sigma = g_\star f$, massless pions $\vec{\Pi}$ and a tower of all possible interactions between these fields which are consistent with the symmetries.

We now have a theory of massless scalars transforming under an unbroken symmetry; our pions form a doublet of $SO(2)$ transforming as $\vec{\Pi} \rightarrow \exp{(i \alpha \sigma_2)} \vec{\Pi}$, corresponding to rotations about the unbroken $SO(3)$ generator. We can complexify by introducing $H \equiv \frac{1}{\sqrt{2}}(\Pi_1 - i \Pi_2)$ to view $SO(2) \simeq U(1)$ as acting $H \rightarrow \exp{(i \alpha)} H$. As it stands, $H$ is an exact Goldstone boson, so cannot pick up a potential to then itself get a vev and break $U(1)$. However, when we gauge a $U(1)$ subgroup of our original $SO(3)$ global symmetry we're introducing explicit breaking of the symmetry and resultingly $H$ becomes a pNGB and can pick up a mass. The tree-level effect of this gauging is simply to upgrade derivatives to gauge covariant derivatives, $\partial_\mu \vec{\Pi} \rightarrow D_\mu \vec{\Pi} = (\partial_\mu-i e A_\mu \sigma_2) \vec{\Pi}$. 

\begin{figure}
	\centering
	\includegraphics[width=0.7\linewidth]{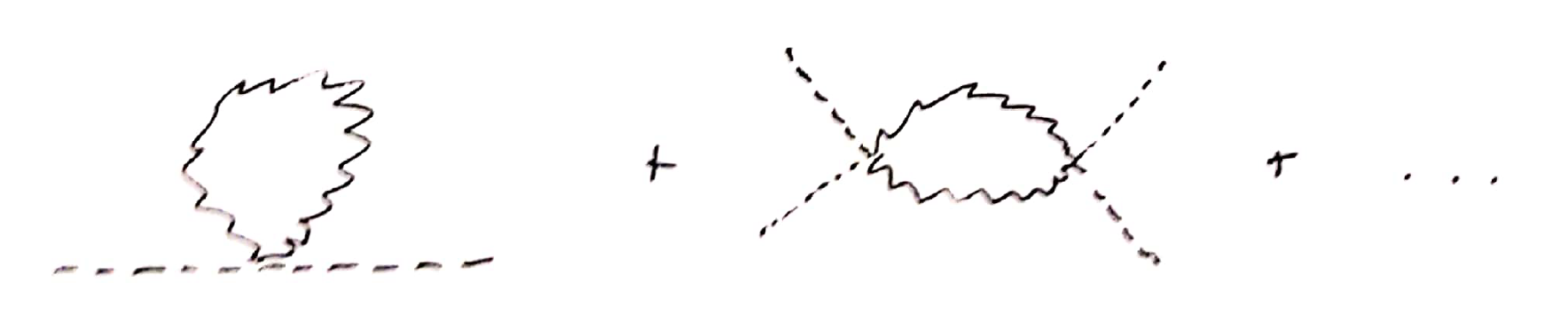}
	\caption{Representative diagrams contributing to the one-loop effective potential for a charged scalar.}
	\label{fig:colemanweinbergabelian}
\end{figure}

Now as a result of $SO(3)$ rotations no longer being an exact symmetry, $H$ is free to pick up a potential, which in general will be radiatively generated as a result of this gauging. It is easy to draw diagrams in which loops of our $U(1)$ gauge bosons generate a nonzero mass and quartic for $H$, as in Figure \ref{fig:colemanweinbergabelian}, and in a realistic model the corrections from loops of top quarks will be especially important. These loop diagrams come along with an obvious cutoff: Above the compositeness scale $f$, the fields $\vec{\Phi}$ no longer exist in the spectrum. There is a beautiful general method for deriving the radiatively-generated potential for $H$ by resumming the one-loop diagrams with various numbers of external $H$ legs, known as the Coleman-Weinberg potential \cite{Coleman:1973jx}.

Unfortunately, absent any of other structure, finding $v/f \ll 1$ still requires some degree of cancellation between various contributions to the potential of $H$. However any amount of $v/f < 1$ will help alleviate the pressure from electroweak precision observables which thus far look empirically as expected for an elementary Higgs field, while still retaining the benefit of forbidding corrections to the Higgs mass above $f$. But one still can't get away from requiring lots of structure near the electroweak scale, which has not (yet) been observed.

There are a variety of important aspects and interesting directions we do not have the space to discuss. Even past understanding the best sorts of group structures to which to apply this strategy, it is clearly important to understand the sorts of field theories which can confine to break $\mathcal{G} \rightarrow \mathcal{H}_1$, as well as the detailed structure of the potential radiatively-generated by SM fields. Then it's important to explore the possibility of a natural structure which dictates $v/f < 1$---while there has been much work on this, we mention in particular the interesting strategy of `collective' symmetry breaking in which a symmetry is broken only by an interplay between different couplings. This class of models is known as the `Little Higgs' \cite{Schmaltz:2005ky,ArkaniHamed:2002qx,ArkaniHamed:2002qy,Kaplan:2003uc,Schmaltz:2002wx,Perelstein:2005ka}, and can be seen as a purely four-dimensional application of the strategy of nonlocal symmetry breaking through `nonlocality in theory space' \cite{ArkaniHamed:2001ca,ArkaniHamed:2001nc}, which is a fascinating topic. The TASI notes by Csaki, Lombardo, Telem \cite{Csaki:2018muy} provide a pedagogical introduction to these topics.

Finally, let me mention that composite Higgs models may be understood as being dual to a novel class of extra-dimensional models known as Randall-Sundrum models \cite{Randall:1999ee,Randall:1999vf}. Unlike in the simple cases we discussed in Section \ref{sec:extradims}, in these models the spacetime does not have a product structure (as did $\mathcal{M}_4\times K$) and the geometry is said to be `non-factorizable'. In this scenario our four-dimensional universe is seen as a brane living on one end of a five-dimensional orbifold of anti-de Sitter space. The minuteness of the electroweak scale compared to the Planck scale is a result of a large fifth-dimensional AdS `warp factor' between the brane we live on (the `IR brane') and the brane on the other end of the space (the `UV brane'). As in Section \ref{sec:led}, this trades the electroweak hierarchy into a geometric hierarchy, but now in AdS we can find novel, natural ways of generating such a hierarchy of scales \cite{Goldberger:1999uk,Goldberger:1999un}. Furthermore, embedding our universe into an AdS spacetime means we can take advantage of the enormously powerful program of AdS/CFT holography, which enables us to study the strong-coupling phenomena of composite Higgs models via their weakly-coupled gravitational duals. Pedagogical introductions to holography can be found in lecture notes from Sundrum \cite{Sundrum:2011ic} and Kaplan \cite{KaplanJ}, and with more background in the textbook by Ammon \& Erdmenger \cite{Ammon:2015wua}. That machinery is not all necessary to appreciate the workings of Randall-Sundrum models, though, and there are a wide variety of great lectures notes aimed at particle theorists, for example those of Sundrum \cite{Sundrum:2005jf}, Csaki \& Tanedo \cite{Csaki:2016kln}, and Gherghetta \cite{Gherghetta:2010cj}.

\chapter{The Loerarchy Problem}

\setlength{\epigraphwidth}{0.5\textwidth}
\epigraph{The great tragedy of science --- the slaying of a beautiful hypothesis by an ugly fact.
}{Thomas Henry Huxley \\ \textit{Biogenesis and Abiogenesis} (1870) \cite{huxley_2011}}
\setlength{\epigraphwidth}{0.6\textwidth}

\section{The `Little Hierarchy Problem'}\label{sec:little}

We've seen in Chapter \ref{sec:classical} a cadre of theories which can produce a light scalar naturally, and there's one feature all the classic approaches have in common: they predict new states with Standard Model charges close to the mass of the Higgs. This is seemingly inevitable simply from the structure of effective field theory---whatever extra structure protects the Higgs mass at UV scales must be broken close to the electroweak scale to allow the Standard Model, which does not have that extra structure. This feature means that smashing together protons at scales much greater than the electroweak scale would surely reveal the physics of whatever mechanism solves the hierarchy problem. And so the Large Hadron Collider was eagerly awaited to tell us which of these ideas was correct.

Yet even years before the LHC turned on, those who could clearly read the tea leaves were realizing that something was amiss with our naturalness expectations (see e.g. the `LEP paradox' \cite{Barbieri:2000gf}, also \cite{Cheng:2003ju}), and exploring the idea that supersymmetry would not show up to solve the hierarchy problem (e.g. `split supersymmetry' \cite{Wells:2003tf,ArkaniHamed:2004fb,Giudice:2004tc}). Perhaps there was something else present which made weak-scale supersymmetry unnecessary for protecting the Higgs.

Of course the LHC has been a fantastic success. It has confirmed for us the existence of a light Higgs resonance that looks SM-like, and made many great measurements of the SM besides. But rather than revealing to us which TeV-scale new physics kept the Higgs light, we've instead had a march of increasingly powerful constraints on new particles which couple to the Standard Model. 

These null results for physics beyond the Standard Model from run 1 of the LHC rapidly popularized the idea that something else might be responsible for stabilizing the Higgs mass at the electroweak scale up to a higher scale where, say, supersymmetry came in. This line of thinking is termed the `Little Hierarchy Problem'---the idea being that one of those classic solutions would appear at $\Lambda \sim 10 \text{ TeV}$ to solve the `Big Hierarchy Problem' and explain why the Higgs mass was not at the Planck scale, leaving a smaller hierarchy of a couple orders of magnitude between $m_H$ and $\Lambda$ unexplained. Perhaps rather than minimal supersymmetry there was another module which provided this last bit of protection. But this has to be a special module to protect the Higgs mass without introducing new colored particles.

\subsection{The Twin Higgs}  \label{sec:twinhiggsintro}\label{sec:twin}

The first such proposal in fact appeared before the LHC had even turned on. The mirror twin Higgs (MTH) \cite{Chacko:2005pe} introduces a second copy of the SM gauge group and states related to ours by a $\mathbb{Z}_2$ symmetry. Since these `twin' states are neutral under the SM gauge group, they are subject only to indirect bounds from precision Higgs coupling measurements. The two sectors are connected solely by Higgs portal-type interactions between the two $SU(2)$ doublet scalars.\footnote{We return in Section \ref{sec:freezetwin} to the prospect of kinetic mixing between the two $U(1)_Y$ factors, which is also allowed by the symmetries.} Subject to conditions on the quartic coupling, the Higgs sector enjoys an approximate $SU(4)$ global symmetry, and the breaking of this symmetry leads naturally to a pseudo-Nambu Goldstone boson. Seemingly magically, this structure is accidentally respected by the quadratically-divergent one-loop corrections to the Higgs potential, and the pNGB continues to be protected through one-loop from large corrections to its mass. The Twin Higgs thus allows the postponement of a solution to the `Big Hierarchy Problem' until scales a loop factor $16 \pi^2 \sim \mathcal{O}(100)$ above the Higgs mass.

While the space of Neutral Naturalness models has now been explored more thoroughly and we will discuss some generalities below, the mirror twin Higgs remains perhaps the most aesthetically pleasing of all these approaches and serves as a useful avatar for this general strategy. As a result, in Chapter \ref{sec:InTheSky} we consider cosmological signatures of the MTH specifically, so we give here a more-detailed introduction to the twin Higgs in particular. This is necessarily slightly more technical than the rest of this chapter, so the reader who is not planning on reading Chapters \ref{sec:InTheSky} or \ref{sec:InTheGround} in detail may skip ahead $\sim 3$ pages to Section \ref{sec:neutral} without loss of continuity.

The scalar potential in this model is best organized in terms of the accidental $SU(4)$ symmetry involving the $SU(2)$ Higgs doublets of the SM and twin sectors, $H_A$ and $H_B$. The general tree-level twin Higgs potential is given by (see e.g. \cite{Craig:2015pha})
\begin{equation}
V(H_A, H_B) = \lambda (|H_A|^2+|H_B|^2-f^2/2)^2+\kappa(|H_A|^4+|H_B|^4)+\sigma f^2|H_A|^2.\label{TwinPot}
\end{equation}
The first term respects the accidental $SU(4)$ global symmetry, as can be seen by writing it in terms of $\mathcal{H} = \left(H_A, H_B\right)^\intercal$, which transforms as a complex $SU(4)$ fundamental. The second term breaks $SU(4)$ but preserves the $\mathbb{Z}_2$ and the final term softly breaks the $\mathbb{Z}_2$. In order for the $SU(4)$ to be a good symmetry of the potential, we require $\kappa,\sigma\ll\lambda$.

However, the gauging of an $SU(2)\times SU(2)$ subgroup constitutes explicit breaking of the $SU(4)$, so we should worry about whether quantum corrections reintroduce large masses for the would-be Goldstones when $SU(4)$ is broken. But writing down the one-loop corrections reveals a fortuitous accidental symmetry. The one-loop effective scalar potential gets the following leading corrections from the gauge bosons at the quadratic level:
\begin{equation}
V_{1-\text{loop}}(H_A,H_B) \supset \frac{9 g_{2,A}^2}{32 \pi^2} \Lambda^2 |H_A|^2 + \frac{9 g_{2,B}^2}{32 \pi^2} \Lambda^2 |H_B|^2 + \text{ subleading} \underset{\mathbb{Z}_2}{\rightarrow} \frac{9 g_{2}^2}{32 \pi^2} \Lambda^2 |\mathcal{H}|^2,
\end{equation}
where we see explicitly that if the $\mathbb{Z}_2$ symmetry is good at the level of the gauge couplings, then these largest one-loop corrections continue to respect the $SU(4)$. It is easy to see from here by power counting that this holds for all the quadratically-divergent pieces so long as the $\mathbb{Z}_2$ is a good symmetry for the interactions involved. 

There is radiative $SU(4)$-breaking at the level of the quartic, since the $\mathbb{Z}_2$ symmetry no longer suffices to form the Higgses into an $SU(4)$ invariant. The coupling $\kappa$ should naturally be of the order of these corrections, the largest of which comes from the Yukawa interactions with the top/twin top, $\kappa\sim 3y_t^4/(8\pi^2)\log(\Lambda/m_{t})\sim 0.1$ for a cut-off $\Lambda\sim 10$ TeV ($y_t$ being the top quark Yukawa coupling and $m_t$ its mass). Requiring $\lambda\gg\kappa$ therefore implies $\lambda\gtrsim 1$. As the SM and twin isospin gauge groups are disjoint subgroups of the $SU(4)$, the spontaneous breaking of the $SU(4)$ coincides with the SM and twin electroweak symmetry breaking. This gives seven Goldstone bosons, six of which are `eaten' by the $SU(2)$ gauge bosons of the two sectors, which leaves one Goldstone remaining. This will acquire mass through the breaking of the $SU(4)$ that is naturally smaller than the twin scale $f$. For future reference, it is convenient to define the real scalar degrees of freedom in the gauge basis as $h_A=\frac{1}{\sqrt{2}}\Re(H_A^0)-v_A$ and $h_B=\frac{1}{\sqrt{2}}\Re(H_B^0)-v_B$, where $\langle H_A^0\rangle =v_A$ and $\langle H_B^0\rangle =v_B$.

The surviving Goldstone boson should be dominantly composed of the $h_A$ gauge eigenstate in order to be SM-like. The soft $\mathbb{Z}_2$-breaking coupling $\sigma$ is required to tune the potential so that the vacuum expectation values (vevs) are asymmetric and that the Goldstone is mostly aligned with the $h_A$ field direction. The (unique) minimum of the Twin Higgs potential (\ref{TwinPot}) occurs at $v_A\approx \frac{f}{2}\sqrt{\frac{\lambda(\kappa-\sigma)-\kappa\sigma}{\lambda\kappa}}$ and $v_B\approx \frac{f}{2}\sqrt{\frac{\sigma+\kappa}{\kappa}}$. The required alignment of the vacuum in the $H_B$ direction occurs if $\sigma\approx\kappa$, which has been assumed in these expressions for the minimum. The consequences of this are that $v_A\approx v/\sqrt{2}$ and $v_B\approx f/\sqrt{2}\gg v$ (where $v$ is the vev of the SM Higgs, although $v_A\approx 174$ GeV is the vev that determines the SM particle masses and electroweak properties), so that the SM-like Higgs $h$ is identified with the Goldstone mode and is naturally lighter than the other remaining real scalar, a radial mode $H$ whose mass is set by the scale $f$. The component of $h$ in the $h_B$ gauge eigenstate is $\delta_{hB}\approx v/f$ (to lowest order in $v/f$). Measurements of the Higgs couplings restrict $f\gtrsim 3v$ \cite{Burdman:2014zta,Craig:2015pha}, and the Giudice-Barbieri tuning of the weak scale associated with this asymmetry is of order $f^2 / 2 v^2$. 

The spectrum of states in the broken phase consists of a SM-like pseudo-Goldstone Higgs $h$ of mass $m_h^2 \sim 8 \kappa v^2$, a radial twin Higgs mode $H$ of mass $m_H^2 \sim 2 \lambda f^2$, a conventional Standard Model sector of gauge bosons and fermions and a corresponding mirror sector. The masses of quarks, gauge bosons, and charged leptons in the twin sector are larger than their Standard Model counterparts by $\sim f/v$, while the twin QCD scale is larger by a factor $\sim \left(1 +\log(f/v)\right)$ due to the impact of the higher mass scale of heavy twin quarks on the renormalisation group (RG) evolution of the twin strong coupling. The relative mass of twin neutrinos depends on the origin of neutrino masses, some possibilities being $\sim f/v$ for Dirac masses and $\sim f^2/v^2$ for Majorana masses from the Weinberg operator. Mixing in the scalar sector implies that the SM-like Higgs couples to twin sector matter with an $\mathcal{O}(v/f)$ mixing angle, as does the radial twin Higgs mode to Standard Model matter. These mixings provide the primary portal between the Standard Model and twin sectors.

The Goldstone Higgs is protected from radiative corrections from $\mathbb{Z}_2$-symmetric physics above the scale $f$. While the mirror Twin Higgs addresses the little hierarchy problem, it does not address the big hierarchy problem, as nothing stabilizes the scale $f$ against radiative corrections. However, the scale $f$ can be stabilized by supersymmetry, compositeness, or perhaps additional copies of the twin mechanism \cite{Asadi:2018abu} without requiring new states beneath the TeV scale. Minimal supersymmetric UV completions can furthermore remain perturbative up to the GUT scale \cite{Chang:2006ra,Craig:2013fga}.

As mentioned, the collider constraints on twin Higgs models are very mild and pertain mostly to a lower bound on the soft breaking of the $\mathbb{Z}_2$. In this respect, the Twin Higgs naturally reconciles the observation of a light Higgs with the absence of evidence for new physics thus far at the LHC. The primary challenge to these models comes from cosmology due to the effects of additional light particles on the cosmic microwave background. We will discuss these issues in Chapter \ref{sec:InTheSky} and propose a natural resolution.

\subsection{Neutral Naturalness or The Return Of The Orbifold} \label{sec:neutral}

More broadly, the twin Higgs is just the simplest example of the more general `Neutral Naturalness' paradigm in which the states responsible for stabilizing the electroweak scale are not charged under (some of) the Standard Model (SM) gauge symmetries \cite{Chacko:2005pe,Chacko:2005un,Burdman:2006tz,Poland:2008ev,Cai:2008au,Craig:2014aea,Batell:2015aha,Curtin:2015bka,Cheng:2015buv,Craig:2016kue,Cohen:2018mgv,Cheng:2018gvu,Serra:2019omd}, thus explaining the lack of expected signposts of naturalness. 

In the symmetry-based solutions to the hierarchy problem discussed in Chapter \ref{sec:classical}, modifications of the Higgs mass were technically natural as a result of a continuous symmetry which commuted with the SM gauge groups. This na\"{i}vely seems necessary to ensure that the necessary degrees of freedom are present and couple to the Higgs with the right strength to cancel divergences. How is the top quark contribution $\propto N_c y_t$ (where $N_c$ is the number of colors) to be canceled if not by another colored particle with the same coupling to the Higgs, which gives the opposite contribution? Well we saw above that the twin Higgs nevertheless works with a quark charged under a \textit{separate} gauge group. Indeed, at one-loop $N_c$ is really just a `counting factor' and we are free to get those three opposite-sign contributions in a variety of ways. To some extent the space of Neutral Naturalness models is an exercise in interesting ways to find that color factor. We'll be able to see that picture more clearly in the language of orbifold projection, which will also give us good reason to expect that these models with na\"{i}vely strange symmetries in fact do have nice UV completions.

In Section \ref{sec:Orbifold} we saw how orbifolds could be useful dynamically---that is, in affording a higher-dimensional, symmetric theory which at low energies looks like a less symmetric, four-dimensional theory due to boundary conditions imposed by the orbifold discrete symmetries. But these theories had interesting properties in their low-energy behavior below the scale of compactification; we didn't need to make reference to their origins in studying them. 

Now we want to understand the variety of low energy theories we can get very generally, but only at the level of the zero-mode spectrum. Rather than decomposing our fields into modes and integrating over the compact manifold and noticing that only those fields invariant under the orbifold symmetry are left with zero-modes, we're going to skip to the answer and look at the spectrum of fields left invariant under our discrete symmetry. We'll find that these theories have enhanced symmetry properties at one loop.

The underlying reason lies in the `orbifold correspondence' in large $N$ gauge theories \cite{Schmaltz:1998bg,Bershadsky:1998cb,Kachru:1998ys}. Given a `mother' field theory, you can create a `daughter' theory by embedding some given discrete symmetry in the symmetries of the mother theory and projecting out states which are not invariant under that discrete symmetry; we call this process `orbifolding'. Then at leading order in large $N$, the correlation functions of the daughter theory match those of the mother theory. This is nothing short of amazing---a theory with no supersymmetry to speak of can nevertheless `accidentally' exhibit supersymmetric behavior at leading order. 

The general case of the orbifold correspondence and how to construct Neutral Naturalness models is beautiful and I do recommend reading \cite{Schmaltz:1998bg,Burdman:2006tz,Craig:2014roa}, but the group theory required for a full discussion would be too large a detour from our main narrative. Fortunately we can get a good sense for what's going on by considering a few explicit examples, which will not require much mathematical machinery.

\subsubsection*{Example 1: Folded Supersymmetry}\addcontentsline{toc}{subsubsection}{Example 1: Folded Supersymmetry}
Let's first consider the example of `Folded Supersymmetry' \cite{Burdman:2006tz} which was the first Neutral Naturalness model constructed explicitly via orbifolding. The idea is precisely to consider a supersymmetric theory and orbifold project onto a theory with no explicit supersymmetry but in which supersymmetric cancellations still occur at one loop. As a pedagogical example, consider an $\mathcal{N}=1$ supersymmetric $U(2N)_C$ gauge theory with $2N$ flavors of left and right fundamental chiral superfields $Q = (\tilde{q},q)$ which enjoy a $U(2N)_{F,L}\times U(2N)_{F,R}$ global flavor symmetry. Let's decree also that the theory respects $R$-symmetry. We will orbifold by the discrete group $\mathbb{Z}_2$ as before, but we must choose how to embed the $\mathbb{Z}_2$ in each of these symmetry groups. That is, our original `mother' theory is invariant under a large symmetry group $U(2N)_C \times U(2N)_{F,L} \times U(2N)_{F,R} \times U(1)_R$, which contains \textit{many} $\mathbb{Z}_2$ subgroups, and we must decide precisely which $\mathbb{Z}_2$ we want to orbifold by. We choose the following embeddings:
\begin{equation}
\mathbf{C} = \begin{pmatrix}
\mathds{1}_N & 0 \\
0 & -\mathds{1}_N 
\end{pmatrix}, \qquad \mathbf{F} = \begin{pmatrix}
\mathds{1}_N & 0 \\
0 & -\mathds{1}_N 
\end{pmatrix}, \qquad \mathbf{R} = (-1)^F,
\end{equation}
where the first matrix is in color-space, the second is in flavor-space, and the third transformation is by fermion number. Each of these generates a $\mathbb{Z}_2$ subgroup of one of the symmetry factors of the mother theory. To see which fields are invariant under this $\mathbb{Z}_2$, we must only act them on the fields and see how they transform. Gauge bosons $A_\mu$ and their superpartner gauginos $\lambda$ are in the adjoint of the gauge group. Indexing $A,B = 1..N$ separately over the two halves of the gauge indices, we have
\begin{align}
A_\mu &= \begin{pmatrix}
A_{\mu,AA} & A_{\mu,AB} \\
A_{\mu,BA} & A_{\mu,BB}
\end{pmatrix} \rightarrow \mathbf{C} \begin{pmatrix}
A_{\mu,AA} & A_{\mu,AB} \\
A_{\mu,BA} & A_{\mu,BB}
\end{pmatrix} \mathbf{C}^\dagger \mathds{1}_R = \begin{pmatrix}
+A_{\mu,AA} & -A_{\mu,AB} \\
-A_{\mu,BA} & +A_{\mu,BB}
\end{pmatrix} \\
\lambda &= \begin{pmatrix}
\lambda_{AA} &\lambda_{AB} \\
\lambda_{BA} & \lambda_{BB}
\end{pmatrix} \rightarrow \mathbf{C} \begin{pmatrix}
\lambda_{AA} &\lambda_{AB} \\
\lambda_{BA} & \lambda_{BB}
\end{pmatrix} \mathbf{C}^\dagger (-\mathds{1})_R = \begin{pmatrix}
-\lambda_{AA} & +\lambda_{AB} \\
+\lambda_{BA} & -\lambda_{BB}
\end{pmatrix}
\end{align}
where the difference here is because of the different $R$ transformations. We see that if we project down to only those states invariant under this transformation, our gauge group dissolves from $U(2N)_C$ down to disconnected pieces $U(N)_C\times U(N)_C$. We see that embedding the $\mathbb{Z}_2$ non-trivially in the $R$-symmetry group means the daughter theory will be non-supersymmetric. The superpartners of our gauge fields are no longer present, but rather the gauginos have been twisted into bifundamentals under the two gauge factors. In the matter sector, letting $a,b=1..N$ similarly index the two halves of the flavor indices and writing the fields as matrices in a combined color and flavor space, we have
\begin{align}
\tilde{q} &= \begin{pmatrix}
\tilde{q}_{Aa} & \tilde{q}_{Ab} \\
\tilde{q}_{Ba} & \tilde{q}_{Bb}
\end{pmatrix} \rightarrow \mathbf{C} \begin{pmatrix}
\tilde{q}_{Aa} & \tilde{q}_{Ab} \\
\tilde{q}_{Ba} & \tilde{q}_{Bb}
\end{pmatrix} \mathbf{F}^\dagger \mathds{1}_R = \begin{pmatrix}
+\tilde{q}_{Aa} & -\tilde{q}_{Ab} \\
-\tilde{q}_{Ba} & +\tilde{q}_{Bb}
\end{pmatrix} \\
q &= \begin{pmatrix}
q_{Aa} & q_{Ab} \\
q_{Ba} & q_{Bb}
\end{pmatrix} \rightarrow \mathbf{C} \begin{pmatrix}
q_{Aa} & q_{Ab} \\
q_{Ba} & q_{Bb}
\end{pmatrix} \mathbf{F}^\dagger (-\mathds{1})_R = \begin{pmatrix}
-q_{Aa} & +q_{Ab} \\
+q_{Ba} & -q_{Bb}
\end{pmatrix}.
\end{align}
Again we see that we have broken supersymmetry. The flavor group has also broken down to $U(N)_{F,L}\times U(N)_{F,L}$ and the invariant states are squarks which are bifundamentals under `diagonal' combinations of the gauge and flavor groups, and quarks which are bifundamentals under the `off-diagonal' combinations. 

\begin{figure}
	\centering
	\begin{subfigure}{.45\textwidth}
		\centering
		\includegraphics[width=0.7\linewidth]{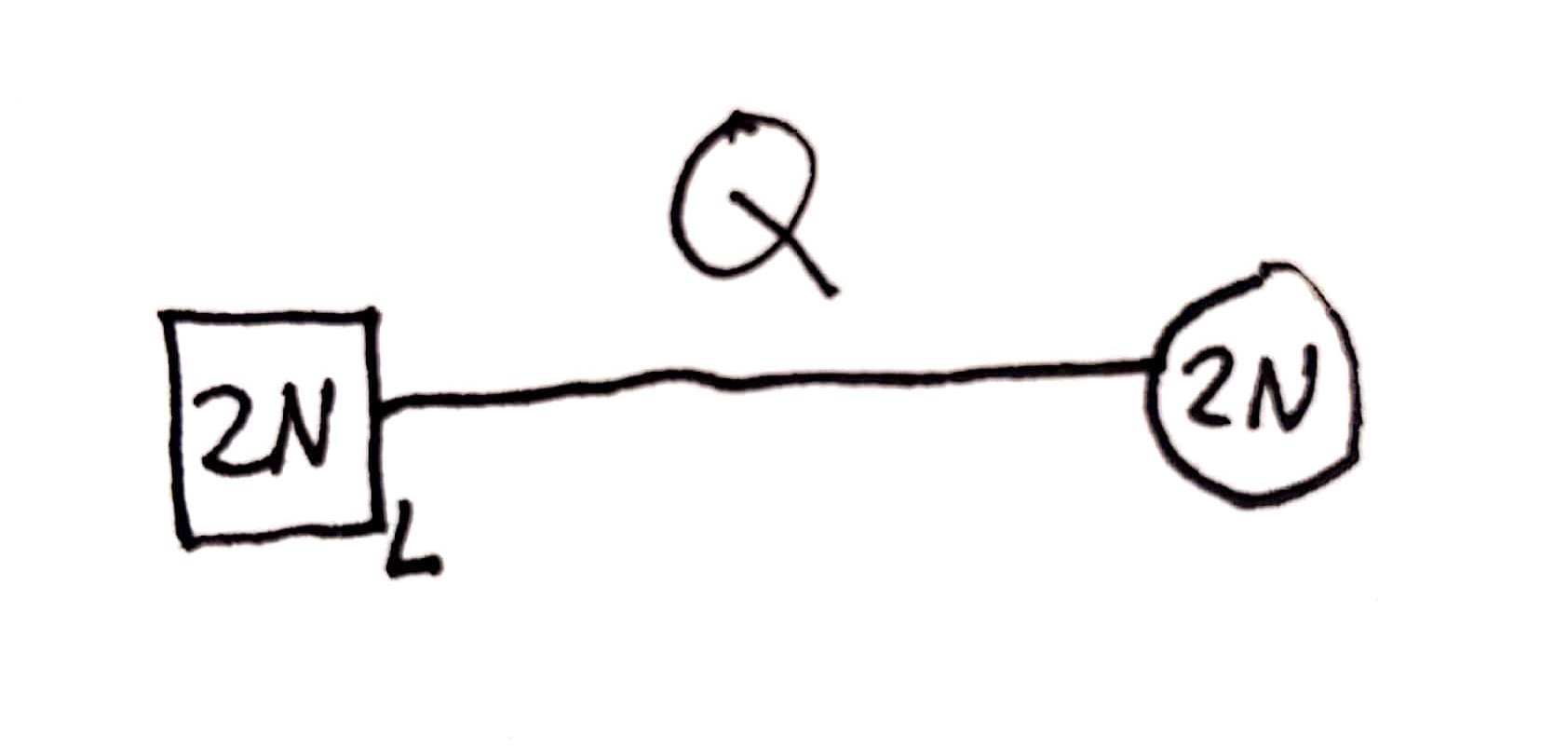}
		\caption{}
		\label{fig:foldedmother}
	\end{subfigure}%
	\begin{subfigure}{.45\textwidth}
		\centering
		\includegraphics[width=0.7\linewidth]{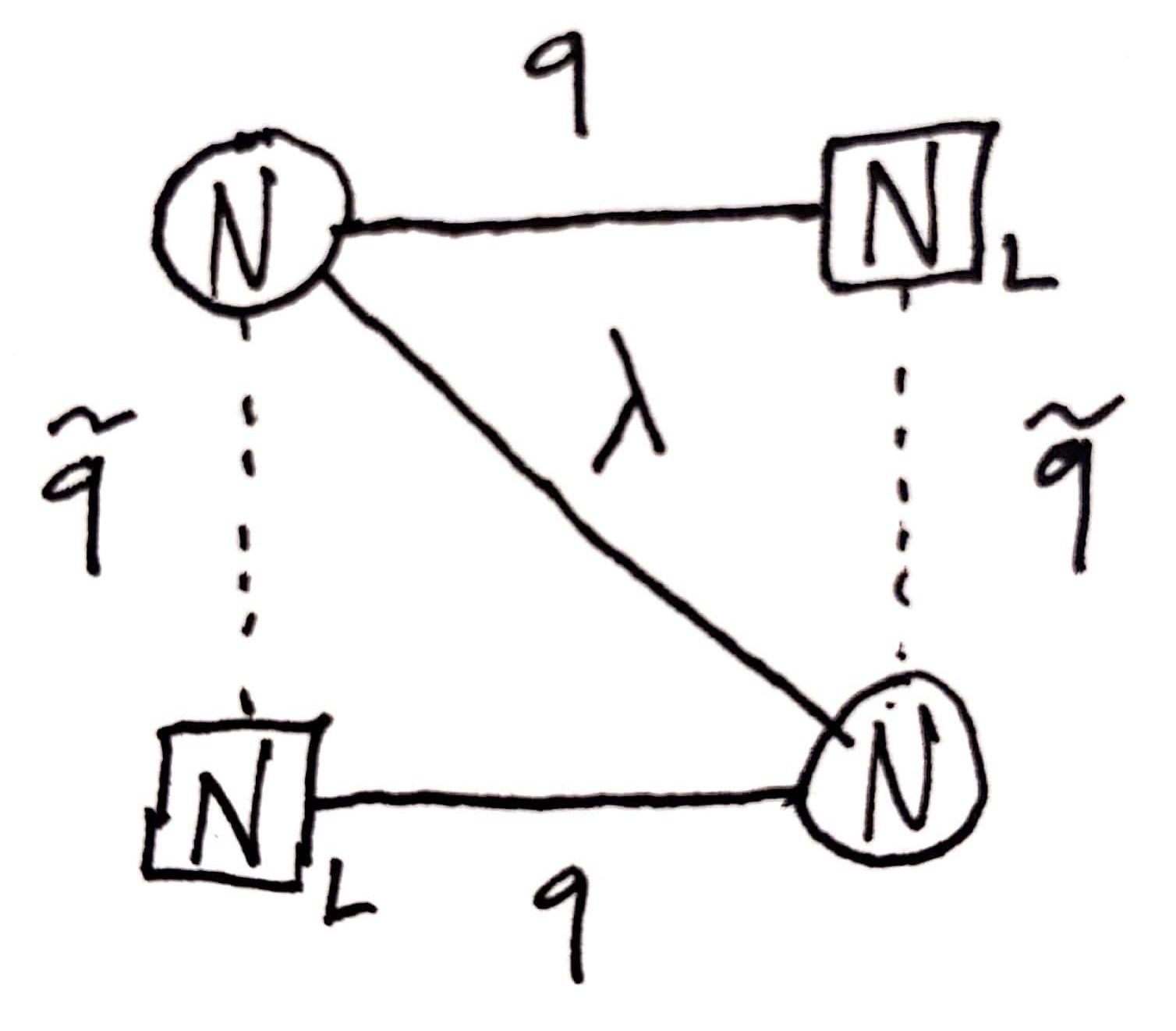}
		\caption{}
		\label{fig:foldeddaughter}
	\end{subfigure}
	\caption{In (a), a quiver diagram for the (left-chiral fields in) the mother theory. A circle corresponds to a gauge symmetry and a square to a flavor symmetry. Lines between denote bifundamentals, which here represents a full chiral supermultiplet. In (b), a quiver diagram for the orbifold daughter theory described in the text. The gauge and flavor groups have been divided into two distinct groups each, and the degrees of freedom in the chiral and vector supermultiplet have been shuffled around. }
	\label{fig:foldedquiver}
\end{figure}

\begin{figure}
	\centering
	\includegraphics[width=1\linewidth]{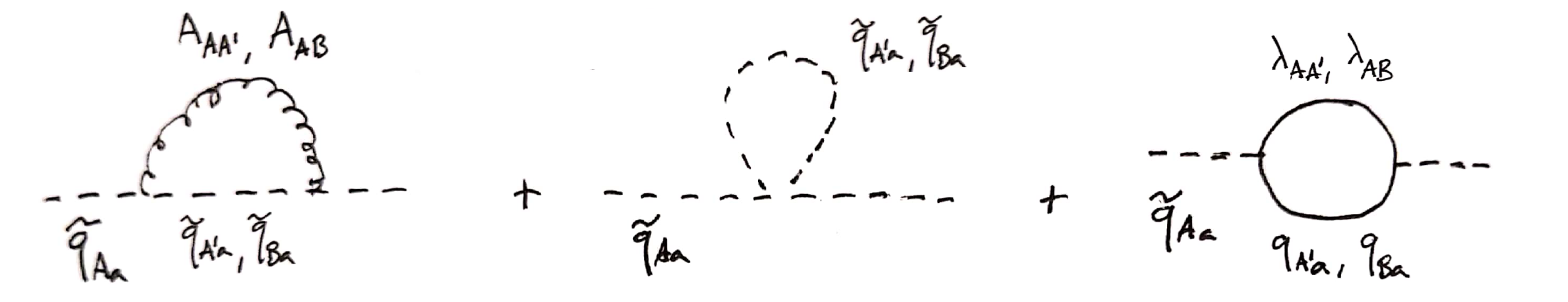}
	\caption{Representative diagrams contribution to the one-loop mass correction of a scalar in the supersymmetric mother theory. The daughter theory projects out some of the internal states of each diagram in a pattern that is non-supersymmetric but still enforces cancellations.}
	\label{fig:foldedsusy}
\end{figure}

It is not too hard to roughly see the magic of how the orbifold-projected theory continues to protect scalar masses. Draw the one-loop the diagrams in the mother theory which would contribute to a calculation of the mass of, say, $\tilde{q}_{Aa}$, as in Figure \ref{fig:foldedsusy}. In the mother theory we know there are no quadratic corrections by supersymmetry. In the daughter theory, half of each sort of diagram will be eliminated by the orbifold projection, so it's clear that there will still be no quadratic divergences. But because it is different internal states that have been eliminated for different classes of diagrams, the daughter theory has no supersymmetry to speak of! Again, as we saw in the example of the twin Higgs, the magic is in that at one-loop we really only need to get the counting right, and so we can use orbifolding to construct theories which do that in clever ways.

The structure of the theory can be succinctly summarized in a `quiver' or `moose' diagram where the various symmetry groups correspond to nodes in a graph and the matter is represented by links between these groups which represent their charges, as in Figure \ref{fig:foldedquiver}.

Despite the fact that our daughter theory has \textit{no} supersymmetry, the orbifold correspondence guarantees that in the $N \rightarrow \infty$ limit the correlation functions have full supersymmetric protection. For finite $N$ the supersymmetric relations are broken, but only by $1/N$ corrections. Going through this explicitly is useful, and pedagogical discussions of it can be found in \cite{Schmaltz:1998bg,Craig:2014aea}.

\subsubsection*{Example 2: The $\Gamma$-plet Higgs}\addcontentsline{toc}{subsubsection}{Example 2: The $\Gamma$-plet Higgs} We can understand the twin Higgs as the simplest example of a $\Gamma$-siblings Higgs which is the orbifold projection $SU(3\Gamma)\times SU(2\Gamma)\times SU(\Gamma)_F/\mathbb{Z}_\Gamma$, where the first two factors are gauged and the last is a flavor symmetry. We focus on the sector which contains the Yukawa interaction of the top quark, as the top partners have the most effect on the naturalness of the Higgs. In the mother theory the matter content is:
\begin{center}
	\begin{tabular}{ |c|c|c|c| } 
		\hline
		 & $SU(3\Gamma)$ & $SU(2\Gamma)$ & $SU(\Gamma)_F$ \\ \hline
		$H$ & - & $\Box$ & $\overline{\Box}$ \\ \hline
		$Q$ & $\Box$ & $\overline{\Box}$ & - \\ \hline
		$u$ & $\overline{\Box}$ & - & $\overline{\Box}$ \\ \hline
	\end{tabular}
\end{center}
where `$\Box$' denotes the fundamental representation and `$\overline{\Box}$' denotes the antifundamental, and we note that the charges allow the operator $y_t H Q u$. The Abelian generalization of the twin Higgs is found by embedding the $\mathbb{Z}_\Gamma$ in these groups as the block-diagonal element $\mathbf{S}_n = \text{diag}\left[\mathds{1}_n,\mathds{1}_n \exp(2\pi i/\Gamma),\dots,\mathds{1}_n \exp(2\pi k i/\Gamma),\dots ,\mathds{1}_n \exp(2\pi (\Gamma-1) i/\Gamma)\right]$, where $n=1,2,3$ corresponds to which $SU(n\Gamma)$ factor the element belongs to and $\mathds{1}_n$ is the $n\times n$ identity matrix. 

The pattern of orbifolding is extremely simple in this example. We write the Higgs field as a matrix of $\Gamma$ columns and $2\Gamma$ rows in blocks of two, with the field in block $i,j$ being $H^{(i)}_{A_j}$ where $i$ simply labels the column and $A_j = 1,2$ is an $SU(2)$ index. Now we can look at how this field transforms under the chosen $\mathbb{Z}_\Gamma$, 
\begin{align}
H &= \begin{pmatrix}
H^{(0)}_{A_0} & \dots & H^{(\Gamma - 1)}_{A_0} \\
\vdots & \ddots & \vdots \\
H^{(0)}_{A_{\Gamma-1}} & \dots & H^{(\Gamma - 1)}_{A_{\Gamma - 1}} \\
\end{pmatrix} \rightarrow \mathbf{S}_2 H \mathbf{S}_1^\dagger \\ &\rightarrow
\begin{pmatrix}
\mathds{1}^{A_0} H^{(0)}_{A_0} \mathds{1} & \dots & \mathds{1}^{A_0}  H^{(\Gamma - 1)}_{A_0} \mathds{1} e^{-2\pi (\Gamma-1) i/\Gamma} \\
\vdots & \ddots & \vdots \\
\mathds{1}^{A_{\Gamma-1}}e^{2\pi (\Gamma-1) i/\Gamma} H^{(0)}_{A_{\Gamma-1}} \mathds{1} & \dots & \mathds{1}^{A_{\Gamma-1}}e^{2\pi (\Gamma-1) i/\Gamma} H^{(\Gamma - 1)}_{A_{\Gamma - 1}} e^{-2\pi (\Gamma-1) i/\Gamma} \\
\end{pmatrix},
\end{align}
and we see immediately that the only invariant elements are those on the diagonal. It is simple to repeat this for the $Q,u$ fields to find the same feature. Then in this example there are no off-diagonal fields at all, and the daughter theory consists of $\Gamma$ SM-like sectors. These sectors are all identical and so have a $S_\Gamma$ rearrangement symmetry, leading to a one-loop quadratic potential of the form
\begin{equation}
V_{1-\text{loop}} \sim c \frac{\Lambda^2}{16 \pi^2} \sum_{i=0}^{\Gamma-1} \left|H^{(i)}_{A_i}\right|^2,
\end{equation}
That is, just as in the twin Higgs, the $SU(2\Gamma)$ symmetry of the scalar potential is respected by the one-loop quadratic corrections as a result of the discrete symmetry despite being explicitly broken by the gauge groups. This clearly opens up a much wider space of Neutral Naturalness models where the Higgs is a pseudo-Nambu Goldstone boson and so receives some protection of its mass without new colored particles at the electroweak scale. 

The general orbifolding approach to constructing models of Neutral Naturalness was fully laid out in \cite{Craig:2014roa,Craig:2014aea}, where they in particular explore a `regular representation' embedding of the discrete group into the continuous symmetries of the mother theory.

This approach of orbifolding to find low-energy models with accidental symmetries is useful also because such models come along with guides for how to UV-complete them. When we wrote down the twin Higgs model above, it was perhaps not obvious that there is a nice UV completion of this theory. But now we see that the twin Higgs is an orbifold projection of a $SU(6)\times SU(4)$ gauge theory by $\mathbb{Z}_2$, so we expect we can uplift this to a five-dimensional UV completion where the twin Higgs emerges \textit{dynamically} at low energies from orbifold boundary conditions. So we can confidently study solely the low-energy effective theory of the zero-modes we've projected out without worrying explicitly about whether a UV completion exists.

\section{The Loerarchy Problem}

Modern particle theorists must now confront a new version of the issue of electroweak naturalness. When originally understood, the pressing problem was understanding what sorts of UV structure could protect a scalar from large mass corrections. Of course this structure needed to be broken to get the SM structure in the IR and there's lots of interesting physics and subtleties on that end as well. But with the structure of the weak scale barely explored, the ways in which this could be done were abundant.

Over the ensuing decades we explored electroweak physics with increasing precision, which has provided invaluable guidance for how the SM structure must appear. Gradually the IR dynamics became more and more constrained to the point where now, as we have emphasized, we have enormous constraints on any appearance of new physics with SM quantum numbers up to mass scales that are often many times the electroweak scale.

We thus have a qualitative change in the electroweak naturalness issue over the past decades. We term the modern, empirical, low-energy puzzle of electroweak naturalness without visible structure around the weak scale as `The Loerarchy Problem', for obvious reasons.\footnote{For readers who do not share my sense of humor, the reasoning is an implied fake etymology for the word `hierarchy' as `high + erarchy', and a retcon of the term `The Hierarchy Problem' as emphasizing the `high' energy, UV aspects of the issue, by which we are comparatively emphasizing the `low' energy, IR aspects, suggesting that a natural parallel term would similarly combine `low + erarchy' to form `loerarchy', which is not a dictionary word.}

In this language, the little hierarchy problem is just one approach toward this problem, which assumes that one of the classic solutions is just out of reach and another module is needed to postpone the appearance of SM charged particles. While it's more than worthwhile to continue looking for and exploring those theories, in the face of increasingly powerful LHC data in excellent agreement with the Standard Model it's worth thinking transversely. As intriguing as the Neutral Naturalness models are, these classes of models rely on one-loop accidents and so do not completely relieve the empirical pressure from a lack of new physics at the energy frontier. Such models are gradually also being constrained by the LHC, so what else are we to do? The \textit{maximalist} interpretation of the LHC data is that Nature may be leading us to the conclusion that \textit{there is no new physics at the weak scale}. With every inverse femtobarn of LHC data without a signal of new physics, the impetus for such a paradigm shift becomes stronger.

But how can we generate a scale \textit{without} additional structure appearing surrounding it? Within the context of effective field theory, decoupling theorems demand that if the RG evolution is to change around a scale $M$ it must be due to fields close to the scale $M$. This was perhaps clearest in dim reg with $\overline{\text{MS}}$, where the beta functions explicitly change solely at mass thresholds, though the Wilsonian approach is more useful for physical intuition. So how are we to break this feature? This undertaking is the maximalist approach to the Loerarchy Problem.

\section{Violations of Effective Field Theory} \label{sec:EFTbar}

\epigraph{There are more things in heaven and earth, Horatio, \\
	Than are dreamt of in your philosophy.}{William Shakespeare \\ \textit{Hamlet}, c. 1600 \cite{shakespeare_hamlet}}

The line of thought we suggest here is that perhaps the apparent violation of EFT expectations at the weak scale is a sign of the breakdown of EFT itself. Depending on how much background in particle physics one has this statement may seem more or less heretical, but the idea is not as radical as it may at first seem---for one reason, the cosmological constant problem has inspired sporadic reexaminations of the validity of effective field theory in our universe for decades.

The cosmological constant problem is the fine-tuning issue with the \textit{other} dimensionful parameter in the Standard Model. Just as with the Higgs mass, there is no protective symmetry in the Standard Model for the vacuum energy, and so the natural expectation is $\Lambda \sim M_\pl^4$, some $120$ orders of magnitude higher than observations suggest. However there's an important difference in the severity of these problems---for the Higgs, as emphasized in Section \ref{sec:input}, the worrisome mass corrections are those from new physics. This is why the severity of the problem has only ratcheted up in recent years, as we have seen nothing to protect the Higgs from BSM mass corrections. But for the vacuum energy, there are finite, calculable, physical contributions in the Standard Model itself! For a start, EWSB by the Higgs yields a contribution $\sim -v^4$, and chiral symmetry-breaking yields a contribution $\sim - \Lambda_{QCD}^4$. How is it that these can be nearly perfectly canceled off in the late universe? 

There have been important attempts to address the cosmological constant problem with a violation of effective field theory, from Coleman's suggestion \cite{Coleman:1988tj} that nonlocality induced by wormholes may allow the early universe to be sensitive to late-time requirements to Cohen-Kaplan-Nelson's suggestion \cite{Cohen:1998zx} that the Bekenstein bound demands an infrared cutoff on the validity of any EFT. From one perspective, our suggestion to extend this philosophy to the hierarchy problem appears natural in light of its apparent need in cosmology. We can point to an even-more-general motivation with the realization that gravity necessarily violates EFT.

\subsection{Gravity and EFT} \label{sec:graveft}

The perturbative quantum field theory of the Einstein-Hilbert Lagrangian \cite{Hilbert1915} is quite clearly an effective field theory of a symmetric two-index tensor field which obeys diffeomorphism invariance and with power counting in $1/M_\pl$ \cite{Donoghue:1994dn}, as we can easily see by writing it out and expanding around flat space $g_{\mu\nu} = \eta_{\mu\nu} + h_{\mu\nu}$:
\begin{equation}
-\mathcal{L}_{EH} = \frac{1}{2 M_\pl^2} \sqrt{g} R \sim \partial^2 h^2 + \frac{1}{M_\pl} \partial^2 h^3 + \frac{1}{M_\pl^2} \partial^2 h^4 + \dots
\end{equation}
where we have only given the schematic form of the operators in terms of the number of derivatives and linearized gravitational fields they contain, as the full expressions quickly become complicated \cite{Feynman:1963ax,DeWitt:1967ub}. Then we may expect that this effective field theory will be a good approximation to infrared gravitational physics until we get to energies close to the Planck scale, at which point the higher-dimensional operators are unsuppressed and we need a UV completion. From this bottom-up approach it's not clear why gravity should be particularly special. But we can get some insight at a very basic level by thinking about gravitational scattering.

Let's compare two effective field theories: the four-Fermi theory of the weak interaction below the weak scale $G_F^{-1/2}$ and the perturbative theory of quantum gravity below the Planck mass. If we consider scattering two leptons at $\sqrt{s} \ll G_F^{-1/2}$, we can make very precise predictions by computing in the four-Fermi theory---the possible final states, the differential cross-section; whatever we'd like. And similarly if we scatter two particles at $\sqrt{s} \ll M_{\pl}$, we can compute to high precision in quantum gravitational corrections what will happen.

Now inversely, imagine scattering two leptons at $\sqrt{s} \gg G_F^{-1/2}$ in the four-Fermi theory. At a scale like $10^{15} G_F^{-1/2}$, the EFT has obviously broken down and we can say essentially nothing about what this process will look like---any calculation we tried would be hopelessly divergent, and we have no idea what sorts of states might exist at energies that large.  

However, what if we scatter two particles at $\sqrt{s} \gg M_\pl$, say a ridiculously trans-Planckian scale like $10^{15} M_\pl$? In this case, in fact, we know what will happen to incredible precision---a black hole will form! This will be a `big' black hole of mass $\gg M_\pl$ with a lifetime of order days. It will be well-described by classical general relativity for some macroscopic time, and then semiclassical GR for an $\mathcal{O}(1)$ fraction of the full lifetime. In fact, unless you are interested in \textit{incredibly} detailed measurements which involve collecting essentially every emitted Hawking quanta (of which there will be $n \sim 10^{30}$ in this case!) and finding their entanglement structure, we know how to describe the evaporation nearly completely.\footnote{I recommend Giddings' Erice lectures \cite{Giddings:2011xs} for more on the perspective of quantum gravitational behavior as a function of the Mandelstam variables.}

So something profoundly weird is going on. The key point being that in gravity, the far UV of the theory is controlled by \textit{classical, infrared} physics. This is obviously a feature that we do not see in other EFTs.

This is not a new idea; it has long been known that gravity contains low-energy effects which cannot be understood in the context of EFT. The fact that black holes radiate at temperatures inversely proportional to their masses \cite{Hawking:1974sw} necessitates some sort of `UV/IR mixing' in gravity---infrared physics must somehow `know about' heavy mass scales in violation of a na\"{i}ve application of decoupling. As a perhaps-more-fundamental \textit{raison d'\^{e}tre} for such behavior, the demand that observables in a theory of quantum gravity must be gauge-(that is, diffeomorphism-)invariant dictates that they must be nonlocal (see e.g. \cite{Torre:1993fq,Giddings:2005id,Donnelly:2015hta,Donnelly:2016rvo,Giddings:2018umg}), again a feature which standard EFT techniques do not encapsulate. In view of this, the conventional position is that EFT should remain a valid strategy up to the Planck scale, at which quantum gravitational effects become important. But once locality and decoupling have been given up, how and why are violations of EFT expectations to be sequestered to inaccessible energies? Indeed, the `firewall' argument \cite{Almheiri:2012rt} evinces tension with EFT expectations in semiclassical gravity around black hole backgrounds at arbitrarily low energies and curvatures, as does recent progress finding the Page curve from semiclassical gravity \cite{Penington:2019npb,Almheiri:2019hni,Almheiri:2019psf,Almheiri:2019qdq,Penington:2019kki,Marolf:2020xie}.

That quantum gravitational effects will affect infrared particle physics is likewise not a new idea. This has been the core message of the Swampland program \cite{Vafa:2005ui}, which has been cataloging---to varying degrees of concreteness and certainty---ways in which otherwise allowable EFTs may conjecturally be ruled out by quantum gravitational considerations. These are EFTs which would look perfectly sensible and consistent to an infrared effective field theorist, yet the demand that they be UV-completed to theories which include Einstein gravity reveals a secret inconsistency. While this is powerful information, the extent to which the UV here meddles with the IR is relatively minor---just dictating where one must live in the space of infrared theories. Even so, they have been found to have possible applications to SM puzzles, including the hierarchy problem \cite{Cheung:2014vva, Ooguri:2016pdq, Ibanez:2017kvh, Ibanez:2017oqr, Hamada:2017yji, Lust:2017wrl, Gonzalo:2018tpb, Gonzalo:2018dxi, Craig:2018yvw, Craig:2019fdy,March-Russell:2020lkq}. 

Let us review briefly the approach to connect the Weak Gravity Conjecture (WGC) \cite{ArkaniHamed:2006dz} to the hierarchy problem. The WGC is one of the earliest and most well-tested Swampland conjectures, its formulation is relatively easy to understand, and it's emblematic of the way one might try to connect the hierarchy problem to Swampland conjectures in general.

The prime motivation for formulating the WGC was the well-known folklore that quantum gravity does not respect global symmetries. The simple argument for this fact is that `black holes have no hair'---the only quantum numbers a black hole has correspond to its mass, spin, and gauge charges. This means that if we make a black hole by smashing together a bunch of neutrons, there is no reason why it cannot decay into solely photons, which violates the global $U(1)_{B-L}$ symmetry of the SM but no gauge symmetries. The authors suggested that the fact that Abelian gauge symmetries smoothly become global symmetries as the gauge coupling vanishes means that something must go wrong with very small gauge couplings as well, so as for the physics to be smooth in this limit.

Another line of thinking for quantum gravity not respecting global symmetries, which connects more closely to the WGC formulation, is based on entropic arguments. If global symmetries could be exact, then we could create a big black hole with an arbitrary global charge, and wait a very very long time while it Hawking evaporates down to the Planck scale. Unlike gauge charges, the global charge does not affect the metric, so the black hole does not shed this charge as it evaporates. We then end up with a Planck-sized black hole with an arbitrarily large global charge, which would necessitate the existence of arbitrarily-many black hole microstates for a fixed-mass black hole. 
But this is a disaster! In calculating a scattering amplitude one has to sum over all possible intermediate states, in principle including black holes. These effects are surely Boltzmann-suppressed by enormous amounts, but if there are an \textit{infinite} number of possible black holes then any nonzero contribution from a single black hole will lead to a divergence. A cuter (albeit somewhat tongue-in-cheek) `hand waving argument' is provided by \cite{Marolf:2003wu}: Were there incredibly large numbers of super-Planckian states which could populate a thermal bath, a vigorous wave of your hand would produce them in Unruh radiation, and you could feel them against your hand as they evaporated.

Now if we have a gauge symmetry, there are no longer Planck-sized black holes with arbitrarily large charge because there is an extremality bound: $g |Q| \leq M/M_{pl}$, with $Q$ the gauge charge and $M$ the mass, in units of the gauge coupling $g$ and the Planck mass. If this were disobeyed a naked singularity would appear, which would violate cosmic censorship \cite{Penrose:1969pc}. But for a very tiny gauge coupling, while we no longer have strictly infinitely-many black holes at each mass, there are still an \textit{enormous} number of black hole microstates for a Planck-sized black hole as $g \rightarrow 0$. So a worry like the hand-waving argument still applies. 

This sort of reasoning motivated Arkani-Hamed, Motl, Nicolis, and Vafa to conjecture that just as quantum gravitational theories must not have exact global symmetries, they must also suffer some physical malady which disallows the limit where Abelian gauge symmetries become global symmetries. Their conjecture has two forms: The `magnetic' conjecture dictates that a quantum gravitational theory with a $U(1)$ gauge symmetry must be enlarged into a theory allowing magnetic monopoles by a cutoff $\Lambda \lesssim g M_{pl}$, which connects to such a description never being valid in the limit $g \rightarrow 0$. The `electric' form of the WGC is that such a theory must contain a particle which is `super-extremal'---it has charge greater than its mass $g q M_{pl} > m$. The existence of such a particle would destabilize the extremal, charged black holes, allowing them to decay (though the extent to which this really soothes our entropic worries is unclear). 

Well this super-extremality bound should look very interesting to us, as it provides an upper bound on a mass scale. While we cannot apply this directly to the Higgs because it is not charged under any unbroken Abelian gauge symmetries, we know that one of the Higgs' jobs is to provide mass to other particles. So if the weak gravity conjecture bound must apply to some state $\phi$ with mass $m_\phi = y v$, with $v$ the Higgs vev, then this still amounts to a bound on the electroweak scale. 

This was suggested in the context of gauged $U(1)_{B-L}$ with very tiny coupling giving an upper bound on the lightest neutrino mass \cite{Cheung:2014vva}, but the magnetic form of the WGC is difficult to deal with in this context. This can be circumvented by introducing a new dark Abelian gauge group $U(1)_X$ and charged states which get (some of) their mass from the Higgs \cite{Craig:2019fdy}. 

The way such a model solves the hierarchy problem is by changing the shape of our prior for the electroweak scale, as mentioned in Section \ref{sec:philosophy}. As na\"{i}ve effective field theorists with no information about quantum gravity, we assumed some sort of flat prior in $\left[-M_{pl},M_{pl}\right]$. But in fact much of this space is ruled out by quantum gravitational constraints, consisting of theories `in the Swampland', meaning that our prior should be reshaped to include only values which can actually be produced by a theory of quantum gravity. This is how much of the connection from Swampland conjectures to the real world has worked schematically: one says that Quantum Gravity demands one live in a subregion of the EFT parameter space.

In theory far more flagrant violations of low-energy expectations are permissible---that is, the extent to which quantum gravitational violation of EFT will affect the infrared of our universe is not at all certain. Of course any proposal to see new effects from a breakdown of EFT must contend with the rampant success of the SM EFT in the IR---though not in the \textit{far} IR, recalling the cosmological constant problem. Certainly a violation of EFT must both come with good reason and be deftly organized to spoil only those observed EFT puzzles. For the former, the need for quantum gravity is obviously compelling. As to the latter, it is interesting to note that the most pressing mysteries involve the relevant parameters in the SM Lagrangian.

Ultimately, our ability to address the hierarchy problem through quantum gravitational violations of EFT is limited by our incomplete understanding of quantum gravity. This motivates finding non-gravitational toy models that violate EFT expectations on their own, providing a calculable playground in which to better understand the potential consequences of UV/IR mixing. In Chapter \ref{sec:newtrail} we pursue the idea that UV/IR mixing may have more direct effects on the SM by considering noncommutative field theory (NCFT) as such a toy model. These theories model physics on spaces where translations do not commute \cite{Snyder:1946qz,Connes:1994yd}, and have many features amenable to a quantum gravitational interpretation---indeed, noncommutative geometries have been found arising in various limits of string theory \cite{Connes:1997cr,Douglas:1997fm,Seiberg:1999vs,Myers:1999ps}.


\chapter{Neutral Naturalness in the Sky}
\label{sec:InTheSky}
\setlength{\epigraphwidth}{0.5\textwidth}
\epigraph{In the beginning the Universe was created. \\
	This has made a lot of people very angry and been widely regarded as a bad move.
}{Douglas Adams \\ \textit{The Restaurant at the End of the Universe} (1980) \cite{adams1980restaurant}}
\setlength{\epigraphwidth}{0.6\textwidth}

\section{Particle Cosmology}

It is an amazing and serendipitous fact that the universe started off hot. As a result of the initially high energies and densities, the details of microscopic physics greatly affected the large-scale evolution of the universe. Since the speed of light is finite, by looking out in the sky at enormous distances we can not only learn about the history of the universe but we can use this information to learn about particle physics. While cosmology doesn't give us probes of arbitrarily high temperatures, there's still a humongous amount to be learned---in part due to further serendipity. The fact that the universe transitions from radiation domination to matter domination shortly before it becomes transparent to photons means that the cosmic microwave background (CMB) encodes information both about the light, radiation-like degrees of freedom as well as the matter density in the early universe. Had radiation domination ended far before recombination, it would be far more difficult to use the CMB to constrain light degrees of freedom like extra neutrinos. Had radiation domination ended far after recombination, there would be little evidence of dark matter in the CMB, which is the strongest evidence for particle dark matter instead of, say, a modification of gravity at large distances. In fact such a `cosmic coincidence' also occurs much later, as there is a very long epoch of matter domination before the universe transitions to dark energy domination. Were there just slightly less dark energy, its effects would be essentially invisible thus far in the history of the universe, and it would be very difficult to measure dark energy at all.

All that is to say that there is enormous value in collaboration between particle physics and cosmology. In this chapter we investigate this connection for the twin Higgs model in particular, though our findings are relevant for general Neutral Naturalness theories as well. In Section \ref{sec:twinhiggsintro} we noted that the energy frontier does not effectively probe these theories. Since they do not introduce new particles with Standard Model charges, it is only precision electroweak measurements made at colliders that constrain them at all. However, such theories are in fact probed very well by cosmology, as they introduce new light degrees of freedom. Despite the fact that these do not directly interact with normal matter, their gravitational effects still contribute to the evolution of the universe, and so the CMB provides a powerful constraint on new light particles.

It is this cosmological effect that provided the biggest obstacle to the original twin Higgs proposal \cite{Chacko:2005pe}, which became an urgent issue after the null results of run 1 of the LHC and the increased interest in models where the lightest states responsible for Higgs naturalness were SM-neutral. The landmark approach taken in \cite{Craig:2015pha} was to pare the model down to a `minimal' version where only those states necessary for Higgs naturalness appeared in the twin spectrum. This revived the twin Higgs as a solution to the little hierarchy problem, and their `fraternal' version brought about many interesting phenomenological possibilities.

The `fraternal twin Higgs' has a twin sector consisting---at energies below its cutoff---solely of the third generation of fermions, and with ungauged twin hypercharge. This brilliantly removes all light particles from the spectrum, so their effects would not cause trouble in the early universe. But this approach leaves perhaps a niggling unpleasant taste for those worried about parsimony. Yes the fraternal twin Higgs introduces fewer new particles than the mirror twin Higgs, so a na\"{i}ve desire to solve problems with few ingredients might suggest that this is a windfall. However, the mirror twin Higgs really consists of only two `ingredients': a $\mathbb{Z}_2$ symmetry and some soft breaking to misalign the resulting Higgs vevs---whereas the fraternal twin Higgs has much more structure.

While those statements seem to be of a very subjective sort, we can ground this unease in physics by considering what's needed to UV-complete such a model. At the cutoff of this model $\Lambda \sim 10 \text{ TeV}$, we need the $\mathbb{Z}_2$ to still be a relatively good symmetry among the largest couplings in the twin sector---that is, the gauge couplings $g_2, g_3$ and the top Yukawa $y_t$---such that the cancellation of contributions from the two sectors to the Higgs potential works well. Yet in other parts of the theory we have done great violence to the structure of the theory, having broken twin hypercharge and removed parts of the spectrum. How are we to ensure firstly that the correct degrees of freedom gain large masses, and secondly that this does not radiatively feed into the remaining light degrees of freedom?

In the midst of this digression, I should mention a terminological confusion. In the literature, the phrase `mirror twin Higgs' often refers to models in which the full collection of twin degrees of freedom are present in the low-energy theory, regardless of how much $\mathbb{Z}_2$-breaking is present. Judicious introduction of such asymmetries has been used to create models which avoid cosmological issues by making all the twin fermions heavy, while still staying within the technical definition of the mirror twin Higgs. But this is an overreliance on a definition; the fraternal twin Higgs is merely a limit of these theories in which the $\mathbb{Z}_2$-breaking is severe enough to push some degrees of freedom above the cutoff. The real distinction between classes of Neutral Naturalness models should be between those which break the $\mathbb{Z}_2$ only minimally and those which do greater violence to the symmetry. It is this distinction which classifies the difficulty involved in finding a UV completion.

Now let me emphasize that this is not to undercut the value of such a model. After all, the Yukawa interactions in the Standard Model badly break the large global symmetries it would otherwise have. Indeed, the fraternal twin Higgs showcased interesting phenomena, pointed to new general experimental probes, and provided a basis for many intriguing lines of research. In fact we will return to this model in Chapter \ref{sec:InTheGround} to study a novel collider search strategy to which it lent credence and which turns out to be a broadly useful probe of many theories of BSM physics. Despite the fact that the vast, vast majority of theory papers written in particle physics will not ultimately be the exact right model of the universe, they still contain value. They may guide experimental searches toward interesting classes of signals to look for, or teach us new things about the range of particle phenomenology or quantum field theories. Regardless, we obviously don't know in advance which model will be correct, so exploring all possible directions is crucial.

Yet when there \textit{is} the possibility for a more parsimonious model, it's certainly worth pursuing that option. This is the philosophy that led to my collaborators and me looking into the prospect of attaining a realistic twin Higgs cosmology that respected the $\mathbb{Z}_2$ symmetry.

\section{Asymmetric Reheating} \label{sec:asymmetric}

\maketitle
\flushbottom
\end{comment}

\subsection{Introduction}

The primary challenge to the mirror Twin Higgs comes not from LHC data, but from cosmology. An exact $\mathbb{Z}_2$ exchange symmetry predicts mirror copies of light Standard Model neutrinos and photons states, which contribute to the energy density of the early universe. In particular, twin neutrinos and a twin photon provide a new source of dark radiation that is strongly constrained by CMB and BBN measurements \cite{Ade:2015xua, Cyburt:2015mya}. While these constraints could be avoided if the two sectors were at radically different temperatures, the Higgs portal couplings required by naturalness keep the two sectors in thermal equilibrium down to relatively low temperatures. 

Constraints on dark radiation in the mirror Twin Higgs have motivated models in which the $\mathbb{Z}_2$ symmetry is approximate \cite{Craig:2014aea, Craig:2014roa,Geller:2014kta, Barbieri:2015lqa, Low:2015nqa,Craig:2015pha,Craig:2016kue,Barbieri:2016zxn,Csaki:2017spo,Liu:2019ixm,Harigaya:2019shz}, in which case the dark radiation component can be made naturally small. These models 
have proved to be a boon for phenomenology. Among other things, they
quite generally motivate looking for 
Higgs decays to long-lived particles at colliders \cite{Clarke:2015ala,Curtin:2015fna,Csaki:2015fba,Pierce:2017taw,Alipour-Fard:2018lsf,Kilic:2018sew,Alipour-fard:2018mre,Li:2019ulz} 
and contain well motivated dark matter (DM) candidates~\cite{Garcia:2015loa,Craig:2015xla,Garcia:2015toa,Farina:2015uea,Freytsis:2016dgf,Farina:2016ndq,Prilepina:2016rlq,Barbieri:2017opf,Hochberg:2018vdo,Cheng:2018vaj,Terning:2019hgj,Earl:2019wjw}. However, such cosmological fixes come at the cost of minimality, as models with approximate $\mathbb{Z}_2$ symmetries require a considerable amount of additional structure near the TeV scale. 

In this work we take an alternative approach and investigate ways in which early universe cosmology can reconcile the mirror Twin Higgs with current CMB and BBN observations. In doing so, we find compelling scenarios that transfer the signatures of electroweak naturalness from high-energy colliders to cosmology. We consider several possibilities in which the energy density of the light particles in the twin sector is diluted by the out-of-equilibrium decay of a new particle after the two sectors have thermally decoupled. Crucially, the new physics in the early universe respects the exact (albeit spontaneously broken) $\mathbb{Z}_2$ exchange symmetry of the mirror Twin Higgs. This symmetry may be used to classify representations of the particle responsible for this dilution. We concentrate on two minimal cases:
In the first, the long-lived particle is $\mathbb{Z}_2$-even and the asymmetry is naturally induced by kinematics. In the second, there is a pair of particles which are exchanged by the $\mathbb{Z}_2$ symmetry and which may be responsible for inflation.\footnote{A third case exists, in which the particle is $\mathbb{Z}_2$-odd. This may additionally be related to the spontaneous $\mathbb{Z}_2$-breaking in the Higgs potential, although we find that a realisation of such a scenario is dependent upon the UV completion of the model.} Moreover, in these cases the new physics does not merely reconcile the existence of a mirror twin sector with cosmological constraints, but predicts contributions to cosmological observables that may be probed in current and future CMB experiments. This raises the prospect of discovering evidence of electroweak naturalness first through cosmology, rather than colliders, and provides natural targets for future cosmological constraints on minimal realizations of neutral naturalness.

The next sections are organized as follows: In Section \ref{sec:thermal} we discuss the thermal history of the mirror Twin Higgs, with a particular attention to the interactions keeping the Standard Model and twin sector in thermal equilibrium and the cosmological constraints on light degrees of freedom. In Section \ref{sec:late} we present a simple model where the out-of-equilibrium decay of a particle with symmetric couplings to the Standard Model and twin sector leads to a temperature difference between the two sectors after they decouple. We turn to inflation in Section \ref{sec:twinflation}, constructing a model of ``twinflation'' in which the softly broken $\mathbb{Z}_2$-symmetry extends to the inflationary sector and leads to two periods of inflation. The first primarily reheats the twin sector, while the second primarily reheats the Standard Model sector. We conclude in Section \ref{sec:conc}.

\subsection{Thermal History of the Mirror Twin} \label{sec:thermal}

The primary challenge to the mirror Twin Higgs comes from cosmology, rather than collider physics. The mirror Twin contains not only states responsible for protecting the Higgs against radiative corrections (such as the twin top), but also a plethora of extra states due to the $\mathbb{Z}_2$ symmetry that are irrelevant to naturalness. The lightest of these, namely the twin photon and twin neutrinos, contribute significantly to the energy density of the early universe around the era of matter-radiation equality, since they have a temperature comparable to that of the Standard Model plasma at all times. This is because the same Higgs portal coupling that makes the Higgs natural also keeps the two sectors in thermal equilibrium down to $\mathcal{O}$(GeV) temperatures. Then the identical particle content in the twin and Standard Model sectors guarantees that they remain at comparable temperatures even after they decouple - for every massive Standard Model species that becomes non-relativistic and transfers its entropy to the rest of the plasma, its twin counterpart does the same within a factor of $f/v$ in temperature.

In this section we undertake a detailed study of the decoupling between the Standard Model and twin sectors as well as the constraints from precision cosmology.

\subsubsection{Twin Degrees of Freedom} \label{Sec:Limits}

\begin{figure}[h!]
\centering
\includegraphics[width=1.0\linewidth]{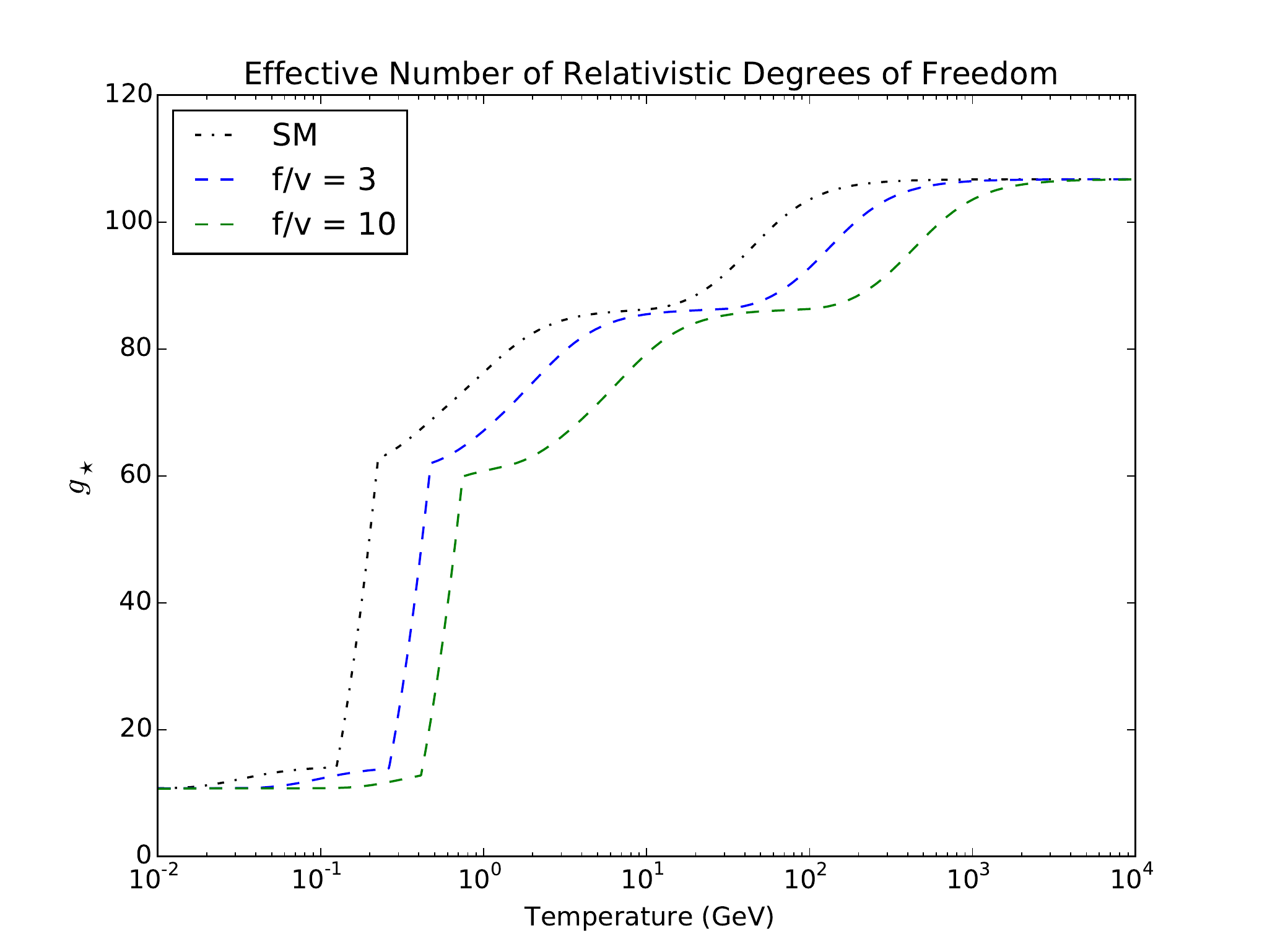}
\caption{The effective number of relativistic degrees of freedom for mirror Twin Higgs models for different values of $f/v$. The dash-dotted line is the for the Standard Model $g^{\text{SM}}_\star(T)$ and the dashed lines are the twin sector degrees of freedom $g^{\text{t}}_\star(T)$. The evolution of $g_\star$ during the QCD phase transition (QCDPT) is not well-understood, so we assign the SM QCDPT a central value of $175$ MeV and a width of $50$ MeV and interpolate linearly between the values of $g_\star$ at $225$ MeV for free partons and at $125$ MeV for pions. Further discussion may be found in \protect\cite{Drees:2015exa}. For the twin sector we use a central value and width which are $(1 + \log(\frac{f}{v}))$ times larger than the SM values. Note that new mass thresholds, expected to appear at energies $\sim 10$ TeV in UV completions of the twin Higgs, have not been included.}
\label{Fig:geff}
\end{figure}

In thermal equilibrium, each relativistic degree of freedom has roughly the same energy density. In general, we express the energy density of the universe $\rho$ during the radiation-dominated era as $\rho \equiv g_{\star} \frac{\pi^2}{30} T^4,$ where we define $g_\star$ through this relation as the effective number of relativistic degrees of freedom and $T$ the temperature of the SM photons. This then determines the evolution of the scale factor through the first Friedmann equation
\bea
H = \frac{1}{\mpl} \left[ \frac{\pi^2}{90} g_\star T^4\right]^{1/2} 
\eea
(assuming spatial flatness), where $\mpl$ is the reduced Planck mass. In general, the energy density of a particular species $i$ may be computed from $\rho_i = g_i \int \frac{d^3p}{(2\pi)^3} f_i(p,T_i) E(p)$, where $g_i$ are the number of internal degrees of freedom, $E(p)$ is the energy as a function of momentum $p$, while $f_i(p,T_i)$ is the phase-space number density and is a Bose-Einstein or Fermi-Dirac distribution if the species is in equilibrium at temperature $T_i$. The number of effective relativistic degrees of freedom may then be defined for each sector separately as $g_{\star}^{\text{SM}}(T)$ and $g_{\star}^{\text{t}}(T) $ satisfying $\rho_{\text{SM}}(T)=\frac{\pi^2}{30}g_{\star}^{\text{SM}}(T)  T^4$ and $\rho_{\text{t}}(T)=\frac{\pi^2}{30}g_{\star}^{\text{t}}(T) T^4$, respectively, where $\rho_{\text{SM}}(T)$ and $\rho_{\text{t}}(T)$ are the total energy densities of SM and twin particles. The values of $g_{\star}(T)$ for the SM and twin sectors are shown in Figure \ref{Fig:geff}, where all species within each sector are in thermal equilibrium. These can then be used to calculate the total number $g_\star$ as a function of temperature, by weighting twin sector energy density by its temperature: $g_\star(T)=g_\star^{\text{SM}}(T)+g_\star^{\text{t}}(\hat{T})(\hat{T}/T)^4$, where $\hat{T}$ is the twin sector photon temperature when the SM photon temperature is $T$.

Likewise, entropy densities for each sector $i$ are defined as $s_i(T)=\frac{2\pi^2}{45}g^{\text{i}}_\star(T) T^3$. We neglect the small differences between the number of relativistic degrees of freedom defined from energy and entropy densities, which are not significant over the range of temperatures of interest here.

\subsubsection{Decoupling} \label{Sec:Decoupling}

In the early universe, the two sectors are thermally linked by interactions mediated by the Higgs, which, through mixing with both $h_A$ and $h_B$ components, allows for SM fermions and weak bosons to scatter off or annihilate into their twin counterparts. However, once the temperature drops sufficiently for this Higgs-mediated interaction to become rare on the expansion time-scale, the sectors decouple and thereafter thermally evolve independently. More precisely, thermal decoupling will occur once the rate at which energy can be exchanged between SM and twin particles (through the Higgs) falls below the Hubble rate. 

Thermal decoupling is traditionally formulated from the Boltzmann equations describing the evolution of single-particle phase space number densities, wherein collisions induce instantaneous changes to the shape of these distributions. When the collisions occur faster than the expansion rate, the phase space probability density functions of the interacting species are expected to relax to an equilibrium distribution (Boltzmann, neglecting quantum statistics, will be applicable to our case). However, once the rate of collisions falls below the expansion rate, collisions become rare on cosmological time scales and the phase space distributions depart from equilibrium. The decoupling temperature is determined as that at which the scattering rate of a participating particle, $\Gamma$, drops below the Hubble rate, assuming that this occurs instantaneously across the entire phase space where the number density is significant. This formulation can be used to determine the time at which a particular species of particle will cease to scatter off twin particles on cosmological time scales.

In the case of interest here, however, both sectors of particles remain thermalised within themselves while the interactions between sectors freeze-out. This implies that the phase space number densities are still Boltzmann distributions throughout decoupling, with a different temperature for each sector. As it is the twin sector temperature that ultimately determines the impact of the light twin degrees of freedom on the cosmological observables (discussed below in Section \ref{Sec:Limits}), we wish to describe the thermal evolution of the two sectors by that of their entire energy or entropy content and the bulk heat flows between them. They may then be identified as thermally decoupled once the rate at which they exchange energy falls below the expansion rate. 

If the SM and twin sector plasmas have temperatures $T$ and $\hat{T}$ respectively, then calling $q$ the net heat flow density from the SM to the twin sector, the rate at which the twin entropy densities $s_{\text{t}}$ and $s_{\text{SM}}$ evolve is determined by
\bea
\frac{ds_{\text{t}}}{dt}+3Hs_{\text{t}}&=&\frac{1}{\hat{T}}\frac{dq}{dt}=\frac{1}{\hat{T}}\Big(\frac{dq_{\text{in}}}{dt}-\frac{dq_{\text{out}}}{dt}\Big)\label{ThermEv}\\
\frac{ds_{\text{SM}}}{dt}+3Hs_{\text{SM}}&=&\frac{-1}{T}\frac{dq}{dt}=-\frac{1}{T}\Big(\frac{dq_{\text{in}}}{dt}-\frac{dq_{\text{out}}}{dt}\Big).
\eea
Here, $H$ is the Hubble rate. The heat flow rate has been decomposed into the sum of the energy transferred into and out of the twin sector by collisions in the second equality in each line, where $\frac{dq_{\text{in}}}{dt}$ and $\frac{dq_{\text{out}}}{dt}$ are both positive.

The rate of heat flow $q$ may be calculated by performing a phase space average of the rate that energy is transferred from the SM to the twin sector through particle interactions. Since the decay rates of top quarks or weak bosons are fast compared to their scattering rate and the Hubble rate, energy transferred to them is instantaneously transferred to the rest of the plasma. Similarly, the scattering rate of lighter fermions off other particles of the same sector (such as photons or gluons) is much faster than their interaction rate with twin fermions. Energy transferred to the lighter fermions therefore quickly diffuses throughout their respective plasmas. The rate of heat flow between sectors may therefore be well approximated by the rate at which energy is transferred from SM particles to twin particles in Higgs mediated interactions. This may occur through elastic scattering of SM particles off twin particles or annihilations of SM particle/antiparticle pairs into twin particles (or the reverse). The energy density transferred to twin particle $i$ from SM particle $j$ in scattering is given by 
\begin{align}
\frac{dq_{ij\rightarrow ij}}{dt}=\frac{g_ig_j}{(2\pi)^6}\int\int \frac{d^3k}{2E_i(k)} \frac{d^3h}{2E_j(h)}&f_i(k,\hat{T})f_j(h,T)\, \nonumber \\ 4E_i(k)E_j(h)&\int v_{rel}(E_i(p)-E_i(k))\frac{d\sigma_{ij\rightarrow ij}}{d\Omega}d\Omega, \hspace{2mm} \label{energytrans}
\end{align}
where $p$ is the outgoing 4-momentum of particle $i$. In the cosmic comoving frame, the phase space number densities $f_i$ and $f_j$ are just Boltzmann factors, although evaluated at the different temperatures of each sector. The factor $g_i$ is the number of internal degrees of freedom of particle $i$, which here includes colour (the cross section should not be colour averaged, as each colour of quark is present in the plasma in equal abundances and each mediates the exchange of energy, so have their contributions summed). Finally, $E_i(k)$ is the on-shell energy of particle $i$ with momentum $k$, while $\frac{d\sigma_{ij\rightarrow ij}}{d\Omega}$ is the differential scattering cross section for species $i$ scattering off $j$ per solid angle $\Omega$ and $v_{rel}$ is the usual relative speed of the incoming particles. As described in \cite{1969PhFl...12..799K}, the factor in the integrand giving the energy transferred per reaction is simply a component of a 4-vector, 
\bea
X=4E_i(k)E_j(h)\int (p-k)v_{rel}\frac{d\sigma_{ij\rightarrow ij}}{d\Omega}d\Omega.
\eea
This may be calculated in the centre-of-mass frame and then boosted back into the cosmic comoving frame where the integrals in (\ref{energytrans}) can be evaluated, similarly to the thermal averaging procedure described in \cite{Edsjo:1997bg}. 

The integral (\ref{energytrans}) may be decomposed into two terms giving the positive and negative energy changes of the twin particle, which respectively contribute to $\frac{dq_{\text{in}}}{dt}$ and $\frac{dq_{\text{out}}}{dt}$. When evaluated in the centre-of-mass frame, these terms correspond to the cases where the scattering angle of the twin particle is respectively less than and greater than the angle between its initial momentum and the total momentum of the system. However, when $T\neq\hat{T}$, we find the integrals involved in this decomposition substantially more arduous than when they are evaluated together.

Energy transferred through annihilations may be similarly calculated as 
\begin{align}
\frac{dq_{j\bar{j}\rightarrow i\bar{i}}}{dt}=\frac{g_j^2}{(2\pi)^6}\int\int \frac{d^3k}{2E_j(k)} \frac{d^3h}{2E_j(h)} &f_j(k)f_j(h)\,\nonumber\\4E_j(k)E_j(h)&\int v_{rel}(E_j(h)+E_j(k))\frac{d\sigma_{j\bar{j}\rightarrow i\bar{i}}}{d\Omega}d\Omega\nonumber\\
-\frac{g_i^2}{(2\pi)^6}\int\int \frac{d^3k}{2E_i(k)} \frac{d^3h}{2E_i(h)}&f_i(k)f_i(h)\,\nonumber\\4E_i(k)E_i(h)&\int v_{rel}(E_i(h)+E_i(k))\frac{d\sigma_{i\bar{i}\rightarrow j\bar{j}}}{d\Omega}d\Omega, \hspace{2mm} \label{HeatFlow}
\end{align}
where $\frac{d\sigma_{j\bar{j}\rightarrow i\bar{i}}}{d\Omega}$ is now the differential annihilation cross section. This rate may be evaluated as described above and is more directly amenable to the factorisation of the integrals observed in \cite{Edsjo:1997bg}. See also \cite{Adshead:2016xxj} for further details of similar calculations. The first term of (\ref{HeatFlow}) is the energy transferred from the SM to the twin sector and contributes to $\frac{dq_{\text{in}}}{dt}$ in (\ref{ThermEv}), while the second term is the energy transferred from the twin sector to the SM and contributes to $\frac{dq_{\text{out}}}{dt}$.

In thermal equilibrium, the rate of energy transferred through collisions into one sector will be balanced by that of energy transferred out of it so that there is negligible net heat flow. This state will be rapidly attained (compared to the age of the universe) if $\frac{d q_{\text{in,out}}}{dt}\gg 3H\hat{T}s_{\text{t}}$. However, as the universe expands and the plasma cools, the energy transfer rates fall faster than the Hubble rate. This is demonstrated in the Figure \ref{Fig:decoupling} below.
Once they drop below the Hubble rate, energy exchange ceases on cosmological time scales and the sectors thermally decouple, thereafter thermodynamically evolving independently.

To determine the decoupling temperature of the sectors, we calculate the rates of positive energy exchange for the twin particles interacting with the SM particles. The cross sections are calculated using a tree-level effective fermion-twin fermion contact interaction that, in the full twin Higgs model, would be UV completed by a SM Higgs exchange (the heavier mass of the radial mode would make its exchange subdominant). The interaction strength is determined by the masses of the fermions through their Yukawa couplings, as well as the mixing angle of the SM-like mass state $h$ with the gauge eigenstate $h_B$, giving a 4-fermion coupling of strength $\frac{m_f m_{\hat{f}}}{m_h^2f^2}$ (here $m_f$ and $m_{\hat{f}}$ are the masses of fermions $f$ and $\hat{f}$). See \cite{Chang:2006ra}, \cite{Barbieri:2016zxn} for a more detailed discussion of the cross sections. This effective interaction is appropriate for the temperatures of interest here and helps to simplify the integrals of (\ref{energytrans}). In order to further simplify the integrations of (\ref{energytrans}) when it is to be decomposed into terms in which the energy exchange is positive and negative, we calculate $\frac{dq_{\text{in}}}{dt}$ under the assumption that the sectors have the same temperature (this ensures that the rate $\frac{dq_{\text{out}}}{dt}$ is identical). This is then combined with the rate of energy transferred from annihilation. A similar calculation of these rates was recently performed in \cite{Barbieri:2016zxn}, for cases where the Yukawa couplings do not respect the $\mathbb{Z}_2$ twin symmetry. 

In Figure \ref{Fig:decoupling} we compare the energy transfer rate to the Hubble rate in order to determine when decoupling occurs. As long as the energy exchange rate exceeds the expansion rate, the sectors will be thermalised and have the same temperature. Decoupling then occurs once this rate drops below the Hubble rate. From Figure \ref{Fig:decoupling}, this occurs at a temperature $\sim 2$ GeV. However, even after the energy exchange rate drops below the Hubble rate, the sectors will remain at the same temperature unless some event that either injects or redistributes entropy occurs within a sector (such as the temperature dropping below a mass threshold). As the heavy quark masses roughly coincide with the decoupling temperature, these do cause the twin sector to be mildly reheated with respect to the SM below decoupling. However, the resulting temperature difference is small and the energy exchange rates are expected to continue to be well-approximated by the rates presented in Figure \ref{Fig:decoupling} beyond decoupling.

The lower plot of Figure \ref{Fig:decoupling} illustrates the decomposition of the energy exchange rates into contributions from interactions involving different SM quarks. The interaction cross sections are proportional to the Yukawa couplings of the interacting fermions. The greatest heat exchange is therefore expected to be mediated by the most massive particles, provided that their abundances are not too Boltzmann suppressed. As expected, at temperatures $\sim 1$ GeV, the bottom quark is the best conduit of thermal equilibration, followed by the charm quark and then the $\tau$ (with colour factors enhancing the former two with respect to the latter). The rate of heat flow that the top quarks and weak bosons can mediate at these temperatures (or below) is negligible because of Boltzmann suppression. The bend in the curves at temperatures $\sim 5$ GeV in the lower plot corresponds to a transition from temperatures where the dominant energy exchange rate is through scatterings to those where it occurs through annihilations, as can be seen in the upper plot. The annihilation rate into twin bottom quarks is the dominant component at high enough energies (again because of the larger Yukawa coupling), but this becomes rapidly threshold suppressed as the temperature drops. As can also be inferred in the upper plot, the energy exchange rate through annihilations involving the twin charmed quarks and tau leptons overtakes that of twin bottom quarks at similar temperature, but are still subdominant to scatterings.

\begin{figure}[h!]
\centering
\includegraphics[width=.8\textwidth]{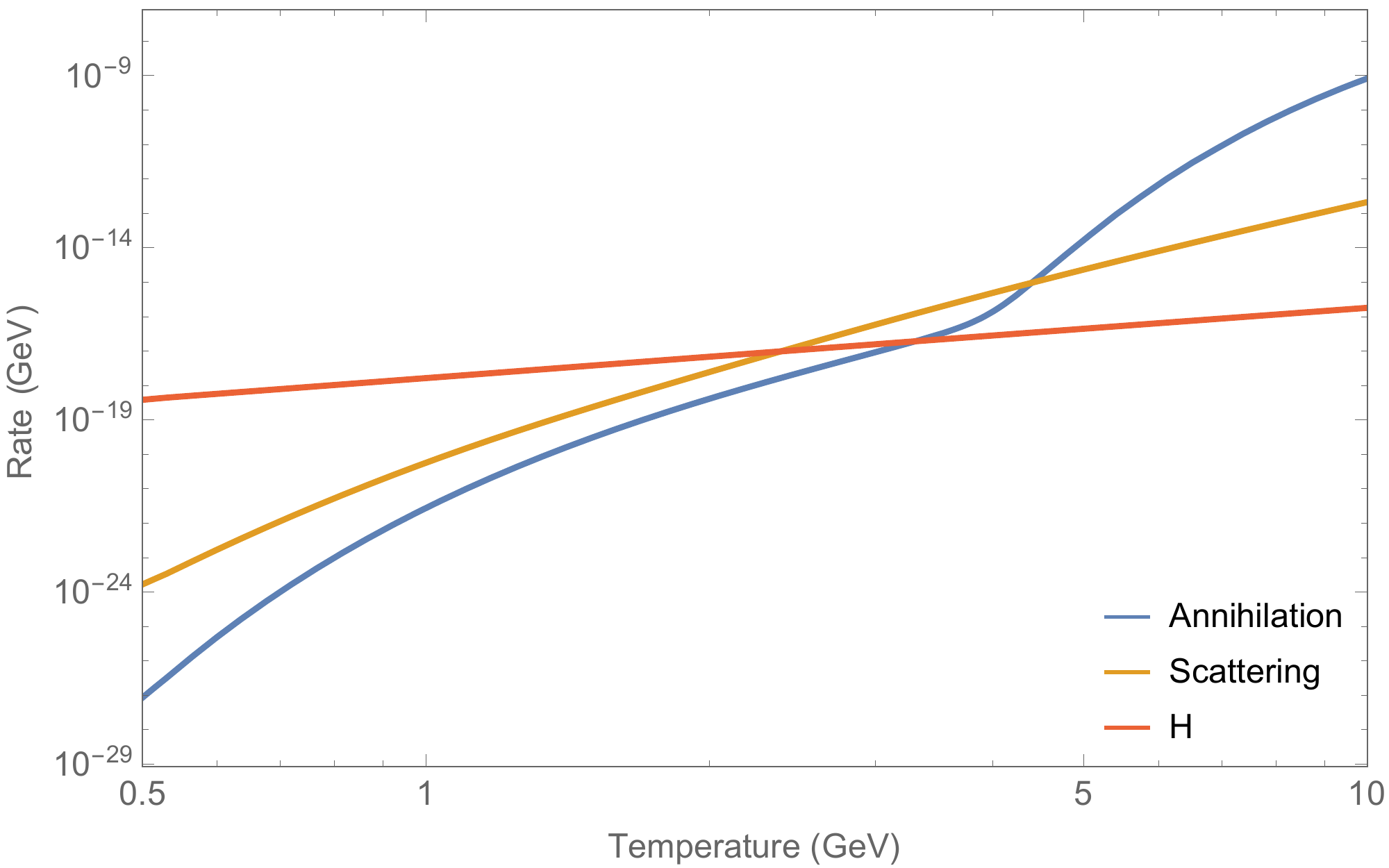}
\centering
\includegraphics[width=.8\textwidth]{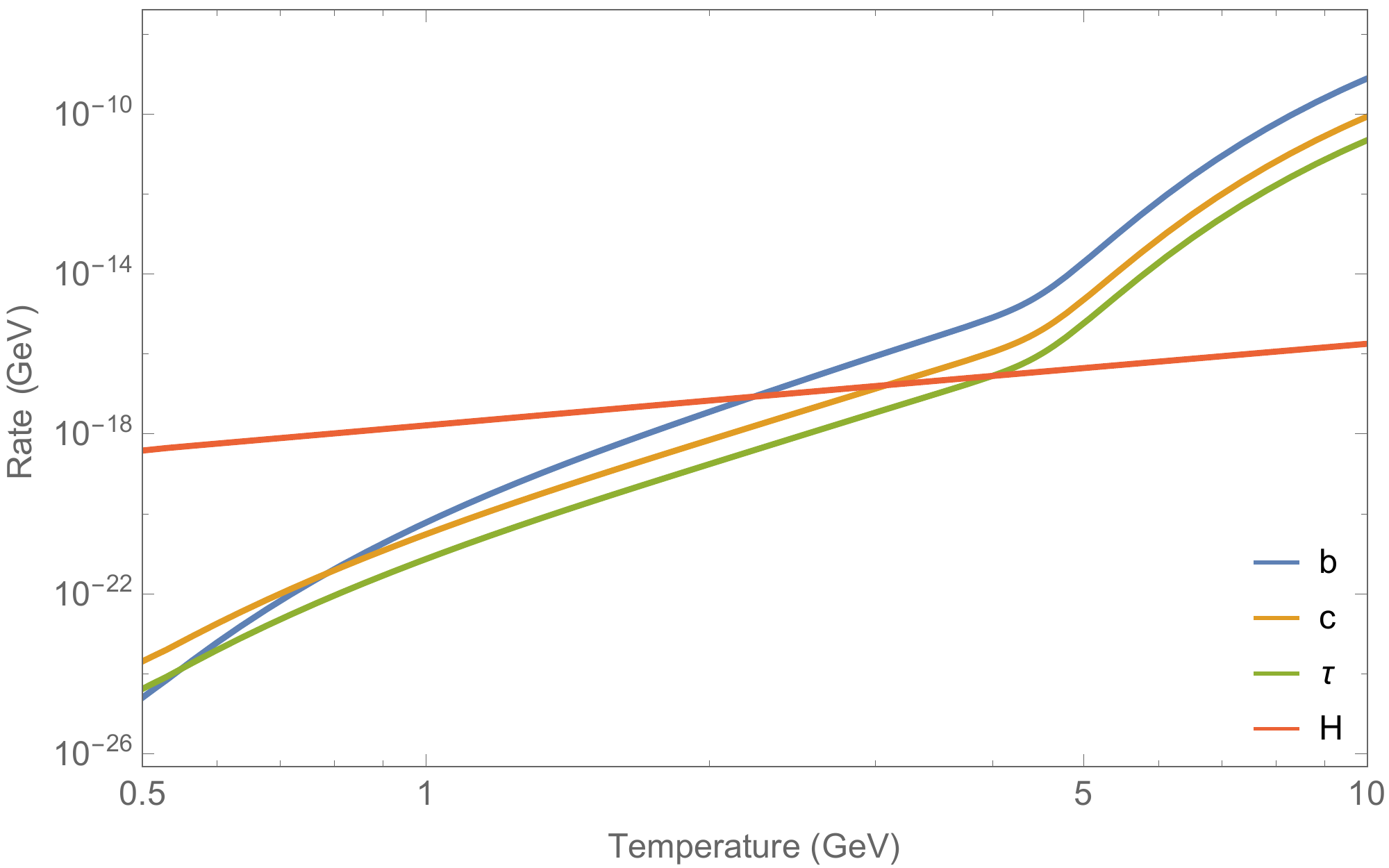}
\caption{Rates of energy density exchange per twin entropy density ($\frac{1}{3s_\text{t}\hat{T}}\frac{dq_{\text{in}}}{dt}$) decomposed into contributions from scattering and annihilation (top) and for interactions involving different species of SM fermions (bottom), along with the Hubble parameter, for $f/v=4$. The decoupling temperature is that where the sum of the energy exchange rates equals the Hubble rate, which occurs at $T_{\text{decoup}}\approx 2$ GeV.}\label{Fig:decoupling}
\end{figure}

The decoupling temperature depends upon $f/v$, which sets both the mass scale of the twin sector and the strength of the Higgs-mediated coupling. As $f/v$ is increased, decoupling occurs earlier because of the greater Boltzmann suppression, although this is only a relatively small effect that, for $f/v=10$, increases the decoupling temperature by only $4$ GeV. 

When the twin sector is colder than the SM  (which will be important for much of what follows) the heat flow is typically dominated by annihilations of SM into twin particles. However, the energy exchange from elastic scattering can be comparable to that from annihilations, as illustrated in Figure \ref{Fig:decoupling}. Although the energy exchange in an annihilation will generally exceed that of a scattering because all of the energy involved in the process must be transferred, the annihilation rate also becomes more Boltzmann or threshold suppressed when the temperature drops below the mass of the heavier twin particles. It is therefore not always clear that energy transfer through annihilations dominates. 

Decoupling is not exactly instantaneous and there is some range of temperatures over which the rate of heat flow freezes-out. The net heat flow rate $\frac{dq}{dt}$ is greater for larger temperature differences between sectors. The generation of a potentially large temperature difference within this brief epoch of sector decoupling, such as those discussed below in Section \ref{sec:late}, may be cut off when the heat flow rate becomes comparable to the Hubble rate. For a given SM temperature $T$, the minimum twin-sector temperature $\hat{T}_{\text{min}}$ during the decoupling period may be roughly estimated as that which satisfies
\bea
H\sim \frac{1}{3s_\text{t}\hat{T}}\frac{dq}{dt}\Big|_{\hat{T}=\hat{T}_{\text{min}}}. \label{HeatFreeze}
\eea
Twin temperatures colder than $\hat{T}_{\text{min}}$ will partially thermalise back to this value. As the participating fermions are not non-relativistic, instantaneous decoupling is not as accurate an approximation as it is, for example, for chemical decoupling of a WIMP, although it is still reliable. 

\begin{figure}[h!]
\centering
\includegraphics[width=.7\textwidth]{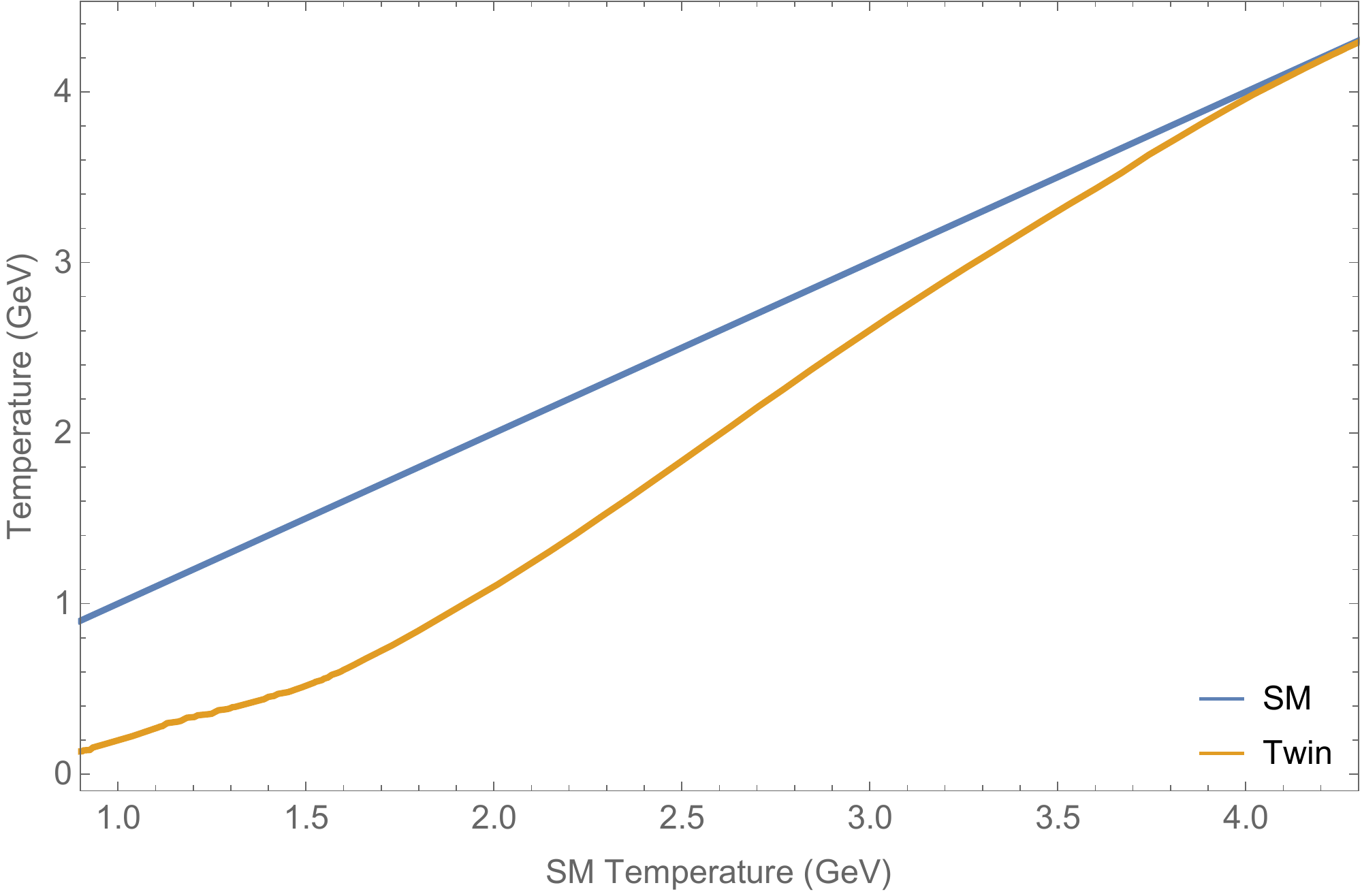}
\caption{Minimum temperature of the twin sector that will not be heated by interactions with a hotter SM plasma, as a function of SM temperature, for $f/v=4$. Also shown is the SM temperature, for reference.}
\label{Fig:decouplingMin}
\end{figure}

In Figure \ref{Fig:decouplingMin}, we show the minimum temperature that the twin sector may have as a function of SM temperature for heat flow to freeze out, estimated using (\ref{HeatFreeze}). Only annihilations have been included in the determination of the minimum temperature, although we have verified that, for these temperatures, the scatterings contribute only $\lesssim 10\%$ to the heat flow. Note that while the energy exchange rate, such as $\frac{1}{\hat{T}}\frac{dq_{\text{in}}}{dt}$ in (\ref{ThermEv}), in scattering processes may be faster, the net energy flow rate, or heat flow ($\frac{1}{\hat{T}}\frac{dq}{dt}$ in (\ref{ThermEv})), which is the difference between energy exchange rates into and out of the sector, is actually dominated by annihilations. Generally, we find that decoupling begins at temperatures $\sim 4$ GeV. The temperature difference can reach an order of magnitude without relaxing once the SM temperature drops to $\sim 1$ GeV.  

While the extent of thermal decoupling is temperature dependent, the maximum temperature difference that will not relax grows quickly as the SM temperature drops. Then we may describe the two sectors as being decoupled if, in a given cosmology, all events that raise the temperature of one sector relative to the other (such as the crossing of a mass threshold and the resulting entropy redistribution, the most significant of which is the confinement of colour) induce temperature differences that are too small to partially relax.

At energies $\lesssim 1$ GeV in Figure \ref{Fig:decoupling}, the reliability of the calculation of the heat flow rate diminishes because of the strengthening of the strong coupling and the eventual confinement of colour. Fortunately, for a cooler twin sector, which will be of interest in subsequent sections, annihilations from the SM dominate other processes over most of the parameter space. These are the least sensitive to higher order corrections and non-perturbative effects because of their higher temperature, and hence energy, compared to the potentially cooler twin sector. The range of temperatures illustrated in Figures \ref{Fig:decoupling} and \ref{Fig:decouplingMin} have been selected to roughly illustrate the duration of decoupling, but may extend below the range where the perturbative calculation of the heat flow rate is valid. For example, at temperatures below the twin sector QCDPT, which occurs at $\sim\left(1+\log(\frac{f}{v})\right)$ higher temperatures than in the SM, the partonic calculation of twin quark/anti-quark pair production must be replaced by a hadronic one. Furthermore, the growth of the twin strong coupling necessitates that the quark-Higgs Yukawa couplings be RG evolved to the scale of the energy exchanged, which can induce an $\mathcal{O}(1)$ change to the cross section, although this has only a relatively small effect on the decoupling temperature. It is nevertheless clear that decoupling is mostly complete by then and that these uncertainties are not large enough to affect this conclusion.

In the standard mirror Twin Higgs cosmology, knowing the decoupling temperature tells us how the temperatures of the two sectors will be related at subsequent times. The sectors separately evolve adiabatically after decoupling, though they redshift in the same way and differences in temperature only arise from events that redistribute entropy. Non-minimal cosmological events that could potentially cause the temperatures of each sector to diverge can therefore only be effective if they leave each sector colder than this approximate decoupling temperature.

\subsubsection{Cosmological Constraints} \label{Sec:Limits}

Given that the twin and Standard Model sectors remain in thermal equilibrium to $\mathcal{O}(\text{GeV})$ temperatures, the simplest mirror Twin Higgs scenario is cosmologically inviable due to the presence of light twin species (photons and neutrinos) with abundances comparable to those of the SM. The cosmological observables through which evidence of light species may be inferred are typically represented by $N_{\text{eff}}$, the ``effective number of neutrino species'' in the early universe; their individual masses, which determine their free-streaming distances; and the ``effective mass'' $m_\nu^{\text{eff}}$, which parameterises their contribution to the present-day energy density of non-relativistic matter. These observables are probed by both the CMB and large scale structure (LSS).



\paragraph{Effective number of neutrinos} \label{Sec:Neff}

The parameter $N_{\text{eff}}$ describes the amount of radiation-like energy density during the evolution of the CMB anisotropies before photon decoupling. It is defined as the effective number of massless neutrinos with temperature as predicted in the standard cosmology that would give equivalent energy density in radiation:
\begin{equation}
\rho_r=\rho_\gamma+\frac{7}{8}\left(\frac{4}{11}\right)^{4/3}N_{\text{eff}}\rho_\gamma,\label{Neffeq}
\end{equation}
where $\rho_r$ is the energy density of radiation and $\rho_\gamma$ is the energy density of photons (the factor of $\Big(\frac{4}{11}\Big)^{4/3}$ arises from the relative reheating of the photons from electron/positron annihilation, which occurs after most of the neutrinos have decoupled, and the factor of $7/8$ is from the opposite spin statistics). A deviation from the Standard Model prediction of $3.046$ \cite{Mangano:2005cc} is denoted by $\Delta N_{\text{eff}}=N_{\text{eff}}-3.046$. 
This definition of radiation, or equivalently, relativistic degrees of freedom, becomes less clear if the new fields have a non-negligible mass, as we discuss further below. 

We here review the CMB physics of dark radiation, summarising the discussion in \cite{Hou:2011ec}. See also \cite{Ade:2015xua} for further review. The angular size and scale of the first acoustic peak is well-measured and this approximately fixes the scale factor at matter-radiation equality $a_{eq}$. If we imagine fixing all other $\Lambda$CDM parameters, extra radiation would delay the epoch of matter-radiation equality. This would have a pronounced effect on the power spectrum in the vicinity of the first acoustic peak through the early Integrated Sachs-Wolfe (eISW) effect. The modes corresponding to this feature entered the horizon close to matter-radiation equality and the evolution of their potentials is highly sensitive to the radiation energy density. However, the impact of a $\Delta N_{\text{eff}}\sim\mathcal{O}(1)$ deviation on the peak height can be simultaneously balanced by increasing the amount of non-relativistic matter, to the extent to which other observations providing independent constraints upon $\Omega_c$ permit (for $\Lambda$CDM$+N_{\text{eff}}$, a variation of $\sim 10\%$ in $\Omega_ch^2$ is consistent with present CMB+BAO measurements \cite{Ade:2015xua}, although these variations must be consistent with other observables).
This degeneracy is not expected to be broken by CMB-S4 \cite{CMB-S4:2016}.

Given that $a_{eq}$ is approximately fixed, the utility of $N_{\text{eff}}$ arises because, in simple extensions of the $\Lambda$CDM model, it approximately corresponds to the suppression of power in the small scale CMB anisotropies that arises from Silk damping. The reason for this is roughly that, although the greater expansion rate induced by the extra radiation reduces the time that CMB photons have to diffuse before decoupling, it also reduces the sound horizon size more severely. As the angular size of the sound horizon is determined by the location of the acoustic peaks and is also well measured, the reduction in the sound horizon must be compensated for by a reduction in the angular diameter distance to the CMB. This effectively raises the angular distance over which photon diffusion proceeds and results in a prediction of smoother temperature anisotropies at small scales. This correspondence with the Silk damping allows $N_{\text{eff}}$ to be approximately factorised from other parameters and constrained independently, providing a direct observational avenue for detecting the presence of new, massless fields \cite{Hou:2011ec} (see \cite{Brust:2013ova} for further implications for model building). This relationship arises because the fixing of $a_{eq}$ implies that $N_{\text{eff}}$ effectively determines the energy density of the universe, and hence the Hubble rate, during CMB decoupling. Note, however, that further extensions of $\Lambda$CDM may complicate this correspondence, in particular deviations from the standard Big Bang Nucleosynthesis prediction of the primordial helium abundance.

The contribution to $N_{\text{eff}}$ (or $\Delta N_{\text{eff}}$)
in the mirror Twin Higgs arises from two sources: the twin photons, which can be treated as massless dark radiation with an appropriate twin temperature $T_{\text{eq}}^{\text{t}}$ at the time of matter-radiation equality, and the twin neutrinos, whose non-zero masses may need to be accounted for. For the twin photons, the contribution to $N_{\text{eff}}$ is simple; their equation of state is always $w = 1/3$ and their energy density is given by $g \frac{\pi^2}{30} \left(T^\text{t}_{\text{eq}}\right)^4$, where $g = 2$. The twin temperature at matter-radiation equality is found from the SM temperature using comoving entropy conservation,
\begin{equation}
\frac{T^\text{t}_{\text{eq}}}{T^\text{SM}_{\text{eq}}} = \left(\frac{g^\text{t}_{\star}(T_{\text{decoup}})}{g^\text{SM}_{\star}({T}_{\text{decoup}})}\right)^{1/3} \left(\frac{g^\text{SM}_{\star}(T^\text{SM}_{\text{eq}})}{g^\text{t}_{\star}({T}^\text{t}_{\text{eq}})}\right)^{1/3},
\end{equation}
where the two sectors have the same number of thermalized degrees of freedom by this time. Here, $T^\text{SM}_{\text{eq}}$ is the SM photon temperature at matter-radiation equality and ${T}_{\text{decoup}}$ is the sector decoupling temperature.

Since neutrinos are massive, their behavior is more complicated. Their equation of state parameter takes on a scale factor dependence which is controlled by their mass. 
In the Standard Model, this sensitivity is negligible because present CMB bounds imply that neutrinos are ultra-relativistic at $a_{\text{eq}}$ to good approximation \cite{Ade:2015xua}. However, the factor by which the twin neutrino masses are enhanced may raise them to order $T^t_{\text{eq}}$ or greater (see Section \ref{sec:twin} for discussion of the scaling of the masses with $f/v$).

To better describe the impact of the extra twin (semi-)relativistic degrees of freedom on the CMB, we choose to define $N_{\text{eff}}$ through the effects of neutrinos at matter-radiation equality, when the impact on the expansion rate of the universe for most of the period relevant for the evolution of the CMB is greatest. Note that, in their presentation of joint exclusion bounds on $N_{\text{eff}}$ and $\sum m_\nu$ (the sum of SM neutrino masses) or $m_\nu^{\text{eff}}$ (effective mass contributing to the present-day non-relativistic matter density of an extra sterile neutrino), the Planck collaboration define $N_{\text{eff}}$ as the value in (\ref{Neffeq}) at temperatures sufficiently high that the neutrinos are fully relativistic. Our values cannot be directly compared with their analysis, although we consider ours to be a reasonable rough estimate that is more representative of the CMB constraints. The ensuing correction from the finite neutrino masses is, in the cases considered in this work, a small effect anyway.

To determine this correction and provide a definition of $N_{\text{eff}}$ that better describes the impact of quasi-relativistic particles on the CMB, we first define the epoch of matter-radiation equality as the time at which the average equation of state parameter of the universe is $\bar{w} = 1/6$ (the equation of state is defined as $\rho=\bar{w} P$, where $\rho$ is energy density and $P$ is pressure). We can express this condition as 
\begin{equation}
\left. \frac{d\ln H}{d\ln a} \right\vert_{a_{eq}}= -\frac{7}{4},
\end{equation}
as in \cite{Dodelson:2005tp}.

Call the quasi-relativistic neutrino energy density $\tilde{\rho}(a)$ with time-evolving equation of state parameter $w(a)$, which is to be balanced against some extra non-relativistic energy density $\Delta \rho_{CDM}(a)\propto a^{-3}$ to keep $a_{eq}$ the same. This amount of non-relativistic energy density $\Delta \rho_{CDM}$ is
\begin{eqnarray}
 \Delta \rho_{CDM} (a_{eq}) = \rho_r(a_{eq}) - \rho_m(a_{eq}) - 2 a_{eq} \left.\frac{d\tilde{\rho}}{da}\right\vert_{a_{eq}} - 7 \tilde{\rho}(a_{eq}),
\end{eqnarray}
where $\rho_r$ and $\rho_m$ are the energy densities of the radiation and non-relativistic matter. 

For a perfect fluid, $\frac{d\tilde{\rho}}{da}=-3(1+w(a))\tilde{\rho}/a$ (neglecting the anisotropic stress that is expected only to contribute to a weak phase shift in the CMB \cite{Baumann:2015rya}), 
this results in a Hubble parameter of
\begin{equation}
H^2(a_{eq}) = \frac{2}{3 M_\text{pl}^2} \left[\rho_r(a_{eq}) + 3 w(a_{eq}) \tilde{\rho}(a_{eq})\right].
\end{equation}
This suggests a definition of the effective number of neutrinos, $N_{\text{eff}}$, via
\begin{eqnarray}
H^2(a_{eq}) &=& \frac{2}{3 M_\text{pl}^2} \left. \left(\rho_\gamma + N_{\text{eff}} \rho_{\nu,m=0}^{th} \right) \right\vert_{a_{eq}} \\
N_{\text{eff}} &\equiv& \sum_i\frac{w_i}{1/3} \frac{\rho_i}{\rho_{\nu,m=0}^{th}},
\end{eqnarray}
where $\rho_i$ is the contribution to the energy density from some species $i$ with equation of state parameter $w_i$ and $\rho_{\nu,m=0}^{th}$ is the energy density of a massless neutrino with a thermal distribution in the standard cosmology. Then $3w$ gives the `relativistic fraction' of the energy density. Note that this is simply a ratio of the pressure exerted by the new fields to that of a massless neutrino. The effectiveness of this approximation was discussed in \cite{Archidiacono:2013cha} in the context of thermal axions (while effective at keeping $a_{eq}$ fixed, changes to odd peak heights subsequent to the first are imperfectly cancelled and require further changes to $H_0$ to compensate - see Section \ref{Sec:meff} below).

Calling $T^i_\nu$ the temperature at which the neutrinos in sector $i$ freeze-out and $a^i_\nu$ the corresponding scale factor, then assuming instantaneous decoupling, the phase space number density for scale factor $a$ is given by a redshifted Fermi-Dirac distribution \cite{Jacques:2013xr}
\begin{equation} \label{Eq:nudis}
f^i_\alpha(p) \approx \left[ 1 + e^{pa/\left(a^i_\nu T^i_\nu\right)} \right]^{-1}
\end{equation}
for the $\alpha$ neutrino mass eigenstate in the $i$ sector ($m_\alpha^i\ll T_\nu^i$, so has been dropped). The energy density and pressure are
\begin{eqnarray} 
\rho^i_{\nu_\alpha} &=& \frac{g_\alpha}{2\pi^2} \int_{0}^{\infty} dp \ p^2 \sqrt{p^2 + \left(m^i_\alpha\right)^2}  f^i_\alpha(p) \label{Eq:nuen}\\
P^i_{\nu_\alpha} &=& \frac{g_\alpha}{2\pi^2} \int_{0}^{\infty} dp \ \frac{p^4}{3\sqrt{p^2 + \left(m^i_\alpha\right)^2}}  f^i_\alpha(p) \label{Eq:nupr},
\end{eqnarray}
where $g_\alpha = 2$ is the number of degrees of freedom for a neutrino species. 

Since the neutrino decoupling temperature depends on the strength of the weak interaction as $T_\nu \propto G_F^{-2/3}$, while $G_F \propto v^2$, then the twin neutrino decoupling temperature $T^\text{t}_\nu$ is related to the SM neutrino decoupling temperature $T^\text{SM}_\nu$ by
\bea
T^\text{t}_\nu = (f/v)^{4/3} T^\text{SM}_\nu.\label{nudec}
\eea
We can then simply use (\ref{Eq:nuen}) and (\ref{Eq:nupr}) at matter-radiation equality to find $\Delta N_{\text{eff}}$ (assuming instantaneous decoupling). We thus obtain
\begin{eqnarray}
H^2(a_{eq}) &=& \frac{2}{3 M_\text{pl}^2} \left. \left(\rho^\text{SM}_\gamma + 3.046 \rho_{\nu,m=0}^{th} + \rho^\text{t}_\gamma + \sum_\alpha 3 w_{\nu_\alpha} \rho^\text{t}_{\nu_\alpha} \right) \right\vert_{a_{eq}}
\end{eqnarray}
and
\begin{eqnarray} 
\Delta N_{\text{eff}} &=& \left(\frac{11}{4}\right)^{4/3} \frac{120}{7 \pi^2 \left(T^\text{SM}\right)^4} \left( \rho^\text{t}_\gamma + \sum_\alpha 3 w^t_{\nu_\alpha} \rho^\text{t}_{\nu_\alpha}\right),\label{CorDelNeff}
\end{eqnarray}
where we now have equation of state parameters $w_{\nu_\alpha}$ for each neutrino, while $\rho^\text{SM}_\gamma$ and $\rho^\text{t}_\gamma$ are the SM and twin photon energy densities, $\rho_{\nu,m=0}^{th}$ and $\rho^\text{t}_{\nu_\alpha}$ are the neutrino energy densities.

\paragraph{Neutrino masses} \label{Sec:meff}

Because they are so weakly interacting, the neutrinos have a long free-streaming scale given by the distance travelled in a Hubble time $v_\nu/H$, with $v_\nu \propto m_\nu^{-1}$ the speed of the neutrino once it becomes non-relativistic. This defines a free-streaming momentum scale $k_{fs} = \sqrt{\frac{3}{2}} \frac{a H}{v_\nu} \propto m_\nu$, above which neutrinos do not cluster. Below this scale, perturbations in the matter density consist coherently of neutrinos and other matter, but well above it only non-neutrino matter contributes to density perturbations. This results in a suppression of the matter power spectrum on large scales which is proportional to the fraction of energy density in the free-streaming matter. Since this occurs at late times when neutrinos are non-relativistic, the energy density is simply $\rho_{\nu_\alpha} = n_{\nu_\alpha} m_{\nu_\alpha}$ for each neutrino species $\alpha$, where $n_{\nu_\alpha}$ is the number density. 
Constraints on the sum of neutrino masses then come from the observations of power on small scales, which is suppressed relative to that expected for massless neutrinos by a factor $\appropto 1-8 f_\nu$, where $f_\nu = \Omega_\nu/\Omega_m$ is the fraction of non-relativistic energy in neutrinos at late times \cite{Lesgourgues:2012uu}.


More generally, inferences of the matter power spectrum constrain the present-day energy density fraction of free-streaming species that do not cluster on small scales and have since become non-relativisitic, $\Omega_\nu=(\sum m_\nu + m_\nu^{\text{eff}})/(94.1\,\text{eV})$, where $\sum m_\nu$ is the sum of SM neutrino masses and $m_\nu^{\text{eff}}$ is the sum of twin neutrino masses weighted by their number density
\begin{equation}
m_\nu^{\text{eff}} = \frac{n^\text{t}_\nu}{n^\text{SM}_\nu} \sum_\alpha m^\text{t}_{\nu_\alpha}.
\end{equation} 
Here $n^{\text{t}}_\nu$ is the number density of a relic twin neutrino flavour and $n^\text{{SM}}_\nu$ is that for a SM neutrino. It is assumed that the neutrinos have been thermally produced as hot relics.

The relic abundance of a neutrino species is given by its number density when it decoupled, diluted by the factor by which the universe has since expanded. The scale factors at which neutrino decoupling occurs in the two sectors, $a^\text{SM}_\nu$ and $a^\text{t}_\nu$ can be determined from (\ref{nudec}), the relative temperatures in the two sectors and comoving entropy conservation, to obtain 
\begin{eqnarray}
a^\text{t}_\nu &=& a^\text{SM}_\nu \left(\frac{v}{f}\right)^{4/3} 
\left( \frac{g^\text{t}_\star\left(T_{\text{decoup}}\right)}{g^\text{SM}_\star\left(T_{\text{decoup}}\right)}\right)^{1/3}
\end{eqnarray}
where the same mass thresholds have been assumed in each sector below their neutrino decoupling temperatures, so that $g^\text{SM}_\star\left(T^\text{SM}_\nu\right)=g^\text{t}_\star\left(T^\text{t}_\nu\right)$. The neutrino number densities are then
\begin{eqnarray}
\frac{n^\text{t}_\nu}{n_\nu^\text{SM}} =\left(\frac{T^\text{t}_\nu a^\text{t}_\nu}{T^\text{SM}_\nu a^\text{SM}_\nu}\right)^3 
=\frac{g^\text{t}_\star\left(T_{\text{decoup}}\right)}{g^\text{SM}_\star\left(T_{\text{decoup}}\right)}.
\end{eqnarray}
For $f/v$ from $3$ to $10$ and using $T_{\text{decoup}} \sim 2 - 6$ GeV from Section \ref{Sec:Decoupling}, we find \\ $g^\text{t}_\star\left(T_{\text{decoup}}\right)/\ g^\text{SM}_\star\left(T_{\text{decoup}}\right) \sim 0.8$ and thus arrive at 
\begin{equation}
m_\nu^{\text{eff}} \approx 0.8 \left(\frac{f}{v}\right)^n \sum_\alpha m^\text{SM}_{\nu_\alpha},
\end{equation}
where $n = 1$ for Dirac masses and $n = 2$ for Majorana masses. 

If they are sufficiently light and hot, the twin neutrinos only affect the CMB as dark radiation and their masses may then only be inferred from tests of the matter power spectrum. However, if heavier and colder, they are better described as a hot dark matter component. Their impact on the CMB is discussed in \cite{Dodelson:1995es}, where the shape of the power spectrum can depend upon the individual neutrino kinetic energies through their characteristic free-streaming lengths. The early Integrated Sachs-Wolfe effect (eISW) is also sensitive to the masses if the neutrinos become non-relativistic during decoupling (thereby affecting the radiation energy density and the growth of inhomogeneities) \cite{Lesgourgues:2012uu}.

There is a significant degeneracy in cosmological fits to the CMB between $\Omega_m$ and $H_0$ (the Hubble constant) \cite{2012JCAP...04..027H}, where raising the non-relativistic matter fraction, such as with nonrelativistic neutrinos, can be accommodated by a decrease in $H_0$ (or equivalently, the dark energy density), which keeps the angular diameter distance to the CMB approximately fixed. This degeneracy can be broken by measurements of the baryon acoustic oscillations (BAOs), which are sensitive to the expansion rate of the late universe and provide an independent measurement of $\Omega_m$ and $H_0$. It is through combination with these results that bounds from Planck on neutrino masses are strongest \cite{Ade:2015xua}.

\paragraph{Bounds}\label{Bounds}

The authors are unaware of any specialised analysis of the present and projected future cosmological constraints on scenarios with both massless dark radiation and additional light, semi-relativistic sterile neutrinos. In the absence of this, we use bounds from \cite{Ade:2015xua} as a rough indication of the present level of sensitivity to these parameters, which we nevertheless expect to be a reliable indication of the (in)viability of this model. The 95\% confidence limits on these parameters are $N_{\text{eff}} = 3.2 \pm 0.5$ and $\sum m_\nu < 0.32 \text{ eV}$ when each are constrained separately with the other fixed. This, of course, overlooks correlations between the impacts of masses and $\Delta N_{\text{eff}}$ on the CMB and LSS. Bounds on an additional sterile neutrino as the only source of dark radiation are also presented with number density, or equivalently, contribution to $\Delta N_{\text{eff}}$, left to float. These are similar to the limit on $\sum m_\nu$. It was found in \cite{DiValentino:2016ikp} that, allowing $\sum m_\nu$ and $m_\nu^{\text{eff}}$ to float independently for a single extra sterile neutrino, the bound mildly relaxes to $m_\nu^{\text{eff}}\lesssim 1$ eV, although the bound may be stronger depending on the combination of data sets chosen (the lensing power spectrum presently prefers higher neutrino masses and raises the combined bounds if included). Other bounds from LSS on $\sum m_\nu$ exist and are potentially stronger than those placed from the CMB, possibly as low as $m_\nu^{\text{eff}}\lesssim 0.05$ eV, again depending on data sets combined (see \cite{Costanzi:2014tna}, \cite{Gariazzo:2015rra}), although these are subject to greater uncertainties in the inference of the power spectra of dark matter halos from galaxies surveys and the Ly$\alpha$ forest.

It must also be noted that the shape of the CMB temperature anisotropies depends upon both the mass of individual neutrino components (through their free-streaming distance) and their contribution to the energy density of the nonrelativistic matter that does not cluster on small scales. However, it is not expected that improvements in bounds on the former will be made from improved measurements of the primary CMB itself, but rather from weak lensing of the CMB, in conjunction with future measurements from DESI of the BAOs to break degeneracy with $\Omega_m$. The lensing spectrum, like inferences of the matter power spectrum made in galaxy surveys, is expected to measure the suppression of small scale power and therefore to strengthen constraints upon $m_\nu^{\text{eff}}$, rather than the individual neutrino masses. 
One of the goals of CMB-S4 will be the detection of neutrino masses, given the present lower bound $\sum m_\nu\gtrsim 0.06$ eV from oscillations. Projected bounds are as low as $\sim 0.02$ eV \cite{CMB-S4:2016}, although this assumes no extra dark radiation or sterile neutrinos. A projection of the joint bound on $N_{\text{eff}}$ (from extra massless dark radiation) and $m_\nu^{\text{eff}}$ combining improved measurements CMB temperature measurements, lensing and BAOs indicates a limit of $m_\nu^{\text{eff}}\lesssim 0.1$ eV at $1\sigma$ \cite{CMB-S4:2016}. Any contribution from additional states to $m_\nu^{\text{eff}}$ may therefore be testable and bounded by the excess of the neutrino mass inference over the minimum neutrino mass, although laboratory measurements or measurements of $\Delta N_{\text{eff}}$ will be required to further ascertain the contribution from the new particles. 

Constraints on $\Delta N_{\text{eff}}$ from improved measurements of the damping tail as part of CMB-S4 are projected to be $\sim 0.02-0.05$ at $1\sigma$ \cite{CMB-S4:2016}. 
In the following sections, we use an optimistic estimate of $0.02$ for its reach in order to identify as much of the potentially testable parameter space as possible.

\begin{figure}[h!]
\centering
\includegraphics[width=1.0\linewidth]{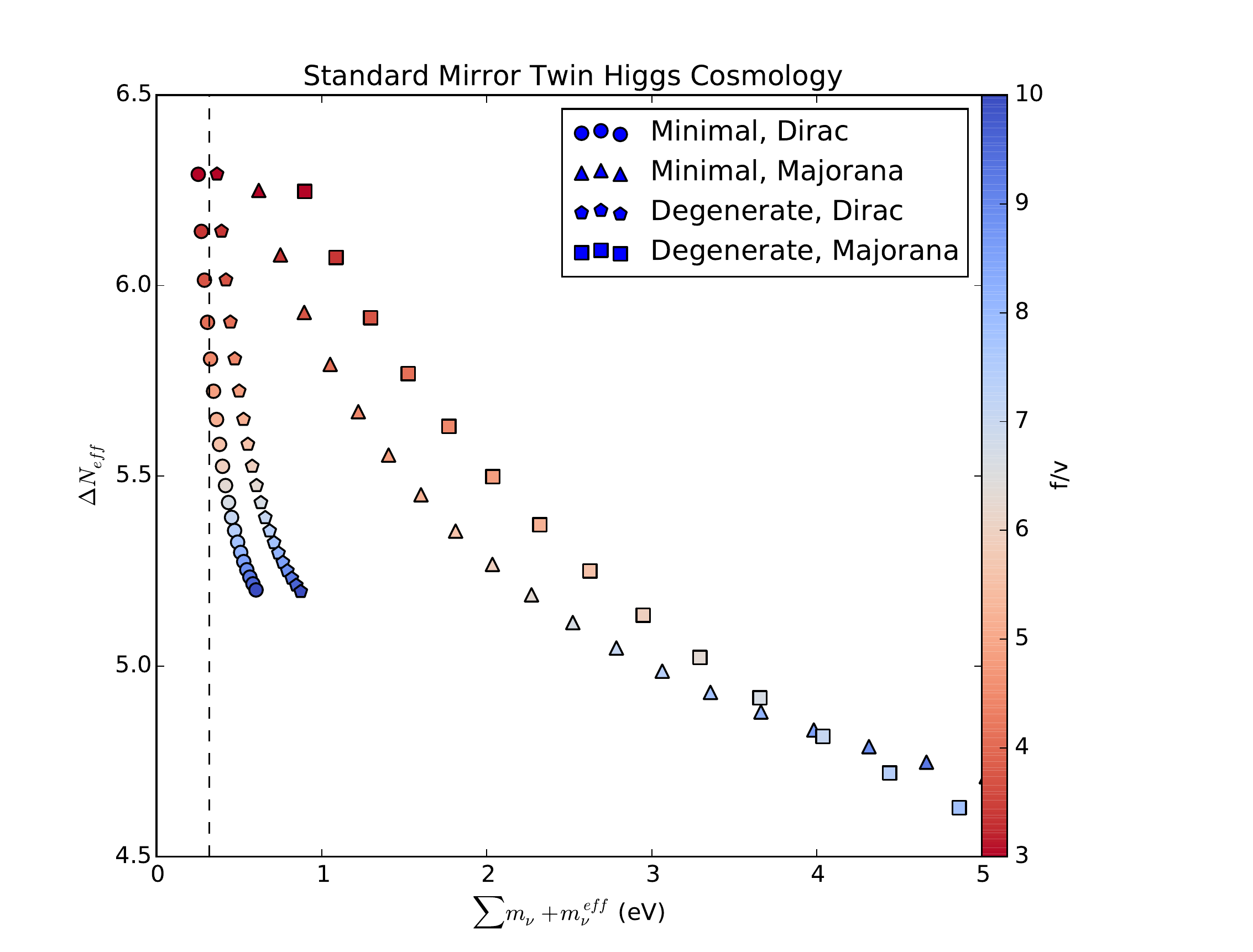}
\caption{Predicted values of $\Delta N_{\text{eff}}$ and $\sum m_\nu + m_\nu^{\text{eff}}$ for minimal and degenerate neutrino mass spectra with both Dirac and Majorana masses for $f/v$ from 3 to 10. The Planck 2015 constraint\protect\cite{Ade:2015xua} is the dashed line; the corresponding $N_{\text{eff}}$ upper bound is well below the bottom of the plot. All points are excluded by the combination of bounds on $\Delta N_{\text{eff}}$ and $\sum m_\nu + m_\nu^{\text{eff}}$.}
\label{Fig:mirrorPred}
\end{figure} 

To estimate the impact of current and projected CMB limits on the mirror Twin Higgs, we consider two scenarios: the minimal Standard Model neutrino mass spectrum of $m_\nu = \left[0.0,0.009 \text{ eV},0.06 \text{ eV}\right]$ and a degenerate spectrum of $m_\nu = \left[0.1\text{ eV} ,0.1\text{ eV},0.1 \text{ eV}\right]/3$ from \cite{Ade:2015xua}. In Figure \ref{Fig:mirrorPred} we plot the predictions of the mirror Twin Higgs for $\Delta N_{\text{eff}}$ and $m_\nu^{\text{eff}}$ for both types of spectra, as well as for both Dirac and Majorana masses (which scale differently with $f/v$). As is plainly evident, the mirror Twin Higgs is ruled out cosmologically, no matter the choices of neutrino masses one makes, if only for the presence of the twin photon.
In the standard cosmology, the twin sector will have roughly the same temperature as the SM, giving $4.6\lesssim\Delta N_\text{eff}\lesssim 6.3$ for $f/v<10$, according to the definition of (\ref{CorDelNeff}). This range depends upon $f/v$ through the twin neutrino decoupling temperature (\ref{nudec}), which determines the extent to which the twin photons are reheated relative to the twin neutrinos after twin electron/positron annihilations. This is sufficiently large that even the cold dark matter fraction cannot be adjusted to keep matter-radiation equality fixed, resulting inevitably in changes to the height and shape of the first acoustic peak. The energy density in neutrinos is predicted to be above the present observational upper bounds for most neutrino mass configurations, with the exception of the minimal values permitted by neutrino oscillation measurements with $f/v\lesssim 6$. We therefore discuss cosmological mechanisms in which the twin radiation is diluted to levels compatible with these observational bounds in the subsequent sections of this paper. 

\subsection{Reheating by the decay of a scalar field} \label{sec:late}

We now turn to simple scenarios that reconcile the mirror Twin Higgs with cosmological bounds, while taking care to respect the softly-broken $\mathbb{Z}_2$ symmetry. We begin with the out-of-equilibrium decay of a particle with symmetric couplings to the Standard Model and twin sectors, in which the desired asymmetry is generated kinematically. That is to say, the dimensionless couplings between the decaying particle and the two sectors are equal, and asymmetric energy deposition into the two sectors is a direct consequence of the asymmetric mass scales. In this respect, the scenario is philosophically similar to $N$naturalness \cite{Arkani-Hamed:2016rle}, albeit with a parsimonious $N=2$ sectors. See also \cite{Reece:2015lch}, \cite{Randall:2015xza} and \cite {Adshead:2016xxj} for other recent related ideas of using long-lived particles for the dilution of dark sectors.

For simplicity, here we will focus on the case of a real scalar $X$ coupled symmetrically to the $A$ and $B$ sector Higgs doublets. Due to the difference in masses between the sectors after electroweak symmetry breaking, simple kinematic effects give $X$ a larger branching ratio into the Standard Model. This occurs over a range of $X$ masses within a few decades of the weak scale. If $X$ decays out-of-equilibrium below the decoupling temperature of the two sectors, this injects different amounts of energy into the two sectors, effectively suppressing the temperature of the twin sector relative to the Standard Model. This relative cooling suppresses the contribution of the light degrees of freedom of the mirror Twin Higgs to below cosmological bounds. Insofar as the asymmetry is driven entirely by kinematic effects arising from $v \ll f$, the resulting temperature inequality between the two sectors is proportional to powers of $v/f$.

The requisite suppression of the twin sector temperature relative to the Standard Model temperature necessitates that the $X$ dominate the cosmology before it decays. Our main discussion will follow the simplest case of an $X$ which dominates absolutely before it decays, comprising all of the energy density of the universe and effectively acting as a `reheaton'. Afterwards, we will discuss the possibility of a `thermal history' for $X$ -- a scenario where $X$ is in thermal equilibrium with the two sectors, then chemically decouples at some high temperature and grows to dominate the cosmology before it decays. This scheme will result in additional stringent constraints on the viable parameter space.

\subsubsection{Asymmetric Reheating} \label{Sec:reheaton}

A $\mathbb{Z}_2$-even scalar $X$ which is a total singlet under the SM and twin gauge groups admits the renormalisable interactions 
\begin{equation} \label{eq:portal}
V \supset \lambda_x X (X + x) \left(\left|H_A\right|^2 + \left|H_B\right|^2\right)+\frac{1}{2}m_X^2X^2,
\end{equation}
where $m_X$ is the mass of $X$ (neglecting corrections from mixing that will be shown below to be tiny), $\lambda_x$ is a dimensionless coupling and $x$ is a dimensionful parameter, which one may imagine identifying as a vacuum expectation value (vev) of $X$ in an UV theory. Note that these interactions preserve the accidental $SU(4)$ symmetry of the Twin Higgs. The $X$ field may additionally possess self-interactions, which we omit here as they do not play a significant role in what follows.

The interactions in (\ref{eq:portal}) allow $X$ to decay into light states in the Standard Model and twin sectors. If $X$ reheats the universe through out-of-equilibrium decays, the reheating temperatures of the two sectors will be determined by its partial decay widths, assuming that the decay products do not equilibrate. In the instantaneous decay approximation, $X$ decays when the Hubble parameter falls to its decay rate $\Gamma_X \sim H$. As we will show in Section \ref{Sec:CMBfx}, in order to evade cosmological constraints we need the $X$ to decay mostly into the SM, so we may estimate $\Gamma_X \sim \Gamma(X \rightarrow \text{SM})$. Then the energy that was contained in the $X$ is transferred into radiation energy density, with the resulting temperature of the radiation given by (see \cite{Kolb:1990vq})
%
\begin{eqnarray}
T \sim 1.2\sqrt{\frac{\Gamma_X \mpl}{\sqrt{g_\star}}}\label{ReheatTemp}
\end{eqnarray}
where $g_\star$ is the effective number of relativistic degrees of freedom, as defined in Section \ref{sec:thermal}, of the particles that are being reheated. Our numerical calculation of the reheating temperature, which will be presented in Section \ref{Sec:cosmosim}, indicates that the approximation $T \sim 0.1\sqrt{\Gamma_X \mpl}$ reliably reproduces the reheating temperature over the range of interest. 

As shown in Section \ref{Sec:Decoupling}, the two sectors thermally decouple when the temperature falls below $T_{\text{decoup}} \sim 1 \text{ GeV}$, so reheating must take place to below this temperature. At even lower temperatures, big bang nucleosynthesis (BBN) places strong constraints on energy injected into the SM at temperatures below $\mathcal{O}(1-10)$ MeV \cite{deSalas:2015glj}.
Requiring that the SM reheating temperature is above $\sim 10$ MeV, these constraints on the SM reheating temperature become constraints on the decay rate of the $X$ into the SM, which in the above approximation becomes
\begin{equation}
5\times 10^{-21} \ \text{GeV} \lesssim \Gamma_X \lesssim 3\times 10^{-16} \ \text{GeV}.\label{ReheatRange}
\end{equation}
This then constrains the couplings $\lambda_x$ and $x$ of the $X$ to the Higgs sector. Importantly, it means that $X$ must couple very weakly, in order to be long-lived enough to reheat to a low temperature, as will be shown below.


The asymmetry in partial widths arises from different effects depending upon the mass of $X$. For masses below the SM Higgs threshold, it is predominantly differences in mass mixing with the two Higgs doublets that produces the asymmetry, where the size of the mixing angles determines the effective coupling of $X$ to the SM and twin particles and therefore its branching fractions. For masses below the twin scale, the relative size of the mixing scales inversely with the vevs in each sector. Thus the hierarchy $v \ll f$ already present in the Higgs sector can automatically gives rise to a hierarchy in partial widths. Note that additional threshold effects can enhance the asymmetry further, in particular when $X$ has mass above threshold for a significant decay channel in the SM, but below the corresponding mass threshold in the twin sector. Decays into on-shell Higgses complicate this picture further.
In what follows, we first give an analytic calculation of the mass mixing effect, then present a more precise calculation of the decay widths into each sector.


To lowest order, $X$ decays via its interactions with the SM and twin Higgs, and only to other fermions and gauge bosons through its mass mixing with the Higgs scalars. Expanding the $X$ potential after the $SU(4)$ is spontaneously broken, the mixing term between $X$ and $h_A$ in the scalar mass matrix is $\sqrt{2}\lambda_x x v_A$, while that between $X$ and $h_B$ is $\sqrt{2}\lambda_x x v_B$. The $h_A$ and $h_B$ components of the $X$ mass eigenstate, which we denote respectively as $\delta_{XA}$ and $\delta_{XB}$, can then be determined. The expressions for the mixing angles are in general complicated, but they simplify in limits $m_X < f$ and $m_X \gg f$:
\begin{equation} \label{Eq:mixing} 
\left(\delta_{XA}, \delta_{XB}\right) \approx 
\begin{cases}
\frac{4 \lambda_x x v_A}{m_X^2 - m_h^2} \left(\frac{1}{\sqrt{2}},\frac{v_A}{f}\right) & m_X < f \\
\frac{\lambda_x x f}{m_X^2} \left(\frac{\sqrt{2} v_A}{f},1\right) & m_X \gg f
\end{cases}
\end{equation}
to lowest order in $(v/f)^2$ and $\kappa/\lambda$. The partial width for the decay of $X$ into SM states (excluding the Higgs) is
\begin{equation}
\Gamma(X \rightarrow \text{SM}) \approx \left|\delta_{XA}\right|^2 \Gamma_h(m_h = m_X),
\end{equation}
where $\Gamma_h(m_h = m_X)$ denotes the decay width of a SM Higgs if it were to have mass $m_X$. Note that the Higgs partial width must be computed using the vev $v_A \approx v/\sqrt{2}$ to determine the masses and couplings of the SM particles. The partial width of the $X$ into twin states is computed the same way using $\delta_{XB}$ and the vev $v_B \approx f/\sqrt{2}$.

From the mixing angles (\ref{Eq:mixing}), it is already apparent over what mass range asymmetric reheating from $X$ decays will work. These give
\begin{equation} 
\frac{\Gamma(X \rightarrow \text{SM})}{\Gamma(X \rightarrow \text{Twin})} \sim
\begin{cases}
f^2/v_A^2 \gg 1 & m_X < f \\
v_A^2/f^2 \ll 1 & m_X \gg f.
\end{cases}
\end{equation}
Thus when the mass of $X$ is less than the twin scale, the Standard Model will be reheated to a higher temperature than the twin sector, but in the large mass limit this mechanism works in the opposite direction and would appear to lead to preferential reheating of the twin sector. 

More precise statements about the relative branching ratios and resulting temperatures require additional care. In addition to decaying through mass mixing, $X$ can decay into the Higgs mass eigenstates themselves if above threshold. 
As the energy is ultimately transferred to the SM and twin sectors, we then need to consider how these states decay and account for the further mixing of the Higgs mass eigenstates into Higgs gauge eigenstates.

For $m_X > 2 m_h$, decay can occur into the lighter (SM-like) Higgs mass eigenstate $h$ with partial width
\begin{equation}
\Gamma(X\rightarrow hh)\approx\frac{\lambda_x^2 x^2}{16\pi m_X}\sqrt{1-\left(\frac{2m_h}{m_X}\right)^2}\label{Eq:Xtohh}.
\end{equation}
Similarly, for $m_X>2m_H$, decays can proceed into $HH$ with a similar partial width, but with the $h$ mass replaced with that of the $H$.
Above the intermediate threshold $m_X > m_h + m_H$, there is also the mixed decay
\begin{equation}
\Gamma(X\rightarrow hH)\approx\frac{\lambda_x^2}{2\pi m_X}\sqrt{1-\left(\frac{m_H+m_h}{m_X}\right)^2}(f\delta_{AX}+2v_A\delta_{BX})^2. 
\end{equation}
Here, $\delta_{AX} \approx-\delta_{hA}\delta_{XA}-\delta_{hB}\delta_{XB}$ is the component of the $h_A$ gauge eigenstate in the $X$ mass eigenstate and $\delta_{BX}\approx\delta_{hB}\delta_{XA}-\delta_{hA}\delta_{XB}$ is the corresponding component of the $h_B$ gauge eigenstate, where $\delta_{hA}$ and $\delta_{hB}$ are, respectively, the components of the SM Higgs in the $h_A$ and $h_B$ gauge eigenstates to zeroth order in $\lambda_x$. Combining all ingredients, this decay width is of order $\lambda_x^4x^2$. Since it is only the total decay width that is constrained to be small by the demand that the SM reheating temperature lie in the required window, this fixes only a product of $\lambda_x$ and $x$. If $x\sim v$, then the mixed decay to $hH$ is effectively second order in the small coupling $\lambda_x^2$ and can be neglected relative to the other partial widths. Conversely if $x\ll v$, then $\lambda_x$ is much larger and this decay cannot be neglected. In what follows we will work in the region of parameter space where mixed decays to $hH$ are negligible.

The rate of heat flow into each sector may be well approximated by adding the decay rates of $X$ into each channel and weighting these by the fraction of energy transferred into the particular sector. Of course, when $X$ decays into Higgs particles, these in turn decay out of equilibrium into both the Standard Model and twin sectors. As the Higgs decays are almost instantaneous, the fraction of energy transferred into each sector is simply that carried by the Higgs decay products multiplied by their branching fractions for each sector. 
The total rate at which $X$ particles are transferred into the SM plasma is 
\begin{multline}
W(X\rightarrow SM)\approx \Gamma(X\rightarrow SM)+\Gamma(X\rightarrow hh)Br(h\rightarrow SM)\\ \hspace{0cm}\hbox{}+  \Gamma(X\rightarrow HH)(Br(H\rightarrow SM)+Br(H\rightarrow hh)Br(h\rightarrow SM)).\label{Xrate}
\end{multline}
The corresponding rate for energy deposition into the twin sectors is simply given by the replacement of ${ SM} \mapsto {\text{Twin}}$. The first term is the rate at which $X$ decays directly into the SM through mass mixing with the Higgs. The second is the fraction of $X$ energy that is transferred into lighter Higgs states that subsequently decay into the SM. The third is the analogous term for decays into the heavy Higgs, where cascade decays of the $H$ into the $h$ and subsequently other SM particles must be included. Note that decays of the heavy Higgs into the light Higgs make up a majority of decay width, because of the large quartic coupling required for the twin Higgs potential. 

Below the $hh$ threshold, it is possible for $X$ to decay via one on-shell and one off-shell Higgs boson. The partial width for off-shell Higgs production was calculated for $X \rightarrow hh^* \rightarrow hb\bar{b}$ and found to be negligible compared to two-body decays through mass mixing and so we omit three-body decay widths in what follows.

\begin{figure}[t]
\centering
\includegraphics[width=0.8\linewidth]{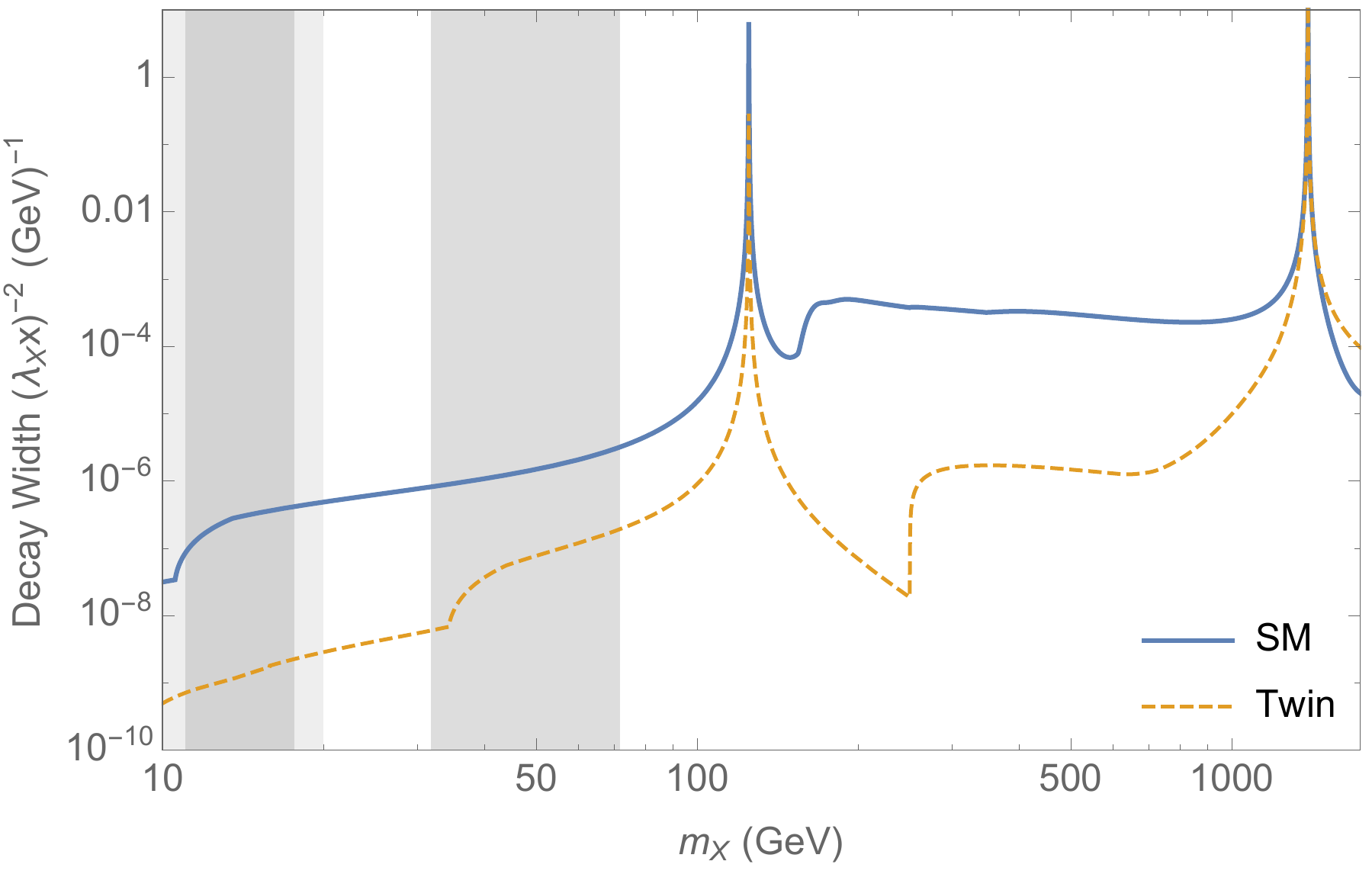}
\caption{The partial widths of the $X$ into the SM (solid blue line) and twin sector (dashed orange) for $f/v = 3$ in units of $(\lambda_x x)^2$. The light gray bands indicate regions of QCD-related uncertainty in the SM calculation, while the darker gray bands indicate the corresponding regions of uncertainty for the twin calculation.}
\label{Fig:decayWidths}
\end{figure}
\begin{figure}[t]
\centering
\includegraphics[width=0.7\linewidth]{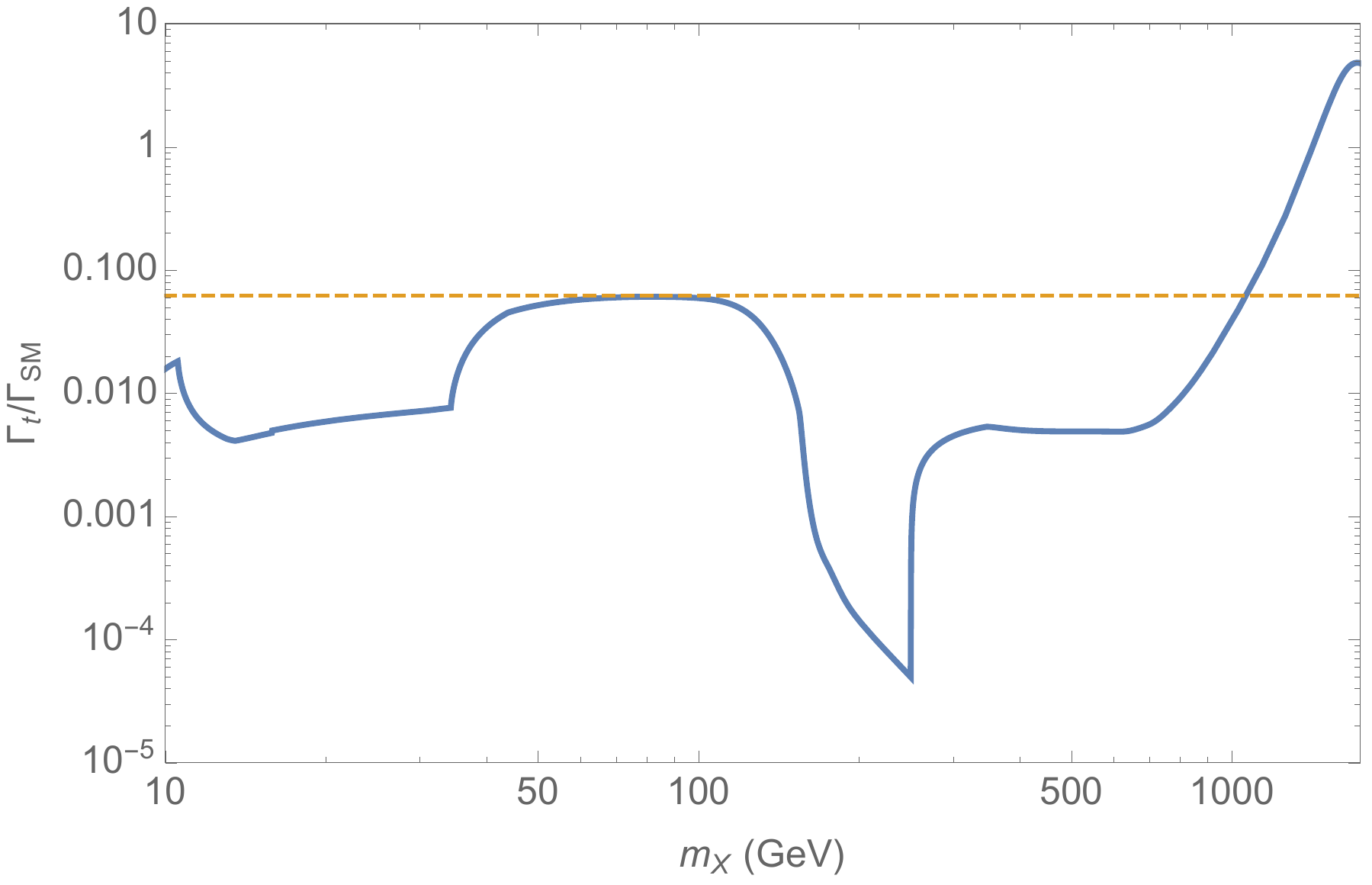}
\caption{The ratio of branching fractions of the $X$ into the SM and twin sectors at $f/v=3$. The dashed line gives the expected $\left(v/f\right)^2$ scaling from the mass mixing; deviations are due to various mass threshold effects.}
\label{Fig:brRatio}
\end{figure}

Ultimately, the complete partial widths for the decay of $X$ into the Standard Model and twin sectors includes the sum of decays into Higgs bosons $h$ and $H$ and direct decays into the fermions and gauge bosons of the two sectors. We compute the latter to an intended level of accuracy of $\sim 10\%$ (including, e.g., NLO QCD corrections to decays into light-flavor quarks), mostly following \cite{Djouadi:2005gi}. The resulting partial widths into the Standard Model and twin sectors are shown as a function of $m_X$ in Figure \ref{Fig:decayWidths} with the ratio of branching fractions displayed in Figure \ref{Fig:brRatio}.  

Over much of the space below the Higgs mass, the branching ratio exhibits the expected $(f/v)^2$ scaling from the mass mixing. Below $\sim 40$ GeV, suppression of the twin partial width arises because the twin bottom quark pair production threshold is crossed. As $m_X$ nears $m_h$, the SM branching fraction grows by $\sim 4$ orders of magnitude as the $WW^*$, $ZZ^*$, and then $WW$ and $ZZ$ decays go above threshold. Since the analogous thresholds are at much higher energies in the twin sector, the enhancement is not paralleled by decays into the twin sector until $m_X$ is close to the twin scale. There is therefore a large range of masses $m_h \lesssim m_X \lesssim m_H$ over which the SM branching fraction dominates by several orders of magnitude.

Above the $X\rightarrow hh$ threshold, the ratio of decay widths is roughly constant in mass up to the $HH$ threshold. The twin sector decay rate is dominated by decays of on-shell light Higgs into twin states, $\Gamma(X \rightarrow \text{Twin}) \approx \Gamma(X\rightarrow hh)\text{Br}(h\rightarrow\text{Twin}) \propto 1/m_X $ as in (\ref{Eq:Xtohh}). If the SM were also predominantly reheated through this channel, then the ratio of branching fractions would again be approximately $\delta_{hA}^2/\delta_{hB}^2\approx (f/v)^2$. However, the SM decay width also receives a larger contribution from decays through mass mixing between the $X$ and the Higgs gauge eigenstates. 

For masses $m_X>2m_h$, decays through mass mixing are dominated by the SM $WW$ and $ZZ$ channels. In this mass region, the decay rate of a Higgs into longitudinally polarized vector bosons scales as $\Gamma(h\rightarrow WW,ZZ)\sim m_X^3$, but the mixing angle scales as $\delta_{AX}^2\sim 1/m_X^4$ (as in (\ref{Eq:mixing})), resulting in the same $\sim 1/m_X$ scaling and thus a roughly constant ratio in this range of masses. 
Near $m_X \sim 1 \ \text{TeV}$, decays into twin vector bosons through mass mixing begin to dominate, and there is no favourable asymmetry in the branching fractions, as discussed in this section. Even at higher masses, the effects of heavy Higgs decays into light Higgs do not compensate sufficiently, as this partial width scales with $m_X$ in the same way as the partial width for longitudinally polarised weak bosons.

The constraint on the decay width from the required reheating temperature (\ref{ReheatRange}) translates into a constraint on the size of the coupling $\lambda_x x$. For $m_X\gtrsim m_h$, this gives $10^{-8.5}\, \text{GeV} \lesssim \lambda_x x\lesssim 10^{-6}$ GeV, while for lower masses, this range increases to $10^{-7}\, \text{GeV} \lesssim \lambda_x x\lesssim 10^{-5.5}$ GeV at $m_X\sim 20$ GeV.

The gray bands in Figure \ref{Fig:decayWidths} highlight regions where our analytic estimates of the partial widths encounter enhanced uncertainties arising from the bottom and charm thresholds in both sectors. Over most of these ranges, we estimate the size of these uncertainties to be either $\sim 10\%$ or confined to very small subregions. The thicknesses of these bands have been chosen conservatively, and ultimately the branching ratios should be accurate to within a factor of $\pm \Lambda_{QCD}$ of the bottom and charm mass thresholds. In particular, the prescription of \cite{Drees:1989du} has been followed for approximating the bottom partial width close to the open flavour threshold. Resonant decay into gluons from bottomonia mixing has been neglected, although these resonant mass ranges are expected to be only $\sim$ MeV wide at the CP-even, spin-0 bottomonia masses $m_X=m_{{\chi_b}_i}$ (see \cite{Drees:1989du} and \cite{Baumgart:2012pj}). It should be noted, however, that at temperatures above that of the QCD phase transition, the quark decay products behave differently compared to that expected in a low temperature environment. In particular, for hot enough temperatures, the $b$ or $c$ quarks may not hadronise and the partonic partial widths may more reliable. The applicability of the treatment of the flavour thresholds used here may therefore not be valid if the decay occurs in the hot early universe. However, it is only very close to the threshold itself (within several GeV) that this uncertainty becomes significant. Finally, quark masses have been neglected in the gluon partial width. For $m_X$ close to the flavour thresholds, this approximation breaks down, but the gluon branching fraction is only $\sim 10\%$ and so the error does not contribute to the uncertainty of the total width by more than this order (it is this uncertainty that is responsible for most of the extension of the length of the gray bands about the flavour threshold).

Close to the charm threshold, the analogous uncertainties are even more poorly understood. Below the charm threshold, hadronic decays of a light scalar are highly uncertain (see \cite{Clarke:2013aya} for discussion). We avoid these regions altogether by restricting our considerations to $m_X$ roughly above the twin charm threshold. Note that below the SM charm threshold, the smaller decay rate of a Higgs-like scalar necessitates larger couplings $\lambda_X x$ for $X$ to have a lifetime within the required reheating window. The larger couplings then imply potentially stronger constraints from invisible mesonic decays. See \cite{Baumgart:2012pj, Clarke:2013aya, Haisch:2016hzu} for further discussion and recent analysis of the pertinent experimental constraints.

Taken together, the results in Figures \ref{Fig:decayWidths} and \ref{Fig:brRatio} bear out the expectation that a scalar $X$ with symmetric couplings to the Standard Model and twin sectors may nonetheless inherit a large asymmetry in partial widths from the hierarchy between the scales $v$ and $f$. Across a wide range of masses $m_X$, the asymmetry is proportional to (or greater than) $v^2/f^2$, tying the reheating of the two sectors to the hierarchy of scales.

Before proceeding to our computation of cosmological observables, we comment on an alternative variation on the reheating mechanism presented here that involves having $X$ odd under the twin parity. This permits two renormalisable interactions with the Higgses to give a Higgs potential of the form:
\bea
\mathcal{V} \supset m_0^2 \left( |H_A|^2 + |H_B|^2 \right) + \lambda_0 \left( |H_A|^4 + |H_B|^4 \right) + \epsilon X^2 \left( |H_A|^2 + |H_B|^2 \right) + \tilde{\epsilon} X \left( |H_A|^2 - |H_B|^2 \right).\label{OddPot}
\eea
If $X$ then acquires a vev at some scale, it may be possible to arrange for the resulting spontaneous breaking of the $\mathbb{Z}_2$ to give that required in the Higgs potential. However, we find that, in order for $X$ to be long-lived and reheat the universe, its couplings to the Higgs must be highly suppressed and therefore that the resulting vev of $X$ required to explain the soft $\mathbb{Z}_2$-breaking in the Higgs potential must be many orders of magnitude above the twin scale. If this is to be identified with the characteristic mass scale of $X$, then a UV-completion of the twin Higgs is required for anything further to be said of the prospects of this possibility. However, if such a UV completion has similar structure to the couplings in (\ref{OddPot}), then asymmetric reheating may require a cancellation between the odd and even couplings of $X$ to the Higgs potential in order to suppress its twin-sector branching fraction (because the odd coupling appears with opposite signs in the coupling between $X$ and the $h_A$ and $h_B$ states). We do not consider this possibility further.

\subsubsection{Imprints on the CMB} \label{Sec:CMBfx}

For appropriate values of $m_X$, the out-of-equilibrium decay of $X$ reheats the two sectors to different temperatures and effectively dilutes the energy density in the twin sector. We obtain an analytic estimate of the effects of the $X$ decay on the number of light degrees of freedom observed from the CMB by approximating both the decay of $X$ and the decoupling of species as instantaneous in Section \ref{Sec:instant}. We then demonstrate that this estimate is reliable over most of the parameter space of interest with a numerical calculation in Section \ref{Sec:cosmosim}. In Section \ref{Sec:neutrinos} we consider neutrino masses and their joint constraints with $N_{\text{eff}}$.


\paragraph{Analytic estimate of $N_{\text{eff}}$} \label{Sec:instant}

If $X$ dominates the energy density of the universe and then decays, depositing energy $\rho_\text{SM}$ and $\rho_\text{t}$ into the SM and twin sectors respectively, then the temperature ratio is determined by 
\begin{equation}
\frac{\rho_\text{t}}{{\rho_\text{SM}}} = \frac{g^\text{t}_{\star}(T^\text{t}_{\text{reheat}})}{g^\text{SM}_{\star}(T^\text{SM}_{\text{reheat}})} \left( \frac{T^\text{t}_{\text{reheat}}}{T^\text{SM}_{\text{reheat}}}\right)^4 \approx \frac{\Gamma(X \rightarrow \text{Twin})}{\Gamma(X \rightarrow \text{SM})},\label{EnergyRatio}
\end{equation}
where $T^\text{SM}_{\text{reheat}}$ and $T^\text{t}_{\text{reheat}}$ are the reheating temperatures for each sector, while $g^\text{SM}_\star$ and $g^\text{t}_\star$ are the SM and twin effective number of relativistic degrees of freedom, respectively. We have assumed that the two sectors are cool enough that they have already decoupled. We point out that not only does the number of effective degrees of freedom in each sector need to be evaluated at the temperature of that sector, but that $g^\text{t}_\star$ and $g^\text{SM}_\star$ differ as functions of temperature due to the differences in the spectra of the sectors, as seen in Figure \ref{Fig:geff}. As is well-known \cite{Kolb:1990vq}, reheating is a protracted process that occurs over a time-scale given by the lifetime of the reheaton. During this time, the temperature of the plasma cools slowly because, while the energy is being replenished by the decay of the reheaton, it is simultaneously diluted and redshifted with the expansion of the universe. It is assumed in (\ref{EnergyRatio}) that any primordial energy density in either sector is subdominant.

The temperatures of both sectors then redshift in the same way, so the only additional differences between their temperatures arise from changes to the effective number of degrees of freedom in each sector. By conservation of comoving entropy within each sector, each evolves as $T^i_{eq}/T^i_\text{reheat} = \left( g^i_{\star}(T^i_\text{reheat})/g^i_{\star}(T^i_{eq}) \right)^{1/3}a(T_\text{reheat})/a(T_{eq})$ where $T^i_{eq}$ is the temperature of the sector at matter-radiation equality, which the CMB probes as explained in Section \ref{Sec:Limits}, and $a(T)$ is the scale factor as a function of temperature. In the mirror Twin Higgs model, the two sectors have the same number of light degrees of freedom at recombination (three neutrinos and a photon, assuming that the neutrinos are still relativistic), so 
\begin{equation}
\left(\frac{T^\text{t}_{eq}}{T^\text{SM}_{eq}}\right)^4 = \left(\frac{T^\text{t}_{\text{reheat}}}{T^\text{SM}_{\text{reheat}}}\right)^4 \left(\frac{g^\text{t}_{\star}(T^\text{t}_{\text{reheat}})}{g^\text{SM}_{\star}({T}_{\text{reheat}})}\right)^{4/3}= \frac{\Gamma(X \rightarrow \text{Twin})}{\Gamma(X \rightarrow \text{SM})} \left(\frac{g^\text{t}_{\star}(T^\text{t}_{\text{reheat}})}{g^\text{SM}_{\star}({T}_{\text{reheat}})}\right)^{1/3}.\label{reheat}
\end{equation}
As our range of reheat temperatures encompasses the QCD phase transitions of both sectors, the factors of $g_{\star}$ can be important.

Given the temperatures of the two sectors after $X$ decays, we can obtain a simple estimate of the contribution to $N_{\text{eff}}$ that neglects the impact of masses of the twin neutrinos discussed in Section (\ref{Sec:Neff}), 
\begin{align} 
(\Delta N_{\text{eff}})_{m_\nu = 0} &= \frac{4}{7} \left(\frac{11}{4}\right)^{4/3} g^\text{SM}_{\star}(T^\text{SM}_{eq}) \frac{\rho_\text{t}(T^\text{t}_{eq})}{\rho_\text{SM}(T^\text{SM}_{eq})} \\
&\approx 7.4 \times \frac{\text{Br}(X \rightarrow \text{Twin})}{\text{Br}(X \rightarrow \text{SM})} \left(\frac{g^\text{t}_{\star}(T^\text{t}_{\text{reheat}})}{g^\text{SM}_{\star}(T^\text{SM}_{\text{reheat}})}\right)^{1/3}.
\end{align}
In this limit the most recent Planck data give a $2\sigma$ bound of $\Delta N_{\text{eff}} \lesssim 0.40$ assuming pure $\Lambda$CDM+$N_{\text{eff}}$ \cite{Ade:2015xua}. This translates into the requirement $\frac{\rho_\text{t}(T^\text{t}_{eq})}{\rho_\text{SM}(T^\text{SM}_{eq})} \approx \frac{\Gamma(X \rightarrow \text{Twin})}{\Gamma(X \rightarrow \text{SM})}  \lesssim 0.05$, ignoring possible differences in $g_{\star}$. 

Of course, as discussed in Section \ref{sec:thermal}, the twin neutrino masses are relevant at the temperature of matter-radiation equality, so we can obtain a more meaningful estimate of $\Delta N_{\text{eff}}$ using the results of Section \ref{Sec:Neff} evaluated at the twin temperature determined above:
\begin{eqnarray} \label{Eq:deltaNeff}
\Delta N_{\text{eff}} &=& \left(\frac{11}{4}\right)^{4/3} \frac{120}{7 \pi^2 \left(T^\text{SM}_{eq}\right)^4} \left( \rho^\text{t}_\gamma\left(T^\text{t}_{eq}\right) + \sum_\alpha 3 w^t_{\nu_\alpha}\left(T^\text{t}_{eq}\right) \rho^\text{t}_{\nu_\alpha}\left(T^\text{t}_{eq}\right)\right) \\
T^\text{t}_{eq} &=& T^\text{SM}_{eq} \left(\frac{\Gamma(X \rightarrow \text{Twin})}{\Gamma(X \rightarrow \text{SM})}\right)^{1/4} \left(\frac{g^\text{t}_{\star}(T^\text{t}_{\text{reheat}})}{g^\text{SM}_{\star}({T}^\text{SM}_{\text{reheat}})}\right)^{1/12}\label{TwinReh}
\end{eqnarray}
with $T^\text{SM}_{eq} \approx 0.77 \text{ eV}$ \cite{Ade:2015xua} the photon temperature. While the right-hand side of this equality has implicit dependence on  $T^\text{t}_{eq}$ through $g^\text{t}_{\star}$, this is only important if the reheating occurs between the SM and twin QCDPTs and the neglecting of the factors of $g_\star$ is otherwise reliable. With the further inclusion of Standard Model neutrino masses or an extra sterile neutrino, the bound described above weakens to $\Delta N_{\text{eff}} \lesssim 0.7$. As discussed in Section \ref{Bounds}, we are not aware of any analyses specific to our model involving both pure dark radiation and three sterile neutrinos with masses of order the photon decoupling temperature of the CMB and possibly cooler temperatures. In the absence of such an analysis, we use the inequality $\Delta N_{\text{eff}} \lesssim 0.7$ to indicate where the present CMB measurements are likely to constrain the light degrees of freedom of this model, leaving a more detailed analysis of the CMB constraints as future work. In this case, the bound on the decay width ratio is $\frac{\Gamma(X \rightarrow \text{Twin})}{\Gamma(X \rightarrow \text{SM})}  \lesssim 0.09$. The next generation of CMB experiments are projected to strengthen this constraint to $\Delta N_{\text{eff}} \lesssim 0.02$ at the $1\sigma$ level \cite{Abazajian:2013oma}.

\paragraph{Numerical Calculation of $N_{\text{eff}}$} \label{Sec:cosmosim}

A more precise study of the effect of $X$ decay on the number of effective neutrino species at recombination may be performed by numerically solving a system of differential equations for the entropy in $X$ and the two sectors as a function of time. Following the analysis of Chapter 5.3 of \cite{Kolb:1990vq} we have
\begin{gather}
H=\frac{1}{a}\frac{da}{dt}=\sqrt{\frac{1}{3M_{Pl}^2}(\rho_X+\rho_{SM}+\rho_t)}\\
\frac{d\rho_X}{dt}+3H\rho_X=-\Gamma_X\rho_X \label{Eq:XEnDec}\\
\rho_i=\frac{3}{4}\left(\frac{45}{2\pi^2 g^i_{\star}}\right)^{1/3}S_i^{4/3}a^{-4}\\
S_i^{1/3}\frac{dS_i}{dt}=\left(\frac{2\pi^2 g^i_{\star}}{45}\right)^{1/3}a^4\Big(\rho_X\Gamma_{X\rightarrow i}+\frac{dq_{j\rightarrow i}}{dt}\Big),\label{Eq:XHeat}
\end{gather} 
where $S_i$ are comoving entropy densities and it has been assumed that $X$ is cold by the time it decays so that $\rho_X = m_X n_X$ with number density $n_X$ (this is reliable as we only consider $m_X>10$ GeV, which is above the decoupling temperature of $\sim 1$ GeV). The rate of heat flow from sector $j$ to $i$ per proper volume, $\frac{dq_{j\rightarrow j}}{dt}$, is defined in (\ref{HeatFlow}). To account for the temperature-dependence of the effective number of relativistic degrees of freedom in each sector, these equations are solved iteratively in the profiles of $g^i_\star(T^i)$. 

The equations are solved in three stages: before, during and after the decoupling of the SM and twin sectors. The ratio $f/v$ is fixed to $4$ for this analysis. Initial conditions were chosen with $\rho=10^{-12}\rho_X$, for combined SM and twin energy densities $\rho$. However, it is only the requirement that the initial energy density of $X$ dominates over that of the SM and twin sectors that is important for simulating the cosmology over the times of interest here, as the entirety of the latter is then generated by the subsequent decay. The results close to the decoupling and reheating epochs are otherwise insensitive to the initial conditions and ultimately match onto the standard outcome \cite{Kolb:1990vq} expected by equating the Hubble rate with the decay rate of X. The sectors are assumed to be in thermal equilibrium and sharing entropy until a temperature of $10$ GeV, below which they are evolved separately with the heat flows $\frac{dq_{i\rightarrow j}}{dt}$ switched on. Elastic scatterings were neglected from the heat flow rate to accelerate the computation. It was verified for the results found below that their contribution to the heat flow was always $\lesssim 10\%$ while the heat flow was itself not dominated by the Hubble rate. Heat flow was switched off again once the twin temperature reaches $0.1$ GeV, by which time thermal decoupling is long-since complete, and the sectors are subsequently evolved separately. Again, although the strengthening of the colour force and the QCDPT make the perturbative tree-level computation of the scattering rates unreliable at temperatures below $\sim 1$ GeV, as found in Section \ref{Sec:Decoupling} and also in the results below, the sectors decouple above these temperatures. Notably, the impact of $X$ on the expansion rate causes decoupling to occur at slightly hotter temperatures than expected from the analysis of Section \ref{Sec:Decoupling} for the decoupling in the standard cosmology.


\begin{figure}[h!]
\centering
\includegraphics[width=0.8\linewidth]{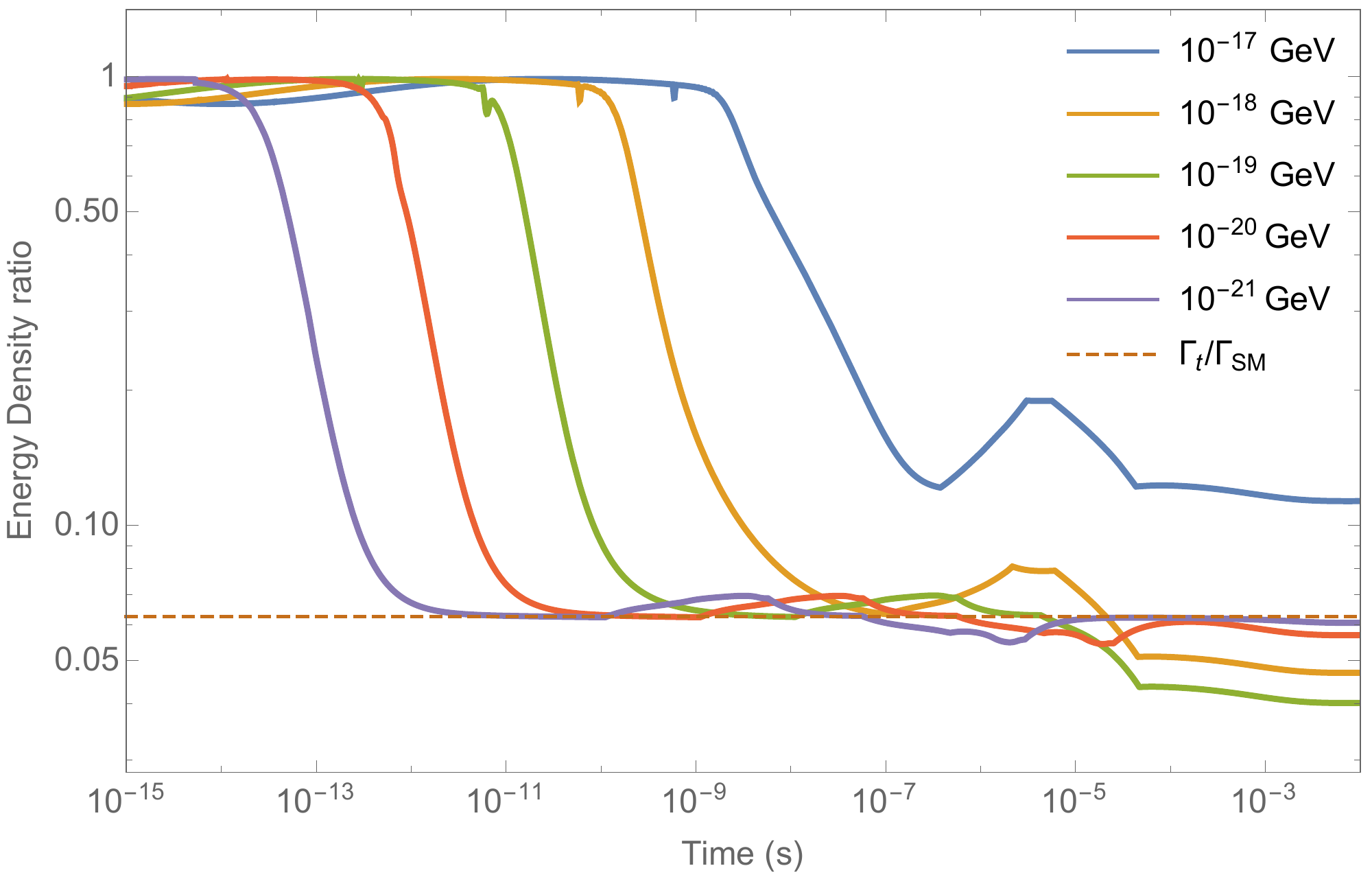}
\caption{Ratio of twin to SM energy densities throughout decoupling and reheating, for different decay rates $\Gamma_X$. The dashed line corresponds to the prediction of from the ratio of decay widths, here selected to be $1/16$.}
\label{Fig:enRatio}
\end{figure} 

The ratio of energy densities in each sector determines $N_\text{eff}$, from (\ref{Eq:deltaNeff}). A plot of this ratio over time is shown in Figure \ref{Fig:enRatio}, with the expectation under the approximations of the previous section shown as well. 
This approximation is reliable as long as the lifetime of $X$ is much longer than the temperature at which decoupling concludes, here $\sim 1$ GeV. The larger asymptotic value of the ratio of the blue line arises because the lifetime lies close to the decoupling period, so that a significant fraction of the energy is transferred while the sectors are thermalised or partially thermalised and does not contribute toward asymmetric reheating. Equivalently, as will be discussed below, insufficient time elapses between decoupling and reheating for the twin energy density to dilute and be repopulated by the decays to the level predicted by (\ref{EnergyRatio}). The subsequent bump represents the period between the reheating of the twin sector by its QCD phase transition followed by that of the SM. The green and orange lines correspond to reheating temperatures that lie between SM and twin QCD phase transitions. In these cases, the reheating of the SM from the subsequent SM QCD phase transition raises its energy density relative to the twin sector above that expected from the ratio of branching fractions. As this occurs after the lifetime of the reheaton, the estimate of the reheating temperatures presented in (\ref{reheat}) is still good as subsequent changes in the ratio due to the evolution of $g_{\star}$ are accounted for in our analysis of the reheating scenarios. 


%
The steep drop in the energy density ratio corresponds to the brief period during which the energy density of the twin sector present at decoupling dilutes and redshifts, which continues until it reaches a comparable size to the energy density that is being replenished by reheating. If the twin-sector branching fraction is highly suppressed, as can occur in the ``valley'' region in Figure \ref{Fig:brRatio} with $m_h\lesssim m_X\lesssim 2m_h$, then a longer time is required for this to happen, especially close to the decay epoch where the diminishing of the $X$ population also contributes to a reduced reheating rate. These effects can prolong the time required for the energy density ratio to converge to the asymptotic prediction of (\ref{EnergyRatio}).

\begin{figure}[h!]
\centering
\includegraphics[width=0.8\linewidth]{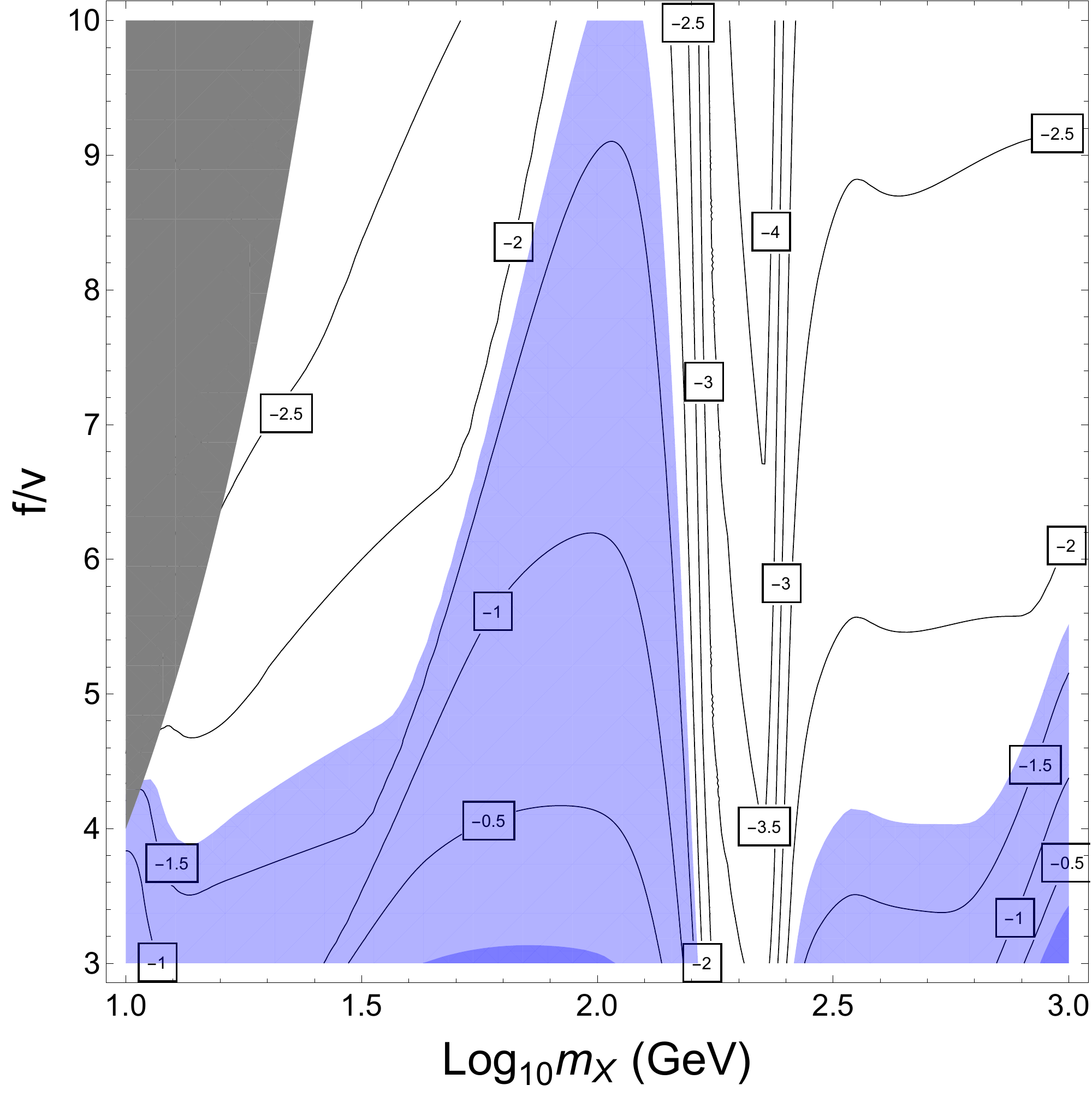}
\caption{Contours of $\log_{10}\Delta N_{\text{eff}}$ as a function of $m_X$ and $f/v$, for $T^{\text{SM}}_{\text{reheat}}=0.7$ GeV. The dark blue region is in tension with Planck, while the light blue region will be tested by CMB-S4. Gray regions are where the $X$ mass is below the twin charm threshold and our calculation of the twin sector partial width is unreliable.}
\label{Fig:XNeffmin}
\end{figure}

Contour plots of $\Delta N_{\text{eff}}$ as a function of $m_X$ and $f/v$ appear in Figure \ref{Fig:XNeffmin}, along with current and predicted bounds using the analytic results of Section \ref{Sec:instant}. The minimum neutrino mass configuration with Dirac masses has also been assumed, although the results are relatively insensitive to this provided that the twin neutrino masses are not well above the eV scale. A SM reheating temperature of $0.7$ GeV has been assumed. At this temperature, we have verified using the numerical calculation of Section \ref{Sec:cosmosim} that the twin sector reheating temperature is always roughly above the twin neutrino decoupling temperature over the parameter space of the figure, ensuring that the neutrinos thermalise once produced in the decays and hence that the predictions of Section \ref{Sec:instant} are valid. A treatment of the case in which the twin neutrinos are produced below their decoupling temperature is beyond the scope of this analysis, but would involve the computation of the phase space spectrum of the neutrino decay products of the $X$. 

Also, as discussed in Section \ref{Sec:Decoupling}, a large temperature difference may partially relax back if reheating occurs close to sector decoupling. However, a reliable calculation of the heat flow at the temperatures of interest here must incorporate non-perturbative effects. We do not perform such a computation, but note that, at a slightly higher SM reheating temperature of $2$ GeV where this computation is more reliable, $\Delta N_{\text{eff}}$ in Figure \ref{Fig:XNeffmin} can be raised by up to an order of magnitude in the region with $f/v\lesssim 4$ and $150\,\text{GeV}\lesssim m_X\lesssim 200\,\text{GeV}$, notably where the twin sector partial width is suppressed relative to the SM by several orders of magnitude. The resulting $\Delta N_{\text{eff}}$ prediction is, nevertheless, still out of observable reach. At the lower SM reheating temperature assumed in Figure \ref{Fig:XNeffmin}, it is expected that decoupling will be further advanced and the enhancement in $\Delta N_{\text{eff}}$ would be weaker.


We emphasize that, if the lifetime of $X$ is sufficiently close to the time of decoupling, or equivalently, that the reheating temperature is sufficiently close to the decoupling temperature, then the residual twin energy density left-over may be comparable to or greater than that regenerated by reheating. Consequently, the suppression in $\Delta N_{\text{eff}}$ would be less than that predicted in (\ref{reheat}). In this respect, the projection of Figure \ref{Fig:XNeffmin} should be regarded as a lower bound on $\Delta N_{\text{eff}}$. In the regions of high suppression, such as the ``valley'' region, the full asymmetry may not be generated before the complete decay of $X$ when the reheating temperature is of similar order as the decoupling temperature. In particular, for the reheating temperature chosen here of $0.7$ GeV and branching fraction ${\text{Br}}(X\rightarrow \text{Twin})\sim 10^{-5}$, the numerical calculation of the energy density ratio saturates at $\sim 4\times 10^{-5}$. We do not include this effect in Figure \ref{Fig:XNeffmin} as its only impact is to mildly shift the unobservably small $\Delta N_{\text{eff}}=10^{-4}$ contour. Lower reheating temperatures would agree with the prediction of (\ref{EnergyRatio}) were it not for the caveat that the twin neutrinos may be produced out of equilibrium. However, this minimum value at which $\Delta N_{\text{eff}}$ is saturated can grow significantly with hotter reheating temperatures upon which it is highly dependent. 

CMB-S4 observations will be able to probe a large portion of the most natural parameter space, save the region $m_h \lesssim m_X \lesssim 2 m_h$ where decays into the Standard Model dominate well beyond the ratio $f^2/v^2$, as previously discussed. Significantly, precision Higgs coupling measurements at the LHC are unlikely to probe the mirror Twin Higgs model beyond $f \sim 4 v$, so that the observation of additional dark radiation may be the {\it first} signature of a mirror Twin Higgs.

\paragraph{Neutrino Masses} \label{Sec:neutrinos}

In addition to the bounds on $N_{\text{eff}}$, we must also respect the bounds on neutrino masses. The analysis remains nearly the same as in Section \ref{Sec:meff}, but now with the twin neutrinos at a lower temperature, as determined above. As mentioned above, for large enough $f/v$ and SM reheating temperature sufficiently close to the lower bound, the reheating temperature of the twin sector may be below the twin neutrino decoupling temperature and the resulting energy density would be more difficult to compute. For simplicity, we choose $\lambda_x x$ large enough such that the twin reheating temperature is always above the twin neutrino decoupling temperature. 

As before, we compute $m_\nu^{\text{eff}}$ as 
\begin{equation}
m_\nu^{\text{eff}} = \frac{n^\text{t}_\nu}{n^\text{SM}_\nu} \sum_\alpha m^\text{t}_{\nu_\alpha}.
\end{equation}  
In relating the scale factors at neutrino decoupling in each sector, we now have to use the above temperature ratio to find, analogously to Section \ref{Sec:meff}, that 
%
%
\begin{equation}
m_\nu^{\text{eff}} = \left( \frac{\Gamma_\text{t}}{\Gamma_\text{SM}} \right)^{3/4}  \left( \frac{g^\text{t}_\star\left(T^\text{t}_\text{reheat}\right)}{g^\text{SM}_\star\left(T^\text{SM}_\text{reheat}\right)}\right)^{1/4} \left(\frac{f}{v}\right)^n \sum_\alpha m^\text{SM}_{\nu_\alpha},
\end{equation}
where, again, $n = 1$ for Dirac masses and $n = 2$ for Majorana masses. Interestingly, if the branching ratios scale as $\Gamma_\text{t}/\Gamma_\text{SM} = (v/f)^2$, then we have $m_\nu^{\text{eff}} \propto (f/v)^{-3/2 + n}$, so the contribution grows with $f/v$ for Majorana masses, but is suppressed for Dirac masses.

\begin{figure}[h!]
\centering
\includegraphics[width=0.9\linewidth]{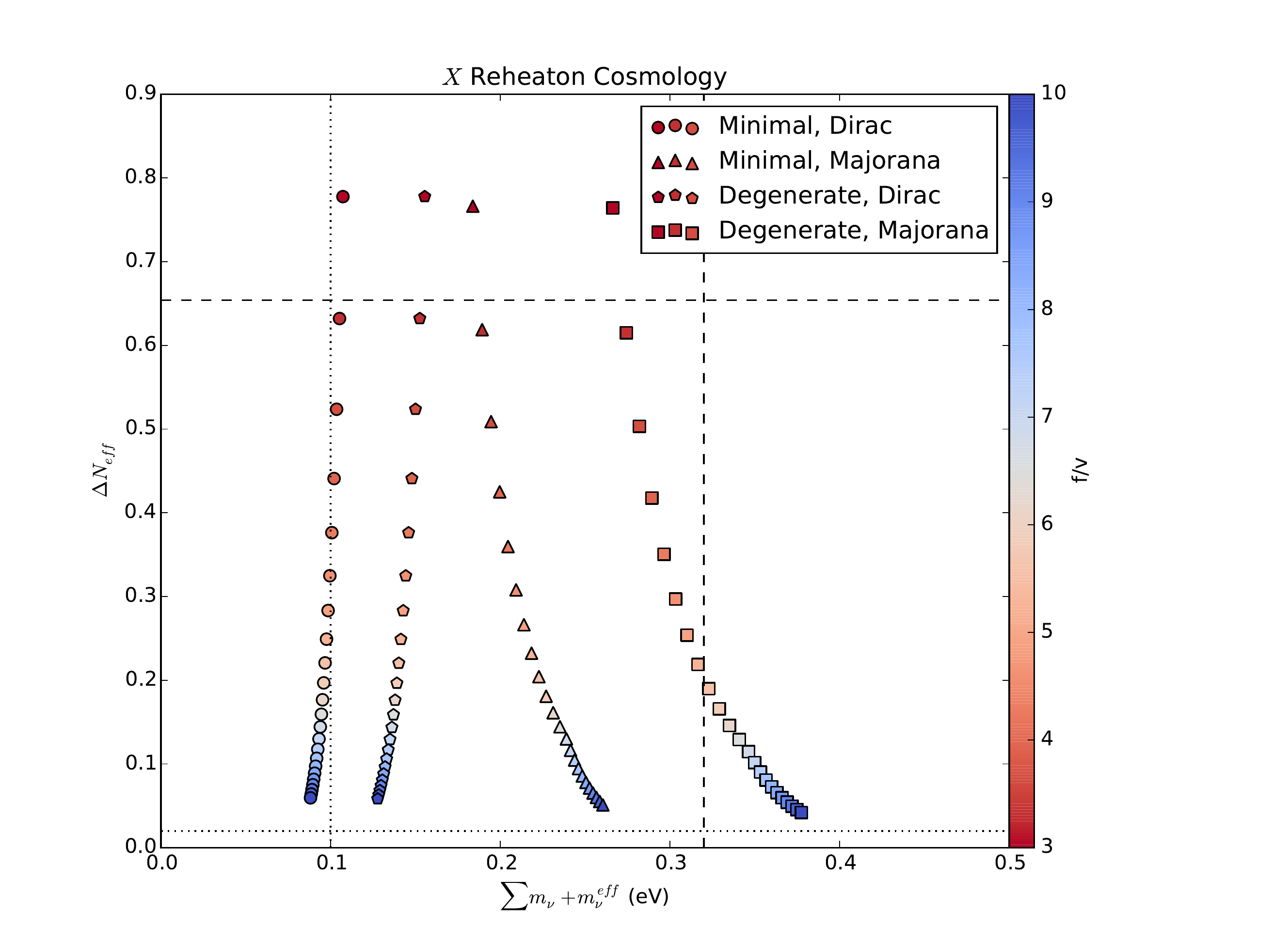}
\caption{Predicted values of $\Delta N_{\text{eff}}$ and $\sum m_\nu + m_\nu^{\text{eff}}$ for minimal and degenerate neutrino mass spectra with both Dirac and Majorana masses for $f/v$ from 3 to 10. The Planck 2015 \protect\cite{Ade:2015xua} bounds on $\sum m_\nu$ and $N_{\text{eff}}$, as discussed in Section \protect\ref{Bounds}, are represented by the dashed lines, and the projected CMB-S4 constraints are given by the dotted lines. It has been assumed that $\frac{\Gamma_\text{t}}{\Gamma_\text{SM}} \sim (v/f)^2$.  Note however, that, from Figure \ref{Fig:XNeffmin}, this scaling of the partial widths holds only for the mass range $50\,\text{GeV}\,\lesssim m_X\lesssim 120\,\text{GeV}$, outside of which the twin partial width is more suppressed and the model is only testable through $\Delta N_{\text{eff}}$ over a smaller range in $f/v$.}
\label{Fig:asymPred}
\end{figure}

As before, we consider the minimal mass spectrum of $m_\nu = \left[0.0,0.009,0.06 \text{ eV}\right]$ and a degenerate spectrum of $m_\nu = \left[0.1\text{ eV} ,0.1\text{ eV},0.1 \text{ eV}\right]/3$. In Figure \ref{Fig:asymPred} we plot the predictions of the $X$ reheating for $\Delta N_{\text{eff}}$ and $m_\nu^{\text{eff}}$ for both spectra and both Dirac and Majorana masses using the approximations of Section \ref{Sec:Limits}, for $f/v$ from 3 to 10 and assuming the $\frac{\Gamma_\text{t}}{\Gamma_\text{SM}} \sim (v/f)^2$ scaling; there are regions in the space of $m_X$ where the suppression of $m_\nu^{\text{eff}}$ would be much higher. 

Dashed lines indicate the rough locations of present experimental limits from Planck 2015, and projected bounds from CMB-S4. As mentioned in Section \ref{Sec:meff}, we are unaware of any study of bounds on both $m_\nu^{\text{eff}}$ and $\Delta N_{\text{eff}}$ treated jointly. In the absence of this, we show present and projected constraints on $N_{\text{eff}}$ and $\sum m_\nu$ from \cite {Ade:2015lrj} and \cite{CMB-S4:2016}, ignoring correlations, as described in Section \ref{Bounds}. 

\subsubsection{Thermal Production} \label{Eq:thermal}
In our discussion up to this point, we have been agnostic about the origin of the cosmic abundance of $X$ and have operated under the assumption that it absolutely dominates the cosmology before it decays. Here, we consider the possibility that $X$ was thermally produced through freeze-out and subsequently dominates the universe as a relic before decaying. This thermal history is viable, but places strong constraints on the mass and couplings of the $X$.

The energy density of relativistic species redshifts as $\rho_r \propto a^{-4} \propto T^4$, while the energy density of non-relativistic, chemically decoupled matter scales as $\rho_m \propto a^{-3}$. The energy density contained in the $X$ can therefore only grow relative to the energy density in the thermal bath once it becomes non-relativistic. We found in Section \ref{Sec:instant} that by recombination, $\rho_\text{t}/\rho_\text{SM} \lesssim 0.09$ is needed to evade current bounds on $\Delta N_\text{eff}$. Thus we need to have the energy density in the $X$ dominate over the SM and twin plasmas by more than this factor when it decays. 
If $X$ becomes non-relativistic instantaneously at the moment that its temperature reaches some fraction $c\sim \mathcal{O}(0.1)$ of its mass, then, as 
%
%
$T \propto 1/a$ and $\rho_X$ is $\sim 1/g_\star$ of the total energy density, the mass is required to satisfy $m_X\gtrsim 10/c\times g_\star\left(T = m_X\right)T^\text{SM}_{X \text{reheat}}$. 
Since the SM reheating temperature is strongly constrained to be above BBN, this effectively puts a lower limit on the mass of the $X$. Importantly, $X$ must freeze-out when relativistic or its energy density will be further Boltzmann suppressed. The lower limit on the mass of the $X$ becomes an upper limit on the $X$'s couplings - if it couples too strongly to the thermal bath, then it won't freeze out early enough to be hot.

%
%
In fact the situation is somewhat less favorable than the above analysis suggests, because it is relevant operators that must keep $X$ in thermal equilibrium. For an $X$ with the interactions introduced in Section \ref{Sec:reheaton}, the annihilations 
have rates that scale with temperature as $\Gamma \sim n_X \left\langle \sigma v \right \rangle \sim T$ for $T\gtrsim m_X,m_h$ (where $n_X$ is the number density of $X$ and $\left\langle \sigma v \right \rangle$ is its thermally averaged annihilation cross section). However, in a radiation-dominated universe, $H \sim T^2$. Thus, at high enough temperatures, $X$ is not in thermal equilibrium with the plasma and it is only once the universe cools enough that it may thermalise. 
Then, as the temperature drops, $XX \rightarrow q\bar{q}$ annihilations become suppressed by the Higgs mass and subsequent Boltzmann suppression causes $X$ to freeze-out. Note that the rates of these annihilation processes are controlled by the coupling $\lambda_x$, independently of $x$, which is unconstrained by itself (other processes mediated by $\lambda_x x$ are found to be subdominant in the ensuing analysis, for the range of $\lambda_x$ over which thermal production is successful). If the coupling is too weak to begin with, then the $X$ never thermalises and thermal production cannot happen. Thermal production therefore requires a careful balancing of parameters - small coupling $\lambda_x$ is preferred for $X$ to freeze-out hot and as early as possible, but the coupling is bounded from below by the requirement that $X$ reach thermal equilibrium. This combination of constraints severely restricts the size of the parameter space over which thermal production is viable to cases in which the coupling is selected so that $X$ enters and departs from thermal equilibrium at close to the same temperature. 

To obtain numerical predictions for this scenario, the calculation of Section \ref{Sec:cosmosim} was modified to account for the time after the freeze-out of $X$ before it becomes non-relativistic. During this period we use (\ref{Eq:nudis}) and (\ref{Eq:nuen}) for the energy density of the $X$, approximating decays as being negligible, before switching over to (\ref{Eq:XEnDec}) when the temperature drops below the mass of the $X$. The approximation that the $X$ does not decay appreciably while it is relativistic must be good if there is to be sufficient time for it to grow to dominate between becoming non-relativistic and decaying. The decay width of $X$ was fixed to $5\times10^{-21}\,\text{GeV}$, corresponding to a reheating temperature close to the $\sim 10$ MeV lower limit, in order to maximise the amount of time over which the energy density of $X$ may grow relative to the SM plasma, thereby providing the greatest possible reheating.

\begin{figure}[!h]
\centering
\includegraphics[width=.8\linewidth]{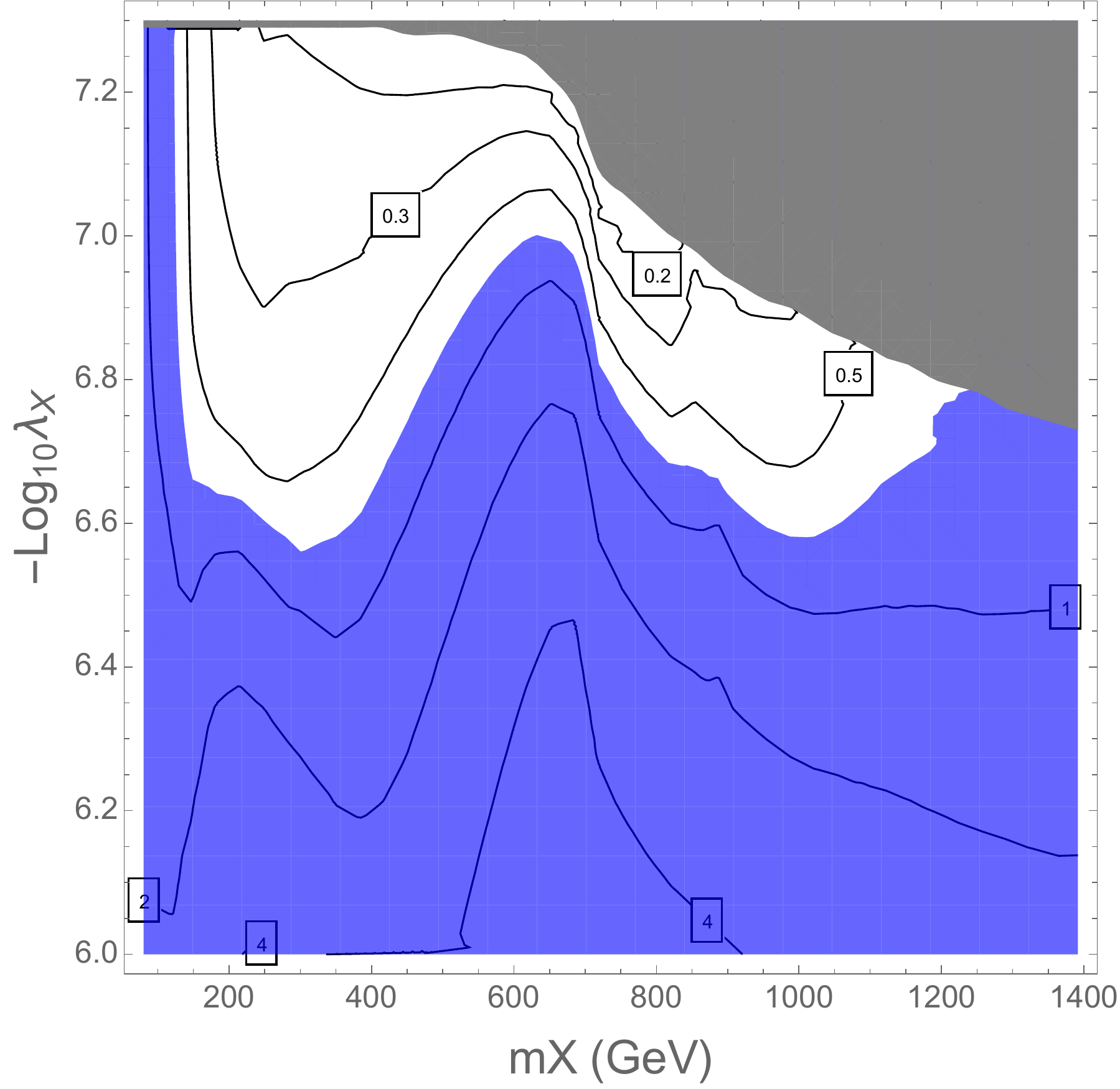}
\caption{Parameter space where thermal production of $X$ gives a large enough relic abundance to dilute the twin sector, for $f/v=4$. In the gray region, the coupling is too weak for $X$ to ever reach thermal equilibrium. The blue region is in tension with recent Planck measurements of $\Delta N_{\text{eff}}$, whereas all of the white region will be tested by CMB-S4. Predictions presented here for $\Delta N_{\text{eff}}$ close to the gray boundary are more uncertain because of the high sensitivity of the freeze-out temperatures to the coupling.}
\label{Fig:freezeout}
\end{figure}

The predictions for $\Delta N_\text{eff}$ from a thermally produced $X$ are shown in Figure \ref{Fig:freezeout} for the small regions of parameter space where this is viable, with $f/v=4$. We find that the dominant annihilation channels over this region are $XX\rightarrow t\bar{t}$ and $XX\rightarrow b\bar{b}$, mediated by the light Higgs, as well as their twin analogues, mediated by the heavy Higgs. As expected, the primordial energy density in the twin sector is too large compared to that generated by the $X$ for the asymmetric reheating to be effective when $m_X$ is too light ($\lesssim 100$ GeV in this case). Similarly, when the coupling is too strong, the X is held in equilibrium for longer and freezes-out underabundant compared to the twin energy density. However, when the coupling is too weak (the gray region), $X$ never thermalises to begin with (close to the boundary with this region, $X$ freezes-out almost immediately after thermalising). The peak in the contours occurs because of the ``$H$-funnel'' in which the twin Higgs resonantly enhances annihilations into twin quarks. All of this region will be testable by CMB-S4. 

\subsection{Twinflation} \label{sec:twinflation}


As an alternative to the model presented above of late, out-of-equilibrium decays of a $\mathbb{Z}_2$-symmetric scalar, one may imagine that the field driving primordial inflation reheats only the Standard Model to below the decoupling temperature of the two sectors. Production of the twin particles then ceases at some time after the temperature drops below the decoupling temperature during reheating.


To make this consistent with a softly-broken $\mathbb{Z}_2$ symmetry, we extend the inflationary sector and introduce a `twinflaton' that couples solely to the twin sector. The combined inflationary and twinflationary sectors respect the $\mathbb{Z}_2$ symmetry. However, if the two sectors are entirely symmetric then one generally expects both inflationary dynamics to happen coincidentally, which would result in identical reheating.
We therefore rely on soft $\mathbb{Z}_2$-breaking to give an asymmetry between the two sectors that causes the twinflationary sector to dominate the universe first. With the right arrangement we can end up with two distinct periods of inflation - a first caused by the ``twinflaton'' and a second that then reheats the Standard Model to below the decoupling temperature, having diluted the sources of twin-sector reheating from the first period.

One simple mechanism for $\mathbb{Z}_2$-breaking which is well-suited for introducing asymmetry to inflationary sectors is to introduce an additional $\mathbb{Z}_2$-odd scalar field $\eta$ (as was done in \cite{Berezhiani:1995am}). This admits linear and quadratic interactions to antisymmetric and symmetric combinations of the inflationary sector fields, respectively. When $\eta$ acquires a vev, this introduces an asymmetry in the fields to which it was coupled, dependent on the combination of its vev and its couplings. If $\eta$ is coupled to both the inflationary sectors and the Higgs sectors, it could be the sole source of $\mathbb{Z}_2$-breaking in a twinflationary theory. One may generally imagine that, in some UV completion, the mechanism that softly breaks the symmetry in the Higgs potential could also be the origin of the soft breaking of the inflationary sector.

Cosmologically, this possibility may have similar observational signatures as the model discussed in Section \ref{sec:late}, where the amount of twin-sector dark radiation is determined by the partial widths of the inflaton of the second inflationary epoch. If this dominantly couples to the SM, then $\Delta N_{\text{eff}}$ will be suppressed which, while successfully resolving the cosmological problems of the Mirror Twin Higgs, may also be observationally inaccessible. However, additional, distinctly inflationary signatures may make this potentially testable by other cosmological observations.

The mechanism of twinflation completes a catalog of models of asymmetric reheating by late decays, which may be indexed by representations of the twin parity: the case of a $\mathbb{Z}_2$-even particle, in which a kinematic asymmetry in the partial widths provides the reheating asymmetry, the case of a $\mathbb{Z}_2$-odd particle, which can also provide the spontaneous $\mathbb{Z}_2$-breaking required in the Higgs potential, and the case where two distinct, long-lived particles couple to each sector, which may also be related to inflation.

\subsubsection{Toy Model}

As a toy model we here consider `twinning' the simple $\varphi^2$ chaotic inflation scenario. The inflationary dynamics in this case are easy to understand and we have the additional benefit that this inflationary model has been considered in the literature before as `Double Inflation' (see \cite{Silk:1986vc}, \cite{Polarski:1992dq} and \cite{Wands:2002bn}). We furthermore specialize to `double inflation with a break', where there are two distinct periods of inflation which produces a step in the power spectrum, and we consider the constraints that this places on our model. In this case, it is assumed that each inflaton field couples and therefore decays dominantly into the sector to which it belongs. 
We will comment briefly on the case without a break and the additional signals one could look for in that case. 

The potential of the inflationary sector for inflaton $\varphi_A$ and twinflaton $\varphi_B$ is
\begin{equation}
V = \frac{1}{2} m_A^2 \varphi_A^2 + \frac{1}{2} m_B^2 \varphi_B^2,
\end{equation} 
where $m_A \neq m_B$ may arise from soft $\mathbb{Z}_2$- breaking, perhaps related to the soft $\mathbb{Z}_2$-breaking in the Higgs potential. In order for the `twinflation' to occur first, we require that the energy of the $B$ field initially dominates the energy density of the universe. We take the initial positions of the fields to be the same and $m_B^2 \gg m_A^2$.\footnote{Note that merely giving the twin field a much larger initial condition does not instigate twinflation. The dynamics of the subdominant field in this case are such that it will track the dominant field and both will reach the critical value at the same time. This is easily confirmed numerically.} Call $\varphi_A(0) = \varphi_B(0) = n \sqrt{2} M_\text{pl} = n \varphi_c$, where $\varphi_c$ is the critical value at which inflation stops and $m_B = r m_A = r m$ with $n,r > 1$. The inflationary dynamics are then those of slowly-rolling scalar fields. At some point in the early universe we imagine that the slow-roll approximation holds for both fields and the inflationary sector dominates the universe. 
%
%
The dominating field then slow-rolls down its potential for $\frac{n^2-1}{2}$ $e$-folds, while the lighter field's velocity is suppressed by approximately $\frac{\varphi_A}{r^2 \varphi_B}$. Solving the system numerically reveals that the motion of $\varphi_A$ during this period can be neglected entirely.

After $\varphi_B$ reaches the critical value $\sqrt{2} \mpl$, it stops slow-rolling and begins oscillating around the minimum of its potential. For there to be two distinct periods of inflation, there must be a period where these oscillations dominate the universe, which requires that the energy densities of each inflaton $\rho_A$ and $\rho_B$ satisfy  $\rho_B(\varphi_c) = r^2 m^2 M_\text{pl}^2 \ > \ \rho_A(\varphi(0)) = n^2 m^2 M_\text{pl}^2 $ and therefore $r>n$. For a $\varphi_B^2$ potential, the energy in these oscillations redshifts as $\rho_B \sim a^{-3}$. Eventually, the energy density in $\varphi_B$ drops below that of $\varphi_A$ and a new epoch of inflation, driven by $\varphi_A$, begins. This provides a further $\frac{n^2 - 1}{2}$ $e$-folds of inflation to give $n^2 - 1$ in total, while the $B$-sector energy density is diluted away.

%
%


Note that in order for our toy model to reheat below the decoupling temperature of the two sectors, reheating must occur well after the end of inflation. If, during the coherent oscillation of an inflaton, it becomes the case that the inflaton decay width $\Gamma \sim H$, then reheating will occur and result in temperature $T_{\text{reheat}} \sim 0.1 \sqrt{\Gamma M_\text{pl}}$. However, if $\Gamma \gg H$ when inflation ends, then all of the energy in the inflaton is immediately transferred and we instead have reheating temperature $T_{\text{reheat}} \sim 0.1 \sqrt{m_\alpha M_\text{pl}}$ for an inflaton of mass $m_\alpha$. But in order for $T_{\text{reheat}}\lesssim 1$ GeV, it is required that $m_\alpha \lesssim 10^{-7}$ eV, so this possibility that the inflaton is short lived is not viable.
The procedure of twinning inflationary potentials may be generalised to other, more realistic models, provided that this constraint upon the reheating temperature can be satisfied.


\subsubsection{Observability}

One could always make a twinflationary scenario consistent with observational constraints by letting the second inflationary period of inflation last long enough. In our toy model, this would correspond to setting $n$ high enough that the momentum modes which left the horizon during the first inflation have not yet re-entered the horizon - such a scenario would look exactly like single-field chaotic inflation.

Alternatively, we may also allow for $n$ small enough that all the momentum modes that left the horizon during the second inflation are currently sub-horizon. In this case, fluctuations at large enough wavenumbers (equivalently, small enough length scales) are `processed' (cross the horizon) at a different inflationary energy scale than those that were processed earlier, giving a step in the power spectrum. While Planck has measured the primordial power spectrum for modes with $10^{-4} \ \text{Mpc}^{-1} \lesssim k \lesssim 0.3 \ \text{Mpc}^{-1}$ (where the lower bound is set by the fact that smaller modes have not yet re-entered the horizon), proposed CMB-S4 experiments will increase this range \cite{CMB-S4:2016} somewhat, as will be discussed further below. We wish to show that the power spectrum of our toy model is not ruled out and, furthermore, may be observed in the coming decades. 

The height of the step in the primordial power spectrum is determined by the energy scale of each period of inflation, so modes crossing the horizon in the second inflationary period should be suppressed by a factor of $r^2 > n^2 \gtrsim 25$ compared to those exiting in the first period. This degree of suppression is ruled out by Planck for the range of modes over which it has reconstructed the power spectrum \cite{Ade:2015lrj}. A computation of the primordial power spectrum for double inflation was given in \cite{Polarski:1992dq}. It was found that significant damping does not occur for modes which cross outside the horizon during the first inflationary period, re-enter during the inter-inflationary period and again cross the horizon during the second inflationary period. It is only those scales which first cross the horizon during the second inflationary period that are significantly damped (although other features in the shape, such as oscillations, may be present for modes that are subhorizon during the intermediate period).



The relation of this characteristic scale to present-day observables is easily done using the framework given in \cite{Hong:2015oqa}. Let the subscripts $a,b,c,d,e$ respectively correspond to the beginning of the first inflationary period, the end of that period, the beginning of the second inflationary period, the end of that period, and the beginning of radiation domination. During the coherent oscillation periods, the inflaton acts as matter and the energy density falls as $\rho \propto a^{-3}$. Let $k_i$ be the momentum whose mode is horizon-size at the $i$ epoch; $k_i = a_i H_i$. 
\begin{figure}[h]
\centering
\includegraphics[width=1.0\linewidth]{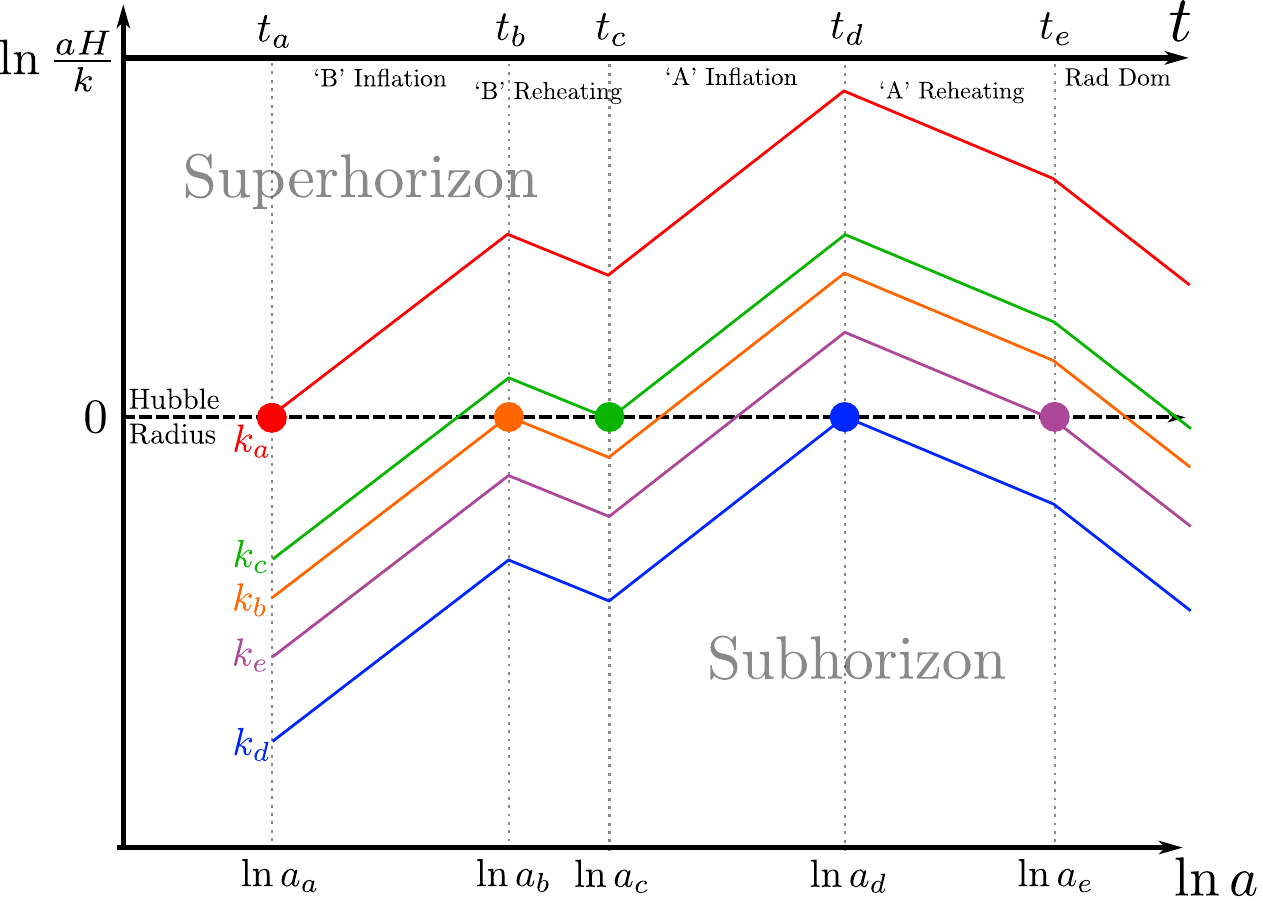}
\caption{Schematic evolution of the characteristic scales in Twinflation, as seen by comparing wavenumbers to the Hubble radius over time. Note that the time axis is not a linear scale.}
\label{Fig:scales}
\end{figure}
The scales $k_i$ can be related using the number of $e$-folds in each period, which are themselves determined from the first Friedmann equation. Denoting $N_{ij} = \ln \frac{a_j}{a_i}$, we have $k_a = e^{-N_{ab}} n k_b$, $k_b =e^{\frac{1}{2} N_{bc}} k_c$
%
%
and similarly for the other characteristic modes, where, in particular, slow-roll inflation predicts that $N_{ab} = N_{cd} = \frac{n^2-1}{2}$. 
%
%
The evolution of the characteristic momentum scales is shown schematically in Figure \ref{Fig:scales}. Finally, $k_e$ can be determined using the conservation of comoving entropy:
\begin{eqnarray}
k_e &=& \frac{\pi g_{\star}^{1/3}(T_0) g_\star^{1/6}(T_{\text{reheat}}) T_0 T_{\text{reheat}}}{3 \sqrt{10} M_\text{pl}},
\end{eqnarray}
where $T_0$ and $a_0$ are the temperature and scale factor today and $T_{\text{reheat}}$ is the reheating temperature (which is sufficiently low that only SM particles are produced). We work explicitly with the convention $a_0 = 1$. The characteristic modes associated with the break can then be determined.

As mentioned above, \cite{Polarski:1992dq} shows that damping occurs for modes that exit the horizon only during the second inflationary period, so we should take the characteristic damping scale to be the smallest such scale, which here corresponds roughly to $k_b$
This can be determined as
\begin{eqnarray} \label{char}
k_b &=& n  e^{\frac{1}{2} N_{bc} - N_{cd} + \frac{1}{2} N_{de}} k_e \nonumber \\
&=& n \left(\frac{r}{n}\right)^{1/3} \exp\left(-\frac{n^2-1}{2}\right)  \left[ \frac{\frac{1}{2}m^2 M_\text{pl}^2}{\frac{\pi^2}{30}g_\star(T_{\text{reheat}}) T_{\text{reheat}}^4}\right]^{1/6} \frac{\pi g_{\star}^{1/3}(T_0) g_\star^{1/6}(T_{\text{reheat}}) T_0 T_{\text{reheat}}}{3 \sqrt{10}  M_\text{pl}}
\end{eqnarray} 
where $k_c$ only differs by the factor of $\left(r/n\right)^{1/3}$ (which is roughly close to unity). Once again, between $k_b$ and $k_c$ are oscillatory features, so $k_b$ should merely be taken as the rough characteristic scale of the damping.


\begin{figure}[h]
\centering
\includegraphics[width=1.0\linewidth]{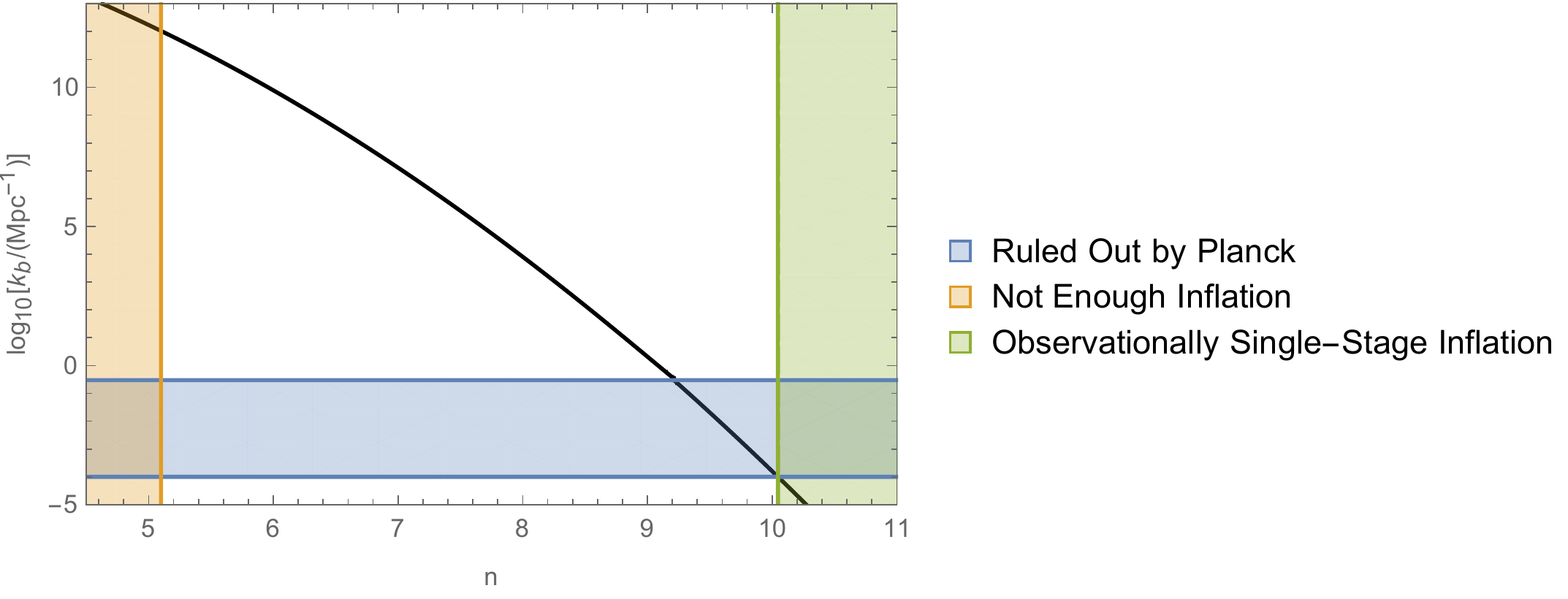}
\caption{The prediction for the characteristic suppression scale as a function of the initial values of the fields. The mapped regions should be interpreted not as having hard boundaries, but rather fuzzy endpoints where they break down. Here we have used $T_{\text{reheat}} = 10 \ \text{MeV}$ and $r = 2n$.}
\label{Observable}
\end{figure}

Now the characteristic damping scale is determined by $m$, $n$, $r$, and $T_{\text{reheat}}$. Our observational bound on $k_b$ is that Planck has not seen this suppression on momentum scales at which it has been able to reconstruct the primordial power spectrum from the angular temperature anisotropy power spectrum, which is roughly $k \lesssim 0.3 \ \text{Mpc}^{-1}$. We have constraints on the reheating temperature from rethermalization of the twin sector or interrupted big bang nucleosynthesis $10\, \text{MeV} \lesssim T_{\text{reheat}} \lesssim 1 \ \text{GeV}$, on having a period of intermediate matter domination between the two inflations $r > n$ and on the total number of $e$-folds $n^2 - 1 \gtrsim 25$ to solve cosmological problems. Note that we require fewer $e$-folds of inflation than is typically assumed in the standard cosmology. Since the low reheating temperature gives fewer $e$-folds from reheating up to today, less inflation is needed to explain the large causal horizon and flatness.

The normalization of the spectrum provides a further constraint, the most recent measurement of which come from Planck \cite{Ade:2015lrj}. The scalar power spectrum at $k_\star = 0.05 \  \text{Mpc}^{-1}$ is measured to be $\mathcal{P}_{\mathcal{R}}(k_\star) = e^{3.094 \pm 0.034} \times 10^{-10}$. Then for $k_\star < k_c$ (i.e. $k_\star$ having left the horizon during the first period of inflation and not re-entered before the second, so no deviation from single-field inflation would be seen at this scale), the spectrum of \cite{Polarski:1992dq} yields the constraint 
\begin{equation} \label{norm}
2.03 \times 10^{-6} = \frac{r^2 m^2}{M_\text{pl}^2} \ln\left(\frac{k_b}{k_\star}\right)\left(\ln\frac{k_b}{k_\star} + \frac{n^2}{2}\right).
\end{equation}

The characteristic scale (\ref{char}) depends much more strongly on $n$ than it does on any of the other parameters. In Figure \ref{Observable}, we give a rough idea of the scale as a function of $n$, having set $T_{\text{reheat}} = 10 \ \text{MeV}$ and $r = 2n$, while $m$ is chosen to satisfy the normalization condition. We also show the constraint on $k_b$ set by Planck. Note again that the region described as ``observationally single-stage inflation" \textit{does} still provide a solution to the problem of reconciling cosmology with the mirror Twin Higgs.

CMB-S4 will improve the constraint on $k_b$ through its improved measurement of polarization anisotropies \cite{CMB-S4:2016}. With only precision measurements of temperature anisotropies, the un-lensed power spectrum cannot be so easily reconstructed from the lensed spectrum. The effects of gravitational lensing of CMB place an upper limit on the size of primordial temperature anisotropies that can be measured \cite{Hu:2003vp}, which Planck has saturated. However, the polarization anisotropy power spectrum allows the removal of lensing noise from the temperature spectrum so that higher primordial modes can be detected. The polarization power spectrum itself also gives us another window into the high-$\ell$ modes of the primordial power spectrum, as the signal does not become dominated by polarized foreground sources until higher scales near $\ell \sim 5000$. CMB-S4 is projected to make cosmic variance limited measurements of both the temperature and polarization anisotropy power spectra up to the modes where they become foreground-contaminated and so provide additional information on the shape of the primordial power spectrum \cite{CMB-S4:2016}. The map from measurements of angular modes $\ell$ to contraints on spatial modes $k$ depends on the evolution of the power spectrum between inflation and the CMB, so forecasting constraints requires careful study. However, these improvements will not test most of the parameter space presented in Figure \ref{Observable}, where the step is predicted on extremely small distance scales.

We have discussed a twinflationary model of double inflation with a break for simplicity, but there is a parametric regime where double inflation without a break gives the required amount of asymmetric reheating into the Standard Model. With two periods of inflation, the second period dilutes the energy density of the heavier field sufficiently that there is no observable signal of it produced in reheating. However, even with only one period, inflation can continue for long enough after the inflaton turns the corner in field space such that, at late times, the fraction of the inflaton in the $B$ state relative to the $A$ state is small enough that the expected energy densities that are transferred into each sector satisfy $\rho_B/\rho_A < 0.1$. This occurs as long as $r \gtrsim 1.2$, assuming that the mixing angle of the slow-rolling field with the $\varphi_A$ and $\varphi_B$ fields entirely determines the fraction of its energy that reheats each sector. There is thus a much larger range of $r$ where this toy model of inflation passes $N_{\text{eff}}$ bounds than our above analysis shows. The resulting imprint on the CMB could resemble that of the long-lived decay model of Section \ref{sec:late}, with $\Delta N_{\text{eff}}$ again being related to the ratio of branching fractions, although this is dependent upon the UV completion of the Twin Higgs.


When there is only one period of inflation, the step is smoothed out and less pronounced and it is necessary to locate the feature numerically. Furthermore, having multiple degrees of freedom available allows for non-trivial evolution of momentum modes after they become super-horizon, which does not occur in single-field inflation but may be calculated from the full solution to the field equations \cite{Wands:2002bn}. While a twinned potential leading to two periods of inflation generally predicts a step in the power spectrum, when there is no break the predictions, and thus constraints, this prediction become more model-dependent. Therefore we leave detailed predictions in that case for future study using realistic models and merely state that the range of $r = 1$ to $n$ interpolates between the single field spectrum and that with a step, as one would expect. 

There are also at least two other detectable effects one might expect in double inflation without a break and in general realistic twinflationary models. Interactions between inflaton fields may produce primordial non-Gaussianities, while the presence of additional oscillating degrees of freedom may produce isocurvature perturbations. These do not appear in our toy model because the heavy field is exponentially damped during the second inflation. CMB-S4 is projected to improve Planck's bounds on non-Gaussianities by a factor of $\sim 2$ and on isocurvature perturbations by perhaps an order of magnitude (though model-independent projections have not been made), so may be able to detect or place useful constraints on realistic twinflationary models \cite{CMB-S4:2016}.

We have introduced twinflation as a mirror Twin Higgs model which suppresses the cosmological effects of twin light degrees of freedom. It extends the mirror symmetry to the inflationary sector. The soft $Z_2$ symmetry-breaking of the Higgs sector may be used in the inflationary sector to cause distinct periods of inflation. There exists a parametric region where this is cosmologically indistinct from single-stage inflation, but also another in which it may be observable. As the direct product of inflation and the Mirror Twin Higgs, this is in some sense a minimal solution.

\subsection{Conclusion} \label{sec:conc}

In this section we have considered scenarios in which cosmology provides meaningful insight on solutions to the electroweak hierarchy problem. In particular, we have demonstrated several simple mechanisms in which the cosmological history of a mirror Twin Higgs model is reconciled with current CMB constraints and provides signatures accessible in future CMB experiments. In the case of out-of-equilibrium decays, we have found that decays of $\mathbb{Z}_2$-even scalars sufficiently dilute the energy density in the twin sector without the addition of any new sources of $\mathbb{Z}_2$-breaking. In much of the parameter space, the residual contribution to $\Delta N_{\text{eff}}$ is directly proportional to the ratio of vacuum expectation values $v^2/f^2$ parameterizing the mixing between Standard Model and twin sectors (as well as the tuning of the electroweak scale), and may be within reach of CMB-S4 experiments. In the case of twinflation, we have found that a (broken) $\mathbb{Z}_2$-symmetric inflationary sector may successfully dilute the energy density in the twin sector, as well as potentially leave signatures in the form of a step in the primordial power spectrum or in departures of primordial perturbations from adiabaticity and Gaussianity. In both cases, these models raise the tantalizing possibility that signatures of electroweak naturalness may first emerge in the CMB, rather than the LHC.

There are a variety of possible directions for future work. Here we have focused on the cosmological consequences of late-decaying scalars and twinned inflationary sectors without specifying their origin in a microscopic model. It would be interesting to construct complete models (where, e.g., supersymmetry or compositeness protect the scale $f$ from UV contributions) in which the existence and couplings of late-decaying scalars arise as intrinsic ingredients of the UV completion. Likewise, we have considered only a toy model of twin chaotic inflation; it would be interesting to see if twinflation may be realized in complete inflationary models that match the observed spectral index and constraints on the tensor-to-scalar ratio.

While we have taken care to ensure that our scenarios respect the well-measured cosmological history beneath $T \sim$ 1 MeV, we have not addressed the origin of the observed baryon asymmetry. In the case of out-of equilibrium decays, there are a number of possibilities. It is plausible that a somewhat larger baryon asymmetry is generated through various conventional mechanisms and diluted by late decays. Alternatively, the decay mechanism itself may possibly be expanded to generate a baryon asymmetry or some other late decay may generate the baryon asymmetry below $\sim 1$ GeV. In the case of twinflation, inflationary dilution of pre-existing baryon asymmetry requires that baryogenesis occur in association with reheating or via another mechanism at temperatures below $\sim 1 \ \text{GeV}$. It would be worthwhile to study models for the baryon asymmetry consistent with these scenarios. Steps in this direction have been taken in \cite{Farina:2016ndq}, which attempted to relate this to asymmetric dark matter in the twin sector.

Likewise, any investigation of dark matter, be it related directly to the twin mechanism or otherwise, must also address implications of the dilution. Previous work attempting to construct dark matter candidates in the twin sector \cite{Barbieri:2016zxn,Garcia:2015loa,Craig:2015xla, Garcia:2015toa, Farina:2015uea, Freytsis:2016dgf,Farina:2016ndq, Prilepina:2016rlq}) has relied upon explicit $\mathbb{Z}_2$-breaking that is not present in the mirror model. Dark matter may alternatively be unrelated to the Twin Higgs mechanism, such as a a WIMP in some minimal extension of the electroweak sector that freezes-out as an overabundant thermal relic and is then diluted to the observed density during reheating. Alternatively, it may be that the dark matter abundance is produced directly during reheating. It would be interesting to study extensions of our scenarios that incorporate dark matter candidates directly related to the mechanism of dilution.

Finally, we have only approximately parameterized Planck constraints and the reach of CMB-S4 on twin neutrinos and twin photons. Ultimately, more precise constraints and forecasts may be obtained via numerical CMB codes. This strongly motivates the future study of CMB constraints on scenarios with three sterile neutrinos and additional dark radiation whose temperatures differ from the Standard Model thermal bath.

\section{Freeze-\textit{Tw}in Dark Matter}
	\label{sec:freezetwin}

\title{Freeze-\textit{Tw}in Dark Matter}

\begin{abstract}
\noindent
The mirror twin Higgs (MTH) addresses the little hierarchy problem by relating every Standard Model (SM) particle to a twin copy, but is in tension with cosmological bounds on light degrees of freedom. Asymmetric reheating has recently been proposed as a simple way to fix MTH cosmology by diluting the twin energy density. We show that this dilution sets the stage for an interesting freeze-in scenario where both the initial absence of dark sector energy and the feeble coupling to the SM are motivated for reasons unrelated to dark matter production. We give the twin photon a Stueckelberg mass and freeze-in twin electron and positron dark matter through the kinetic mixing portal. The kinetic mixing required to obtain the dark matter abundance is of the loop-suppressed order expected from infrared contributions in the MTH.
\end{abstract}

\maketitle
\end{comment}
\subsection{Introduction}

In this work, we build on a MTH framework where `hard' breaking of the $\mathbb{Z}_2$ is absent. In \cite{Craig:2016lyx,Chacko:2016hvu}, it was realized that late-time asymmetric reheating of the two sectors could arise naturally in these models if the spectrum were extended by a single new state.
This asymmetric reheating would dilute the twin energy density and so attune the MTH with the cosmological constraints. 
This dilution of twin energy density to negligible levels would seem to hamper the prospect that twin states might constitute the dark matter, and generating dark matter was left as an open question. This presents a major challenge toward making such cosmologies realistic. 
However, we show that asymmetric reheating perfectly sets the stage for a MTH realization of the `freeze-in' mechanism for dark matter production~\cite{Asaka:2005cn,Gopalakrishna:2006kr,Asaka:2006fs,Page:2007sh,Hall:2009bx,Chu:2011be,Bernal:2017kxu,Dvorkin:2019zdi}.

Freeze-in scenarios are characterized by two assumptions: 1) DM has a negligible density at some early time and 2) DM interacts with the SM so feebly that it never achieves thermal equilibrium with the SM bath.\footnote{We note that the feeble connection between the two sectors may originate as a small dimensionless coupling or as a small ratio of mass scales, either of which deserves some explanation.} This second assumption is motivated in part by the continued non-observation of non-gravitational DM-SM interactions.  
Both assumptions stand in stark contrast to freeze-out scenarios. 

Freeze-twin dark matter is a particularly interesting freeze-in scenario because both assumptions are fulfilled for reasons orthogonal to dark matter considerations: 1) the negligible initial dark matter abundance is 
predicted by the asymmetric reheating already necessary to resolve the MTH cosmology, and 2) the kinetic mixing necessary to achieve the correct relic abundance is of the order expected from infrared contributions in the MTH. 
To allow the frozen-in twin electrons and positrons to be DM, we need only 
break the $\mathbb{Z}_2$ by a relevant operator to give a Stueckelberg mass to twin hypercharge.
Additionally, the twin photon masses we consider can lead to dark matter self-interactions at the level relevant for small-scale structure problems~\cite{Tulin:2017ara}. 

The next sections are organized as follows: In Section \ref{sec:mth}, we review the MTH and 
its cosmology in models with asymmetric reheating, and
in Section \ref{sec:kin} we introduce our extension. 
In Section \ref{sec:freezein}, we calculate the freeze-in yield for twin electrons and discuss the parameter space to generate dark matter and constraints thereon. 
We discuss future directions and conclude in Section \ref{sec:conclu}. For the interested reader, we include some discussion of the irreducible IR contributions to kinetic mixing in the MTH in Appendix \ref{sec:KinMixapp}.

\subsection{The Mirror Twin Higgs \& Cosmology}\label{sec:mth}

The mirror twin Higgs  
framework \cite{Chacko:2005pe} introduces a twin sector $B$, which is a `mirror' copy of the Standard Model sector $A$, related by a $\mathbb{Z}_2$ symmetry. 
Upgrading the $SU(2)_{A}\times SU(2)_{B}$ gauge symmetry of the scalar potential to an $SU(4)$ global symmetry adds a Higgs-portal interaction between the $A$ and $B$ sectors:
\begin{equation}
\label{eq:MTHV}
V = \lambda \left( |\mathcal{H}|^2 - f^2/2\right)^2,
\end{equation}
where $\mathcal{H} = \begin{pmatrix} H_A \\ H_B \end{pmatrix}$ is a complex $SU(4)$ fundamental consisting of the $A$ and $B$ sector Higgses in the gauge basis. The SM Higgs is to be identified as 
a pseudo-Goldstone mode arising from the breaking of $SU(4)\rightarrow SU(3)$ when $\mathcal{H}$ acquires a vacuum expectation value (vev) $\langle \mathcal{H} \rangle = f/\sqrt{2}$. Despite the fact that the global $SU(4)$ is explicitly broken by the gauging of $SU(2)_{A}\times SU(2)_{B}$ subgroups, the $\mathbb{Z}_2$ is enough to ensure that the quadratically divergent part of the one-loop effective action respects the full $SU(4)$.  The lightness of the SM Higgs is then understood as being protected by the approximate accidental global symmetry up to the UV cutoff scale $\Lambda \lesssim 4 \pi f$, at which point new physics must come in to stabilize the scale $f$ itself.

We refer to twin particles by their SM counterparts primed with a superscript ', and we refer the reader to \cite{Chacko:2005pe,Craig:2015pha} for further discussion of the twin Higgs mechanism.

The thermal bath history in the conventional MTH is fully dictated by the Higgs portal in Eq.~\eqref{eq:MTHV}  
which keeps the SM and twin sectors in thermal equilibrium down to temperatures $\mathcal{O}(\text{GeV})$. 
A detailed calculation of the decoupling process was performed in \cite{Craig:2016lyx} by tracking the bulk heat flow between the two sectors as a function of SM temperature. It was found that for the benchmark of $f/v=4$, decoupling begins at a SM temperature of $T \sim 4 \text{ GeV}$ and by $\sim 1 \text{ GeV}$, 
the ratio of twin-to-SM temperatures may reach $\lesssim 0.1$  without rebounding.
While heat flow rates become less precise below $\sim 1 \text{ GeV}$ due to uncertainties in hadronic scattering rates, especially close to color-confinement, decoupling between the two sectors is complete by then for $f/v \gtrsim 4$.
For larger $f/v$, the decoupling begins and ends at higher temperatures. 

As mentioned above, one class of solutions to this $\Neff$ problem uses hard breaking of the $\mathbb{Z}_2$ at the level of the spectra~\cite{Craig:2015pha,Barbieri:2016zxn,Csaki:2017spo,Liu:2019ixm,Harigaya:2019shz} while keeping a standard cosmology. An alternative proposal is to
modify the cosmology with asymmetric reheating to dilute the energy density of twin states. 
For example, \cite{Chacko:2016hvu} uses late, out-of-equilibrium decays of right-handed neutrinos, while \cite{Craig:2016lyx} uses those of a scalar singlet.
These new particles respect the $\mathbb{Z}_2$, but dominantly decay to SM states due to the already-present soft $\mathbb{Z}_2$-breaking in the scalar sector.
In \cite{Chacko:2016hvu}, this is solely due to extra suppression by $f/v$-heavier mediators, 
while in \cite{Craig:2016lyx}, the scalar also preferentially mass-mixes with the heavier twin Higgs. \cite{Craig:2016lyx} also presented a toy model of `Twinflation', where a softly-broken $\mathbb{Z}_2$-symmetric scalar sector may lead to inflationary reheating which asymmetrically reheats the two sectors to different temperatures.
In any of these scenarios, the twin sector may be diluted to 
the level where it evades Planck bounds \cite{Aghanim:2018eyx} on extra radiation, yet is potentially observable with 
CMB Stage IV~\cite{Abazajian:2016yjj}. 

We will stay agnostic about the particular mechanism at play, and merely assume that by $T \sim 1 \text{ GeV}$, the Higgs portal interactions have become inefficient and some mechanism of asymmetric reheating has occurred such that the energy density in the twin sector has been largely depleted, $\rho_{\text{twin}} \approx 0$.\footnote{If asymmetric reheating leaves some small $\rho_{\text{twin}}>0$, then mirror baryon asymmetry can lead to twin baryons as a small subcomponent of dark matter \cite{Chacko:2018vss}.} This is consistent with the results of the decoupling calculation in \cite{Craig:2016lyx} given the uncertainties in the rates at low temperatures, and will certainly be true once one gets down to $\text{few } \times  10^2 \text{ MeV}$. 

One may be concerned that there will be vestigial model-dependence from irrelevant operators induced by the asymmetric reheating mechanism which connect the two sectors. However, these operators will generally be suppressed by scales above the reheating scale, as in the example studied in~\cite{Craig:2016lyx}. Prior to asymmetric reheating, the two sectors are in thermal equilibrium anyway, so these have little effect. After the energy density in twin states has been diluted relative to that in the SM states, the temperature is far below the heavy masses suppressing such irrelevant operators, and thus their effects are negligible. So we may indeed stay largely agnostic of the cosmological evolution before asymmetric reheating as well as the details of how this reheating takes place. We take the absence of twin energy density as an initial condition, but emphasize that there are external, well-motivated reasons for this to hold in twin Higgs models, as well as concrete models that predict this occurrence naturally.

\subsection{Kinetic Mixing \& A Massive Twin Photon} \label{sec:kin}

In order to arrange for freeze-in, we add to the MTH kinetic mixing between the SM and twin hypercharges and a Stueckelberg mass for twin hypercharge.  
At low energies, these reduce to such terms for the photons instead, parametrized as
\al{ \label{eq:addl}
\mathcal{L} \mathrel{+}= - \frac{\epsilon}{2} F_{\mu\nu} F'^{\mu\nu} - \half m_{\gamma'}^2 A'_\mu A'^\mu.
}
This gives each SM particle of electric charge $Q$ an effective twin electric charge $\epsilon Q$.\footnote{Note that twin charged states do not couple to the SM photon. Their coupling to the SM Z boson has no impact on freeze-in at the temperatures under consideration. Furthermore, the miniscule kinetic mixing necessary for freeze-in has negligible effects at collider experiments. See Ref.~\cite{Chacko:2019jgi} for details.} The twin photon thus gives rise to a `kinetic mixing portal' through which the SM bath may freeze-in light twin fermions in the early universe.

The Stueckelberg mass constitutes soft $\mathbb{Z}_2$-breaking,\footnote{While we are breaking the $\mathbb{Z}_2$ symmetry by a relevant operator, the extent to which a Stueckelberg mass is truly \textit{soft} breaking is not clear. Taking solely Eq. (\ref{eq:addl}), we would have more degrees of freedom in the twin sector than in the SM, and in a given UV completion it may be difficult to isolate this $\mathbb{Z}_2$-asymmetry from the Higgs potential. 
One possible fix may be to add an extremely tiny, experimentally allowed Stueckelberg mass for the SM photon as well \cite{Goldhaber:2008xy}, though we note this may be in violation of quantum gravity \cite{Reece:2018zvv,Craig:2018yld} 
or simply be difficult to realize in UV completions without extreme fine-tuning. We will remain agnostic about this UV issue and continue to refer to this as `soft breaking', following \cite{Chacko:2019jgi}.} but has no implications for the fine-tuning of the Higgs mass since hypercharge corrections are already consistent with naturalness \cite{Craig:2015pha}. We will require $m_{\gamma'}> m_{e'}$, to prevent frozen-in twin electron/positron annihilations, and $m_{\gamma'}> 2 m_{e'}$, to ensure that resonant production through the twin photon is kinematically accessible.
Resonant production will allow a much smaller kinetic mixing to generate the correct relic abundance, thus avoiding indirect bounds from supernova cooling. We note that while taking $m_{\gamma'} \ll f$ does bear explanation, the parameter is technically natural. 

On the other hand, mixing of the twin and SM $U(1)$s preserves the symmetries of the MTH EFT, so quite generally one might expect it to be larger than that needed for freeze-in. However, it is known that in the MTH a nonzero $\epsilon$ is not generated through three loops \cite{Chacko:2005pe}. While such a suppressed mixing is phenomenologically irrelevant for most purposes, here it plays a central role. In Appendix \ref{sec:KinMixapp}, we discuss at some length the vanishing of infrared contributions to kinetic mixing through few loop order. If nonzero contributions appear at the first loop order where they are not known to vanish, kinetic mixing of the order $\epsilon \sim 10^{-13} - 10^{-10}$ is expected. 

The diagrams which generate kinetic mixing will likely also generate higher-dimensional operators. These will be suppressed by (twin) electroweak scales and so, as discussed above for the irrelevant operators generated by the model-dependent reheating mechanism, freeze-in contributions from these operators are negligible.

\subsection{Freezing-\textit{Tw}in Dark Matter} \label{sec:freezein}

As we are in the regime where freeze-in proceeds while the temperature sweeps over the mass scales in the problem, it is not precisely correct to categorize this into either ``UV freeze-in'' or ``IR freeze-in''. Above the mass of the twin photon, freeze-in proceeds through the marginal kinetic mixing operator, and so a naive classification would say this is IR dominated. However, below the mass of the twin photon, the clearest approach is to integrate out the twin photon, to find that freeze-in then proceeds through an irrelevant, dimension-six, four-Fermi operator which is suppressed by the twin photon mass. Thus, at temperatures $T_{\text{SM}} \lesssim m_{\gamma'}$, this freeze-in looks UV dominated. This leads to the conclusion that the freeze-in rate is largest at temperatures around the mass of the twin photon. Indeed, this is generally true of freeze-in --- production occurs mainly at temperatures near the largest relevant scale in the process, whether that be the largest mass out of the bath particles, mediator, and dark matter, or the starting thermal bath temperature itself in the case that one of the preceding masses is even higher.

As just argued, freeze-in production of dark matter occurs predominantly at and somewhat before $T \sim m_{\gamma'}$. 
We require $m_\gamp \ll 1 \text{ GeV}$ so that most of the freeze-in yield comes from when $T <1 \text{ GeV}$, which
allows us to retain `UV-independence' in that we need not care about how asymmetric reheating has occurred past providing negligible density of twin states at $T=1 \text{ GeV}$. Specifically, we limit ourselves to $m_{\gamma'} < 2 m_{\pi^0}$, both for this reason and to avoid uncertainties in the behavior of thermal pions during the epoch of the QCD phase transition. However, we emphasize that freeze-in will remain a viable option for producing a twin DM abundance for even heavier dark photons. But the fact that the freeze-in abundance will be generated simultaneously with asymmetric reheating demands that each sort of asymmetric reheating scenario must then be treated separately. Despite the additional difficulty involved in predicting the abundance for larger twin photon masses, it would be interesting to explore this part of parameter space. In particular, it would be interesting to consider concrete scenarios with twin photons in the range of tens of GeV \cite{Batell:2019ptb}.

To calculate the relic abundance of twin electrons and positrons, we use the Boltzmann equation for the number density of $e'$:
\al{
\dot{n}_\ep+3Hn_\ep=\sum_{k,l} -\avg{\sigma v}_{\ep \ebp \to kl} \prn{n_\ep n_\ebp-n_\ep^\text{eq} n_\ebp^\text{eq}},
}
where $\avg{\sigma v}_{\ep \ebp \to kl}$ is the thermally averaged cross section for the process $\ep \ebp \to kl$, the sum runs over all processes with SM particles in the final states and $\ep \ebp$ in the initial state, and $n_\ep^\text{eq}$ is the equilibrium number density evaluated at temperature $T$. As we are in the parametric regime in which resonant production of twin electrons through the twin photon is allowed, $2 \rightarrow 2$ annihilation processes $\bar{f} f \to \gamp \to \ebp \ep$, with $f$ a charged SM fermion, entirely dominate the yield. 

In accordance with the freeze-in mechanism, $n_\ep$ remains negligibly smaller than its equilibrium number density throughout the freeze-in process, and so that term is ignored. It is useful to reparametrize the abundance of $\ep$ in terms of its yield, $Y_\ep = n_\ep/s$ where $s=\frac{2\pi^2}{45} g_{\ast s}T^3$ is the entropy density in the SM bath. Integrating the Boltzmann equation using standard methods, we then find the yield of $\ep$ today to be
\al{
\label{eq:Yepexact}
Y_\ep&=\int_{0}^{T_i} dT \frac{\prn{45}^{3/2}}{\sqrt{2}\pi^3 \sqrt{g_\ast}g_{\ast s}} \frac{M_{Pl}}{T^5} \prn{\frac{1}{T}+\frac{\dd_T g_{\ast s}}{3 g_{\ast s}}} \nonumber \\
&\times \sum_{\bar{f}f \to \ebp \ep} \avg{\sigma v}_{\bar{f}f \to \ebp \ep} n_\ebp^\text{eq} n_\ep^\text{eq},
}
where $T_i = 1 \text{ GeV}$ is the initial temperature of the SM bath at which freeze-in begins in our setup, $g_\ast(T)$ is the number of degrees of freedom in the bath, and $M_{Pl}$ is the reduced Planck mass. We will calculate this to an intended accuracy of 50\%. To this level of accuracy, we may assume Maxwell-Boltzmann statistics to vastly simplify the calculation \cite{Blennow:2013jba}. As a further simplification, we observe that the $\partial_T g_{\star s}$ term is negligible compared to $1/T$ except possibly during the QCDPT - where uncertainties on its temperature dependence remain \cite{Drees:2015exa} - and so we ignore that term. 
The general expression for the thermally averaged cross section of the process $12 \to 34$ is then
\al{
\avg{\sigma v}n_1^\text{eq}n_2^\text{eq}&=\frac{T^4}{2^9 \pi^5 s_{34}} \int^\oo_{\text{Max} \prn{\frac{m_1+m_2}{T},\frac{m_3+m_4}{T}}} dx x^2 \\ 
&\times \sqrt{\left[1,2\right]} \sqrt{\left[3,4\right]} K_1 (x) \int d\prn{\cos \theta} \magn{\MM}^2_{12 \to 34}, \nonumber
}
where $s_{34}$ is 1 if the final states are distinct and 2 if not, $x=\sqrt{s}/T$, $\sqrt{\left[i,j\right]}=\sqrt{1-\prn{\frac{m_i+m_j}{xT}}^2}\sqrt{1-\prn{\frac{m_i-m_j}{xT}}^2}$, and $\magn{\MM}^2_{12 \to 34}$ is the matrix element squared summed (not averaged) over all degrees of freedom. 
To very good approximation, the yield results entirely from resonant production, and so we may analytically simplify the matrix element squared for $\bar{f}f \to \ebp \ep$ using the narrow-width approximation 
\al{
\int d\prn{\cos \theta} \magn{\MM}^2_{\bar{f}f \to \ebp \ep} &\approx \frac{256 \pi^3 \alpha^2 \epsilon^2}{3} \prn{2m_f^2+m_\gamp^2}  \\ 
&\times \frac{ \prn{2m_{e'}^2+m_\gamp^2}}{\Gamma_\gamp m_\gamp^2 T} \delta \prn{x-m_\gamp/T}. \nonumber
}
$\Gamma_\gamp$ is the total decay rate of the twin photon.

For the range of $m_\gamp$ we consider, the twin photon can only decay to twin electron and positron pairs. Thus, its total decay rate is 
\al{
\label{eq:Gamgamp}
\Gamma_{\gamp}=\frac{\alpha \prn{m_\gamp^2+2m_\ep^2}}{3m_\gamp} \sqrt{1-\frac{4m_\ep^2}{m_\gamp^2}}.
}
Its partial widths to SM fermion pairs are suppressed by $\epsilon^2$, and so contribute negligibly to its total width.

The final yield of twin electrons is then
\al{
\label{eq:Yep}
Y_\ep \! \approx \! \frac{3m_\gamp^2}{2\pi^2} \frac{\prn{45}^{3/2}M_{Pl}}{\sqrt{2} \pi^3} \! \sum_f \! \int_{T_f}^{T_i} \! dT \Gamma_{\gamp \to \bar{f} f} \frac{K_1(\frac{m_\gamp}{T})}{\sqrt{g_\ast} g_{\ast s}T^5}, \!
}
where $T_f=\QCD$ for quarks, $T_f=0$ for leptons, $\Gamma_{\gamp \to \bar{f} f}$ is the partial decay width of the twin photon to $f \bar{f}$, and the sum is over all SM fermion-antifermion pairs for which $m_\gamp > 2 m_f$.

Since we have approximated the yield as being due entirely to on-shell production and decay of twin photons, the analytical expression for the yield in Eq.~\eqref{eq:Yep} exactly agrees with the yield from freezing-in $\gamp$ via `inverse decays' $\bar{f}f \to \gamp$, as derived in~\cite{Hall:2009bx}. We have validated our numerical implementation of the freeze-in calculation by successfully reproducing the yield in similar cases found in \cite{Blennow:2013jba,Chang:2019xva}. We have furthermore checked that reprocessing of the frozen-in dark matter \cite{Chu:2011be,Forestell:2018dnu} through $e'\bar{e'} \rightarrow e'\bar{e'}e'\bar{e'}$ is negligible here,\footnote{To be conservative, we calculate the rate assuming all interactions take place at the maximum $\sqrt{s} \simeq m_\gamp$ and find that it is still far below Hubble. We perform the calculation of the cross section using MadGraph \cite{Alwall:2014hca} with a model implemented in Feynrules \cite{Alloul:2013bka}.} as is the depletion from $e' \bar{e'} \to \nu' \bar{\nu'}$.

An equal number of twin positrons are produced as twin electrons from the freeze-in processes. Requiring that $\epsilon$ reproduce the observed DM abundance today, we find
\al{
\epsilon = \sqrt{\frac{\Omega_\chi h^2 \rho_{\text{crit}}/h^2}{2m_\ep \Tilde{Y}_\ep s_0}},
}
where $\Omega_\chi h^2 \approx 0.12$, $\rho_{\text{crit}}/h^2 \approx 1.1 \times 10^{-5} \text{GeV}/\text{cm}^3$, and $s_0 \approx 2900/\text{cm}^3$ \cite{Tanabashi:2018oca}. $\Tilde{Y}_\ep$ is the total yield with the overall factor of $\epsilon^2$ removed.
\begin{figure}[t!]
\includegraphics[width = \columnwidth]{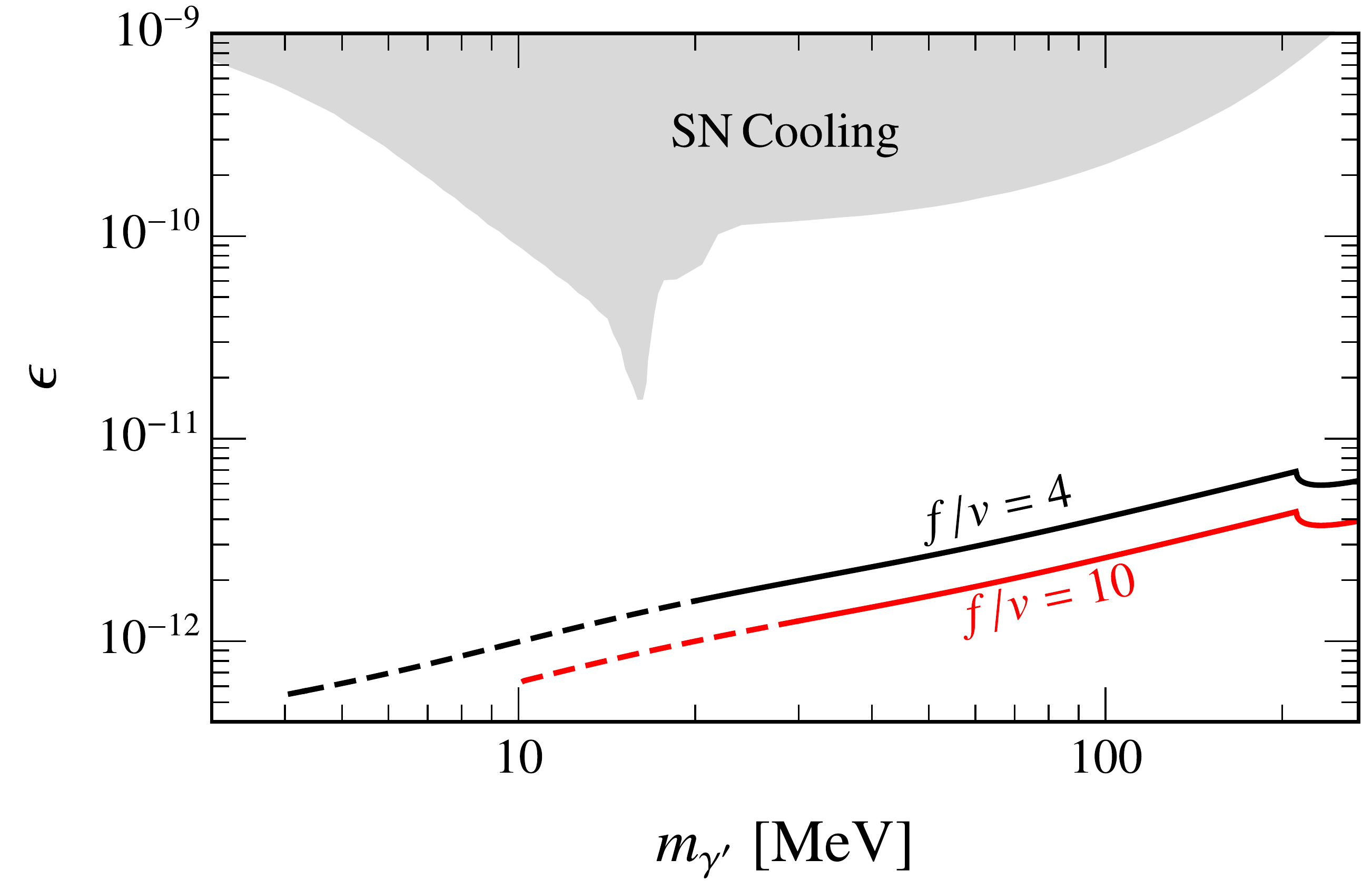}
\caption{\label{fig:mAvseps} Contours in the plane of twin photon mass $m_\gamp$ and kinetic mixing $\epsilon$ which freeze-in the observed DM abundance for two values of $f/v$. The dip at high masses corresponds to additional production via muon annihilations. In the dashed segments, self-interactions occur with $\sigma_{\text{elastic}}/m_{e'}\gtrsim 1 \text{cm}^2/\text{g}$. Also included are the combined supernova cooling bounds from \protect\cite{Chang:2016ntp,Hardy:2016kme}.} 
\end{figure}
This requisite kinetic mixing appears in Fig.~\ref{fig:mAvseps} as a function of the twin photon mass $m_{\gamma'}$ for the two benchmark $f/v$ values 4 and 10. In grey, we plot constraints from anomalous supernova cooling. To be conservative, we include both, slightly different bounds from \cite{Chang:2016ntp,Hardy:2016kme}. The dashed regions of the lines show approximately where self-interactions through Bhabha scattering are relevant in the late universe, $\sigma_{\text{elastic}}/m_{e'}\gtrsim 1 \text{ cm}^2/\text{g}$. Self-interactions much larger than this are constrained by the Bullet Cluster~\cite{Clowe:2003tk,Markevitch:2003at,Randall:2007ph} among other observations. Interestingly, self-interactions of this order have been suggested to fix small-scale issues, and some claimed detections have been made as well. We refer the reader to \cite{Tulin:2017ara} for a recent review of these issues. 

As mentioned above and discussed further in Appendix \ref{sec:KinMixapp}, the level of kinetic mixing required for freeze-in is roughly of the same order as is expected from infrared contributions in the MTH. It would be interesting to develop the technology to calculate the high-loop-order diagrams at which it may be generated. In the context of a complete model of the MTH where kinetic mixing is absent in the UV, $\epsilon$ is fully calculable and depends solely on the scale at which kinetic mixing is first allowed by the symmetries. Calculating $\epsilon$ would then predict a minimal model at some $m_{\gamma'}$ to achieve the right dark matter relic abundance, making this effectively a zero-parameter extension of MTH models with asymmetric reheating. Importantly, even if $\epsilon$ is above those shown in Fig.~\ref{fig:mAvseps}, that would simply point to a larger value of $m_\gamp$ which would suggest that the parameter point depends in more detail on the mechanism of asymmetric reheating. We note that in the case that the infrared contributions to $\epsilon$ are below those needed here, the required kinetic mixing may instead be provided by UV contributions and the scenario is unaffected.

\subsection{Conclusion} \label{sec:conclu}

The mirror twin Higgs is perhaps the simplest avatar of the Neutral Naturalness program, which aims to address the increasingly severe little hierarchy problem. Understanding a consistent cosmological history for this model is therefore 
crucial, and an important step was taken in \cite{Craig:2016lyx,Chacko:2016hvu}. As opposed to prior work, the cosmology of the MTH was remedied without hard breaking of the $\mathbb{Z}_2$ symmetry by utilizing asymmetric reheating to dilute the twin energy density. Keeping the $\mathbb{Z}_2$ as a good symmetry should simplify the task of 
writing high energy completions of these theories, but low-scale reheating may slightly complicate cosmology at early times. These works left as open questions how to set up cosmological epochs such as dark matter generation and baryogenesis in such models. We have here found that at least one of these questions has a natural answer. 

In this work, we have shown that twin electrons and positrons may be frozen-in as dark matter following asymmetric reheating in twin Higgs models. This requires extending the mirror twin Higgs minimally with a single free parameter: the twin photon mass. Freezing-in the observed DM abundance pins the required kinetic mixing to a level expected from infrared contributions in MTH models. In fact, the prospect of calculating the kinetic mixing generated in the MTH could make this an effectively parameter-free extension of the MTH. Compared to generic freeze-in scenarios, it is interesting in this case that the ``just so'' stories of feeble coupling and negligible initial density were already present for reasons entirely orthogonal to dark matter.

This minimalism in freeze-twin dark matter correlates disparate signals which would allow this model to be triangulated with relatively few indirect hints of new physics. If deviations in Higgs couplings are observed at the HL-LHC or a future lepton collider, this would determine $f/v$ \cite{Ahmed:2017psb,Chacko:2017xpd,Alipour-Fard:2018lsf,Alipour-fard:2018mre}, which would set the dark matter mass. An observation of anomalous cooling of a future supernova through the measurement of the neutrino `light curve' might allow us to directly probe the $m_{\gamma'},\epsilon$ curve \cite{Chang:2016ntp,Hardy:2016kme}, though this would rely on an improved understanding of the SM prediction for neutrino production.\footnote{We thank Jae Hyeok Chang for a discussion on this point.} Further astrophysical evidence of dark matter self-interactions would point to a combination of $f/v$ and $m_{\gamma'}$. All of this complementarity underscores the value of a robust experimental particle physics program whereby new physics is pursued via every imaginable channel. 

\chapter{Neutral Naturalness in the Ground} \label{sec:InTheGround}
\setlength{\epigraphwidth}{0.5\textwidth}
\epigraph{There is nothing like looking, if you want to find something. You certainly usually find something, if you look, but it is not always quite the something you were after.}{J.R.R. Tolkien \\ \textit{The Hobbit}, 1937 \cite{tolkien_hobbit}}
\setlength{\epigraphwidth}{0.6\textwidth}
\section*{Long-Lived Particles}

When we introduced models of Neutral Naturalness in Section \ref{sec:neutral}, we were motivated by the lack of signals of new, SM-charged particles at colliders. However, experimentalists are very clever, and it is in fact possible to observe the effects of such particles if you know where to look. 

In particular, Neutral Naturalness models usually exhibit a `hidden valley' type phenomenology, wherein a dark sector is connected to the SM only through some heavy states. At a collider, a high energy collision can transfer energy from our sector to these heavy dark sector states, which may then decay just as heavy SM particles do. If the dark sector contains absolutely stable particles, then the energy may cascade into those states, which simply leave the detector. But without a symmetry dictating that stability, dark sector states may be destabilized by the effects of that same high-energy link to the SM. That connection may induce higher-dimensional operators which allow the decay of the dark sector states back into SM states. Since this decay channel comes from interactions of heavy states, the width may be highly suppressed, leaving to a macroscopically long lifetime, even when the scale set by the particle's mass is microscopically small. This leads to the appearance of SM particles out of nowhere inside a detector a macroscopic distance away from the interaction point in a collider.

\begin{figure}
	\centering
	\includegraphics[width=0.6\linewidth]{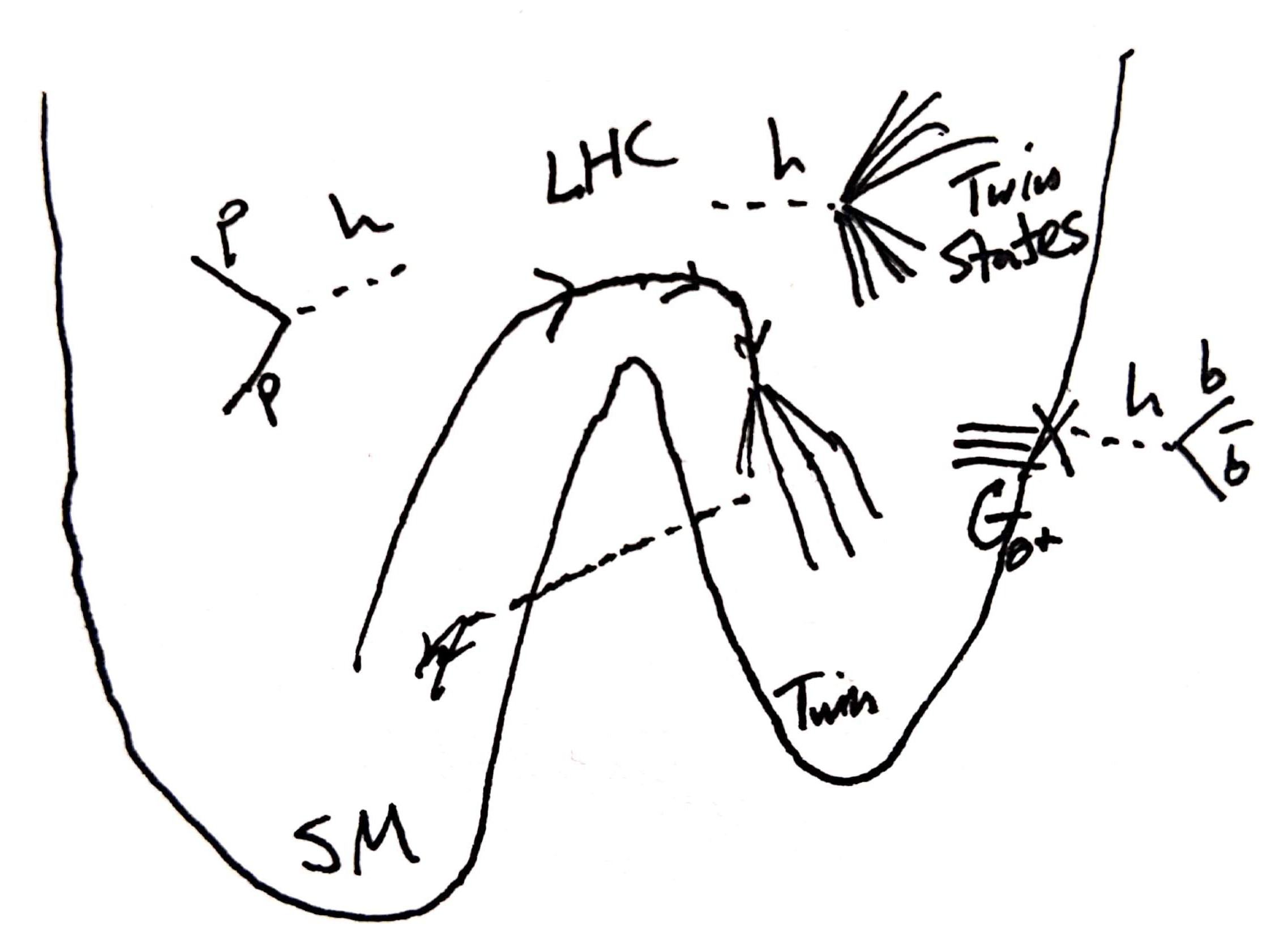}
	\caption{Schematic depiction of hidden valley phenomenology. When another sector is connected to the SM only through heavy particles, it is possible to produce particles from that sector in high energy collisions. If the other sector has phenomenology similar to QCD, this may lead to the production of many states in a dark shower. If (some of) the lowest-lying states in the dark sector may decay back to SM particles, these will generically be displaced decays because they must go through the higher-dimensional operator. In the Twin Higgs, this may be production of either Higgs leading to dark showers with many twin glueballs, some of which mix with the Higgs and decay into SM states. Figure adapted from \protect\cite{Strassler:2008bv}.}
	\label{fig:hiddenValley}
\end{figure}

In fact there are many reasons an unstable particle may be long-lived, and we see a multitude of examples in the SM itself. In analogy with the hidden valley phenomenology, charged pions are long-lived compared to the QCD scale because their decay must take place through a heavy W boson. Neutrons are similarly very long-lived as a result of their small mass splitting with protons. Protons themselves are \textit{very} long-lived as a result of an approximate global symmetry (see Section \ref{sec:bottop} for some little-appreciated subtleties in this reasoning). The SM Higgs is long-lived compared to the electroweak scale because its leading decay mode proceeds through the small bottom quark Yukawas. Given the genericity of long-lived particles in our sector, it's entirely reasonable to imagine that some dark sector will have similar phenomenology. So even aside from our Neutral Naturalness motivation, such searches are generically useful and interesting things to look for.

In this section we forecast how well an electron-positron collider will be able to probe Higgs decays to long-lived particles, using the parameters for some machines which have been proposed by the community and search strategies of our own design. Such forecasting is crucially important at the present time, as the community is still discussing what the next collider is that we will build\footnote{Note that I have no idea what the `present time' is for you, the reader, but I am confident this statement remains true regardless.}. It's clearly necessary to know what sort of physics program we expect we can carry out before we build a machine to do it. And these studies are used to motivate different types of detectors, their detailed design features and how trigger bandwidth is allocated.

\title{Long Live the Higgs Factory: Higgs Decays to Long-Lived Particles at Future Lepton Colliders}

\maketitle

\begin{abstract}
We initiate the study of exotic Higgs decays to long-lived particles (LLPs) at proposed future lepton colliders, focusing on scenarios with displaced hadronic final states. Our analysis entails a realistic tracker-based search strategy involving the reconstruction of displaced secondary vertices and the imposition of selection cuts appropriate for eliminating the largest irreducible backgrounds. The projected sensitivity is broadly competitive with that of the LHC and potentially superior at lower LLP masses. In addition to forecasting branching ratio limits, which may be freely interpreted in a variety of model frameworks, we interpret our results in the parameter space of a Higgs portal Hidden Valley and various incarnations of neutral naturalness, illustrating the complementarity between direct searches for LLPs and precision Higgs coupling measurements at future lepton colliders.
\end{abstract}
\end{comment}

\section{Introduction}
Following the discovery of the Higgs boson in 2012 \cite{Aad:2012tfa, Chatrchyan:2012xdj}, the precision study of its properties has rapidly become one of the centerpieces of the physics program at the LHC. The expansion of this program beyond the LHC has become one of the key motivators for proposed future accelerators, including lepton colliders such as CEPC \cite{CEPCStudyGroup:2018rmc, CEPCStudyGroup:2018ghi}, FCC-ee \cite{Gomez-Ceballos:2013zzn}, the ILC \cite{Behnke:2013xla, Baer:2013cma}, and CLIC \cite{Aicheler:2012bya, deBlas:2018mhx} that would operate in part as Higgs factories.  

The potential gains of a precision Higgs program pursued at both the LHC and future colliders are innumerable. Confirmation of Standard Model predictions for Higgs properties would mark a triumphant validation of the theory and illuminate phenomena never before seen in nature. The observation of deviations from Standard Model predictions, on the other hand, would point the way directly to additional physics beyond the Standard Model. Such deviations could take the form of changes in Higgs couplings to itself or other Standard Model states, or they could appear as exotic decay modes not predicted by the Standard Model. The latter possibility has been extensively explored for {\it prompt} exotic decay modes in the context of both the LHC (see e.g.~\cite{Curtin:2013fra, deFlorian:2016spz}) and future Higgs factories \cite{Liu:2016zki}. 

However, an equally compelling possibility is for new physics to manifest itself in exotic decays of the Higgs boson to long-lived particles (LLPs). Such signals were first considered in the context of Hidden Valleys \cite{Strassler:2006im, Strassler:2006ri, Han:2007ae} and subsequently found to arise in a variety of motivated scenarios for physics beyond the Standard Model, including solutions to the electroweak hierarchy problem \cite{Craig:2015pha} and models of baryogenesis \cite{Cui:2014twa}; for an excellent recent overview, see \cite{Curtin:2018mvb}. The search for exotic Higgs decays into LLPs necessarily involves strategies outside the scope of typical analyses. The non-standard nature of these signatures raises the compelling possibility of discovering new physics that has been heretofore concealed primarily by the novelty of its appearance.

There is a rich and rapidly growing program of LLP searches at the LHC. A variety of existing searches by the ATLAS, CMS, and LHCb collaborations (e.g.~\cite{CMS:2014hka, Aaboud:2018arf, Aaij:2017mic}; for a recent review see \cite{Lee:2018pag}) constrain Higgs decays into LLPs at roughly the percent level across a range of LLP lifetimes. Significant improvements in sensitivity are possible in future LHC runs with potential advances in timing \cite{Liu:2018wte}, triggers \cite{Gershtein:2017tsv, CMS:2018qgk, LHCb:2018hiv}, and analysis strategies \cite{Csaki:2015fba, Curtin:2015fna}. Most notable among these is the possible implementation of a track trigger \cite{Gershtein:2017tsv, CMS:2018qgk}, which would significantly lower the trigger threshold for Higgs decays into LLPs and potentially allow sensitivity to branching ratios on the order of $10^{-6}$ in zero-background scenarios. 

While studies of prompt exotic Higgs decays at future colliders \cite{Liu:2016zki} have demonstrated the potential for significantly improved reach over the LHC, comparatively little has been said about the prospects for constraining exotic Higgs decays to long-lived particles at the same facilities.\footnote{A notable exception is CLIC, for which a study of tracker-based searches for Higgs decays to LLPs has been recently performed \cite{Kucharczyk:2625054}. For preliminary studies of other non-Higgs LLP signatures at future lepton colliders, see e.g.~\cite{Antusch:2016vyf}. For studies of LLP signatures at future electron-proton colliders, see \cite{Curtin:2017bxr}.} In this work we take the first steps towards filling this gap by studying the sensitivity of $e^+ e^-$ Higgs factories to hadronically-decaying new particles produced in exotic Higgs decays with decay lengths ranging from microns to meters. For the sake of definiteness we restrict our attention to circular Higgs factories operating at or near the peak rate for the Higgsstrahlung process $\eehz$, namely CEPC and FCC-ee, while also sketching the corresponding sensitivity for the $\sqrt{s} = 250$ GeV stage of the ILC. While essentially all elements of general-purpose detectors may be brought to bear in the search for long-lived particles, the distribution of decay lengths for a given average lifetime make it advantageous to exploit detector elements close to the primary interaction point. We thus focus on signatures that can be identified in the tracker. In order to provide a faithful forecast accounting for realistic acceptance and background discrimination, we employ a realistic (at least at the level of theory forecasting) approach to the reconstruction and isolation of secondary vertices. 

A key question is the extent to which future Higgs factories can improve on the LHC sensitivity to Higgs decays to LLPs, insofar as the number of Higgs bosons produced at the LHC will outstrip that of proposed Higgs factories by more than two orders of magnitude. Higgs decays to LLPs are sufficiently exotic that appropriate trigger and analysis strategies at the LHC should compensate for the higher background rate and messier detector environment. As we will see, there are two natural avenues for improved sensitivity at future lepton colliders: improved vertex resolution potentially increases sensitivity to LLPs with relatively short lifetimes, while lower backgrounds and a cleaner detector environment improves sensitivity to Higgs decays into lighter LLPs whose decay products are collimated. 

This chapter is organized as follows: In Section 2 we present a simplified signal model for Higgs decays into pairs of long-lived particles, which in turn travel a macroscopic distance before decaying to quark pairs. We further detail the components of our simulation pipeline and lay out an analysis strategy aimed at eliminating the majority of Standard Model backgrounds. In Section 3 we translate this analysis strategy into the sensitivity of future lepton colliders to long-lived particles produced in Higgs decays as a function of the exotic Higgs branching ratio and the mass and decay length of the LLP. While these forecasts are generally applicable to any model giving rise to the signal topology, we additionally interpret the forecasts in terms of the parameter space of several motivated models in Section 4. We summarize our conclusions and highlight avenues for future development in Section 5.

\section{Signal and analysis strategy} \label{sec:signal}

Exotic decays of the Higgs to long-lived particles encompass a wide variety of intermediate and final states. The decay of the Higgs itself into LLPs can proceed through a variety of different topologies. Perhaps the most commonly-studied scenario is the decay of the Higgs to a pair of LLPs, $h \rightarrow XX$, though decays involving additional visible or invisible particles (such as $h \rightarrow X + {\rm invisible}$ or $h \rightarrow XX + {\rm invisible}$) are also possible. The long-lived particles in turn may have a variety of decay modes back to the Standard Model, including $X \rightarrow \gamma \gamma, jj, \ell \bar \ell,$ or $jj \ell,$ including various flavor combinations. These decay modes may also occur in the company of additional invisible states. Moreover, a given long-lived particle may possess a range of competing decay modes, as is the case for LLPs whose decays back to the Standard Model are induced by mixing with the Higgs.

Our aim here is to be representative, rather than comprehensive, as each production and decay mode for a long-lived particle is likely to require a dedicated search strategy. For the purposes of this study, we adopt a simplified signal model in which the Higgs decays to a pair of long-lived scalar particles $X$ of mass $m_X$, which each decay in turn to pairs of quarks at an average ``proper decay length'' $c \tau$.\footnote{Of course, ``proper decay length'' is a bit of a misnomer, but we use it as a proxy for $c$ times the mean proper lifetime $\tau$.} Both the mass $m_X$ and proper decay length $c \tau$ are treated as free parameters, though they may be related in models that give rise to this topology. For the sake of definiteness, for $m_X > 10 \text{ GeV}$ we take a branching ratio of $0.8$ to $b \bar b$ and equal branching ratios of $0.05$ to each of $u \bar u, d \bar d, s \bar s, c \bar c$, though the precise flavor composition is not instrumental to our analysis. For $m_X \leq 10 \text{ GeV}$ we take equal branching ratios into each of the lighter quarks. We further restrict our attention to Higgs factories operating near the peak of the $\eehz$ cross section, for which the dominant production process will be $\eehz$ followed by $h \rightarrow XX$. The associated $Z$ boson provides an additional invaluable handle for background discrimination. Here we develop the conservative approach of focusing on leptonic decays of the $Z$, though added sensitivity may be obtained by incorporating hadronic decays.

Given the signal, there are a variety of possible analysis strategies sensitive to Higgs decays to long-lived particles, exploiting various parts of a general-purpose detector. Tracker-based searches are optimal for decay lengths below one meter, with sensitivity to shorter LLP decay lengths all the way down to the tracker resolution. Timing information using timing layers between the tracker and electromagnetic calorimeter offers optimal coverage for slightly longer decay lengths, while searches for isolated energy deposition in the electromagnetic calorimeter, hadronic calorimeter, and muon chambers provides sensitivity to decay lengths on the order of meters to tens of meters. In principle, instrumenting the exterior of a general-purpose detector with large volumes of scintillator may lend additional sensitivity to even longer lifetimes. In this work we will focus on tracker-based searches at future lepton colliders, as these may be simulated relatively faithfully and ultimately are among the searches likely to achieve zero background while retaining high signal efficiency. 

We define our signal model in \texttt{FeynRules} \cite{Alloul:2013bka} and generate the signal $\eehz \rightarrow XX + \ell \bar \ell $ at $\sqrt{s} = 240$ GeV using \texttt{MadGraph 5} \cite{Alwall:2014hca}. Where appropriate, we will also discuss prospects for Higgs factories operating at $\sqrt{s} = 250$ GeV (potentially with polarized beams) such as the ILC by rescaling rates with the appropriate leading-order cross section ratios. In order to correctly simulate displaced secondary vertices, the decay of the LLP $X$ and all unstable Standard Model particles is then performed in \texttt{Pythia 8} \cite{Sjostrand:2007gs}.

In addition to the signal, we consider some of the leading backgrounds to our signal process and develop selection cuts aimed at achieving a zero-background signal region. The most significant irreducible backgrounds from Standard Model processes include $\eehz$ with $Z \rightarrow \ell \bar \ell$ and $h \rightarrow b \bar b$ as well as $e^+ e^- \rightarrow ZZ \rightarrow \ell \bar \ell + b \bar b$. Unsurprisingly, there are a variety of other Standard Model backgrounds, but they are typically well-controlled by imposing basic Higgsstrahlung cuts, and we do not simulate them with high statistics. In addition to irreducible backgrounds from hard collisions, there are possible backgrounds from particles originating away from the interaction point, including cosmic rays, beam halo, and cavern radiation; algorithmic backgrounds originating from effects such as vertex merging or track crossing; and detector noise. Such backgrounds are well beyond the scope of the current study, and will require dedicated investigation with full simulation of the proposed detectors.

Correctly emulating the detector response to LLPs using publicly-available fast simulation tools is notoriously challenging. In particular, we have found that the default clustering algorithms in the detector simulator \texttt{Delphes} \cite{deFavereau:2013fsa} tends to cluster calorimeter hits from different secondary vertices into the same jets, significantly complicating the realistic reconstruction of secondary vertices. As such, we develop an analysis strategy using only ingredients from the \texttt{Pythia} output, although we do further run events through \texttt{Delphes} and utilize \texttt{ROOT} \cite{BRUN199781} for analysis. 

We implement two distinct tracker-based analyses with complementary signal parameter space coverage, which we denote as the `large mass' and `long lifetime' pipelines. We shall eventually see that the former will be effective for $m_X \gtrsim 10 \text{ GeV}$ down to proper decay lengths $c\tau \gtrsim 1 \mu \text{m}$, while the latter is able to push down in $m_X$ by a factor of a few though is only fully effective for $c\tau\gtrsim 1 \text{ cm}$. Full cut tables for both irreducible backgrounds and a variety of representative signal parameter points appear in Tables \ref{tab:large_mass} and \ref{tab:long_life}, respectively. 

\begin{center}
\captionof{table}{\label{tab:large_mass} Cut flow of the `large mass' analysis for the CEPC with entries of acceptance $\times$ efficiency. The top set of rows gives the cut flow on 500k $Z(b \bar b)Z(\ell \bar \ell)$ events and 100k $h(b \bar b)Z(\ell \bar \ell)$ background events, which are used to confirm our analysis is in the no-background regime. The next sets of rows give cut flows on 5k signal events at representative parameter points, where the different columns are labeled by $m_X/\text{GeV},c\tau/\text{m}$. The full row labels are given in the top set of rows and the labels below are abbreviations for the same cuts or selections.}
    \begin{tabular}{|p{6cm}||l|l|}
 \hline
 \text{Cut/Selection} & \text{$ZZ$ Background} & \text{$hZ$ Background} \\
\hline \hline
 \text{Dilepton Invariant Mass} & 0.97 & 0.98 \\
\hline
 \text{Recoil Mass} & 0.006 & 0.94 \\
\hline
 \text{Displaced Cluster ($\geq$ resolution)} & 0.004 & 0.94 \\
\hline
 \text{Invariant Charged Mass (6 GeV)} & 0 & 0.00005 \\
\hline
 \text{Invariant `Dijet' Mass} & 0 & 0.00005 \\
\hline
 \text{Pointer Track} & 0 & 0.00001 \\
\hline
    \end{tabular}
    
    \begin{tabular}
    {|l||p{1.38cm}|p{1.38cm}|p{1.38cm}||p{1.38cm}|p{1.38cm}|p{1.38cm}|}
\hline
$m_X, c \tau$ & $\text{7.5, }10^{-4}$ & $\text{7.5, }10^{-2}$ & $\text{7.5, }10^0$ & $\text{10, }10^{-4}$ & $\text{10, }10^{-2}$ & $\text{10, }10^0$ \\
\hline \hline
 $M_{\ell\ell}$ & 0.97 & 0.97 & 0.97 & 0.97 & 0.98 & 0.97 \\
\hline
 $M_{\text{recoil}}$ & 0.93 & 0.93 & 0.93 & 0.93 & 0.94 & 0.93 \\
\hline
 $|\vec{d}_{\text{cluster}}|$ & 0.93 & 0.93 & 0.41 & 0.93 & 0.94 & 0.50 \\
\hline
 $M_{\text{charged}}$ & 0.27 & 0.28 & 0.08 & 0.55 & 0.55 & 0.21 \\
\hline
  $M_{\text{cluster}}$ & 0.27 & 0.28 & 0.08 & 0.55 & 0.55 & 0.21 \\
\hline
 Pointer & 0.25 & 0.28 & 0.08 & 0.50 & 0.55 & 0.21 \\
\hline
    \end{tabular}
    \begin{tabular}
    {|l||p{1.38cm}|p{1.38cm}|p{1.38cm}||p{1.38cm}|p{1.38cm}|p{1.38cm}|}
\hline
$m_X, c \tau$  & $\text{25, }10^{-4}$ & $\text{25, }10^{-2}$ & $\text{25, }10^0$ & $\text{50, }10^{-4}$ & $\text{50, }10^{-2}$ & $\text{50, }10^0$ \\
\hline \hline
 $M_{\ell\ell}$ & 0.97 & 0.97 & 0.98 & 0.97 & 0.98 & 0.97 \\
\hline
 $M_{\text{recoil}}$ & 0.92 & 0.92 & 0.93 & 0.92 & 0.92 & 0.93 \\
\hline
 $|\vec{d}_{\text{cluster}}|$ & 0.92 & 0.92 & 0.80 & 0.92 & 0.92 & 0.93 \\
\hline
 $M_{\text{charged}}$ & 0.76 & 0.77 & 0.57 & 0.82 & 0.85 & 0.81 \\
\hline
 $M_{\text{cluster}}$ & 0.76 & 0.77 & 0.57 & 0.76 & 0.80 & 0.76 \\
\hline
 Pointer & 0.73 & 0.76 & 0.57 & 0.75 & 0.77 & 0.76 \\
\hline
    \end{tabular}
\end{center}

\begin{center}
\captionof{table}{\label{tab:long_life} Cut flow of the `long lifetime' analysis for the CEPC with entries of acceptance $\times$ efficiency. The top set of rows gives the cut flow on 500k $Z(b \bar b)Z(\ell \bar \ell)$ events and 100k $h(b \bar b)Z(\ell \bar \ell)$ background events, which are used to confirm our analysis is in the no-background regime. The next sets of rows give cut flows on 5k signal events at representative parameter points, where the different columns are labeled by $m_X/\text{GeV},c\tau/\text{m}$. The full row labels are given in the top set of rows and the labels below are abbreviations for the same cuts or selections.}

    \begin{tabular}{|p{6cm}||l|l|}
\hline
 \text{Cut/Selection} & \text{$ZZ$ Background} & \text{$hZ$ Background} \\ 
\hline \hline
 \text{Dilepton Invariant Mass} & 0.97 & 0.98 \\
\hline
 \text{Recoil Mass} & 0.006 & 0.94 \\
\hline
 \text{Displaced Cluster ($\geq 3$ cm)} & 0.004 & 0.62 \\
\hline
 \text{Charged Invariant Mass (2 GeV)} & 0 & 0.002 \\
\hline
 \text{`Dijet' Invariant Mass} & 0 & 0.002 \\
\hline
 \text{Pointer Track} & 0 & 0.001 \\
\hline
 \text{Isolation} & 0 & 0.00005 \\
\hline
    \end{tabular}
    \begin{tabular}{|l||p{1.37cm}|p{1.37cm}|p{1.37cm}||p{1.37cm}|p{1.37cm}|p{1.37cm}|}
    \hline
$m_X, c \tau$ & $\text{2.5, }10^{-4}$ & $\text{2.5, }10^{-2}$ & $\text{2.5, }10^0$ & $\text{7.5, }10^{-4}$ & $\text{7.5, }10^{-2}$ & $\text{7.5, }10^0$ \\
\hline \hline
 $M_{\ell\ell}$ & 0.97 & 0.97 & 0.98 & 0.97 & 0.97 & 0.97 \\
\hline
 $M_{\text{recoil}}$ & 0.93 & 0.93 & 0.93 & 0.93 & 0.93 & 0.93 \\
\hline
$|\vec{d}_{\text{cluster}}|$ & 0.21 & 0.89 & 0.15 & 0.41 & 0.89 & 0.41 \\
\hline
$M_{\text{charged}}$ & 0 & 0.40 & 0.05 & 0 & 0.74 & 0.34 \\
\hline
$M_{\text{cluster}}$ & 0 & 0.40 & 0.05 & 0 & 0.74 & 0.34 \\
\hline
Pointer & 0 & 0.40 & 0.05 & 0 & 0.74 & 0.34 \\
\hline
 Isolation & 0 & 0.33 & 0.045 & 0 & 0.51 & 0.33 \\
\hline

    \end{tabular}

    \begin{tabular}{|l||p{1.37cm}|p{1.37cm}|p{1.37cm}||p{1.37cm}|p{1.37cm}|p{1.37cm}|}
\hline \hline
$m_X, c \tau$ & $\text{15, }10^{-4}$ & $\text{15, }10^{-2}$ & $\text{15, }10^0$ & $\text{50, }10^{-4}$ & $\text{50, }10^{-2}$ & $\text{50, }10^0$ \\
\hline \hline
$M_{\ell\ell}$ & 0.97 & 0.98 & 0.97 & 0.97 & 0.98 & 0.97 \\
\hline
$M_{\text{recoil}}$ & 0.93 & 0.93 & 0.93 & 0.92 & 0.92 & 0.93 \\
\hline
$|\vec{d}_{\text{cluster}}|$ & 0.59 & 0.87 & 0.65 & 0.64 & 0.69 & 0.92 \\
\hline
$M_{\text{charged}}$ & 0.001 & 0.71 & 0.63 & 0 & 0.10 & 0.91 \\
\hline
$M_{\text{cluster}}$ & 0.001 & 0.71 & 0.63 & 0 & 0.09 & 0.90 \\
\hline
Pointer & 0.001 & 0.65 & 0.60 & 0 & 0.08 & 0.84 \\
\hline
Isolation & 0.0002 & 0.42 & 0.58 & 0 & 0.05 & 0.77 \\
\hline
    \end{tabular}

\end{center}

As a first step in either analysis, we select Higgsstrahlung events by requiring that our events have an opposite sign electron (muon) pair in the invariant mass range $70 \leq M_{ee} \leq 110$ GeV ($81 \leq M_{\mu\mu} \leq 101$ GeV) and with recoil mass $M_{\text{recoil}}^2 \equiv \big( (\sqrt{s},\vec{0})^\mu - p^\mu_{\ell\ell}\big)^2$ in the range $120 \leq M_{\text{recoil}} \leq 150$ GeV, with $p^\mu_{\ell\ell}$ the momentum of the lepton pair. This allows us to limit our background considerations to the irreducible backgrounds mentioned above and cuts down severely on the $e^+ e^- \rightarrow ZZ$ background, as seen in Tables \ref{tab:large_mass} and \ref{tab:long_life}.

We next identify candidate secondary vertices using a depth-first `clustering' algorithm, which roughly emulates that performed in the CMS search \cite{CMS:2014wda}. We perform this clustering using \textit{all} particles in the event because at later points in the analysis we need this truth-level assignment of neutral particles to clusters, but we expect that this (admittedly unrealistic) inclusion does not significantly modify the performance of this algorithm. Beginning with a single particle as the `seed' particle for our algorithm, we look through all other particles in the event and create a `cluster' of particles consisting of the seed particle and any others whose origins are within $\ell_\text{cluster} = 7 \ \mu \text{m}$ (the projected tracker resolution of CEPC \cite{CEPCStudyGroup:2018ghi}) of the seed particle. We then add to that cluster any particles whose origins are within $\ell_\text{cluster}$ of any origins of particles in the cluster, and do this step iteratively until no further particles are added to the cluster. We then choose a new seed particle which has not yet been assigned to a cluster and begin this clustering process again. We repeat this process until all particles in the event have been assigned to clusters. We assign to each cluster a location $\vec{d}_\text{cluster}$ which is the average of the origins of all charged particles in the cluster. To ensure that our events contain displaced vertices, we impose a minimum bound on the displacement from the interaction point $|\vec{d}_\text{cluster}| > d_\text{min}$, and clusters satisfying this requirement constitute candidate secondary vertices. For our `large mass' analysis we set $d_\text{min}$ to be the impact parameter resolution ($\simeq 5 \ \mu\text{m}$ for both CEPC and FCC-ee \cite{CEPCStudyGroup:2018ghi}), and so retain sensitivity to very short $X$ lifetimes. For our `long lifetime' analysis we set $d_\text{min}= 3 \text{ cm}$, which removes the vast majority of clusters coming from $B$ hadron decays in background events, as seen in Table \ref{tab:long_life}. An upper bound $|\vec{d}_\text{cluster}| < r_\text{tracker}$ is imposed by the outer radius of the tracker, where $r_\text{tracker} = 1.81 \text{m}$ for CEPC and $r_\text{tracker} = 2.14 \text{m}$ for FCC-ee are proposed. 

At this point an experimental analysis might sensibly examine dijets containing candidate secondary vertices and impose an upper bound on the dijet invariant mass to remove backgrounds coming from Standard Model $H$ or $Z$ decays. As discussed above we are limited to \texttt{Pythia} objects, but to mock up the (small) penalty to signal of such a selection we implement a selection on the total invariant mass of the clusters $M_{\text{cluster}}^2 \equiv (\sum_{i \in \text{ cluster}} p_i^\mu)^2$. Since this is truth-level information, to turn it into an analog for the dijet invariant mass we apply a Gaussian smearing with a standard deviation of $10 \text{ GeV}$ to account for the dijet resolution.
We then select only candidate secondary vertices with $M_{\text{cluster}} < m_h/2$. This has no effect on background in our simulation pipeline as the background candidate secondary vertices are the result of hadronic decays, so the invariant masses of these clusters are not analogs for dijet invariant masses. As emphasized above, the imposition of this cut is strictly to account for possible selections that might appear in a more realistic experimental analysis.

While the total invariant mass of the clusters is not an experimental observable, the invariant mass of charged particles in the clusters $M_{\text{charged}}=(\sum_{i \text{  charged}} p_i^\mu)^2$ is experimentally accessible. For our `large mass' analysis we select candidate secondary vertices with $M_{\text{charged}}> 6 \text{ GeV}$, which gets rid of nearly all clusters from hadronic decays, as seen in Table \ref{tab:large_mass}. For the `long lifetime' analysis, while the increased displacement requirement removes $b$ hadrons it still allows $c,s$ hadrons, and so we select $M_{\text{charged}}> 2 \text{ GeV}$ to address this, which Table \ref{tab:long_life} shows is again very effective.

Next we select the cluster closest to the beamline which passes the above selection requirements as our secondary vertex for the event. Choosing the closest one preferentially selects $X$ decay clusters over hadronic decay clusters in the jets to which the $X$ decays, though this can be fooled by a non-zero fraction of `back-flowing' quarks in $X$ decays (quarks with momenta pointing toward the beamline). 

To remove displaced vertices coming from the decays of charged $b$ hadrons we implement a `pointer track' cut in both analyses as follows. For the cluster selected as the secondary vertex, we consider a sphere of radius $r = 0.5 \text{ mm}$ around the position $\vec{d}_\text{cluster}$. We look for any charged particles whose origins are outside this sphere and whose momenta (at the point at which they were created) point into it, and veto the event if there are any such particles. The main effect of this cut is to remove clusters which were produced from the decay of a charged hadron. The sphere size has been chosen to maximize this effect, though this allows a small effect on signal due to geometric coincidence. Since this cut is only on charged particles, roughly $\sim 30\%$ of background clusters are unaffected. For this cut we ignore the effect of the magnetic field in the tracker, which should not highly impact the trajectories on short scales.

For the `long lifetime' analysis we further implement an `isolation' cut to remove neutral hadronic background decays. Given the cluster selected as the secondary vertex, we consider the plane perpendicular to the sum of momenta of charged particles in the cluster which passes through $\vec{d}_\text{cluster}$. We project the paths of prompt (vertex within $3 \ \mu \text{m}$ of the primary vertex, the planned CEPC vertex resolution \cite{CEPCStudyGroup:2018ghi}) charged particles onto this plane (again ignoring the magnetic field) and veto the event if any come within $R = 10 \text{ cm}$ of the position of the secondary vertex. This radius was chosen to maximally reduce background, and does have a deleterious effect on short decay lengths $\lesssim 10 \text{ mm}$, as can be seen in Table \ref{tab:long_life}. This cut is not perfectly effective at rejecting background due to the non-negligible presence of jets whose prompt components have neutral fraction $1$.
 
\section{Results and discussion} \label{sec:results}

To confirm that our analysis pipelines put us in the zero-background regime we run both the `long lifetime' and the `large mass' analyses on 500k $e^+ e^- \rightarrow Z(b \bar b)Z(\ell \bar \ell)$ events and 100k $e^+ e^- \rightarrow h(b \bar b) Z(\ell \bar \ell)$ background events. For both pipelines we find that zero $e^+ e^- \rightarrow Z(b \bar b)Z(\ell \bar \ell)$ events remain, while for $e^+ e^- \rightarrow h(b \bar b) Z(\ell \bar \ell)$ we find efficiencies of $5\times 10^{-5}$ and $1\times 10^{-5}$ respectively. We then run each analysis on 5k signal events to get acceptance $\times$ efficiencies for each $(m_X, c\tau)$ point, for a selection of points with   $m_X = 2.5$ from GeV to $50$ GeV and $c\tau$ from $1 \ \mu\text{m}$ to $50$ m. In Table \ref{tab:long_life} we give a cut table for both backgrounds and some representative signal parameter points for the `long lifetime' analysis, and in Table \ref{tab:large_mass} we do the same for the `large mass' analysis.

In the zero-background regime, Poisson statistics rules out model points which predict 3 or more signal events to $95\%$ confidence (or better) if no signal is detected. We may then find a projected $95\%$ upper limit on branching ratio as 
\begin{equation}
    \text{Br}(h\rightarrow XX)^{95} = \frac{N_{sig}}{\mathcal{L}\times \sigma(e^+ e^- \rightarrow hZ) \times \text{Br}(Z\rightarrow \ell \ell) \times A \times \varepsilon},
\end{equation}
with $N_{sig} = 3$ and $A\times \varepsilon$ the result of our simulations. For both the CEPC and FCC-ee, the most recent integrated luminosity projections \cite{CEPCStudyGroup:2018ghi, FCCee} give $\mathcal{L}\times  \sigma(e^+ e^- \rightarrow hZ) = 1.1 \times 10^6$ Higgses produced.

\begin{figure}
	\centering
	\includegraphics[width=\textwidth]{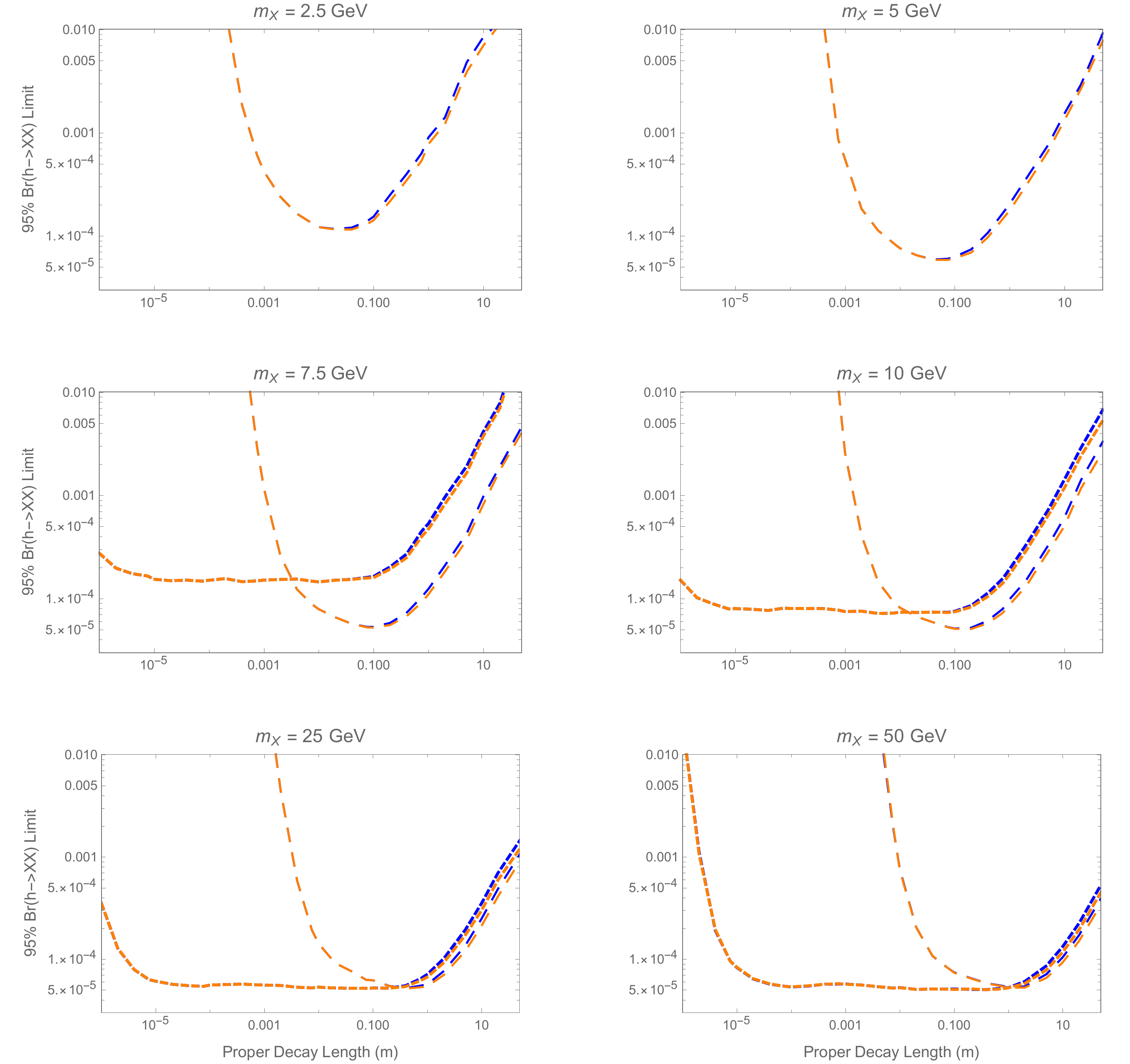}
	\caption{\label{fig:brlimits} Projected 95\% $h \rightarrow XX$ branching ratio limits as a function of proper decay length for a variety of $X$ masses. Blue lines are for CEPC and orange lines are for FCC-ee, and where only one is visible they overlap. The larger dashes are the `long lifetime' analysis and the smaller dashes are the `large mass' analysis.}
\end{figure}

In Figure \ref{fig:brlimits} we show projected 95\% upper limits on $\text{Br}(h\rightarrow XX)$ as a function of $X$ mass and proper decay length. While we plot separate lines for both CEPC and FCC-ee, we only use one set of signal events generated at $\sqrt{s} = 240 \text{ GeV}$ and only account for the difference in tracker radii, so these overlap entirely at smaller lifetimes. 
Approximate limits for the ILC can be obtained by multiplying the above branching ratio limits by a factor of $\sim 1.8$ (i.e.~weakening the limit) to account for the leading order differences in center-of-mass energy, polarization, and integrated luminosity at the $\sqrt{s} = 250$ GeV ILC run, assuming comparable acceptance and efficiency. The ILC limits weaken slightly further for large decay lengths, as its proposed tracker radius is $1.25$ m. Of course, adding the higher-energy ILC runs should significantly improve sensitivity given analyses suitable for the $WW$ fusion production relevant at those energies.

For small masses we are only able to use the `long lifetime' analysis, which requires large displacement from the beamline to cut out the SM $b$ hadron background. As a result we only retain good sensitivity to $X$ decay lengths comparable with the tracker size, though the fact that we only require one displaced vertex (out of two $X$s per signal event) significantly broadens our sensitivity range. This fact also helps us retain efficiency at low masses, as we are able to get down to a projected branching ratio limit of $1\times 10^{-4}$ for $m_X = 2.5$ \text{GeV} despite our $2$ GeV cut on charged invariant mass of the decay cluster. For larger masses this cut has less effect, which allows it to push down to even lower branching ratios $\sim 5 \times 10^{-5}$.

The `large mass' analysis begins working well for masses not far above the $6$ GeV charged invariant mass cut and provides sensitivity to far shorter decay lengths, reaching all the way down to the impact parameter resolution and below. For $m_X = 10$ GeV, where we are aided by the boost factor, we project a limit of $1\times 10^{-4}$ for a proper decay length of $1$ micron. The sensitivity to extremely small decay lengths drops for larger masses, but at $m_X = 50$ GeV we cross below the $10^{-4}$ threshold by $7.5 \ \mu$m. For $X$ masses high enough that the charged invariant mass cut does not remove a large amount of signal events, this analysis projects a branching ratio limit of $\sim 5 \times 10^{-5}$ across roughly the entire range of decay lengths corresponding to the geometric volume of the detector. There is a slight dip in sensitivity for $c\tau \sim 1 \text{ mm}$, where the pair of dijets from the two $X$ decays are most likely to overlap and trigger the cut on `pointer' tracks.

The notable region of this parameter space to which our analyses do not provide good sensitivity is the low mass ($m_X \lesssim 6 \text{ GeV}$) and short proper decay length ($c\tau \lesssim 1 \text{ cm}$) regime. The difficulty is that, from the perspective of the tracker, the $X$ here looks more and more like a neutral SM hadron. An analysis making use of the impact parameter distribution of particles in clusters may help here \cite{CMS:2014wda}, but we leave this to future work. Taking advantage of calorimeter data to distinguish between clusters in single jets versus dijets is also likely to provide good sensitivity, but we again leave this to future exploration.

Broadly speaking, our results suggest a peak sensitivity of ${\rm Br}(h \rightarrow XX) \sim 5 \times 10^{-5}$, weakening to $\sim 10^{-4}$ for lower-mass LLPs. Significant additional improvement could be expected with the inclusion of hadronic $Z$ decays, but this requires further study to ensure the control of corresponding Standard Model backgrounds. These limits are competitive with LHC forecasts based on conventional Higgs triggers \cite{Csaki:2015fba, Curtin:2015fna}, noting that these latter forecasts assume zero background. However, the lepton collider limits are potentially superseded by an efficient CMS track trigger \cite{Gershtein:2017tsv, CMS:2018qgk} for higher-mass LLPs, again assuming zero background is achievable with high signal efficiency across a range of lifetimes. In this respect, the primary strengths of the Higgs factories in searching for exotic Higgs decays to LLPs are the potential to push down to shorter decay lengths and lighter LLPs. In particular, the relatively clean and low-background environment of lepton colliders should enable efficient LLP searches even when the LLP decay products become collimated, which remains a weakness of the corresponding LHC searches.

\section{Signal interpretations} \label{sec:interpretations}

While the bounds presented in the previous section apply to any scenario in which the Higgs decays into pairs of long-lived particles which in turn decay (at least in part) into pairs of quarks,  it is also useful to interpret these bounds in the context of specific models that relate the Higgs branching ratio to LLPs (and the LLP lifetime) to underlying parameters. This illustrates the potential for LLP searches at future lepton colliders to constrain motivated scenarios for physics beyond the Standard Model and allows us to explore the potential complementarity between LLP searches and precision Higgs coupling measurements. To this end, we consider the implications of the LLP limits presented here in the context of both the original Higgs portal Hidden Valley model and a variety of models of neutral naturalness.

\subsection{Higgs portal}

As a general proxy model for Higgs decays into LLPs, we first consider the archetypal Higgs portal Hidden Valley \cite{Strassler:2006ri}. This entails the extension of the Standard Model by an additional real singlet scalar $\phi$, which couples to the Standard Model through the Higgs portal \cite{Silveira:1985rk, McDonald:1993ex, Burgess:2000yq} via
\begin{eqnarray} \nonumber
\mathcal{L} \supset - \frac{1}{2} (\partial_\mu \phi)^2 - \frac{1}{2} M^2 \phi^2 - A |H|^2 \phi - \frac{1}{2} \kappa |H|^2 \phi^2 \\ -\frac{1}{3!} \mu \phi^3 - \frac{1}{4!} \lambda_\phi \phi^4 - \frac{1}{2} \lambda_H |H|^4.
\end{eqnarray} 
If $\phi$ respects a $\mathbb{Z}_2$ symmetry under which $\phi \rightarrow - \phi$, this additionally sets $\mu = A = 0$, such that the singlet scalar only couples to the Standard Model via the quartic interaction $|H|^2 \phi^2$. After electroweak symmetry breaking, in unitary gauge $H = \left(0, \frac{1}{\sqrt{2}} (h+v) \right)$, but the CP-even scalars $h$ and $\phi$ do not mix. Nonetheless, the quartic interaction nonetheless provides a significant portal for the production of $\phi$, as $\phi$ may be pair produced via the decay $h \rightarrow \phi \phi$ for  $m_\phi < m_h / 2$. Of course, $\phi$ is stable if the $\mathbb{Z}_2$ symmetry is exact, rendering it a potential (albeit highly constrained) dark matter candidate \cite{He:2008qm, Gonderinger:2009jp, Mambrini:2011ik}.

This model gives rise to long-lived particle signatures \cite{Strassler:2006ri} if the $\mathbb{Z}_2$ is broken by a small amount, such that $A \neq 0$ but e.g. $A^2/M^2 \ll \kappa $. The relative smallness of $A$ is technically natural, as the $\mathbb{Z}_2$ symmetry is restored when $A \rightarrow 0$. This then leads to mass mixing between the CP even scalars. As long as $A$ is small compared to $M$ and $v$, the mass eigenstates consist of an SM-like Higgs $h_{\rm SM}$ and a mostly-singlet scalar $s$, related to the gauge eigenstates by 
\begin{eqnarray}
h_{\rm SM} &=& h \cos \theta + \phi \sin \theta \\
s &=& - h \sin \theta + \phi \cos \theta,
\end{eqnarray}
where $\theta \ll 1$ is the mixing angle. There are now two parametrically distinct processes: pair production of the scalar $s$ via Higgs decays, governed by the size of the $\mathbb{Z}_2$-preserving coupling $\kappa$, and decay of the $s$ scalar back to the Standard Model, governed by the size of the $\mathbb{Z}_2$-breaking coupling $A$. In the limit of small mixing, the former process is of order
\begin{equation}
\Gamma(h \rightarrow ss) \approx \frac{\kappa^2 v^2}{32 \pi m_h} \sqrt{1 - 4 \frac{m_s^2}{m_h^2}},
\end{equation}
where we are neglecting subleading corrections proportional to $\lambda_H \sin^2 \theta$. The latter process proceeds into whatever Standard Model states $Y$ are kinematically available, with partial widths
\begin{equation}
\Gamma(s \rightarrow YY) = \sin^2 \theta \times \Gamma (h_{\rm SM}[m_s] \rightarrow YY),
\end{equation}
where $h_{\rm SM}[m_s]$ denotes a Standard Model-like Higgs of mass $m_s$. This naturally leads to a scenario in which the $s$ scalars may be copiously produced via Higgs decays but travel macroscopic distances before decaying back to Standard Model particles. 

This scenario may be constrained not only by direct searches for Higgs decays to LLPs (with the scalar $s$ playing the role of the LLP), but also by precision Higgs coupling measurements. Higgs coupling deviations in this scenario arise from two parametrically distinct effects: tree-level deviations proportional to $\theta^2$ due to Higgs-singlet mixing, and one-loop deviations proportional to $\kappa$ due to $s$ loops. Both effects result in a universal modification of Higgs couplings, which is best constrained at lepton colliders via the precision measurement of the $\eehz$ cross section \cite{Englert:2013tya, Craig:2013xia}. The net deviation in the $\eehz$ cross section due to these effects in the limit of small mixing is
\begin{equation}
\frac{\delta \sigma_{hZ}}{\sigma_{hZ}^{\rm SM}} \approx - \theta^2 - {\rm Re} \left. \frac{d \mathcal{M}_{hh}}{d p^2} \right|_{p^2 = m_h^2},
\end{equation}
where the radiative correction \cite{Craig:2013xia}

\begin{eqnarray} 
 \left. \frac{d \mathcal{M}_{hh}}{d p^2} \right|_{p^2 = m_h^2} = - \frac{1}{16 \pi^2} \frac{\kappa^2 v^2}{2 m_h^2} \hspace{3.3cm} \\ \times \left( 1 + \frac{4 m_s^2}{m_h^2} \sqrt{\frac{m_h^2}{m_h^2- 4 m_s^2}} \tanh^{-1} \left[ \sqrt{\frac{m_h^2}{m_h^2 - 4 m_s^2}} \right] \right) \nonumber
 \end{eqnarray}

 is approximated at $\theta = 0$. Either effect can dominate depending on the relative size of $A/M$ and $\kappa$.

\begin{figure}
	\centering
	\includegraphics[width=8cm]{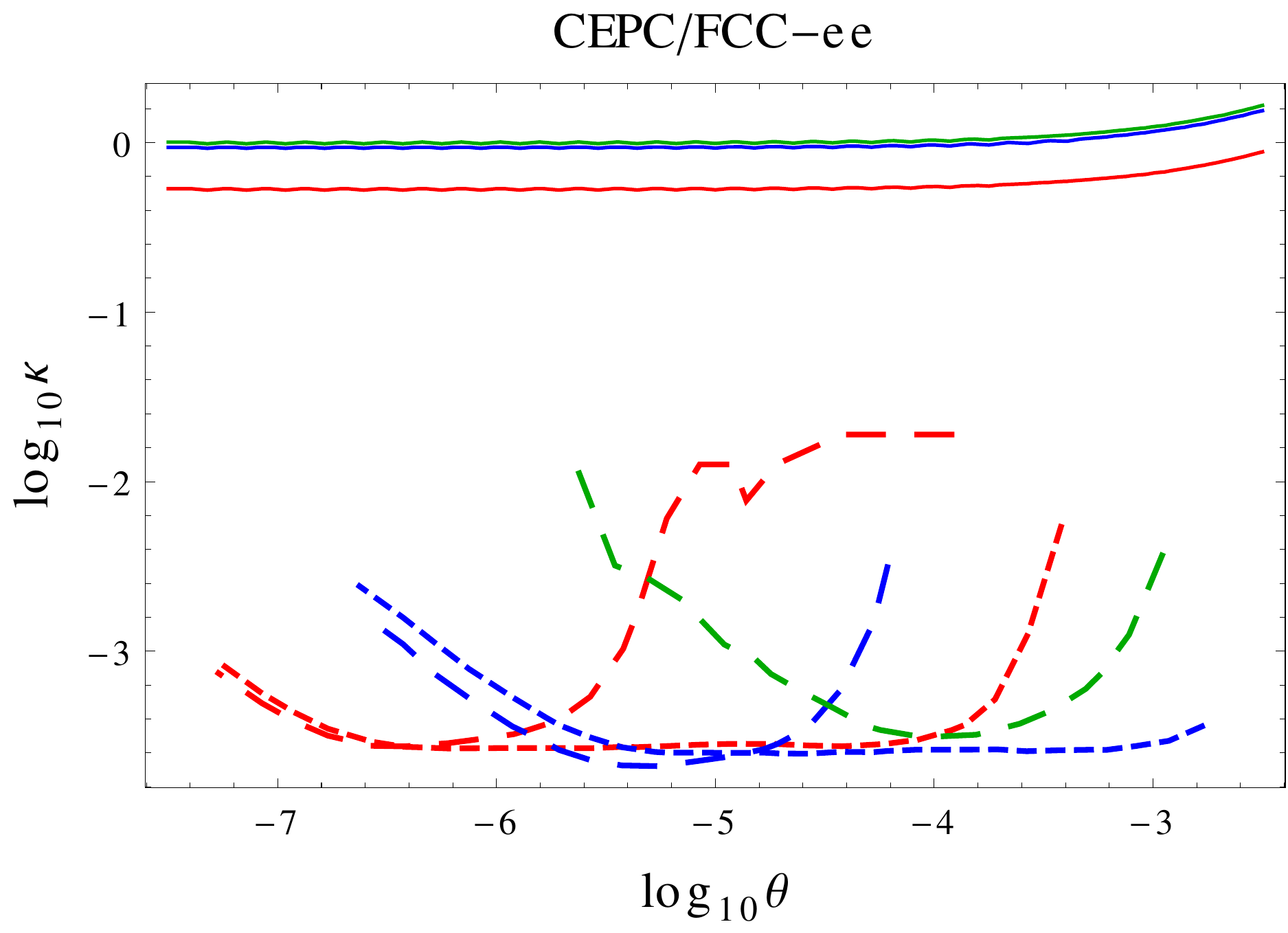}
	\caption{\label{fig:bounds} Projected 95\% limits on the Higgs portal Hidden Valley model in the $\kappa, \theta$ plane for three choices of $m_s$; green lines correspond to $m_s = 2.5$ GeV, blue to $m_s = 10$ GeV, and red to $m_s = 50$ GeV.  The solid lines are the projected lower limits from precision Higgs measurements, taking the CEPC projections \protect\cite{CEPCStudyGroup:2018ghi} for definiteness. The dashed lines are projected limits from this work, which are essentially identical for CEPC and FCC-ee.  Long dashes are from the `long lifetime' analysis and short dashes from the `large mass' analysis. }
\end{figure}

Constraints from a direct search for Higgs decays to LLPs and precision Higgs measurements as a function of the underlying parameters $\theta$ and $\kappa$ are shown in Figure \ref{fig:bounds} for the illustrative benchmarks $m_s = 2.5, 10$, and 50 GeV. Unsurprisingly, in the regime where $s$ is long-lived, the bounds from precision Higgs coupling measurements are modest and direct searches provide the leading sensitivity.

\subsection{Neutral naturalness}

Higgs decays to LLPs are also motivated by naturalness considerations, arising frequently in models of neutral naturalness that address the hierarchy problem with SM-neutral degrees of freedom \cite{Chacko:2005pe, Craig:2014aea}. In these models, partially or entirely SM-neutral partner particles that couple to the Higgs boson are charged under an additional QCD-like sector. Confinement in the additional QCD-like sector leads to a variety of bound states that couple to the Higgs and may be pair-produced in exotic Higgs decays with predictive branching ratios. The bound states with the same quantum numbers as the physical Higgs scalar typically decay back to the Standard Model by mixing with the Higgs. These decays occur on length scales ranging from microns to kilometers, making them a motivated target for LLP searches at colliders \cite{Craig:2015pha, Curtin:2015fna}. 

For simplicity, here we will restrict our focus to scenarios with the sharpest predictions for the Higgs branching ratio to LLPs. In these cases, the LLPs in question are typically glueballs of the additional QCD-like sector, of which the $J^{PC} = 0^{++}$ is typically the lightest. The coupling of the SM-like Higgs to these LLPs is predominantly due to top partner loops, for which the scales and couplings are directly related to the naturalness of the parameter space. In the Fraternal Twin Higgs \cite{Craig:2015pha}, the entirely SM-neutral fermionic partners of the top quark induce Higgs couplings to twin gluons, which then form glueballs; the $0^{++}$ states are the lightest in the twin QCD spectrum only if the other twin quarks are sufficiently heavy. In addition, there are tree-level deviations in Higgs couplings due to the pseudo-goldstone nature of the SM-like Higgs. In Folded SUSY \cite{Burdman:2006tz}, the scalar top partners carry electroweak quantum numbers, leading to radiative corrections to standard Higgs decays as well as the existence of exotic decay modes. Loops of the scalar top partners again induce Higgs couplings to twin gluons, and without light folded quarks the $0^{++}$ glueball is generically the lightest state in the folded QCD spectrum. While there are no tree-level Higgs coupling deviations in this case, the electroweak quantum numbers of the scalar top partners induce significant corrections to the branching ratio $h \rightarrow \gamma \gamma$. Finally, in the Hyperbolic Higgs \cite{Cohen:2018mgv} (see also \cite{Cheng:2018gvu}), the scalar top partners are entirely SM-neutral, and induce couplings to $0^{++}$ glueballs that are generically the lightest states in the hyperbolic QCD spectrum. As with the Fraternal Twin Higgs, however, there are also tree-level Higgs coupling deviations due to mass mixing among CP-even neutral scalars.

In each of these scenarios, the branching ratio of the SM-like Higgs can be parameterized as follows:
\begin{eqnarray} \nonumber
{\rm Br}(h \rightarrow 0^{++} 0^{++}) \approx  \left( 2 v^2 \frac{\alpha_s^\prime(m_h)}{\alpha_s(m_h)} \left[ \frac{y^2}{M^2} \right] \right)^2 \\ \times {\rm Br}(h \rightarrow gg)_{\rm SM} \times \sqrt{1 - \frac{4 m_0^2}{m_h^2}} 
\end{eqnarray}
Here $\alpha_s^\prime$ denotes the coupling of the additional QCD-like sector (whether twin, folded, or hyperbolic), which is necessarily of the same order as the SM QCD coupling $\alpha_s$, and $m_0$ is the mass of the glueball, which is determined in terms of the QCD-like confinement scale. Adopting the schematic notation of \cite{Curtin:2015fna}, the parameter $\left[ \frac{y^2}{M^2} \right]$ encodes the model-dependence of the Higgs coupling to pairs of gluons in the QCD-like sector, with
\begin{equation}
\left[ \frac{y^2}{M^2} \right] \approx \left \{ \begin{array}{ll} 
- \frac{1}{2v^2} \frac{v^2}{f^2} & {\rm ~Fraternal~Twin~Higgs} \\
\frac{1}{4v^2} \frac{m_t^2}{m_{\tilde t}^2} & {\rm ~Folded~SUSY} \\
\frac{1}{4 v^2} \frac{v}{v_{\mathcal{H}}} \sin \theta & {\rm ~Hyperbolic~Higgs}
\end{array} \right.
\end{equation}
For the Fraternal Twin Higgs, $f$ denotes the overall twin symmetry-breaking scale $f^2 = v^2 + v^{\prime 2} $ in terms of the SM weak scale $v$ and the fraternal weak scale $v^\prime$. For Folded SUSY, $m_{\tilde t}$ denotes the mass of the scalar top partners, neglecting possible mixing effects. For the Hyperbolic Higgs, $v_{\mathcal{H}}$ is the hyperbolic scale and $\tan \theta \approx \frac{v}{v_{\mathcal{H}}}$ encodes tree-level mixing effects. In each case, the scales appearing in the effective coupling are related to the fine-tuning of the model, drawing a direct connection between the Higgs exotic branching ratio and the naturalness of the weak scale.

In each case, the $0^{++}$ glueballs of the additional QCD-like sector decay back to the Standard Model by mixing with the SM-like Higgs, with a partial width to pairs of SM particles $Y$ given by
\begin{eqnarray} \nonumber
\Gamma(0^{++} \rightarrow YY) = \left( \frac{1}{12 \pi^2} \left[ \frac{y^2}{M^2} \right] \frac{v}{m_h^2 - m_0^2} \right)^2 \left(4 \pi \alpha_s^B F_{0^{++}}^S \right)^2 \\ \times \Gamma(h_{SM}[m_0] \rightarrow YY),
\end{eqnarray}
where $4 \pi \alpha_s^B F_{0^{++}}^S \approx 2.3 m_0^3$ and, as before, $h_{\rm SM}[m_0]$ denotes a Standard Model-like Higgs of mass $m_0$.

\begin{figure}
	\centering
	\includegraphics[width=\textwidth]{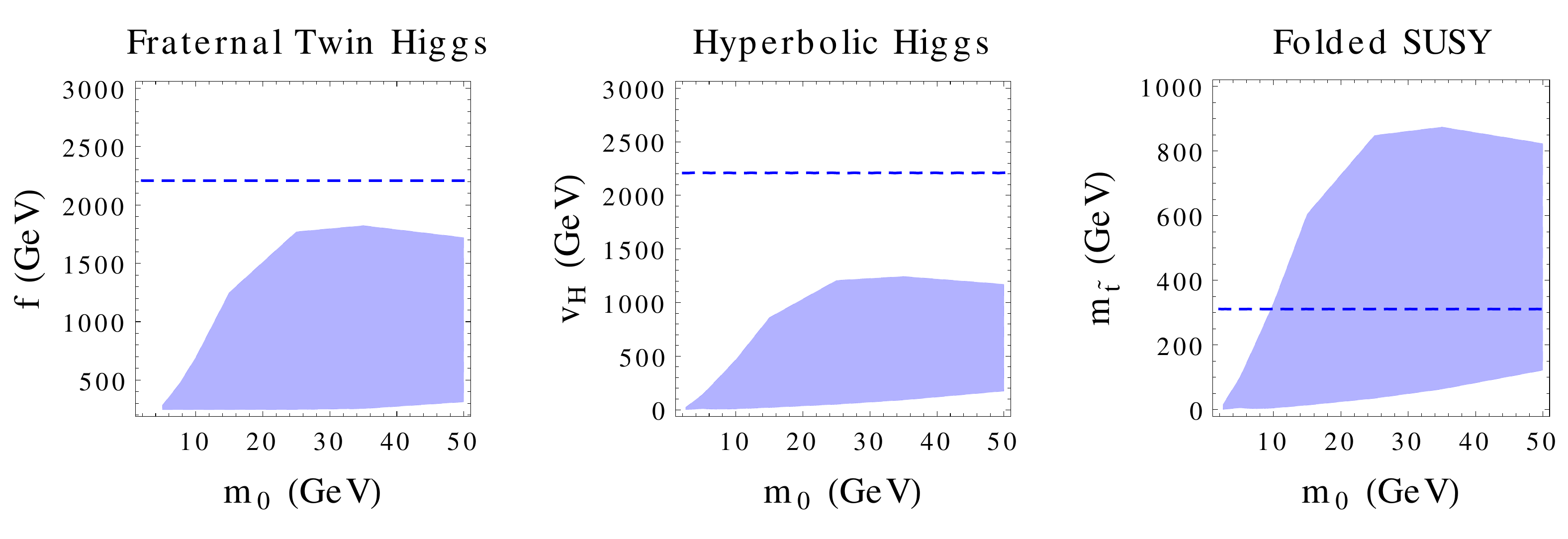}
	\caption{\label{fig:nnplot} Projected 95\% limits on the underlying scale as a function of the LLP mass $m_0$ in three models of neutral naturalness: the Fraternal Twin Higgs ($f$), the Hyperbolic Higgs ($v_{\mathcal{H}}$), and Folded SUSY ($m_{\tilde t}$). The blue dashed line denotes the limit coming from precision Higgs coupling measurements, taking for definiteness the CEPC projections from \protect\cite{CEPCStudyGroup:2018ghi}. For the Fraternal Twin Higgs and Hyperbolic Higgs, the dominant indirect constraint is from $\sigma_{Zh}$, while for Folded SUSY it is from ${\rm Br}(h \rightarrow \gamma \gamma)$. The shaded region denotes the projected limits from direct LLP searches obtained in this work.}
\end{figure}

Constraints on each model from a direct search for Higgs decays to LLPs and precision Higgs measurements are shown in Figure \ref{fig:nnplot} as a function of the LLP mass $m_0$ and the relevant scale ($f, v_\mathcal{H}$, and $m_{\tilde t}$, respectively).   For precision Higgs measurements we use the CEPC projections from \cite{CEPCStudyGroup:2018ghi}. In the Fraternal Twin Higgs and Hyperbolic Higgs, the dominant indirect constraint is from $\sigma_{Zh}$, while for Folded SUSY it is from ${\rm Br}(h \rightarrow \gamma \gamma)$. For both the Fraternal Twin Higgs and the Hyperbolic Higgs, tree-level Higgs coupling deviations make precision Higgs measurements the strongest test of the model. However, the sensitivity of LLPs searches provides valuable complementarity in the event that Higgs coupling measurements yield a discrepancy from Standard Model predictions. In particular, the size of an observed Higgs coupling deviation would single out the relevant overall mass scale ($f$ or $v_{\mathcal{H}}$), providing a firm target for LLP searches that would then validate or falsify these models as an explanation of the deviation. Note also that in the Fraternal Twin Higgs there may be additional contributions to the Higgs branching ratio into LLPs coming from the production of twin bottom quarks, which could lead to sensitivity in the LLP search comparable to that of Higgs couplings.  In the case of Folded SUSY, the absence of tree-level Higgs coupling deviations and the relatively weaker constraints on ${\rm Br}(h \rightarrow \gamma \gamma)$ make the LLP search the leading test of this model at Higgs factories.

\section{Conclusion} \label{sec:conclusions}

The exploration of exotic Higgs decays is an integral part of the physics motivation for future lepton colliders. New states produced in these exotic Higgs decays may themselves decay on a variety of length scales, necessitating a range of search strategies. While considerable attention has been devoted to the reach of future lepton colliders for promptly-decaying states produced in exotic Higgs decays, the reach for long-lived particles is relatively unexplored. 

In this paper we have made a first attempt to study the reach of proposed circular Higgs factories such as CEPC and FCC-ee (as well as approximate statements for the $\sqrt{s} = 250$ GeV run of the ILC) for long-lived particles produced in exotic Higgs decays, focusing on the pair production of LLPs and their subsequent decay to pairs of quarks. We have developed a realistic tracker-based search strategy motivated by existing LHC searches that entails the reconstruction of displaced secondary vertices. Rather than relying on existing public fast simulation tools, which do not necessarily give a sensible parameterization of signal and background efficiencies for long-lived particle searches, we have implemented a realistic approach to clustering and isolation. This allows us to characterize some of the leading irreducible Standard Model backgrounds to our search and determine reasonable analysis cuts necessary for a zero-background analysis. We obtain forecasts for the potential reach of CEPC and FCC-ee on the Higgs branching ratio to long-lived particles with a range of lifetimes. The projected reach is competitive with LHC forecasts and potentially superior for lower LLP masses and shorter lifetimes. In addition to our branching ratio limits, which may be freely interpreted in a variety of model frameworks, we interpret our results in the parameter space of a Higgs portal Hidden Valley and various incarnations of neutral naturalness, demonstrating the complementarity between direct searches for LLPs and precision Higgs coupling measurements.

There are a variety of directions for future work. While we have attempted to investigate some of the leading irreducible backgrounds and impose realistic cuts, we have not attempted to estimate possible backgrounds coming from cosmic rays; algorithmic, detector, or beam effects; or other contributions. Our tracker-based analysis has focused on Higgs decays to pairs of hadronically-decaying LLPs, but a comprehensive picture of exotic Higgs decays would also suggest the investigation of Higgs decays to various LLP combinations as well as the consideration of additional LLP decay modes.  Moreover, tracker-based searches for displaced vertices are but one of many possible avenues to discover long-lived particles. Analogous searches based on timing or on isolated energy deposition in outer layers of the detector (including either the electromagnetic or hadronic calorimeter, the muon chambers, or potentially instrumented volumes outside of the main detector) would be valuable for building a complete picture of LLP sensitivity across a range of lifetimes. 

More broadly, it is an ideal time to study the potential sensitivity of future Higgs factories to long-lived particles, as the results are likely to inform the design of detectors for these proposed colliders. This is a necessary step in motivating the physics case of future Higgs factories and ensuring that they enjoy optimal coverage of possible physics beyond the Standard Model.

\section{Introduction}\label{sec:intro}

At its heart, the electroweak hierarchy problem is a question of how an infrared (IR) scale can emerge from an ultraviolet (UV) scale without fine-tuning of UV parameters. Given the sensitivity of the Standard Model Higgs mass to UV scales, the expectation of effective field theory (EFT) is that the two should coincide. Conventional solutions to the hierarchy problem introduce both symmetries that control UV contributions to the Higgs potential and dynamics that generate IR contributions, leading to considerable structure at the weak scale and correspondingly sharp experimental tests. Ongoing exploration of the weak scale has given no evidence for these solutions, despite their theoretical soundness. 

In the face of increasingly powerful LHC data in excellent agreement with the Standard Model, it's worth taking seriously the possibility that Nature may be leading us to the conclusion that \textit{there is no new physics at the weak scale}. While this is often taken to suggest the existence of considerable fine-tuning in the Higgs potential, here we pursue an alternative idea. Perhaps the apparent violation of EFT expectations at the weak scale is a sign of the breakdown of EFT itself. We'll use the broad term `UV/IR mixing' to denote any effects that the UV has on low-energy physics which goes past that expected in EFT.

In this work we pursue the idea that such UV/IR mixing may have more direct effects on the SM by considering noncommutative field theory (NCFT) as a toy model. These theories model physics on spaces where translations do not commute \cite{Snyder:1946qz,Connes:1994yd}, and have many features amenable to a quantum gravitational interpretation---indeed, noncommutative geometries have been found arising in various limits of string theory \cite{Connes:1997cr,Douglas:1997fm,Seiberg:1999vs,Myers:1999ps}.\footnote{Noncommutative branes arising in gauge theory matrix models have also been found to contain emergent gravitational effects, and so have been suggested as novel quantum theories of gravity \cite{Rivelles:2002ez,Yang:2004vd,Yang:2006hj,Yang:2006mn,Steinacker:2007dq,Steinacker:2008ri,Steinacker:2008ya,Grosse:2008xr,Klammer:2008df,Steinacker:2009mp}. We do not pursue this perspective here, but refer the reader to \cite{Steinacker:2010rh} for a review of this approach.} 

This noncommutativity bears out the general expectation that the general-relativistic notion of spacetime should break down in a theory of quantum gravity \cite{DeWitt:1962cg}. Its realization here leads directly both to UV/IR mixing in the form of a violation of decoupling and to nonlocal effects in interactions. This gives rise to many interesting effects, but particularly fascinating for our purposes is that UV divergences present in the S-matrix elements of QFTs on commutative spaces can be transmogrified into \textit{new infrared poles} in the corresponding field theory on noncommutative space \cite{Minwalla:1999px}. An effective field theorist living in a noncommutative space would have no way to understand the appearance of this infrared scale; its existence is intrinsically linked to the geometry of spacetime and to the far UV of the theory. Such an effective field theorist would see a surprising lack of new physics accompanying this pole to explain its presence. 

It is clear from the outset that the direct application of NCFT to understand the hierarchy problem is immediately hindered by the Lorentz invariance violation which is inherent to these theories. Precisely how fatal this might be is not entirely clear; results regarding the extent to which `generic' Lorentz violation is empirically ruled out \cite{Collins:2004bp} are partly circumvented here by the fact that the Lorentz violation is \textit{not} generic, but comes as part of some larger structure. In this case the novel effects of UV/IR mixing in fact only appear in nonplanar loop diagrams \cite{Filk:1996dm} and care is required when interpreting EFT constraints on Lorentz violation---a point we will emphasize in Section \ref{sec:review}. Even so, it is difficult to imagine that observed properties of the weak scale and the wide range of constraints on Lorentz violation leave room for NCFT to be directly relevant to puzzles of the Standard Model.

Thus we make no claim about having solved the hierarchy problem. The value of this work is in the exploration of this toy model of UV/IR mixing, which possesses the intriguing feature that ultraviolet dynamics generate a scale whose lightness would be baffling to an effective field theorist. As this is the only model (of which we are aware) with this feature---and this feature, at the level of words, increasingly matches the experimental situation with the Higgs---it's worth understanding its appearance in as much detail as possible.

To make this work self-contained for the contemporary particle theorist, we begin with an  extensive introduction. In Section \ref{sec:review}, we review quantum field theory on noncommutative spaces with an emphasis on the violation of EFT expectations. In Section \ref{sec:phi4} we use this technology to go over the classic result of \cite{Minwalla:1999px} which first identified this emergent infrared pole in a Euclidean $\phi^4$ theory. We compute also the effect in dimensional regularization to evince the regularization-independence of the UV/IR mixing effects.

In Section \ref{sec:yukawa} we ask how general the effect of UV/IR mixing is within NCFT, which leads us to study noncommutative Yukawa theory in detail. We find that the scalar propagator again develops a new infrared pole at one loop, in contrast with previous work. Intriguingly, the pole in this case is accessible in $s$-channel scattering in the Lorentzian theory, making Yukawa theory a promising setting for probing phenomenological consequences of UV/IR mixing.

In Section \ref{sec:wesszumino} we upgrade our model to the softly-broken Wess-Zumino model to study the interplay between UV-finiteness and UV/IR mixing effects. When the fermion is kept in the spectrum of the theory below the cutoff, the lack of UV sensitivity of the field theory removes the light pole. As the fermion is taken above the cutoff, an effective theorist again sees effects past those observed in Wilsonian EFT. These results are expected, but this model affords us a concrete demonstration that UV/IR mixing can only have interesting low-energy effects if the field theory is UV sensitive, and puts this naturalness strategy in stark contrast to conventional approaches. Of course, this also makes addressing the hierarchy problem with UV/IR mixing a potentially Pyrrhic victory: to generate an IR scale, the field theory alone cannot be fully predictive.

Finally, in Section \ref{sec:lessons} we examine the appearance of the emergent light pole in NCFT from more general arguments, so as to ascertain the relative importance of nonlocality and Lorentz-violation for these effects. The conclusion is inevitably that in this case the two are inexorably linked, and no strong conclusion about the possibility of finding a light pole in a theory with only one or the other is available. However, we provide some direction toward future explorations into both of these possibilities. We wrap up in Section \ref{sec:conclusions}.

\section{Noncommutative Field Theory}\label{sec:review}
In this section we review the salient features of the formulation of noncommutative field theories and the standard formalism for studying their perturbative physics. Useful general references for this background include \cite{Szabo:2001kg,Douglas:2001ba}. Readers familiar with NCFT may wish to skip to Section \ref{sec:phi4}, but we emphasize that our interest is necessarily non-perturbative in the parameter controlling the noncommutativity, unlike much of the earlier phenomenological literature.

Physics on noncommutative spaces involves the introduction of a nonzero commutator between position operators
\begin{equation}\label{eqn:ncdef}
\left[\hat{x}_\mu,\hat{x}_\nu\right] = i \theta_{\mu\nu},
\end{equation}

\noindent where we will refer to $\theta_{\mu\nu} = - \theta_{\nu\mu}$ as the noncommutativity tensor, and we emphasize that it is covariant under Lorentz transformations. So while it does break Lorentz invariance, it only does so in the way that turning on a magnetic field in your lab chooses a preferred frame, and it can indeed be thought of as simply a background field. This basic definition is reminiscent of the introduction of a nonzero commutator in passing from classical mechanics to quantum mechanics. Indeed much of the structure is precisely analogous, including importantly the construction of noncommutative versions of familiar commutative theories via a quantization map. At an even more basic level, the above nonzero commutator induces an uncertainty relation
\begin{equation}
\Delta \hat{x}_\mu \Delta \hat{x}_\nu \geq \frac{\left|\theta_{\mu\nu}\right|}{2},
\end{equation}

\noindent which immediately makes apparent the presence of UV/IR mixing in this theory. If you attempt to create a wavepacket which is very small in one direction it will necessarily be elongated in another, and so we see already the non-trivial mixing of UV and IR modes. This clearly violates the separation of scales which is baked in to EFT. Thus purely from the defining relation of noncommutative geometry, we see already an indication that noncommutative theories should violate EFT expectations. 

Field theories on this space may be conveniently formulated in terms of fields that are functions of {\it commuting} coordinates imbued with a new field product, known as a Groenewold-Moyal product (or star-product), with position-space representation
\begin{equation}\label{eqn:starprod}
f(x) \star g(x) = \left. \exp\left(\frac{i}{2} \theta_{\mu\nu} \partial_y^\mu \partial_z^\nu\right) f(y) g(z) \right|_{y = z = x} = f(x) \exp\left(\frac{i}{2} \overleftarrow{\partial}^\mu \theta_{\mu\nu} \overrightarrow{\partial}^\nu\right) g(x).
\end{equation}
We derive this procedure in Appendix \ref{sec:NCFTderive}. It is important to observe that this is a nonlocal product, since it contains an infinite series of derivative operators. So we see again that one of the tenets of EFT has been violated by our basic definition of field theory on noncommutative spaces.

With this in hand we may now write down noncommutative versions of familiar theories \textit{in terms of commuting coordinates}, which will then allow us to use normal QFT methods to analyze them. First note that this noncommutative quantization will not affect the quadratic part of the tree-level action due to momentum conservation and the antisymmetry of the noncommutativity tensor. For the interacting part of the action the effects of noncommutative quantization are not so trivial, but are easy to analyze classically. As an example, for a simple $\phi^n$ theory we find
\begin{equation}
\mathcal{L}^{(NC)}_\text{int} = \frac{\lambda}{n!} \overbrace{\phi(x)\star\phi(x)\star\cdots\star\phi(x)}^{\text{n copies}}.
\end{equation}

Note, importantly, that the star-product has endowed our vertices with a notion of ordering, as it is only cyclically invariant. If we now Fourier transform the action to momentum space, we find that we can account for the effects of quantization on the tree-level action with a simple modification of the momentum-space vertex factor:
\begin{equation}
\tilde{V}\left(k_1,\dots,k_n\right) = \delta\left(k_1 + \dots + k_n\right)\exp\left(\frac{i}{2} \sum\limits_{i<j}^{n}k_i^\mu k_j^\nu \theta_{\mu\nu}\right).
\end{equation}

A word of caution is in order. We can now express the action in momentum space as 
\begin{equation}
\mathcal{S}^{(NC)}_\text{int} = \frac{\lambda}{n!}\int \left(\prod_{i}^{n} \text{d}^4k_i \right)\delta\left(k_1 + \dots + k_n\right)\phi(k_1)\phi(k_2)\dots\phi(k_n)\exp\left(\frac{i}{2} \sum\limits_{i<j}^{n}k_i^\mu k_j^\nu \theta_{\mu\nu}\right),
\end{equation}
and so---as good effective field theorists---we may be tempted to expand the exponential for small momenta $\sim \left|k^2\right|\left|\theta\right| \ll 1$. Indeed, doing so would give us a series of irrelevant operators which would correct the leading interaction. However, once the theory is truncated at some finite order in $\theta$, we are left with a perfectly local EFT. In other scenarios where an infinite series of operators appears, this is a valid approximation procedure and allows one to calculate the leading corrections a theory predicts. But here our definition of NCFT introduces UV/IR mixing which we expect to violate EFT expectations. Truncating the series removes these effects entirely, and a theory so defined no longer has anything to do with NCFT---at least not in the effects we will be interested in, which are nonperturbative in $\theta$ as we shall see explicitly in the following sections. There has been much work expended on these `noncommutative-inspired' theories, but they do not contain UV/IR mixing, and do not capture the most striking and most interesting features of physics on a noncommutative space, from our perspective.\footnote{We are not the first to issue a warning of this sort---see e.g. \cite{AmelinoCamelia:2002au,Khoze:2004zc} in the context of connecting noncommutativity to the real world, and \cite{Tomboulis:2015gfa} which discusses the general case of nonlocal interactions.}

With that in mind, we may now proceed to do perturbative quantum field theory calculations, but we must worry about keeping track of all the phases from each of the vertices. In fact there is another simplification that occurs, as found by Filk \cite{Filk:1996dm}, which allows us to simplify the process of finding the phase factor for a diagram to a graph-topological statement. Filk proved two simple rules for the phase factors:
\begin{enumerate}
	\item An internal line which ends on two different vertices can be contracted while keeping the ordering of the other lines fixed. 
	\begin{equation}
	\tilde{V}\left(k_1,\dots,k_{n_1},p\right)\tilde{V}\left(-p,k_{n_1+1},\dots,k_{n_2}\right) = \tilde{V}\left(k_1,\dots,k_{n_2}\right)\delta(k_1 + \dots + k_{n_1} + p)
	\end{equation} 
	\item A loop which doesn't cross any lines can be eliminated. Note that the fixed ordering of the lines at a vertex means that we can now meaningfully speak of lines which do or don't cross each other. 
	\begin{equation}
	\tilde{V}\left(k_1,\dots,k_{n_1},p,k_{n_1+1},\dots,k_{n_2},-p\right) = \tilde{V}\left(k_1,\dots,k_{n_1},k_{n_1 + 1},\dots,k_{n_2}\right) \quad \text{if } \sum\limits_{i = n_1 + 1}^{n_2}k_i = 0
	\end{equation}
\end{enumerate}

The proof of these facts relies only on the antisymmetry of $\theta^{\mu\nu}$ and the fact that each vertex contains a momentum-conserving delta function. We may make use of this to simply find the phase factor of any Feynman diagram. Using the first rule, we can reduce any diagram to a single vertex, which is a rosette of the external lines and closed loops. The second rule allows us to eliminate loops which don't cross other lines. 

If the graph was planar (including, importantly, any tree-level graph), then by definition all loops can be eliminated. So all contributions to phase factors from internal lines cancel, and we're only left with an overall phase corresponding to the ordering of the external lines, which has remained fixed throughout the reduction process.

For a nonplanar graph, in this representation it is easy to see that we only pick up phase factors from lines which cross. The loop gives vertex legs with $\pm p^\mu$, and for an external line which doesn't cross this loop, both loop legs will be on the same side of it in the cyclic ordering, and so the two terms will cancel in the sum. Only for an external line which crosses it are the $\pm p$ on different sides, and so the antisymmetry of $\theta$ will make the two negative signs cancel to give a coherent phase for this vertex. Thus we define $I_{ij}$, the intersection matrix of an oriented graph: 
\begin{equation}
\begin{split}
I_{ij} = \begin{cases}
1& \text{line \emph{j} crosses \emph{i} from right} \\
-1 & \text{line \emph{j} crosses \emph{i} from left} \\
0 & \text{line \emph{j} does not cross \emph{i}}
\end{cases}
\end{split}
\end{equation}

\noindent Then for any graph $\mathcal{G}$, the contribution $\Gamma(\mathcal{G})$ of the phase factors is just 
\begin{equation}
\Gamma(\mathcal{G}) = \tilde{V}\left(\left\lbrace\text{external momenta}\right\rbrace\right)\times\exp\left(\frac{i}{2} \sum\limits_{ij}I_{ij} k_i \wedge k_j \right),
\end{equation}
where we've defined $k_i \wedge k_j \equiv k_i^\mu \theta_{\mu\nu}  k_j^\nu$.

In what follows we will omit the overall external phase when evaluating diagrams, as it will not be important for our purposes. We have now simplified perturbative field theory on noncommutative spaces down to the simple task of marking line-crossings, at least at the level of writing down integrands of amplitudes. The triviality of this task for tree-level graphs leads to the interesting feature that tree-level amplitudes on noncommutative spaces are the same as on commutative manifolds, and it is only at loop-level that we find deviations. We will see in the next section that the loop integration will bring surprising features. 

An important issue for the interpretation of NCFTs is that of their unitarity. There is no problem in Euclidean space, 
but for Lorentzian spacetimes with noncommutativity in the time directions (`timelike' or `space-time' noncommutativity when $- k^\mu \theta_{\mu\rho} \theta^{\rho \nu} k_\nu \equiv k \circ k < 0$ is allowed), one may find a breakdown of unitarity by taking cuts of one-loop diagrams \cite{Gomis:2000zz,Bassetto:2001vf}.\footnote{Though it is interesting to note that the special case of `lightlike' noncommutativity is also unitary \cite{Aharony:2000gz,SheikhJabbari:2010nc}.} This may be interpreted physically as being due to the production of tachyonic states, which if added to the Fock space of the theory result in a formal restoration of the cutting relations whilst making the nonunitarity explicit \cite{AlvarezGaume:2001ka}. 

This failure of unitarity is well-understood from the stringy perspective. Spatial noncommutativity appears from a background magnetic field and the field theory limit to a spacelike NCFT is smooth \cite{Seiberg:1999vs}. In the case of timelike noncommutativity, however, approaching the field theory limit forces an electric field to supercritical values whence pair-production of charged strings destabilizes the vacuum \cite{Seiberg:2000ms}. Study of string theories with timelike noncommutativity (e.g. `noncommutative open string theory' \cite{Seiberg:2000ms,Gopakumar:2000na}) is outside our scope, but there are at least some hints of similar UV/IR mixing effects as those in the NCFT \cite{Torrielli:2002ev}. We note in passing that there are further interesting connections between NCFTs and string theories---not only do particles on noncommutative spaces act in many ways like rods of size $L \sim p \theta$ (see e.g. \cite{SheikhJabbari:1999vm,Bigatti:1999iz,Seiberg:2000gc,Girotti:2001dh,Acatrinei:2002sb}), mimicking the behavior of extended objects, but there have been many hints in the spacelike theories that the curious IR effects in the NCFT are reproducing effects from closed strings, despite the fact that these have been decoupled (e.g. \cite{Minwalla:1999px,Arcioni:2000bz,Rajaraman:2000dw,Fischler:2000fv,VanRaamsdonk:2000rr,Kiem:2000wt,Armoni:2001uw,Torrielli:2002ev,Armoni:2003va,Lopez:2003uq}).

Within the realm of field theory, there have long been suggestions that this difficulty is pointing to the need for a modified definition of quantum field theories on timelike noncommutative spaces (for some early references, see \cite{Gomis:2000gy,Bahns:2002vm,Bozkaya:2002at,Liao2002,Rim:2002if,Denk:2003jj,Fischer:2003jh,Liao:2004sw}). From this perspective, the issue is that such field theories are non-local in time, which renders nonsensical the normal time-ordering involved in the perturbative Dyson series (at the least). That is, our effective definition of these theories above via the diagrammatic expansion may be too na\"{i}ve. An interesting line of work is to formulate a modification of the standard quantum field theory machinery to non-local-in-time theories which avoids the unitarity issue by construction. We note that the same UV/IR mixing effects of interest in the two-point function have been seen to persist in at least some of these approaches (e.g. \cite{Bozkaya:2002at}). For some recent work on the formulation and properties of nonlocal field theories, see e.g. \cite{Barnaby:2007ve,Salminen:2011ut,Biswas:2014yia,Tomboulis:2015gfa, Addazi:2015dxa,Chin:2018puw}. 

Below we will begin in Euclidean space, where $k \circ k \geq 0$ is guaranteed for any $\theta_{\mu\nu}$, but will then venture into Lorentzian signature. All of our calculations and the general features we find, including finding new infrared poles, will hold robustly in spacelike noncommutative theories. However we will comment also on how these features are modified when timelike noncommutativity is turned on, taking license from the aforementioned hints that unitary completions/reformulations of timelike NCFT may retain the UV/IR mixing exhibited in the na\"{i}ve approach. 

\section{Real Scalar \texorpdfstring{$\phi^4$}{Phi4} Theory}\label{sec:phi4}
In this section we review the perturbative physics of the noncommutative real scalar $\phi^4$ theory at one loop, which was first studied in detail by Minwalla, Van Raamsdonk, and Seiberg in \cite{Minwalla:1999px}.\footnote{Some early results in this model may also be found in \cite{Arefeva:1999gex,Chepelev:1999tt}.}

In four Euclidean dimensions the action on noncommutative space becomes
\begin{equation}
S = \int \text{d}^4x \left(\half \partial_\mu \phi \partial^\mu \phi + \half m^2 \phi^2 + \frac{g^2}{4!} \phi\star\phi\star\phi\star\phi\right),
\end{equation}
where we have already used the fact that the quadratic part of the noncommutative action is the same as the commutative theory to eliminate the star product there.  Our object of interest will be the one-loop correction to the two-point function. In the commutative theory this is given by a single Feynman diagram, but the noncommutative theory contains both a planar diagram and a nonplanar diagram. 

\begin{equation*}
- \Gamma^{(2)}_1 = \includegraphics[width=0.25\linewidth,valign=c]{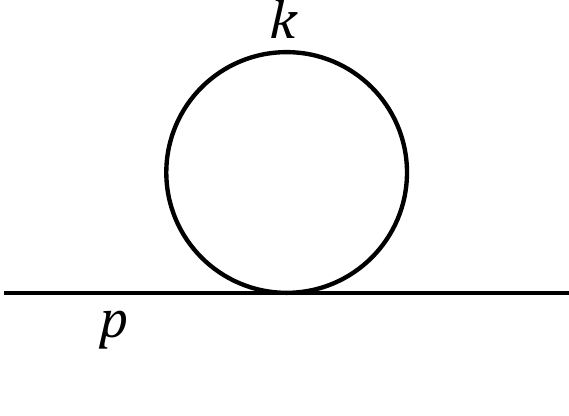} + \includegraphics[width=0.25\linewidth,valign=c]{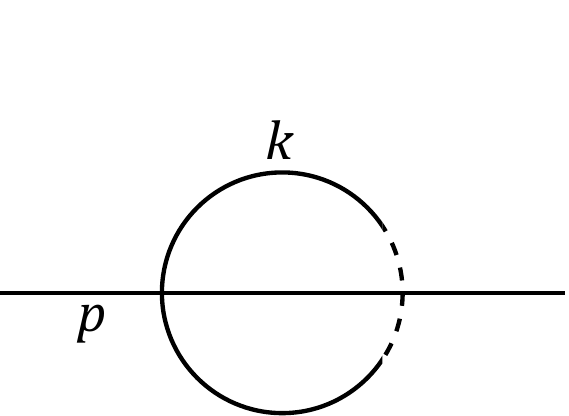}
\end{equation*} 

\noindent The expressions for these two diagrams now differ---not only in symmetry factor but also due to the phase in the integrand. We find
\begin{align}
\begin{split}
\Gamma^{(2)}_{1,\text{planar}} &= \frac{g^2}{3\left(2\pi\right)^4} \int \frac{\text{d}^4k}{k^2 + m^2} \\
\Gamma^{(2)}_{1,\text{nonplanar}} &= \frac{g^2}{6\left(2\pi\right)^4} \int \frac{\text{d}^4k}{k^2 + m^2} e^{i k^\mu \theta_{\mu\nu} p^\nu}.
\end{split}
\end{align}
We may already see that something interesting should happen, as in the nonplanar diagram the phase mixes the internal and external momenta. One may intuit that the rapidly oscillating phase in the UV of the loop integration will dampen the would-be divergence, and indeed we will see that nonplanar diagrams are finite. However, unlike in the case where the vertex factor vanishes rapidly for large Euclidean momenta and so ensures UV-finiteness \cite{Chin:2018puw}, here the damping is in some sense `marginal'. This fact will be responsible for the interesting feature we will find presently. 

The simplest method to evaluate noncommutative diagrams is to use Schwinger parameters, recalling the identity $\frac{1}{k^2+m^2} = \int_{0}^{\infty}\text{d}\alpha \ e^{-\alpha\left(k^2 + m^2\right)}$. The presence of the phase in the nonplanar diagram means we must complete the square before going to spherical coordinates to get a Gaussian integral. This means that after the momentum integrals we end up with 
\begin{align}\label{eqn:phi4gaussianint}
\begin{split}
\Gamma^{(2)}_{1,\text{planar}} &= \frac{g^2}{48\pi^2} \int \frac{\text{d}\alpha}{\alpha^2} e^{-\alpha m^2} \\
\Gamma^{(2)}_{1,\text{nonplanar}} &= \frac{g^2}{96\pi^2} \int \frac{\text{d}\alpha}{\alpha^2} e^{-\alpha m^2-\frac{p\circ p}{4\alpha}}
\end{split}
\end{align}
where again $p \circ q = - p^\mu\theta^2_{\mu\nu} q^\nu$. Moving to Schwinger space trades large-$k$ divergences for small-$\alpha$ divergences, which we now smoothly regulate by multiplying the integrands by $\exp\left(-1/(\Lambda^2 \alpha)\right)$ so that the small $\alpha$ region will be driven to zero. Note that a term of this form already exists in the expression for the nonplanar diagram. After introducing the regulator, we can evaluate the integrals to find 
\begin{align}
\begin{split}
\Gamma^{(2)}_{1,\text{planar}} &= \frac{g^2}{48\pi^2} \left(\Lambda^2 - m^2 \log\left(\frac{\Lambda^2}{m^2}\right) + \mathcal{O}(1)\right) \\
\Gamma^{(2)}_{1,\text{nonplanar}} &= \frac{g^2}{96\pi^2} \left(\Lambda_\text{eff}^2 - m^2 \log\left(\frac{\Lambda_\text{eff}^2}{m^2}\right) + \mathcal{O}(1)\right),
\end{split}
\end{align}
where we've defined 
\begin{equation}
\Lambda_\text{eff}^2 \equiv \frac{1}{1/\Lambda^2 + p \circ p/4},
\end{equation}
which is the effective cutoff of the nonplanar diagram. 

The first thing to note is that it seems the UV divergence of the nonplanar diagram has disappeared---the graph is finite in the limit $\Lambda\rightarrow \infty$, and so appears to have been regulated by the noncommutativity of spacetime. In fact the effect is more subtle, as alluded to earlier, and now the UV and IR limits of this amplitude do not commute. If we first take an infrared limit $p\circ p \rightarrow 0$ we find that $\Lambda_{\text{eff}} \rightarrow \Lambda$ and the ultraviolet divergence of the commutative theory reappears. If we take the UV limit $\Lambda \rightarrow \infty$ first we find an IR divergence $\frac{1}{p \circ p}$, so the noncommutativity has transmogrified the UV divergence into an IR one.\footnote{We note here that the failure of a `correspondence principle' between commutative and noncommutative theories as $\theta^{\mu\nu} \rightarrow 0$ is clearly intrinsically linked to the appearance of UV/IR mixing. This failure doesn't violate Kontsevich's proof of the existence of deformation quantization for any symplectic manifold \cite{Kontsevich:1997vb}, as that is confined solely to `formal' deformation quantization---that is, the production of a formal power series expansion of the algebra of observables in terms of the deformation parameter. As was noted in Section \ref{sec:review} and is now on prime display, the physics of the theory with nonperturbative $\theta$-dependence is starkly different from that of any truncation.}

Turning to the question of renormalizability, one may na\"{i}vely ask if we can absorb all UV divergences into a finite number of counterterms. Under this criterion, it is clear that this procedure works in the noncommutative theory at least when the commutative version is renormalizable. In the current case, we may absorb the UV divergences of this correction to the two-point function into a redefinition of the physical mass, $M^2 = m^2 + \frac{g^2 \Lambda^2}{48 \pi^2} - \frac{g^2 m^2}{48 \pi^2}\log\frac{\Lambda^2}{m^2}$, and so write down a one-particle irreducible quadratic effective action which has a finite $\Lambda\rightarrow\infty$ limit: 
\begin{align} \label{eqn:1PIaction}
S^{(2)}_{1\text{PI}} = \int \frac{\text{d}^4p}{(2 \pi)^4} \ \half \Bigg( p^2 + M^2 + &\frac{g^2}{96\pi^2\left(\frac{p\circ p}{4} + \frac{1}{\Lambda^2}\right)}  \\
 - &\frac{g^2 M^2}{96 \pi^2} \log \frac{1}{M^2\left(\frac{p \circ p}{4} + \frac{1}{\Lambda^2}\right)} + \dots + \mathcal{O}(g^4)\Bigg)\phi(p)\phi(-p).\nonumber
\end{align}

However, in the $\Lambda \rightarrow \infty$ limit one finds that at one loop the propagator now has \textit{two} poles. The first is a standard radiative correction to the free pole, but the second has appeared ex nihilo at one loop:
\begin{align}
\begin{split}
p^2 &= - m^2 + \mathcal{O}(g^2) \\
p \circ p &= - \frac{g^2}{24 \pi^2 m^2} + \mathcal{O}(g^4),
\end{split}
\end{align}
where we have assumed that $\theta^{\mu\nu}$ is full rank. The former is to be interpreted as the on-shell propagation of the particles associated to our fundamental field $\phi$. If $\theta^{\mu\nu}$ has only one eigenvalue $1/\Lambda_\theta^2$---with $\Lambda_\theta$ thought of as the scale associated with the breakdown of classical geometry---we have $p \circ p = \frac{p^2}{\Lambda_\theta^4}$. We see that the new pole appears at $p^2 \propto g^2 \frac{\Lambda_\theta^4}{m^2}$, and so if our field $\phi$ lives in the deep UV of the theory, our new pole appears at parametrically low energy scales. To the extent that poles are particles, we appear to have generated a new light particle from ultraviolet dynamics.

The interpretation of the new pole can be sharpened by considering more carefully the criteria for renormalizability in Wilsonian EFT. In a Wilsonian picture, we upgrade our Lagrangian parameters to running parameters, and define our theory at the scale $\Lambda$ as 
\begin{equation}
S_{Wilson}(\Lambda) = \int \text{d}^4x \left(\half Z(\Lambda) \partial_\mu \phi \partial^\mu \phi + \half Z(\Lambda) m^2(\Lambda) \phi^2 + \frac{Z^2(\Lambda) g^2(\Lambda)}{4!} \phi\star\phi\star\phi\star\phi\right).
\end{equation}
It is immediately apparent from the above calculation that we cannot write the action at a lower scale $\Lambda_0 < \Lambda$ in this same form by choosing appropriate definitions for $Z(\Lambda),m(\Lambda),g(\Lambda)$---there's nowhere to put the $\frac{1}{p \circ p}$ term!\footnote{There has been much work on understanding renormalizability of NCFTs, especially with an eye toward finding a mathematically well-defined four-dimensional quantum field theory with a non-trivial continuum limit. Renormalizability has been proven for modifications of NCFTs where the free action is supplemented by an additional term which adjusts its long-distance behavior. Such an action is manufactured either by requiring it manifest `Langman-Szabo' duality \cite{Langmann:2002cc} $p_\mu \leftrightarrow 2 (\theta^{-1})_{\mu\nu}x^\nu$ \cite{Grosse:2004yu,Grosse:2012uv} or by adding a $1/p\circ p$ term to the free Lagrangian \cite{Gurau:2008vd}, the latter of which directly has the interpretation of adding `somewhere to put the $1/p\circ p$ counterterm'. For recent reviews of these and related efforts we refer the reader to \cite{Grosse:2016yjo,Ydri:2016dmy}. It would be interesting to understand fully the extent to which the physics of these schemes agrees with the interpretation of the IR effects as coming from auxiliary fields \cite{Minwalla:1999px,VanRaamsdonk:2000rr}.}	

Stated more precisely, for Wilsonian renormalizability we require that we can define the running couplings such that correlation functions computed from this action converge uniformly to their $\Lambda \rightarrow \infty$ limits. However, this requirement is flatly violated by the noncommutation of the UV and IR limits of the diagrams. For any finite value of $\Lambda$, the effective action of Equation \ref{eqn:1PIaction} differs significantly from its limiting value for small momenta $p \circ p \ll \frac{1}{\Lambda^2}$. This is the precise sense in which the violation of Wilsonian EFT appears in this one-loop correction.

This brings up the question of how an effective field theorist would describe the universe if they unknowingly lived on a noncommutative space. A consistent Wilsonian interpretation can be regained by including a degree of freedom which can absorb the new infrared dynamics of the quadratic effective action. Since we need this to involve the $\phi$ momentum, this new particle must mix linearly with the $\phi$ field. We manufacture its tree-level Lagrangian such that the problematic inverse $p \circ p$ term in the quadratic effective action of $\phi$ is replaced with its $\Lambda \rightarrow \infty$ value for all values of $\Lambda$, to satisfy our precise condition for Wilsonian renormalizability. To see how this works, we add to our tree level Wilsonian action
\begin{equation}\label{eqn:auxaction}
\Delta S(\Lambda) = \int \text{d}^4x \left(\half \partial\chi \circ \partial\chi + \half \frac{\Lambda^2}{4} (\partial \circ \partial \chi)^2 + i \frac{1}{\sqrt{24 \pi^2}} g \chi \phi\right).
\end{equation} 
Since $\chi$ appears quadratically, we may integrate it out exactly at tree level to find a contribution to the effective action 
\begin{equation}
\Delta S_{1\text{PI}}(\Lambda) = \int \frac{\text{d}^4p}{(2\pi)^4} \half \left(-\frac{g^2}{96\pi^2\left(\frac{p\circ p}{4} + \frac{1}{\Lambda^2}\right)} + \frac{g^2}{24\pi^2 p\circ p}\right)\phi(p)\phi(-p)
\end{equation}

This precisely subtracts off the problematic term in the original 1PI quadratic effective action and adds back its $\Lambda \rightarrow \infty$ limit, as we had wanted. Ignoring the logarithmic term,\footnote{Discussion of the interpretation of logarithmic singularities as being due to auxiliary fields propagating in extra dimensions may be found in \cite{VanRaamsdonk:2000rr}.} we are left with an effective action which is manifestly independent of the cutoff $\Lambda$, and so satisfies our criterion for Wilsonian renormalizability.\footnote{In Equation \ref{eqn:auxaction}, the four-derivative quadratic action of the auxiliary field can be rewritten as two fields with two-derivative actions, one of which is of negative norm and may be thought of as the `Lee-Wick partner' of the positive norm state \cite{Lee:1969fy}, viz.
\begin{equation}
\mathcal{L} = \half \partial \chi' \circ \partial \chi' - \half \partial \tilde{\chi} \circ \partial \tilde{\chi} - \half \frac{4}{\Lambda^2} \tilde{\chi}^2 + i \frac{1}{\sqrt{24 \pi^2}} g\left(\chi' - \tilde{\chi}\right)\phi , \qquad \chi' \equiv \chi + \tilde{\chi} 
\end{equation}
One may then wonder if the lightness of the new IR pole may be understood through the regularization performed by the Lee-Wick field, as is done for the Higgs in the `Lee-Wick standard model' \cite{Grinstein:2007mp}. However, in that theory the Higgs is kept light because every particle comes with a Lee-Wick partner, and so all diagrams contributing to corrections to the Higgs mass are made finite. The presence of the Higgs' Lee-Wick partner alone is not enough to keep it light. Here, the lightness of $\chi$ can be understood diagrammatically as being simply due to the fact that its only interaction is linear mixing with $\phi$, and so any correction to its two-point function is absorbed into that of the two-point function of $\phi$. A further issue with the Lee-Wick rewriting is that the seeming perturbative unitary of the theory is normally guaranteed by the Lee-Wick partner being heavy and unstable. But as we take the $\Lambda\rightarrow\infty$ limit in our Wilsonian action, we see that the Lee-Wick partner becomes massless as well, in accordance with the result that this theory is non-unitary \cite{Gomis:2000zz}.} We discuss the generalization of this procedure in Appendix \ref{app:auxfield}.

Now while we have written down an action which identifies the new observed IR pole with a field and in doing so gives our effective action a Wilsonian interpretation, the extent to which $\chi$ can be taken seriously as a fundamental degree of freedom is unclear.\footnote{We note that in matrix models containing dynamical noncommutative geometries it has been argued that emergent infrared singularities should be associated with the dynamics of the geometry (see e.g. \cite{VanRaamsdonk:2001jd,Steinacker:2010rh}). As our field theories are formulated on fixed noncommutative backgrounds, this interpretation is unavailable to us.} The new pole is inaccessible in Euclidean space---so one does not immediately conclude there is a tachyonic instability---and relatedly, when we na\"{i}vely analytically continue this result to Lorentzian spacetime this new pole is inaccessible in the $s$-channel.\footnote{Note that this peculiar connection regarding (in)accessibility is due to the Lorentz violation.   While the normal pole which is inaccessible in Euclidean signature becomes accessible for timelike momenta in Lorentzian signature, the Wick rotation affects the noncommutative momentum contraction differently. When taking $x_4 \rightarrow - i x_0$, one also rotates $\theta_{4\nu} \rightarrow -i \theta_{0\nu}$ such that Equation \ref{eqn:ncdef} continues to hold for the same numerical $\theta_{\mu\nu}$. For the simplest configuration of full-rank noncommutativity with $\theta_{\mu\nu}$ block-off-diagonal and only one eigenvalue $1/\Lambda_\theta^2$, the Euclidean $p \circ p = p^2/\Lambda_\theta^4$ becomes a Lorentzian $p \circ p = (p_0^2 - p_1^2 + p_2^2 + p_3^2)/\Lambda_\theta^4$. So a noncommutative pole which is inaccessible in the Euclidean theory becomes accessible in the Lorentzian theory for spacelike momenta, while a noncommutative pole which can be accessed in the Euclidean theory becomes accessible in the $s$-channel in Lorentzian signature.} However, its presence is still enough to break unitarity for this theory \cite{Gomis:2000zz}, and in fact may still be interpreted as being due to the presence of tachyons \cite{AlvarezGaume:2001ka}. As discussed in Section \ref{sec:review}, it is possible this may be resolved if analytical continuation is adjusted for nonlocal-in-time theories, or it may be that a UV theory cures this apparent violation. 

Separately, it is not obvious much has been gained by attributing the new pole to a new, independent field, past acting as a formal tool to regain a notion of renormalizability. Since the only interaction of $\chi$ above is linear mixing, its action is not renormalized---any divergences are instead absorbed into the running of $\phi$ parameters---and so no interactions are generated. Furthermore one is obstructed from integrating out the heavy field $\phi$ to come up with an effective action of $\chi$ at low energies by the fact that the kinetic terms of $\chi$ are non-standard, which prevents diagonalization of the quadratic terms in the Lagrangian. Thus it seems it is intrinsically linked with the heavy scalar which begat it.


There are further obstructions to asking that this specific mechanism be responsible for the lightness of an observed particle such as the Higgs. Prime among these is the modified dispersion relation of the new field, $p \circ p = \mathcal{O}(g^2)$, which means that the free propagation of this field would be Lorentz violating.\footnote{This dispersion relation means that $\chi$ only propagates in noncommutative directions, and so attempts to use hidden extra-dimensional noncommutativity to avoid four-dimensional Lorentz violation constraints seem a phenomenological nonstarter.} We will explore these issues further in the next sections, as in the Yukawa theory of Section \ref{sec:yukawa} the new pole will appear with the opposite sign and so will offer the prospect of appearing as an $s$-channel pole. 

We emphasize that a new infrared scale whose lightness is unexplained in the context of Wilsonian effective field theory is an exciting feature that makes further exploration of UV/IR mixing an interesting pursuit. The fact that it here appears as the scale of a pole in a propagator makes the connection to the hierarchy problem captivating, but asking that this toy model---where Lorentz violation is at the fore---literally solve the problem for us would be too much. We proceed without further hindrance in exploring NCFT so as to learn more about the appearance and effects of UV/IR mixing here.

\subsection{Dimensional Regularization} \label{sec:dimreg}

A good question to ask is whether, or to what extent, these effects are an artifact of our choice of regularization. To demonstrate their physicality, we repeat the calculation of the one-loop correction to the two-point function now in dimensional regularization. We set up our integral in $d=4-\epsilon$ dimensions, having defined $g^2 = \tilde{g}^2 \tilde{\mu}^\epsilon$, and we again go to Schwinger space:
\begin{align}
\begin{split}
\Gamma^{(2)}_{1,\text{planar}} &= \frac{\tilde{g}^2 \tilde{\mu}^\epsilon}{3\left(2\pi\right)^d} \int \text{d}^dk \ \text{d}\alpha \ e^{-\alpha(k^2 + m^2)} \\
\Gamma^{(2)}_{1,\text{nonplanar}} &= \frac{\tilde{g}^2 \tilde{\mu}^\epsilon}{6\left(2\pi\right)^d} \int \text{d}^dk \ \text{d}\alpha \ e^{-\alpha(k^2 + m^2) + i k^\mu \theta_{\mu\nu} p^\nu}.
\end{split}
\end{align}

After completing the square in the nonplanar integral, the momentum integral and the Schwinger integral may then be performed analytically, with the results: 
\begin{align} \label{eqn:dimregint}
\begin{split}
\Gamma^{(2)}_{1,\text{planar}} &= \frac{\tilde{g}^2 \tilde{\mu}^\epsilon}{3\left(4\pi\right)^{d/2}} (m^2)^{\frac{d}{2}-1} \Gamma(1 - \frac{d}{2}) \\
\Gamma^{(2)}_{1,\text{nonplanar}} &= \frac{\tilde{g}^2 \tilde{\mu}^\epsilon}{6 \left(4\pi\right)^{d/2}} 2^{\frac{d}{2}} (m^2)^{\half({\frac{d}{2}-1})} \left(\sqrt{p \circ p}\right)^{1 - \frac{d}{2}} K_{\frac{d}{2}-1}\left(m\sqrt{p\circ p}\right).
\end{split}
\end{align}
If we expand the planar graph in the limit $\epsilon \rightarrow 0$, which should be thought of as probing the ultraviolet, we recover
\begin{equation}
\Gamma^{(2)}_{1,\text{planar}} = - \frac{\tilde{g}^2 m^2}{3(4 \pi)^2} \left[ \frac{2}{\epsilon} + \ln\frac{\mu^2}{m^2}\right],
\end{equation}
where in $\overline{\text{MS}}$ we would subtract off the pole and find the renormalization group evolution of $m$ from the logarithmic term, as usual. 

The question of dimensional regularization for the nonplanar diagram is a subtle one \cite{Huffel:2002pv}. If we first take the $\epsilon \rightarrow 0$ limit of Equation \ref{eqn:dimregint}, we see this manifestly has no divergences, and we are simply left with the finite, $\epsilon^0$ term
\begin{equation} \label{eqn:dimregepsfirst}
\Gamma^{(2)}_{1,\text{nonplanar}} = \frac{g^2 m^2}{6(4\pi)^2}\left[\frac{4}{m^2 p\circ p}- \ln \frac{4}{m^2 p \circ p} - 1 + 2 \gamma \right],
\end{equation}
which we have expanded near $p \circ p \rightarrow 0$ to manifest the IR divergence. We have again transmogrified our UV divergence into an IR pole. We now expect to see that the IR limit does not commute with the above UV limit. To do so, we expand Equation \ref{eqn:dimregint} around $p \circ p \rightarrow 0$ to find
\begin{equation}\label{eqn:dimregpopfirst}
\Gamma^{(2)}_{1,\text{nonplanar}} = \frac{\tilde{g}^2 m^2}{6 (4\pi)^2} \frac{\pi^{\epsilon/2}\tilde{\mu}^\epsilon}{m^\epsilon} \Gamma\left(-1 + \frac{\epsilon}{2}\right) + \frac{\tilde{g}^2}{24 \pi^2}\tilde{\mu}^\epsilon \pi^{\epsilon/2} \Gamma\left(1 - \frac{\epsilon}{2}\right) p \circ p^{-1 + \epsilon/2} + \mathcal{O}(p \circ p).
\end{equation}

If we were to now blindly take the $\epsilon\rightarrow 0$ limit of this expression, we would again get Equation \ref{eqn:dimregepsfirst}, contrary to our expectations. However, we notice that if the dimension of spacetime over which we had performed the integral was particularly low $\epsilon > 2$, then we have incorrectly kept the second term in Equation \ref{eqn:dimregpopfirst}, as that term would be at least $\mathcal{O}(p \circ p)$. If we were to work in $d < 2$, expand in $p \circ p \rightarrow 0$ and so ignore that term, and \textit{then} analytically continue back to $d = 4$, we would instead find the $\epsilon^{-1}$ pole
\begin{equation} \label{eqn:dimregnppole}
\Gamma^{(2)}_{1,\text{nonplanar}} = - \frac{\tilde{g}^2 m^2}{6(4 \pi)^2} \left[ \frac{2}{\epsilon} + \ln\frac{\mu^2}{m^2}\right],
\end{equation}
and now we recover the UV divergence that was present in the commutative theory, so that once again we find the UV and IR limits don't commute.

The key to understanding clearly this seemingly ambiguous dimensional regularization procedure is that while  $\Gamma^{(2)}_{1,\text{nonplanar}}(p \circ p) \sim \int \text{d}^dq \ \text{d}\alpha \ e^{-\alpha(q^2 + m^2) - \frac{p\circ p}{4 \alpha}}$ is convergent in $d>2$ for $p \circ p > 0$, at $p \circ p = 0$ it is only convergent for $d < 2$. Since it is a property of dimensional regularization that if an integral converges in $\delta$ dimensions, it converges to the same value in $d < \delta$ dimensions \cite{Collins:1984xc}, we may thus perform the integral at $d < 2$ for \textit{all} $p \circ p$ and correctly find Equation \ref{eqn:dimregint}. It is only when taking the IR limit that we must remember the integral was performed in $d < 2$ dimensions, and so our expansion to get Equation \ref{eqn:dimregnppole} is unambiguously correct. Thus our conclusion that the UV and IR limits of the two-point function do not commute here is robust. 

It is thus clear that the UV/IR mixing we have observed in this model is not an artifact of a choice of regularization, and is in fact a physical feature of this noncommutative field theory.

\section{Yukawa Theory}\label{sec:yukawa}
\subsection{Motivation: Strong UV/IR Duality}

We observed in our first example that the UV divergences of the real $\phi^4$ commutative theory are transmogrified into infrared poles in the noncommutative theory.\footnote{While we only presented the calculation of the one-loop correction to the two-point function, \cite{Minwalla:1999px} goes through corrections to the two- and $n$-point functions for $\phi^n$ with $n=3,4$ and finds the same features in all cases.} It is natural to ask whether this ``strong UV/IR duality'' \cite{RuizRuiz:2002hh} is a common feature of \textit{all} noncommutative theories.

The answer is no, and the simplest counterexample is provided in the case of a complex scalar field with global $U(1)$ symmetry and self-interaction \cite{RuizRuiz:2002hh}. In the quantization of the scalar potential we have two quartic terms which are noncommutatively-inequivalent due to the ordering non-invariance, so the general noncommutative potential is 
\begin{equation}\label{eqn:complexphi4}
V = m^2 |\phi|^2 + \frac{\lambda_1}{4} \phi^* \star \phi \star \phi^* \star \phi + \frac{\lambda_2}{4} \phi^* \star \phi^* \star \phi \star \phi,
\end{equation}
where $\lambda_1$ and $\lambda_2$ are now different couplings. By doodling some directed graphs, one sees simply that the one-loop correction to the scalar two-point function contains planar graphs with each of the $\lambda_1, \lambda_2$ vertices, but the only nonplanar graph has a $\lambda_2$ vertex. There is thus no necessary connection of the ensuing nonplanar IR singularity to the UV divergence in the $\theta\rightarrow 0$ limit, as the coefficients are unrelated (and in particular, we are free to turn off the IR singularity at one loop by setting $\lambda_2 = 0$).

Another important counterexample is that of charged scalars, the simplest example of which is noncommutative scalar QED, which was first constructed in \cite{Arefeva:2000vow}. There is a very rich and interesting structure of gauge theories on noncommutative spaces, a full discussion of which is far beyond the scope of this paper. We refer the reader to \cite{Armoni:2000xr,Martin:2000bk,Chaichian:2001py,Chaichian:2001mu,Chaichian:2004yw,Khoze:2004zc,Arai:2007dm} for discussions of some features relevant to SM model-building. We here satisfy ourselves with the simplest case, for which we have the noncommutative Lagrangian\footnote{It is important to note that many fundamental concepts which one normally thinks of as depending upon Lorentz invariance still hold on noncommutative spaces, due to a `twisted Poincar\'{e} symmetry' \cite{Oeckl:2000eg,Wess:2003da,Chaichian:2004yh,Chaichian:2004za}. This includes the unitary irreducible representations, so it is sensible to speak of a vector field.} 
\begin{equation}
\mathcal{L} = \frac{1}{4g^2} F_{\mu\nu} \star F^{\mu\nu} + (D_\mu \phi)^* \star (D^\mu \phi) + V\left(\phi,\phi^*\right),
\end{equation}
where even though we're quantizing $U(1)$ we have $F_{\mu\nu} = \partial_\mu A_\nu - \partial_\nu A_\mu - i \left[A_\mu \stackrel{*}{,} A_\nu\right]$ due to the noncommutativity, where $\left[\cdot \stackrel{*}{,} \cdot\right]$ is the commutator in our noncommutative algebra. The vector fields transform as $A_\mu \mapsto U \star A_\mu \star U^\dagger + i \partial_\mu U \star U^\dagger$, where $U(x)$ is an element of the noncommutative $U(1)$ group, which consists of functions $U(x) = \left(e^{i\theta(x)}\right)_\star$, which is the exponential constructed via power series with the star-product.

The potential and the covariant derivative both depend on the representation we choose for the scalar. In contrast to commutative $U(1)$ gauge theory, where we merely assign $\phi$ a charge, our only choices now are to put $\phi$ in either the fundamental or the adjoint of the gauge group. Note that an adjoint field smoothly becomes uncharged in the commutative limit. Such a field $\phi$ transforms as  $\phi \mapsto U \star \phi \star U^\dagger$. The covariant derivative is thus $D^\mu \phi = \partial_\mu \phi - i g \left[A_\mu \stackrel{*}{,}\phi\right]$. The gauge-invariant potential then includes both quartic terms in Equation \ref{eqn:complexphi4}, in addition to others such as $\phi^* \star \phi \star \phi \star \phi$, since the adjoint complex scalar is uncharged at the level of the global part of the gauge symmetry. Strong UV/IR duality then should not hold here either.

The situation is even worse if $\phi$ is in the fundamental, where it transforms as $\phi \mapsto U \star \phi$ and $\phi^* \mapsto \phi^* \star U^{-1}$ with covariant derivative $D^\mu \phi = \partial^\mu \phi - i A_\mu \star \phi$. It is easy to see in this case that the $\lambda_2$ interaction term is no longer gauge invariant, and a charged scalar may only self-interact through $V = \lambda_1 \phi^* \star \phi \star \phi^* \star \phi$. Purely from gauge invariance we thus see that a fundamental scalar has no nonplanar self-interaction diagrams in the one-loop correction to its two-point function, and so there is no remnant of strong UV/IR duality to speak of.\footnote{Noncommutative QED also has strange behavior in the gauge sector that runs counter to strong UV/IR duality---the photon self-energy correction gains an infrared singularity from nonplanar one-loop diagrams, even though the commutative quadratic power-counting divergence is forbidden by gauge-invariance. The theory is constructed in detail in \cite{Hayakawa:1999zf}, while more physical interpretation is given in \cite{Matusis:2000jf}, and the possible relation to geometric dynamics in the context of matrix models is discussed in \cite{VanRaamsdonk:2001jd}.}

The question is then whether there are other examples where this strong UV/IR duality \textit{does} occur, or whether it is perhaps a peculiar feature of real $\phi^n$ theories on noncommutative spaces. To answer this, we will study in detail another case of especial phenomenological significance: Yukawa theory. Noncommutative Yukawa theory was first studied in \cite{Anisimov:2001zc}.\footnote{Aspects of noncommutative Yukawa theory have also been studied recently in d=3 in \cite{Bufalo:2016mui}, and with a modified form of noncommuativity in \cite{Bouchachia:2015kxa}.} Our result on the presence of strong UV/IR mixing differs, for reasons we will explain henceforth.

\subsection{Setup}

For reasons that will soon become clear, we will now work directly in Minkowski space, and begin with a commutative theory of a real scalar $\varphi$ and a Dirac fermion $\psi$ with Yukawa interaction: 
\begin{equation}
\mathcal{L^{(\text{C})}} = -\half \partial_\mu \varphi \partial^\mu \varphi - \half m^2 \varphi^2 + i \overline{\psi} \slashed{\partial} \psi  - \overline{\psi} M \psi + g \varphi \overline{\psi} \psi.
\end{equation}

When constructing a noncommutative version of this theory, the quadratic part of the action does not change. However, ordering ambiguities appear for the interaction term, and we in fact find two noncommutatively-inequivalent interaction terms which generically appear:
\begin{equation}\label{eqn:yukint}
\mathcal{L^{(\text{NC})}_{\text{int}}} = g_1 \varphi \star \overline{\psi} \star \psi + g_2  \overline{\psi} \star \varphi \star \psi.
\end{equation}

\begin{figure}[!ht]
	\begin{subfigure}[b]{0.4\textwidth}
		\includegraphics[width=\textwidth]{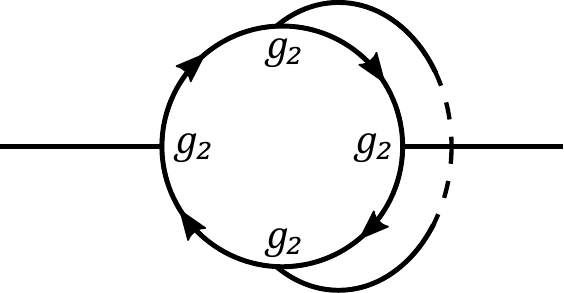}
	\end{subfigure}
	\hfill
	\begin{subfigure}[b]{0.4\textwidth}
		\includegraphics[width=\textwidth]{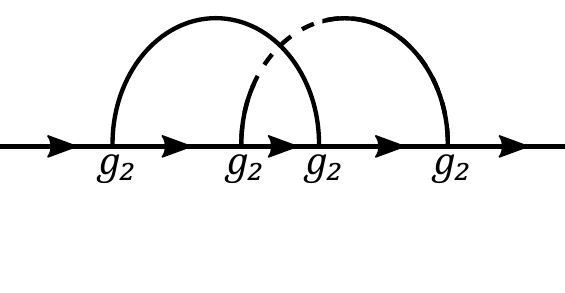}
	\end{subfigure}
	\caption{Representative leading nonplanar corrections to the self-energies in the noncommutative Yukawa theory of \protect\cite{Anisimov:2001zc}. Fermion lines have arrows and dashing denotes nonintersection.} \label{fig:yuk2loop}
\end{figure}

These terms are inequivalent because the star product is only cyclically invariant. In the analysis of \cite{Anisimov:2001zc}, only the $g_2$ interaction was included. As a result, it was concluded that this theory contains no nonplanar diagrams at one loop, and the first appear at two loops as in Figure \ref{fig:yuk2loop}. This immediately tells us that the one-loop quadratic divergence of the scalar self-energy will not appear with a one-loop IR singularity, and so rules out the putative strong UV/IR duality of the theory they studied.

However, we must ask whether we actually have the freedom to choose $g_1$ and $g_2$ independently. To address that question, we must understand the role of discrete symmetries in noncommutative theories. For ease of reference we here repeat our definition of the noncommutativity parameter 
\begin{equation}\tag{\ref{eqn:ncdef}}
\left[x_\mu,x_\nu\right] = i \theta_{\mu\nu}
\end{equation}
It is manifest that the noncommutativity tensor does not transform homogeneously under either parity or time-reversal, but only under their product: $PT: x_\mu \rightarrow - x_\mu \Rightarrow PT: \theta_{\mu\nu} \rightarrow \theta_{\mu\nu}$. So while any Lagrangian with full-rank noncommutativity unavoidably violates both $P$ and $T$, it may preserve $PT$. 

Since both $\varphi$ and the scalar fermion bilinear are invariant under all discrete symmetries, these symmetries na\"{i}vely play no further role in this theory. However, the time-reversal operator is anti-unitary, and thus negates the phase in the star-product: 
\begin{equation}
(PT)^{-1} \left(f(x) \star g(x)\right) PT =  g(x) \star f(x).
\end{equation}

\noindent Armed with this, we may now apply CPT to our interaction Lagrangian, to find

\begin{equation}
(CPT)^{-1} \mathcal{L^{(\text{NC})}_{\text{int}}} CPT = g_1   \overline{\psi} \star\varphi \star \psi  + g_2 \varphi \star \overline{\psi} \star \psi.
\end{equation}
Comparing with Equation \ref{eqn:yukint}, we see that our interactions have been re-cycled! Requiring that our interactions preserve CPT amounts to imposing
\begin{equation}
(CPT)^{-1} \mathcal{L^{(\text{NC})}_{\text{int}}} CPT= \mathcal{L^{(\text{NC})}_{\text{int}}} \quad \Longrightarrow \quad  g_1 = g_2
\end{equation}

And so the theory of \cite{Anisimov:2001zc} appears to violate CPT.\footnote{We note that while the CPT theorem has only been proven in NCFT without space-time noncommutativity \cite{Chaichian:2002vw,Franco:2004gx,AlvarezGaume:2003mb,Soloviev:2006ah}, the difficulty in the general case is related to the issues with unitarity discussed in Section \ref{sec:review}, and we expect it should hold in a sensible formulation of the space-time case as well.} When we instead include both orderings of interactions the nonplanar diagrams now occur at the first loop order. Furthermore, with both couplings set equal the planar and nonplanar diagrams will have the same coefficients, which reopens the question of strong UV/IR duality for this theory. In the following we will keep $g_1$ and $g_2$ distinguished merely to evince how the different vertices appear, but in drawing conclusions about the theory we will set them equal.\footnote{We should note that in the construction of noncommutative QED it has been argued that it is sensible to assign $\theta$ the anomalous charge conjugation transformation $C: \theta^{\mu\nu} \rightarrow - \theta^{\mu\nu}$ (\cite{SheikhJabbari:2000vi} and many others since). The argument is that charged particles in noncommutative space act in some senses like dipoles whose dipole moment is proportional to $\theta$, and so charge conjugation should naturally reverse these dipole moments. Here, however, our particles are uncharged, and thus we have no basis for arguing in this manner. Furthermore, such an anomalous transformation makes charge conjugation relate theories on \textit{different} noncommutative spaces $\mathcal{M}_\theta \rightarrow \mathcal{M}_{-\theta}$. The heuristic picture of the CPT theorem (that is, the reason we care about CPT being a symmetry of our physical theories) is that after Wick rotating to Euclidean space, such a transformation belongs to the connected component of the Euclidean rotation group \cite{Witten:2018lha}, and so is effectively a symmetry of spacetime. So it is at the least not clear that defining a CPT transformation that takes one to a different space accords with the reason CPT should be satisfied in the first place.}

\subsection{Scalar Two-Point Function}

First we consider the planar diagrams, of which there are two: \\
\begin{equation*}
-i\Gamma^{2,s,p}_1(p) \quad = \quad  \includegraphics[width=0.25\linewidth,valign=c]{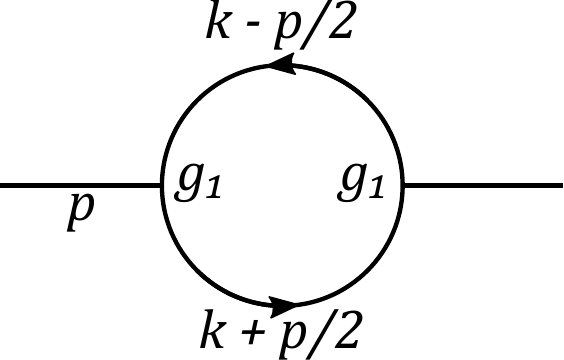} \quad + \quad \includegraphics[width=0.25\linewidth,valign=c]{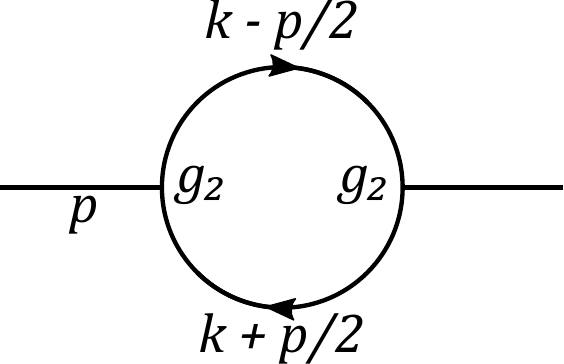}
\end{equation*} 

The `symmetrization' of the momenta of the internal propagators is an important calculational simplification. This calculation is textbook save for our Schwinger-space regularization, so we will be brief and merely point out the salient features. The sum of these diagrams gives
\begin{equation}
\Gamma^{(2),s,p}_1(p) = i(-1)\left((i g_1)^2 + (i g_2)^2\right) \int \frac{\text{d}^4k}{(2\pi)^4} \frac{(-i)^2 \text{Tr}\left[\left(M -  \slashed{k}-\slashed{p}/2\right)\left(M - \slashed{k} +  \slashed{p}/2\right)\right]}{\left((k+p/2)^2 + M^2\right)\left((k-p/2)^2 + M^2\right)}.
\end{equation}

To evaluate this, we must now introduce two Schwinger parameters $\alpha_1, \alpha_2$ and then switch to `lightcone Schwinger coordinates' which effects the change $\int_0^\infty \text{d} \alpha_1 \int_0^\infty \text{d} \alpha_2 \rightarrow \int_0^\infty \text{d} \alpha_+ \int_{-\alpha_+}^{+\alpha_+} \text{d} \alpha_-$. Regulating the integral by $\exp\left[-1/\sqrt{2} \alpha_+ \Lambda^2 \right]$, we may then evaluate and isolate the divergences as $\Lambda \rightarrow \infty$ to find

\begin{equation} \label{eqn:scalaroneloopplanar}
\Gamma^{(2),s,p}_1(p) = - \frac{(g_1^2 + g_2^2)}{2 \pi^2} \left[\Lambda^2 - \frac{6  M^2 + p^2}{4} \log\left(\frac{\Lambda^2}{M^2 + p^2/4}\right) + \dots \right]
\end{equation} 

\noindent Turning now to the nonplanar diagrams, there are again two 
\begin{equation*}
-i \Gamma^{(2),s,np}_1 \quad = \quad  \includegraphics[width=0.25\linewidth,valign=c]{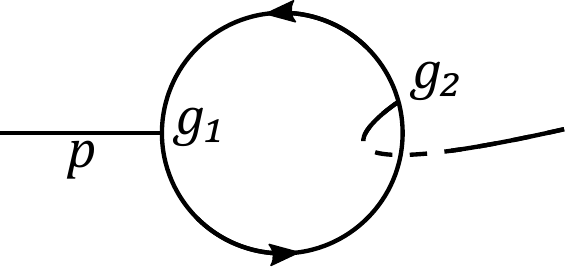} \quad + \quad \includegraphics[width=0.25\linewidth,valign=c]{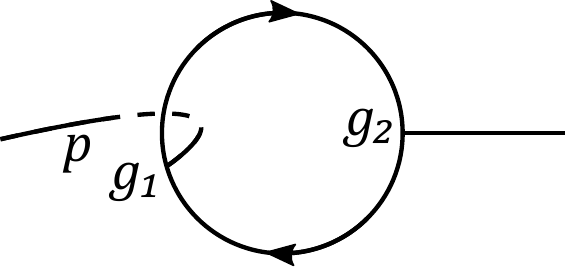}
\end{equation*} 

Each now has one $g_1$ vertex and one $g_2$ vertex, which makes it clear why the analysis of \cite{Anisimov:2001zc} found no such diagrams. The two diagrams will come with opposite phase factors, $e^{i p \wedge k}$ and $e^{i k \wedge p}$, so we can compute one and then find the other by taking $p \mapsto -p$. In this case it's obvious that after completing the square we will only be left with terms which are quadratic in $p$, and so the two diagrams give the same contribution. We can thus compute both terms at the same time.

The phase factor in the integrand will modify our change of variables, as it did in the $\phi^4$ case, to give again an effective cutoff for this diagram due to the noncommutativity. We find
\begin{multline}
\Gamma^{(2),s,np}_1(p) = \frac{g_1 g_2}{\pi^2} \int \text{d}q \text{d}\alpha_1 \text{d}\alpha_2 q^3\left( M^2 - q^2 + \frac{\alpha_1 \alpha_2}{\left(\alpha_1 + \alpha_2\right)^2} p^2+\frac{p \circ p }{4 (\alpha_1 + \alpha_2)^2}\right) \\ \times e^{- \left(\alpha_1 + \alpha_2\right)\left(q^2 + M^2\right) - \frac{\alpha_1\alpha_2}{\alpha_1 + \alpha_2} p^2-\frac{p \circ p }{4 (\alpha_1 + \alpha_2)}}.
\end{multline}
 
We can now follow the same steps to regulate and integrate this, and again find a closed-form expression for the pieces which contain divergences.  Note that unlike the $\phi^4$ calculation, we can already see that the nonplanar expression will not merely be given by $\Lambda \rightarrow \Lambda_{\text{eff}}$, as the change of variables has here modified the numerator of the integrand to give an extra piece to the momentum polynomial multiplying the exponential. And so integration gives us
\begin{multline}  \label{nonplanarscalar}
\Gamma^{(2),s,np}_1(p) = \frac{g_1 g_2}{1920 \pi^2} \left[3\left(640 M^2 + p^4 p \circ p + 40 (4 M^2 + p^2) p \circ p \Lambda_{\text{eff}}^2\right) K_0\left(\frac{\sqrt{4 M^2 + p^2}}{\Lambda_\text{eff}}\right) \right. \\ + \left. 20 \sqrt{4 M^2 + p^2} \Lambda_{\text{eff}} \left(- 96 + p^2 p\circ p + 12 p \circ p \Lambda_{\text{eff}}^2\right) K_1\left(\frac{\sqrt{4 M^2 + p^2}}{\Lambda_\text{eff}}\right)  \right].
\end{multline}

We must now think slightly more carefully about what we want to add to the quadratic effective action to find a Wilsonian interpretation of this theory. We may isolate the IR divergence that appears when the cutoff is removed by first taking the limit $\Lambda \rightarrow \infty$ with $p \circ p$ held fixed, and then expanding around $p \circ p = 0$. We may then ask that this same divergence appears at any value of $\Lambda$. To account for this IR divergence, we must add to our  effective action 
\begin{equation}
\Delta S_{1\text{PI}}(\Lambda) = - \half \int \frac{\text{d}^4p}{(2 \pi)^4} \frac{g_1 g_2}{2 \pi^2} \left(\Lambda_{\text{eff}}^2 - \frac{4}{p\circ p}\right)\varphi(p)\varphi(-p),
\end{equation}
which can easily be done through the addition of an auxiliary scalar field as was done in Section \ref{sec:phi4} and is discussed in more generality in Appendix \ref{app:auxfield}. After having added this to our action, for small $p \circ p$ the scalar two-point function now behaves as $\Gamma_1^s(p) = - \frac{2 g_1 g_2}{\pi^2 p \circ p} + \dots$ for any value of $\Lambda$. The new pole in this case has the opposite sign as that in \ref{eqn:auxaction}, and so will be accessible in Euclidean signature, clearly signaling a tachyonic instability. While this puts the violation of unitarity in this theory on prime display, it also means that this pole will be accessible in the $s$-channel in the Lorentzian theory if we allow for timelike noncommutativity. 

We emphasize that any conclusions about the Lorentzian theory with timelike noncommutativity are speculative and dependent upon a solid theoretical understanding of a unitary formulation of the field theory, and in principle such a formulation could find radically different IR effects than this na\"{i}ve approach. However, it was found in \cite{Bozkaya:2002at} that a modification of time-ordering to explicitly make the theory unitary (at the expense of microcausality violation) leaves the one-loop correction to the self-energy unchanged in $\phi^4$ theory, and the same might be expected to hold true for Yukawa theory. This makes it worthwhile to at least briefly consider the potential phenomenological consequences of the new pole.

At low energies, the propagator is here modified to $m^2 + (p_i + p_j)^2 - \frac{2 g_1 g_2}{\pi^2} \frac{1}{(p_i + p_j)\circ (p_i + p_j)}$. If we consider scattering of fermions through an $s$-channel $\varphi$ and take the simple case of a noncommutativity tensor which in the lab frame has one eigenvalue $1/\Lambda_\theta^2$ with $m^2 \gg \Lambda_\theta^2$, then the emergent pole appears at $s = \frac{2 g_1 g_2}{\pi^2} \frac{1 - \beta^2}{1+\beta^2} \frac{\Lambda^4_\theta}{m^2}$. Here $s = - (p_i + p_j)^2$ is the invariant momentum routed through the propagator, and $\beta$ is the boost of the $(p_i + p_j)$ system with respect to the lab frame. The Lorentz-violation here then has the novel effect of smearing out the resonance corresponding to the light pole for a particle which is produced at a variety of boosts. This is in contrast to the pole at $m^2$, which gives a conventional resonance at leading order. Of course, we have not constructed a fully realistic theory in any respect, and ultimately it may well be that other Lorentz-violating effects provide the leading constraint. Nonetheless, the lineshape of resonances may be an interesting observable in this framework.

A further feature of this opposite sign of the new pole compared to that in the $\phi^4$ theory is that the unusual momentum-dependence of the two-point function will lead to ordered phases which break translational invariance \cite{Gubser:2000cd,Minwalla:1999px,Steinacker:2005wj,Chen:2001an,Castorina:2003zv}. 
While a Lorentz-violating background field may possibly be very well constrained, the detailed constraint depends on its wavelength and the ways in which it interacts with the SM. But this is another obvious line of exploration for constraining realistic NCFTs.

\subsection{Fermion Two-Point Function}
 
There are again two planar diagrams:
\begin{equation*}
-i\Gamma^{(2),f,p}_1 \quad = \quad  \includegraphics[width=0.25\linewidth,valign=c]{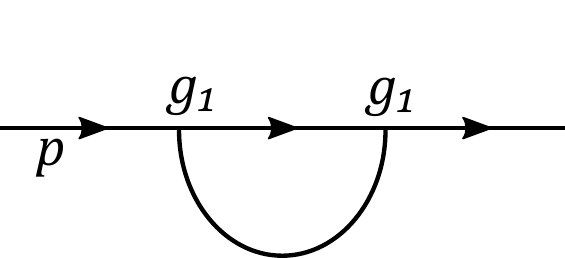} \quad + \quad \includegraphics[width=0.25\linewidth,valign=c]{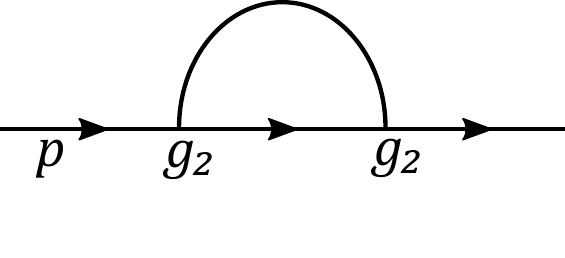}
\end{equation*} 

\noindent No new features appear in the evaluation of these diagrams, so we merely quote the final result:
\begin{equation}
\Gamma^{(2),f,p}_1 = -\frac{g_1^2 + g_2^2}{16 \pi^2} \left(M - \frac{\slashed{p}}{2}\right) \log\frac{4 p^2 \Lambda^2}{m^4 + 2 m^2 (p^2 - M^2) + (M^2 + p^2)^2} + \dots 
\end{equation}

\noindent We also have two nonplanar diagrams, which again mix the two vertices
\begin{equation*}
-i\Gamma^{(2),f,np}_1 \quad = \quad  \includegraphics[width=0.25\linewidth,valign=c]{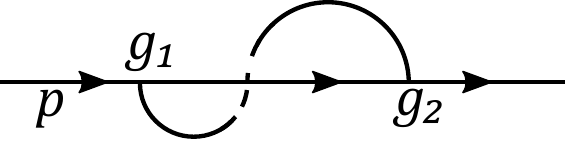} \quad + \quad \includegraphics[width=0.25\linewidth,valign=c]{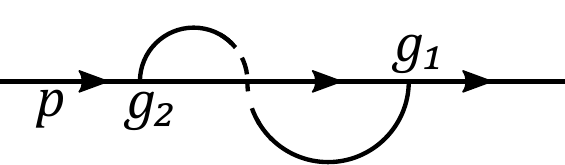}
\end{equation*} 

Here we find that the different phase factors for each diagram, which we saw were inconsequential for the nonplanar corrections to the scalar, have an important role. When we complete the square in each of the two cases, we find that one of the diagrams has an integrand proportional to $\left(M - \slashed{p} \frac{\alpha_2}{\alpha_1 + \alpha_2} - \frac{1}{2}\frac{p_\mu\theta^{\mu\nu}\gamma_\nu}{\alpha_1 + \alpha_2}\right)$ and the other is proportional to $\left(M - \slashed{p} \frac{\alpha_2}{\alpha_1 + \alpha_2} + \frac{1}{2}\frac{p_\mu\theta^{\mu\nu}\gamma_\nu}{\alpha_1 + \alpha_2}\right)$, so the would-be divergence in $p\theta$ will cancel manifestly between the two diagrams. After this everything proceeds as before, and we find 
\begin{equation}
\Gamma^{(2),f,np}_1 = -\frac{g_1 g_2}{8 \pi^2} \left(M - \frac{\slashed{p}}{2}\right) \log\frac{4 p^2 \Lambda_{\text{eff}}^2}{m^4 + 2 m^2 (p^2 - M^2) + (M^2 + p^2)^2} + \dots
\end{equation}

We see that with $g_1 = g_2 \equiv g$, the fermion quadratic effective action also behaves as expected from `strong UV/IR duality'. The logarithmic divergence of the commutative theory has been transmogrified in the nonplanar diagrams into IR dynamics via the simple replacement $\Lambda \rightarrow \Lambda_{\text{eff}}$, and so a $p \circ p \rightarrow 0$ pole will emerge when we remove the cutoff. We discuss the use of an auxiliary field to restore a Wilsonian interpretation here in Appendix \ref{app:auxferm}.

\subsection{Three-Point Function}

The correction to the vertex function constitutes further theoretical data toward the Wilsonian interpretation of the noncommutative corrections. We calculate the one-loop correction in this section and delay the discussion of the use of auxiliary fields to account for them until Appendix \ref{app:aux3pt}. We will find that while we can use the same fields to account for the modifications to both the propagators and the vertices, the physical interpretation of such fields is unclear.

We can compute corrections for each fixed ordering of external lines separately since they're coming from different operators. For simplicity we'll compute the $g_1$ ordering, which we will denote $\Gamma^{\varphi\overline{\psi}\psi}_3(r,p,\ell)$. There are four diagrams in total: one planar diagram with two insertions of the $g_2$ vertex, one nonplanar diagram with two insertions of the $g_1$ vertex, and two nonplanar diagrams with one insertion of each. It is easy to see by looking at the diagrams that the same expressions with $g_1 \leftrightarrow g_2$ compute the correction to the other ordering, $\Gamma_3^{\overline{\psi}\varphi\psi}(r,p,l)$.

The new feature of this computation is that we now need \textit{three} Schwinger parameters, and this presents a problem for our previous computational approach. We won't be able to perform the two finite integrals before expanding in a variable which isolates the divergences when $\alpha_1 + \alpha_2 + \alpha_3 \rightarrow 0$, analogously to what we did in $2d$ Schwinger space. Instead we slice $3d$ Schwinger space such that we can perform the integral which isolates the leading divergences first, and then---as long as we're content only to understand this divergence---we can discard the rest without having to worry about performing the other two integrals.

The planar diagram is 
\begin{equation*}
i\Gamma^{\varphi\overline{\psi}\psi}_{3,p}(p,\ell) \quad = \quad  \includegraphics[width=0.25\linewidth,valign=c]{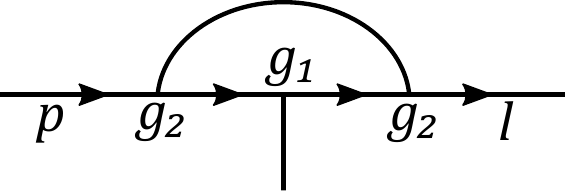},
\end{equation*} 

and corresponds to the expression
\begin{align}
\Gamma^{\varphi\overline{\psi}\psi}_{3,p}(p,\ell) &= -i (i g_1) (i g_2)^2 \times \\ &\int \frac{\text{d}^4k}{(2\pi)^4} \frac{(-i)^3 \left(M - (\slashed{k} + \frac{\slashed{p}}{2} + \frac{\slashed{\ell}}{2})\right)\left(M - (\slashed{k}- \frac{\slashed{p}}{2} - \frac{\slashed{\ell}}{2}\right)}{\left((k+\frac{p}{2}+\frac{\ell}{2})^2 + M^2\right)\left((k-\frac{p}{2}-\frac{\ell}{2})^2 + M^2\right)\left((k+\frac{p}{2}-\frac{\ell}{2})^2 + m^2\right)}. \nonumber
\end{align}
After moving to Schwinger space, integrating over the loop momentum, and introducing a cutoff $\exp\left(-1/\left(\Lambda^2 (\alpha_1 + \alpha_2 + \alpha_3)\right)\right)$, we switch variables to 
\begin{equation}
\alpha_1 = \xi_1 \eta, \qquad 
\alpha_2 = \xi_2 \eta, \qquad
\alpha_3 = (1 - \xi_1 - \xi_2) \eta,
\end{equation}
under which $\int_0^\infty \text{d}\alpha_1 \int_0^\infty \text{d}\alpha_2 \int_0^\infty \text{d}\alpha_3 \rightarrow \int_0^1 \text{d} \xi_1 \int_0^{1-\xi_1} \text{d} \xi_2 \int_0^\infty \text{d} \eta \ \eta^2$. Performing the momentum integral transfers the divergence for large $k$ to a divergence in small $\alpha_1 + \alpha_2 + \alpha_3 = \eta$. This will allow us to find the leading divergent behavior immediately by carrying out the $\eta$ integral and then expanding in $\Lambda \rightarrow \infty$. This yields
\begin{equation}
\Gamma^{\varphi\overline{\psi}\psi}_{3,p}(p,\ell) = \frac{g_1 g_2^2}{16 \pi^2} \log\left(\Lambda^2\right) + \text{ finite},
\end{equation}
where we are unable to determine the IR cutoff of the logarithm, but this suffices for our purposes.

The three nonplanar graphs now each receive a different phase corresponding to which external line crosses the internal line
\begin{align}
i\Gamma^{\varphi\overline{\psi}\psi}_{3,np}(p,\ell) \quad = \quad  &\includegraphics[width=0.25\linewidth,valign=c]{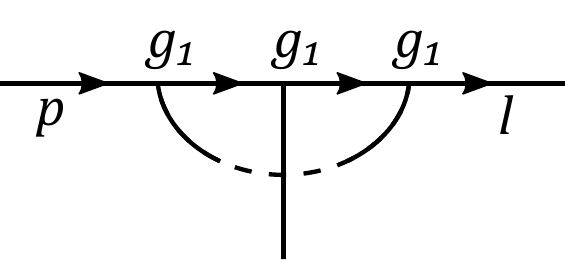} \quad + \quad \includegraphics[width=0.25\linewidth,valign=c]{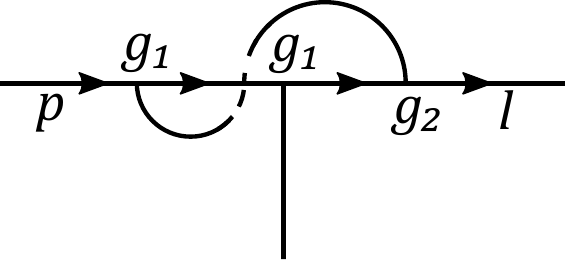} \quad \nonumber \\ &\qquad \qquad + \quad \includegraphics[width=0.25\linewidth,valign=c]{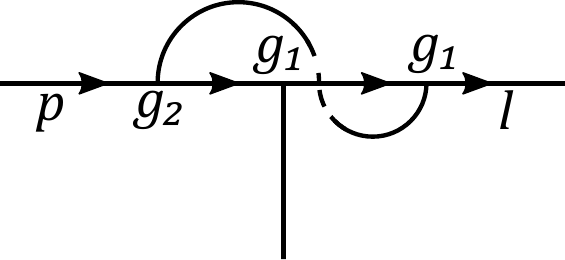},
\end{align} 
where the first gets $\exp\left[-i(k \wedge p + k \wedge \ell + p \wedge \ell)\right]$, the second $\exp\left[-i(k \wedge p + p \wedge \ell/2)\right]$, and the third $\exp\left[-i(k \wedge \ell + p \wedge \ell/2)\right]$. The evaluation of these diagrams proceeds as in the previous examples. If we take the IR limit $p,\ell \rightarrow 0$ of the nonplanar contributions to this ordering of the three-point function and then expand in large $\Lambda$ we find 
\begin{equation}
\lim\limits_{p,\ell\rightarrow 0} \Gamma^{\varphi\overline{\psi}\psi}_{3,np}(p,\ell) = \frac{g_1^2 (g_1 + 2 g_2)}{16 \pi^2} \log\left(\Lambda^2\right) + \text{ finite}.
\end{equation}
However, if we first take the UV limit $\Lambda \rightarrow 0$, and then expand in small momenta, we find
\begin{align}
\lim\limits_{\Lambda\rightarrow \infty} \Gamma^{\varphi\overline{\psi}\psi}_{3,np}(p,\ell) = \frac{g_1^2}{16 \pi^2} \Bigg[g_1 &\log\left(\frac{4}{(p + \ell) \circ (p + \ell)}\right) \\ &+ g_2 \log\left(\frac{4}{p \circ p}\right)+ g_2 \log\left(\frac{4}{\ell \circ \ell}\right)\Bigg] + \text{ finite},\nonumber \label{eqn:3ptfct}
\end{align}
where we again see UV/IR mixing, and we note that each nonplanar diagram has been effectively cutoff by the momenta which cross the internal line. We discuss the use of auxiliary fields to restore a Wilsonian interpretation to this vertex correction in Appendix \ref{app:aux3pt}.

\section{Softly-broken Wess-Zumino Model}\label{sec:wesszumino}
We now turn our attention to the softly-broken noncommutative Wess-Zumino model as a controllable example of the interplay between UV/IR mixing and the finiteness of the field theory. We will restrict ourselves to calculating the one-loop correction to the scalar two-point function. Since the new poles appearing in the quadratic effective action in the scalar and Yukawa theories are intimately related to the quadratic divergences of the commutative theories, we will not be surprised to find that this feature will disappear when both the scalar and the fermion are present in the EFT below the cutoff. By studying the softly-broken theory we can take the fermion above or below the cutoff to smoothly see the relation between the finiteness of the field theory and the effects of UV/IR mixing. The exactly supersymmetric noncommutative Wess-Zumino model was first discussed in detail in \cite{Girotti:2000gc}, and the absence of an infrared pole in a softly-broken theory was first noted in \cite{Matusis:2000jf}. The softly-broken Wess-Zumino model was first considered in \cite{AmelinoCamelia:2002au}.\footnote{Our one-loop results agree with those of \cite{AmelinoCamelia:2002au} save for their claim that logarithmic IR divergences are absent in the exactly supersymmetric theory, which contradicts \cite{Girotti:2000gc}. We will below find a logarithmic IR divergence in the wavefunction renormalization which is independent of the soft-breaking, which is consistent with the expectations of strong UV/IR duality.}

The noncommutative Wess-Zumino theory can be suitably formulated in off-shell superspace as
\begin{equation}
\mathcal{L} = \int \text{d}^4\theta \ Z \Phi^\dagger \Phi + \int \text{d}^2\theta \ \left(\half M \Phi^2 + \frac{1}{6} y \ \Phi \star \Phi \star \Phi\right) + \text{ h.c.},
\end{equation}
where $\Phi$ is a chiral superfield and we have included a wavefunction renormalization factor in the K\"{a}hler potential $Z = 1 + \mathcal{O}(y^2)$. We can introduce soft supersymmetry breaking by promoting this factor to a spurion $Z = 1 + ( \left|M\right|^2 - m^2)  \theta^2 \theta^{\dagger2}$, the only effect of which is to modify the scalar mass spectrum.

Formulating the noncommutative theory including the auxiliary $F$ fields makes it manifest that we have preserved supersymmetry off-shell. This procedure is in fact precisely the same as quantizing after integrating out $F$, and so we end up with a star-product version of the familiar Lagrangian:
\begin{align}
-\mathcal{L}_{\text{NCWZ}} &= Z \partial^\mu \phi^{*} \partial_\mu \phi - i Z \psi^\dagger \bar{\sigma}^\mu \partial_\mu \psi \nonumber\\ 
&+ Z^{-1} m^2 \phi^{*} \phi + \half M \psi \psi + \half M^{*} \psi^\dagger \psi^\dagger \nonumber\\ 
&+ \half Z^{-1} y \phi \star \psi \star \psi + \half Z^{-1} y^{*} \phi^{*} \star \psi^\dagger \star \psi^\dagger  \nonumber\\
&+ \half Z^{-1} y M^{*} \phi \star \phi \star \phi^{*} + \half  Z^{-1} y^{*} M \phi^{*} \star \phi^{*} \star \phi \nonumber\\
&+ \frac{1}{4} Z^{-1} \left|y\right|^2 \phi \star \phi \star \phi^{*} \star \phi^{*}
\end{align}

\noindent where $\phi$ is a complex scalar and $\psi$ is a Weyl fermion. Of course, now that we've introduced supersymmetry breaking we expect to find that there is further renormalization beyond that associated with $Z$, but keeping the manifest factors of $Z$ will allow us to easily compare to our expectations for the supersymmetric limit.

The calculation of the one-loop correction to the two-point function goes much as the previously-demonstrated examples. The presence of the three-scalar interaction gives a new class of diagrams, whose evaluation is routine. The two-component fermions yield slightly different factors than did the Dirac fermions \cite{Dreiner:2008tw}. Finally, it is important to note that the results for the diagrams computed in Section \ref{sec:phi4} cannot be used here, as we must here regulate uniformly using $\exp(-1/(\Lambda^2 (\alpha_1 + \alpha_2)))$ like we did in Section \ref{sec:yukawa}. This may be easily accommodated by writing the integrand in the quartic diagrams as $\frac{1}{k^2 + m^2} \frac{k^2 + m^2}{k^2 + m^2}$.

Adding up all these diagrams and taking the limit where $\Lambda, \Lambda_{\text{eff}}$ are large, we find that the one-loop scalar two-point function may be organized as 
\begin{align}
\Gamma^{(2),s} &\equiv Z p^2 + Z^{-1}(m^2 + \delta m^2) \\
Z &= 1 + \frac{y^2}{32 \pi^2} \log\left[\frac{\Lambda \Lambda_{\text{eff}}}{M^2}\right] + \dots \\
\delta m^2 &= \frac{y^2}{32 \pi^2} \left(M^2 -  m^2\right) \log\left[ \frac{\Lambda \Lambda_{\text{eff}}}{M^2}\right] + \dots,
\end{align}

\noindent where we make manifest the presence of supersymmetric nonrenormalization in the limit $m \rightarrow M$, which acts as a non-trivial check. As expected, the absence of the quadratic UV divergence in the Wess-Zumino model has led to the absence of an infrared pole from the noncommutativity, even as the fermion is made arbitrarily heavy relative to the scalar. However, logarithmic UV/IR mixing still occurs.  

We may repeat this calculation using dimensional regularization and taking note of the issues which arose in Section \ref{sec:dimreg}. Using the same parametrization of the one-loop two-point function as above, the planar diagrams contribute
\begin{align}
Z_{\text{planar}} &= 1 + \frac{y^2}{64 \pi^2} \left(\frac{2}{\epsilon} + \log \frac{\mu^2}{M^2}\right) + \dots \\
\delta m^2_{\text{planar}} &= \frac{y^2}{64 \pi^2} (M^2 - m^2) \left(\frac{2}{\epsilon} + \log \frac{\mu^2}{M^2}\right) + \dots,
\end{align}
as expected. The full form of the nonplanar diagrams is unenlightening, but if we take the IR limit $p \circ p \rightarrow 0$ first, they give precisely the same contribution as the planar diagrams, since the diagram degeneracies are all the same in this case. Taking the UV limit $\epsilon \rightarrow 0$ first (and staying in $d < 2$), we instead find 
\begin{align}
Z_{\text{nonplanar}} &= 1 + \frac{y^2}{64 \pi^2} \log \frac{4}{M^2 p \circ p} + \dots \\
\delta m^2_{\text{nonplanar}} &= \frac{y^2}{64 \pi^2} (M^2 - m^2) \log \frac{4}{M^2 p \circ p} + \dots,
\end{align}
which has precisely the same correspondence with the Schwinger-space regularization as we saw for the $\phi^4$ case.

We thus see clearly the conflict between supersymmetry and the use of UV/IR mixing to explain low-energy puzzles. UV/IR mixing transmogrified UV momentum dependence into IR momentum dependence, and so depended crucially on the sensitivity of our field theory to UV modes. For a theory which is finite as a field theory, the dependence on the UV physics has been removed, and so we see no interesting IR effects. 

Of course, in the presence of a cutoff $\Lambda$ it is also possible to study the behavior of the scalar two-point function when $M^2 \gg \Lambda^2 \gg |M^2-m^2|$ as the fermion is taken above the cutoff while keeping the scalar light. This corresponds to taking $M/\Lambda,M/\Lambda_{\text{eff}} >1$ and then expanding in the limit where $\Lambda, \Lambda_{\text{eff}}$ are large. This gets rid of the nonplanar Yukawa-type diagrams and, as one might expect, results in a return of UV sensitivity in the scalar EFT below the cutoff, foreshadowing a return of the UV/IR mixing effects. The scalar mass-squared in this limit becomes
\begin{equation}
\delta m^2 = \frac{y^2}{256 \pi^2} \left(6 M^2 + 16 \Lambda^2 + 8 \Lambda_{\text{eff}}^2 \right) + \dots.
\end{equation}  
and UV/IR mixing reappears at the quadratic level. So our EFT intuition isn't totally out the window; it's been broken in a controlled way, and we can smoothly interpolate between theories with and without UV/IR mixing by taking the states responsible for finiteness above the cutoff. This sharpens the sense in which UV/IR mixing can do something interesting in the IR as long as the field-theoretic description of our universe is never finite. 

Ultimately, this highlights a central challenge for approaching the hierarchy problem via UV/IR mixing. The hierarchy problem is particularly sharp when the full theory is finite and scale separation is large, in which case the sensitivity of the Higgs mass to underlying scales is unambiguous. But UV/IR mixing effects potentially relevant to the hierarchy problem are absent in this case, and emerge only when finiteness is lost. This tension is not necessarily fatal to UV/IR approaches to the hierarchy problem---ultimately the UV sensitive degrees of freedom are not the ones we would wish to identify with the Higgs---but it bears emphasizing.

Moreover, there is a possible loophole in the general argument that finiteness must be surrendered in order to generate a scale from UV/IR mixing. The presence of interesting effects in the IR here depends solely on the UV sensitivity of the nonplanar diagrams. The `orbifold correspondence' \cite{Kachru:1998ys,Bershadsky:1998cb,Schmaltz:1998bg} provides non-supersymmetric field theories constructed via orbifold truncation of $\mathcal{N}>0$ theories whose planar diagrams agree with those of the supersymmetric theory and so are finite. A noncommutative orbifold field theory \cite{Armoni:1999nj} may then provide a theory which is fully predictive, yet which still generates an infrared scale via UV/IR mixing. Generally, it may be possible that UV/IR mixing appears in such a way that it is the sole effect sensitive to short distances.

\section{Whence UV/IR Mixing?}\label{sec:lessons}
To attempt to formulate a realistic theory which uses UV/IR mixing to solve extant theoretical puzzles, it would be useful to have an understanding of which features of NCFT were responsible for the curious infrared effects discussed above. This would be helpful whether one wishes to test out these ideas in any of the many proposed modifications of NCFT, or to write down other toy models which share some features of NCFT but are based upon different principles.

Qualitatively, the two unusual features involved in the formulation of NCFT are Lorentz invariance violation and nonlocality. However, it is obvious that one may have theories with one or both of these features without the interesting effects we have seen. The answer then is not so simple as pointing to one axiom or another of EFT which has been broken, but depends sensitively on the way in which they are broken. We briefly explore two ways we may better understand the interplay here between nonlocality and Lorentz-violation and how they come together to cause surprising low-energy effects. We first give a general argument based on the way nonlocality appears to postdict the form of the violation of EFT expectations. We then phenomenologically examine the loop integration appearing in our NCFT calculations to diagnose what caused the appearance of the IR pole. This will lead us to discuss an avenue toward investigating (or manufacturing) such effects in nonlocal, Lorentz-invariant theories.

To see how EFT expectations may be violated, consider the peculiar way in which the noncommutative effects in the one-loop action (e.g. Equation \ref{eqn:1PIaction}) induce nonlocality. In Wilsonian EFT, integrating out momentum modes $p \gtrsim \Lambda$ produces a nonlocal theory at those scales, or equivalently on distances $x \lesssim 1/\Lambda$. However, particles on a noncommutative space can be thought of as rods of size $L \sim p \theta$ \cite{SheikhJabbari:1999vm,Bigatti:1999iz,Seiberg:2000gc,Girotti:2001dh,Acatrinei:2002sb}. This tells us that in a NCFT we should expect nonlocality to be present for scales $x \lesssim p \theta$. Comparing the two scales, we see that we should find nonlocal effects past those expected in Wilsonian EFT for $\frac{1}{\Lambda} < p\theta$. Here this momentum-dependent nonlocality occurs in a Lorentz-violating way. This expectation was exactly borne out in the examples above, where we saw that the one-loop effective action in momentum space is nonlocal for $p\circ p \gg 1/\Lambda^2$ \cite{Minwalla:1999px}. 

Purely from this analysis of the form of nonlocality, we may conclude there will be a breakdown of Wilsonian renormalization. After we remove the cutoff, the theory should be nonlocal on all scales $p \circ p > 0$. But if we compute a correlation function at a large-but-finite $\Lambda$, the theory will still be local for momenta $p \circ p < 1/\Lambda^2$, and so will greatly differ from the continuum result. So our surprising discovery of the non-uniform convergence of correlation functions in the examples above is understood easily from this picture. 

While this sort of momentum-dependent nonlocality may seem \textit{ad hoc}, it has been suggested previously for separate purposes. It has been argued \cite{Maggiore:1993rv} that quantum gravity should obey a `Generalized Uncertainty Principle' $\Delta x \gtrsim \frac{\hbar}{\Delta p} + \ell_p^2 \Delta p$, with $\ell_p$ the Planck length, based on the use of Hawking radiation to measure the horizon area of a black hole. This gives precisely the same sort of momentum-dependent nonlocality as we saw above. We refer the reader to \cite{Tawfik:2015rva} for a review of the Generalized Uncertainty Principle, \cite{Konishi:1989wk,Yoneya:2000bt} for similar conclusions within string theory, and \cite{Hossenfelder:2012jw} for a more general review of the appearance of an effective minimal length in quantum gravity. It would be interesting to investigate other field theories which obey such uncertainty principles and determine whether UV/IR mixing causes similar features as appear in NCFT. For theories which violate Lorentz invariance, care must be taken to avoid arguments that even Planck-scale Lorentz violation is empirically ruled out \cite{Collins:2004bp,Polchinski:2011za}.

We may also attempt to phenomenologically diagnose what caused the appearance of the IR pole from the form of the loop integration. The presence of an exponential of momenta was clearly crucial, and this implies a necessity of nonlocality. It's also clear that the modification of the cutoff in the nonplanar diagrams $\Lambda \mapsto \Lambda_{\text{eff}}$, which rendered the diagrams UV finite in a way that brought UV/IR mixing, was a result of the contraction between the loop momentum and the external momentum. Less obviously, one may see that any quadratic term in loop momentum in the exponential would have erased this feature, as after momentum integration one would find an integrand $\sim \frac{1}{1 + \alpha_+}$, and any divergence will have disappeared. Heuristically, the quadratic suppression in loop momentum is too strong and regulates the UV divergence entirely independently of the cutoff, so no UV/IR mixing appears. NCFT disallows such terms as a result of momentum contractions being performed with an antisymmetric tensor, and this particular mechanism seems to imply the necessity of Lorentz invariance violation. However, this argument only considers small deviations from the form of the integral in NCFT. Further discussions of the form of loop integrals with generalizations of the star-product may be found in \cite{Galluccio:2009ss,Ardalan:2010ht}.

Likely a better approach to understand the prospect for finding features similar to that of NCFT in a Lorentz invariant theory is to back up and study formulations of Lorentz invariant extensions of NCFT. This is accomplished by upgrading the noncommutativity tensor $\theta^{\mu\nu}$ from a $c$-number to an operator. This was proposed already by Snyder in 1947 \cite{Snyder:1946qz}, and this approach has been revived a number of times more recently (e.g.  \cite{Doplicher:1994tu,Kase:2002pi,Carlson:2002wj,Heckman:2014xha,Much:2017hcv}). Schematically, this results in an action containing an integral over $\theta^{\mu\nu}$ 
\begin{equation}
S = \int \text{d}^4x \ \text{d}^6\theta \ W(\theta) \ \mathcal{L}(\phi, \partial \phi),
\end{equation}
where $W(\theta)$ is a `weighting function', and the Lagrangian is still defined using the star-product. The challenge in this approach for our purposes is in devising a method for nonperturbative calculations in $\theta$, which as we saw above was necessary to preserve the features of UV/IR mixing.

Searching more generally for Lorentz invariant theories which contain UV/IR mixing will likely allow more promising phenomenological applications. That such theories should exist can be broadly motivated by quantum gravity, as any gravitational theory is expected both to be nonlocal and to have UV/IR mixing. That Lorentz violation should be present is less clear. A particularly interesting line of development is to then understand in detail the class of nonlocal theories that would have UV/IR mixing of a sort similar to that discussed here. Recent work toward placing nonlocal quantum field theories on solid theoretical ground \cite{Tomboulis:2015gfa,Chin:2018puw} is clearly of sharp interest here, though the larger goal is quite distinct. The nonlocality studied in these works is designed to render the field theory UV-finite, and so the nonlocal vertex kernels are chosen precisely to avoid the introduction of new poles by ensuring these are momentum-space entire functions which vanish rapidly in Euclidean directions. The nonlocal vertices of NCFT manage to introduce new poles by oscillating as $p \rightarrow \infty$, which presumably allows for the appearance of new `endpoint singularities' \cite{Eden:1966dnq,Itzykson:1980rh}, though a full examination of the Landau equations in NCFT has not (to our knowledge) been performed. Our interest is thus in a disjoint class of nonlocal theories, where new poles can appear in interesting ways. Classifying the space of such theories and developing an approach to systematically understand their unitarity properties seems well motivated.

\section{Conclusions} \label{sec:conclusions}
The lack of evidence for conventional solutions to the hierarchy problem has placed particle physics at a crossroads. While it is possible that the answer ultimately lies further down the well-trodden path of existing paradigms, the appeal of less-travelled paths grows greater with every inverse femtobarn of LHC data. 

In this work we have ventured to take seriously the apparent failure of expectations from Wilsonian effective field theory regarding the hierarchy problem by investigating a concrete framework---noncommutative field theory---in which Wilsonian EFT itself breaks down. Not only does noncommutative field theory violate Wilsonian expectations, it provides a sharp instance of UV/IR mixing: ultraviolet modes of noncommutative theories can generate an infrared scale whose origin is opaque to effective field theory. To the extent that UV/IR mixing has any relevance to the hierarchy problem, the emergence of an infrared scale seems to be among the most promising effects. Although the real-world applicability of these theories is likely limited by their Lorentz violation, they nonetheless provide valuable toy models for exploring the potential relevance of UV/IR mixing to problems of the Standard Model. 

To this end, we have surveyed existing results on noncommutative theories with an eye towards `strong UV/IR duality'---the transmogrification of UV divergences into infrared poles at the same order. This led us to a detailed analysis of noncommutative Yukawa theory, perhaps the most useful toy model for thinking about the hierarchy problem (insofar as the Yukawa sector of the Standard Model is responsible for the largest UV sensitivity of the Higgs mass, and highlights the relative UV {\it insensitivity} of the fermion masses). In the noncommutative theory, the presence of both inequivalent Yukawa couplings implies the same strong UV/IR duality exhibited by real $\phi^4$ theory: a quadratic divergence in the one-loop correction to the scalar mass from fermion loops gives rise to a simple IR pole, while a logarithmic UV divergence in the one-loop correction to the fermion mass from scalar loops give rise to only a logarithmic IR divergence. Intriguingly, the infrared pole in the scalar two-point function appears accessible in the $s$-channel in the Lorentzian theory, a feature which gives it particular phenomenological relevance. 

We then introduced softly-broken supersymmetry as a way to explore the interplay between (in)finiteness and UV/IR mixing. Choosing soft terms in order to keep the scalar light as the fermion mass is varied concretely illustrates several expected features. Strong UV/IR duality is preserved in the sense that both UV and IR divergences are absent at quadratic order (and persist at logarithmic order) when both the scalar and the fermion are in the spectrum. However, infrared structure reappears as the fermion mass is raised above a fixed cutoff and (quadratic) finiteness is lost. This underlines the sense in which UV/IR mixing may only ever play an interesting role when the field theory is quadratically UV sensitive at all scales, a scenario in which the hierarchy problem is less concrete.

Finally, building on the lessons from the toy models considered here, we have highlighted a variety of interesting lines of exploration in theories featuring nonlocality with or without Lorentz violation that may be of relevance to the hierarchy problem.

While the prospect that UV/IR mixing will solve outstanding theoretical problems in the low-energy universe is possibly fanciful, now is the time for such reveries. The paradigms of the past few decades of particle theory are under considerable empirical pressure, and innovative approaches are needed. At the very least, by pushing the limits of EFT we stand to learn more about the broad spectrum of phenomena possible within quantum field theory.

\begin{comment}\section*{Acknowledgements}
We thank Nima Arkani-Hamed, Matthew Brown, Andy Cohen, Tim Cohen, Brianna Grado-White, Alex Kinsella, Harold Steinacker, Terry Tomboulis, Timothy Trott, and Yue Zhao for valuable discussions. We are grateful to Tim Cohen, Isabel Garcia-Garcia, and Robert McGehee for comments on a draft of this manuscript. This work is supported in part by the US Department of Energy under the Early Career Award DE-SC0014129 and the Cottrell Scholar Program through the Research Corporation for Science Advancement.

\appendix
\section{Wilsonian Interpretations from Auxiliary Fields}\label{app:auxfield}
In this appendix we discuss various generalizations of the procedure introduced in \cite{Minwalla:1999px,VanRaamsdonk:2000rr} to account for the new structures appearing in the noncommutative quantum effective action via the introduction of additional auxiliary fields. 

\subsection{Scalar Two-Point Function}

It is simple to generalize the procedure discussed in Section \ref{sec:phi4} to add to the quadratic effective action of $\phi$ any function we wish through judicious choice of the two-point function for an auxiliary field $\sigma$ which linearly mixes with it. In position space, if we wish to add to our effective Lagrangian 
\begin{equation}
\Delta \mathcal{L}_{\text{eff}} = \half c^2 \phi(x) f(-i\partial) \phi(x),
\end{equation} 
where $f(-i\partial)$ is any function of momenta, and $c$ is a coupling we've taken out for convenience, then we simply add to our tree-level Lagrangian 
\begin{equation}
\Delta \mathcal{L} = \half \sigma(x) f^{-1}(-i \partial) \sigma(x) + i c \sigma(x) \phi(x),
\end{equation}
where $f^{-1}$ is the operator inverse of $f$.  It should be obvious that this procedure is entirely general. As applied to the Euclidean $\phi^4$ model, we may use this procedure to add a second auxiliary field to account for the logarithmic term in the quadratic effective action as
\begin{equation}
\Delta \mathcal{L} = \half \sigma(x) \frac{1}{\log\left[1 - \frac{4}{\Lambda^2 \partial \circ \partial }\right]} \sigma(x) - \frac{g M}{\sqrt{96 \pi^2}} \sigma(x) \phi(x),
\end{equation}
where we point out that the argument of the log is just $4/(\Lambda_{\text{eff}}^2 p \circ p)$ in position space. We may then try to interpret $\sigma$ also as a new particle. As discussed in \cite{VanRaamsdonk:2000rr}, its logarithmic propagator may be interpreted as propagation in an additional dimension of spacetime.

Alternatively, we may simply add a single auxiliary field which accounts for both the quadratic and logarithmic IR singularities by formally applying the above procedure. But having assigned them an exotic propagator, it then becomes all the more difficult to interpret such particles as quanta of elementary fields.

\subsection{Fermion Two-Point Function} \label{app:auxferm}

To account for the IR structure in the fermion two-point function, we must add an auxiliary fermion $\xi$. If we wish to find a contribution to our effective Lagrangian of 
\begin{equation}
\Delta \mathcal{L}_{\text{eff}} = c^2 \bar \psi \mathcal{O} \psi,
\end{equation}
where $\mathcal{O}$ is any operator on Dirac fields, then we should add to our tree-level Lagrangian
\begin{equation}
\Delta \mathcal{L} = - \bar \xi \mathcal{O}^{-1} \xi + c \left( \bar \xi \psi + \bar \psi \xi\right),
\end{equation}
with $\mathcal{O}^{-1}$ the operator inverse of $\mathcal{O}$. In the Lorentzian Yukawa theory of Section \ref{sec:yukawa}, if we add to the Lagrangian
\begin{equation}
\Delta \mathcal{L} =  - \overline{\xi} \frac{M - i \slashed{\partial}/2}{M^2 - \partial^2/4} \left[\log\left(1 - \frac{4}{\Lambda^2 \partial \circ \partial}\right)\right]^{-1} \xi + \frac{g}{2\sqrt{2}\pi} \left( \overline{\xi} \psi + \overline{\psi} \xi\right).
\end{equation}
we again find a one-loop quadratic effective Lagrangian which is equal to the $\Lambda \rightarrow \infty$ value of the original, but now for any value of $\Lambda$.

\subsection{Three-Point Function} \label{app:aux3pt}

We may further generalize the procedure for introducing auxiliary fields to account for IR poles to the case of poles in the three-point effective action. It's clear from the form of the IR divergences in Equation \ref{eqn:3ptfct} that they `belong' to each leg, and so na\"{i}vely one might think this means that the divergences we've already found in the two point functions already fix them. However those corrections only appear in the internal lines and were already proportional to $g^2$, and so they will be higher order corrections. Instead we must generate a correction to the vertex function itself which only corrects one of the legs. 

To do this we must introduce auxiliary fields connecting each possible partition of the interaction operator. However, while an auxiliary scalar $\chi$ coupled as $\chi \varphi + \chi \overline{\psi}\psi$ would generate a contribution to the vertex which includes the $\chi$ propagator with the $\varphi$ momentum flowing through it, it would also generate a new $(\overline{\psi}\psi)^2$ contact operator, which we don't want. To avoid this we introduce \emph{two} auxiliary fields with off-diagonal two-point functions, a trick used for similar purposes in \cite{VanRaamsdonk:2000rr}. By abandoning minimality, we can essentially use an auxiliary sector to surgically introduce insertions of functions of momenta wherever we want them.

We can first see how this works on the scalar leg. We add to our tree-level Lagrangian 
\begin{equation}
\Delta \mathcal{L} = - \chi_1(x) f^{-1}(-i\partial) \chi_2(x) + \kappa_1 \chi_1(x) \varphi(x) + \kappa_2 \chi_2(x) \star \overline{\psi}(x) \star \psi(x).
\end{equation}
Now to integrate out the auxiliary fields we note that for a three point vertex, one may use momentum conservation to put all the noncommutativity between two of the fields. That is, $\chi_2(x) \star \overline{\psi}(x) \star \psi(x) = \chi_2(x)  (\overline{\psi}(x) \star \psi(x)) =  (\overline{\psi}(x) \star \psi(x)) \chi_2(x)$ as long as this is not being multiplied by any other functions of $x$. So we may use this form of the interaction to simply integrate out the auxiliary fields. We end up with 
\begin{equation}
\Delta \mathcal{L}_{\text{eff}} = \kappa_1 \kappa_2 \overline{\psi} \star \psi \star f(-i\partial) \varphi 
\end{equation}
which is exactly of the right form to account for an IR divergence in the three-point function which only depends on the $\varphi$ momentum.

For the fermionic legs, we need to add fermionic auxiliary fields which split the Yukawa operator in the other possible ways. We introduce Dirac fields $\xi,\xi'$ and a differential operator on such fields $\mathcal{O}^{-1}(-i\partial)$. Then if we add to the Lagrangian
\begin{equation}
\Delta \mathcal{L} = - \overline{\xi} \mathcal{O}^{-1}\xi' - \overline{\xi'} \mathcal{O}^{-1} \xi + c_1 (\overline{\xi} \star \psi \star \varphi + \overline{\psi} \star \xi\star \varphi) + c_2 (\overline{\xi} \star \varphi \star \psi  + \overline{\psi} \star \varphi \star \xi) +  c_3(\overline{\xi'} \psi + \overline{\psi} \xi'),
\end{equation}
we now end up with a contribution to the effective Lagrangian
\begin{equation}
\Delta \mathcal{L}_{\text{eff}} = c_1 c_3 \left( \bar \psi \star \mathcal{O} \left(\psi\right) \star \varphi  + \bar \psi \star \mathcal{O}\left(\psi \star \varphi\right)\right) + c_2 c_3 \left( \bar \psi \star \varphi \star \mathcal{O} \left(\psi\right)   + \bar \psi \star \mathcal{O}\left(\varphi \star \psi \right)\right),
\end{equation}
where we have abused notation and now the argument of $\mathcal{O}$ specifies which fields it acts on. These terms have the right form to correct both vertex orderings.

Now that we've introduced interactions between auxiliary fields and our original fields, the obvious question to ask is whether we can utilize the \textit{same} auxiliary fields to correct both the two-point and three-point actions. In fact, using two auxiliary fields with off-diagonal propagators per particle we may insert any corrections we wish. The new trick is to endow the auxiliary field interactions with extra momentum dependence. 

For a first example with a scalar, consider differential operators $f$, $\Phi$, and add to the Lagrangian
\begin{equation}
\Delta \mathcal{L} = - \chi_1 f^{-1}(-i\partial)\chi_2 + \kappa_1 \chi_1 \varphi + \kappa_2 \chi_2 \overline{\psi} \star \psi + g \varphi\Phi(-i\partial) \chi_2.
\end{equation}
We may now integrate out the auxiliary fields and find 
\begin{equation}
\Delta \mathcal{L}_{\text{eff}} = g \kappa_1 \varphi f(\Phi(\varphi)) + \kappa_1 \kappa_2 \overline{\psi} \star \psi \star f(\varphi) 
\end{equation}
where we've assumed that $f$ and $\Phi$ commute. If we take $\Phi = \mathds{1}$ then we have the interpretation of merely inserting the $\chi$ two-point function in both the two-and three-point functions. But we are also free to use some nontrivial $\Phi$, and thus to make the corrections to the two- and three-point functions have whatever momentum dependence we wish. It should be obvious how to generalize this to insert momentum dependence into the scalar lines of arbitrary $n-$point functions.

The case of a fermion is no more challenging in principle. For differential operators $\mathcal{O}, \mathcal{F}$, we add
\begin{multline}
\Delta \mathcal{L} = - \overline{\xi} \mathcal{O}^{-1}\xi' - \overline{\xi'} \mathcal{O}^{-1} \xi + c_1 (\overline{\xi} \star \psi \star \varphi + \overline{\psi} \star \xi \star \varphi) + c_2 (\overline{\xi} \star \varphi \star \psi  + \overline{\psi} \star \varphi \star \xi ) \\ + c_3(\overline{\xi'} \psi + \overline{\psi} \xi') +  \frac{g}{2} \left( \bar \xi \mathcal{O}^{-1}\mathcal{F}\psi + \bar \psi \mathcal{O}^{-1} \mathcal{F} \xi\right),
\end{multline}
and upon integrating out the auxiliary fields we find
\begin{equation}
\Delta \mathcal{L}_{\text{eff}} = g c_3 \bar \psi \mathcal{F} \psi + c_1 c_3 \left( \bar \psi \star \mathcal{O} \left(\psi\right) \star \varphi  + \bar \psi \star \mathcal{O}\left(\psi \star \varphi\right)\right) + c_2 c_3 \left( \bar \psi \star \varphi \star \mathcal{O} \left(\psi\right)   + \bar \psi \star \mathcal{O}\left(\varphi \star \psi \right)\right),
\end{equation}
where the generalization to $n$-points is again clear. Note that in the fermionic case it's crucial that we be allowed to insert different momentum dependence in the corrections to the two- and three-point functions, as these have different Lorentz structures.

Now we cannot quite implement this for the two- and three-point functions calculated in Section \ref{sec:yukawa}, for the simple reason that we regulated these quantities differently. That is, we have abused notation and the symbol `$\Lambda$' means different things in the results for the two- and three-point functions. In order to carry out this procedure, we could simply regulate the two-point functions in $3d$ Schwinger space, though we run into the technical obstruction that the integration method above only calculates the leading divergence, which is not good enough for the scalar case.


\chapter{Conclusion}
\epigraph{Scientists are baffled: What's up with the universe?}{\textit{The Washington Post} Headline \\ November 1, 2019 \cite{achenbach_2019}}

We end the way we began: Declaring it to be an exciting time in particle physics. The picture we have painted above on the state of the field is one of uncertainty---and indeed we have barely even touched on many of the important problems of the Standard Model. Dark matter and neutrino masses, while having had canonical, obvious, beautiful solutions in the context of supersymmetric grand unified theories, are also as yet mysterious. These fields have likewise turned their focus toward alternative mechanisms in the past few years as a result of the lack of observational evidence for their standard solutions. But these facts all make the universe a more exciting place to study. Imagine if we had found weak-scale supersymmetry at the LHC, and our job now was simply to interpret the data in terms of which of the supersymmetric extensions proposed and well-studied in the past decades were correct. Or even worse, if technicolor had really been the answer and we had to watch Nature repeat the same trick she used at the strong scale again at the weak scale. How dreadfully boring!

Yes, yes, this attitude is selfish and a bit flippant, but what we now have is the chance to learn \textit{more} about the universe and about the spectrum of possibilities in physics, and to explore new, radical ideas. 

Let me end with a reminder of another, prior era in which theoretical physicists had thought they had everything figured out, recalled by no less than Max Planck in a 1924 talk at the University of Munich, and bring to your mind the outcome of those predictions:

\begin{quote}
	As I began my university studies [in 1878] I asked my venerable teacher Philipp von Jolly for advice
	regarding the conditions and prospects of my chosen field of study. He described physics to me
	as a highly developed, nearly fully matured science, that through the crowning achievement
	of the discovery of the principle of conservation of energy it will arguably soon take its final
	stable form. It may yet keep going in one corner or another, scrutinizing or putting in order
	a jot here and a tittle there, but the system as a whole is secured, and theoretical physics
	is noticeably approaching its completion to the same degree as geometry did centuries ago.
	That was the view fifty years ago of a respected physicist at the time.\footnote{In fairness to von Jolly (1809-1884), he really was a respected experimental physicist in his day---enough so to have been knighted---and earlier in his life made important contributions to the understanding of gravity and of osmosis \cite{Wells2016InPO}. This attitude was not rare at the time, and he wouldn't be remembered for it were it not for a student of his having played a role in revolutionizing physics.} \\
	--- Max Planck 
	 
	As translated in Wells (2016) \cite{Wells2016InPO} from Planck (1933) \cite{planck1933wege}
\end{quote}


\noindent May the universe continue to surprise us. \hfill $\square$

\appendix

\dsp

\chapter{Kinetic Mixing in the Mirror Twin Higgs}
\label{sec:KinMixapp}

Since kinetic mixing plays a central role in freeze-twin dark matter, we discuss here at some length the order at which it is expected in the low-energy EFT. Of course, there may always be UV contributions which set $\epsilon$ to the value needed for freeze-in. However, if the UV completion of the MTH disallows such terms - for example, via supersymmetry, an absence of fields charged under both sectors, and eventually grand unification in each sector (see e.g. \cite{Berezhiani:2005ek,Falkowski:2006qq,Chang:2006ra,Craig:2013fga,Katz:2016wtw,Badziak:2017syq})- then the natural expectation is for mixing of order these irreducible IR contributions.

To be concrete, we imagine that $\epsilon = 0$ at the UV cutoff of the MTH, $\Lambda \lesssim 4 \pi f$. To find the kinetic mixing in the regime of relevance, at momenta $\mu \lesssim 1 \text{ GeV}$, we must run down to this scale. As we do not have the technology to easily calculate high-loop-order diagrams, our analysis is limited to whether we can prove diagrams at some loop order are vanishing or finite, and so do not generate mixing. Thus our conclusions are strictly always `we know no argument that kinetic mixing of this order is not generated', and there is always the possibility that further hidden cancellations appear. With that caveat divulged, we proceed and consider diagrammatic arguments in both the unbroken and broken phases of electroweak symmetry.

Starting in the unbroken phase, we compute the mixing between the hypercharge gauge bosons. Two- and three-loop diagrams with Higgs loops containing one gauge vertex and one quartic insertion vanish. By charge conjugation in scalar QED, the three-leg amplitude of a gauge boson and a complex scalar pair must be antisymmetric under exchange of the scalars. However, the quartic coupling of the external legs ensures that their momenta enter symmetrically. As this holds off-shell, the presence of a loop which looks like
\begin{center}
\includegraphics[width=0.28\textwidth]{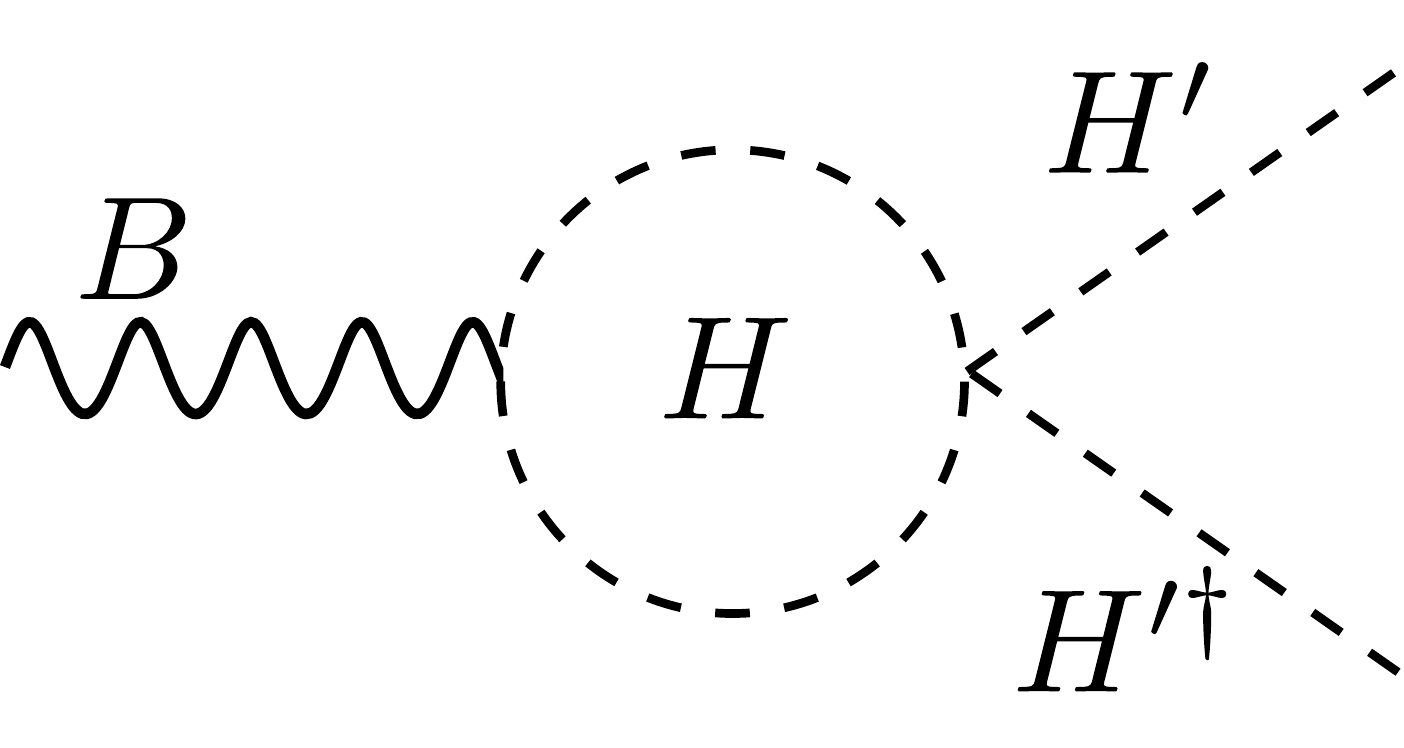}
\end{center}
causes the diagram to vanish. However, at four loops the following diagram can be drawn which avoids this issue:
\begin{center}
\includegraphics[width=0.45\textwidth]{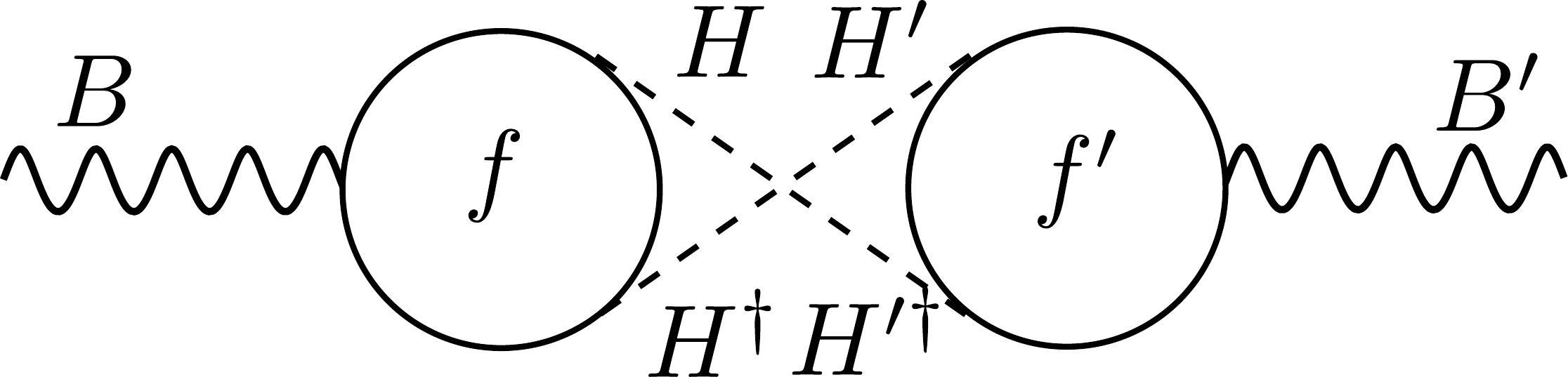}
\end{center}
where the two hypercharges are connected by charged fermion loops in their respective sectors and the Higgs doublets' quartic interaction. This diagram contributes at least from the MTH cutoff $\Lambda \lesssim 4 \pi f$ down to $f$, the scale at which twin and electroweak symmetries are broken. We have no argument that this vanishes nor that its unitarity cuts vanish. We thus expect a contribution to kinetic mixing of $\epsilon \sim g_1^2 c_W^2 / (4 \pi)^8$, with $g_1$ the twin and SM hypercharge coupling and $c_W = \cos \theta_W$ appearing as the contribution to the photon mixing operator. In this estimate we have omitted any logarithmic dependence on mass scales, as it is subleading.

In the broken phase, we find it easiest to perform this analysis in unitary gauge. The Higgs radial modes now mass-mix, but the emergent charge conjugation symmetries in the two QED sectors allow us to argue vanishing to higher-loop order. 
The implications of the formal statement of charge conjugation symmetry are subtle because we have two QED sectors, so
whether charge conjugation violation is required in both sectors seems unclear. However, similarly to the above case, there is a symmetry argument which holds off-shell. The result we rely on here is that in a vector-like gauge theory, diagrams with any fermion loops with an odd number of gauge bosons cancel pairwise. Thus, each fermion loop must be sensitive to the chiral nature of the theory, so the first non-vanishing contribution is at five loops as in:
\begin{center}
\includegraphics[width=0.45\textwidth]{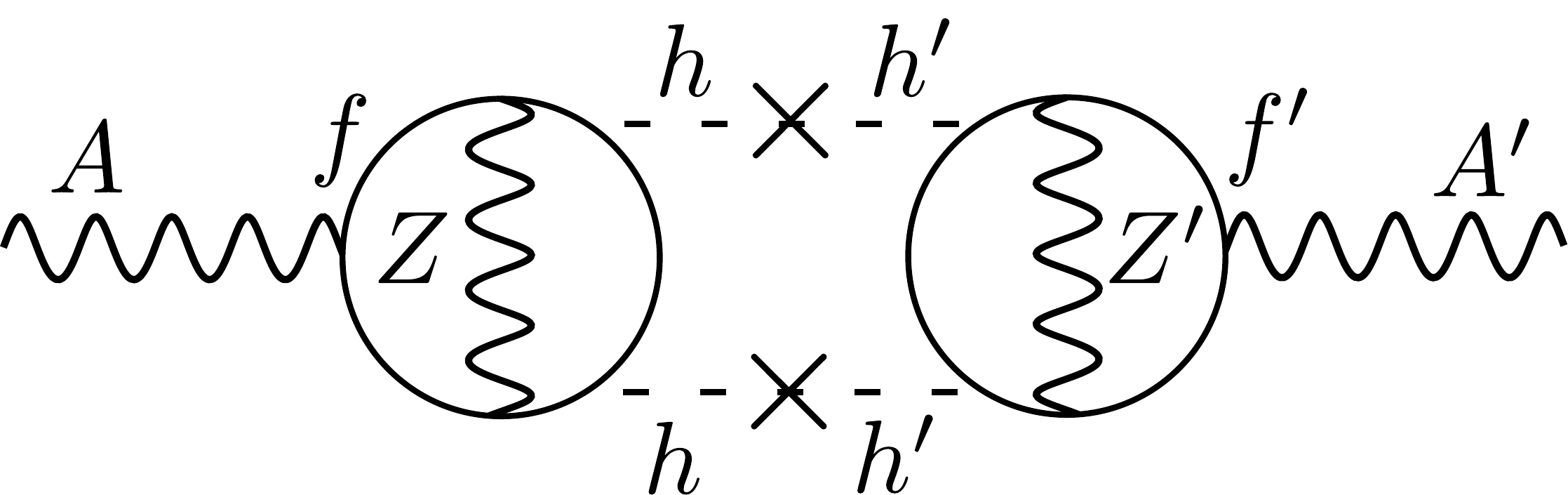}
\end{center}
where the crosses indicate mass-mixing insertions between the two Higgs radial modes which each contribute $\sim v/f$. Thus, both the running down to low energies and the finite contributions are five-loop suppressed. From such diagrams, one expects a contribution $\epsilon \sim e^2 g_A^2 g_V^2 (v/f)^2 / (4 \pi)^{10}$, where with $g_V$ and $g_A$ we denote the vector and axial-vector couplings of the $Z$, respectively. We note there are other five loop diagrams in which Higgses couple to massive vectors which are of similar size or smaller.

Depending on the relative sizes of these contributions, one then naturally expects kinetic mixing of order $\epsilon \sim 10^{-13} - 10^{-10}$. If $\epsilon$ is indeed generated at these loop-levels, then mixing on the smaller end of this range likely requires that it becomes disallowed not far above the scale $f$. However, we note that our ability to argue for higher-loop order vanishing in the broken versus unbroken phase is suggestive of the possibility that there may be further cancellations. We note also the possibility that these diagrams, even if nonzero, generate only higher-dimensional operators. Further investigation of the generation of kinetic mixing through a scalar portal is certainly warranted.


\chapter{How to Formulate Field Theory on a Noncommutative Space}\label{sec:NCFTderive}

In this appendix we provide detail on how to formulate field theories on a space which is defined by Equation \ref{eqn:ncdef}, which we repeat here for convenience:
\begin{equation}
\left[\hat{x}^\mu,\hat{x}^\nu\right] = i \theta^{\mu\nu}.
\end{equation} 
To construct a field theory on this space we must specify the algebra of observables. First we briefly recall the familiar, commutative case. For simplicity, we consider a scalar field theory on flat Euclidean space. We denote by $\text{Alg}\left(\mathbb{R}^d[x],\cdot\right)$ the commutative, $C^{*}$-algebra of Schwartz functions of $d$-dimensional Euclidean space with the standard point-wise product, and this constitutes our algebra of observables. A convenient basis for the vector space is that of plane waves $e^{i p\cdot x}$. 

The case of interest here is noncommutative flat Euclidean space, on which we define $\text{Alg}\left(\mathbb{R}^d_{\theta}[\hat{x}],\cdot\right)$. This now consists of such functions of $d$ variables $\hat{x}^\mu$ related by Equation \ref{eqn:ncdef}, and so is a noncommutative algebra, although we've specified again the normal `point-wise' product. A useful basis will again be that of plane waves, which we may define as the eigenfunctions of appropriately-defined derivatives on the noncommutative space, and which look familiar $e^{i p\cdot \hat{x}}$. To can get a sense for this algebra it is useful to carry out the simple exercise of multiplying two plane waves by simply applying Baker-Campbell-Hausdorff
\begin{equation}\label{eqn:planewaves}
e^{ik\cdot\hat{x}} \cdot e^{i k'\cdot \hat{x}} = \exp \left(ik\cdot\hat{x} + i k'\cdot \hat{x}  - \half  k_\mu k'_\nu \left[\hat{x}^\mu,\hat{x}^\nu\right]\right) = e^{-\frac{i}{2} \theta^{\mu\nu} k_\mu k'_\nu}e^{i\left(k+k'\right)\cdot \hat{x}}.
\end{equation}

As in quantum mechanics, we will wish to study noncommutative versions of familiar commutative theories, and so it will be useful to view $\mathbb{R}^d_\theta$ as a `deformation' of $\mathbb{R}^d$. We then wish to construct a map from our commutative algebra to our noncommutative one which returns smoothly to the identity as $\theta^{\mu\nu} \rightarrow 0$. The standard such choice is the Weyl-Wigner map $\hat{\mathcal{W}}$, which one may roughly think of as merely replacing $x$s with $\hat{x}$s.

The procedure is simply to Fourier transform from commutative space to momenta, and then inverse Fourier transform to noncommutative space. Given a commutative space Schwartz function $f$, we may compose the two operations and write
\begin{equation}
\hat{\mathcal{W}}[f] = \int \text{d}^dx f(x) \hat{\Delta}(x), \qquad \hat{\Delta}(x) = \int \frac{\text{d}^dk}{\left(2\pi\right)^d} e^{i k_i \cdot \hat{x}^i} e^{-i k_i \cdot x^i}.
\end{equation}

Note that this is an injective map of Schwartz functions on $\mathbb{R}^d$ to those on $\mathbb{R}^d_\theta$ which respects the vector space structure but not the structure of the algebra. This property is familiar from quantum mechanics.

We may now construct noncommutative versions of field variables, but we still don't know how to do physics on these spaces. That is, we can write down the Lagrangian for noncommutative $\phi^4$ theory, and we could even determine an action after we formulate a notion of an integral over a noncommutative space. But our familiar results about how to go from the action of a field theory to a calculation for a physical observable most certainly depended implicitly on living on a commutative space, and so it seems we must re-formulate physics from the bottom up.

Fortunately, such a drastic measure may not be necessary, as one may formulate QFT on noncommutative spaces as a simple modification of our normal field theory structure. The core idea is to find an algebra of functions on $\mathbb{R}^d$ which is isomorphic to $\text{Alg}\left(\mathbb{R}^d_{\theta}[\hat{x}],\cdot\right)$ by pushing the noncommutativity into a new field product, known as a Groenewold-Moyal product (or star-product). We diagram the structure we wish to look for in Figure \ref{fig:ncquant}.

\begin{figure}[h!] 
	\centering
	\begin{tikzcd}[sep=large]
	& \text{Alg}\left(\mathbb{R}^d[x],\cdot\right) \arrow[dl,dashed,hook,"\hat{\mathcal{W}}:\mathbb{R}^d{[x]}\rightarrow\mathbb{R}^d_\theta{[\hat{x}]}"'] \arrow[dr,dashed,leftrightarrow, "\text{Id}_{\mathbb{R}^d[x]}"] 
	& \\
	\text{Alg}\left(\mathbb{R}^d_\theta[\hat{x}],\cdot\right) \arrow[rr,leftrightarrow,"\hat{\mathcal{W}} \text{ an isomorphism of algebras}"] &                         & \text{Alg}\left(\mathbb{R}^d[x],\star_\theta\right)
	\end{tikzcd}
	\caption{The relations between the algebra of a commutative field theory, the noncommutative algebra one finds by applying the Weyl-Wigner map, and the noncommutative algebra most useful for field theory making use of the Groenewold-Moyal product.The vertical arrows respect only the vector space structure, and one should think of constructing a new algebra by applying a vector space map and then endowing the vectors with a multiplication operation.}\label{fig:ncquant}
\end{figure}
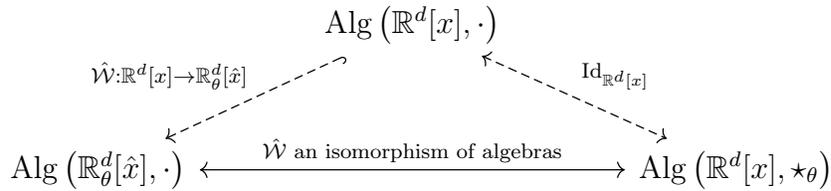

In particular, we may do this by demanding that our quantization map $\hat{\mathcal{W}}$ is upgraded to an isomorphism between $\text{Alg}\left(\mathbb{R}^d_{\theta}[\hat{x}],\cdot\right)$ and an algebra on the vector space of functions of commutative Euclidean space, with a multiplication operation which is chosen to preserve the algebraic structure. That is, we must satisfy 
\begin{equation}
\hat{\mathcal{W}}\left[f\star_\theta g\right] = \hat{\mathcal{W}}\left[f\right] \cdot \hat{\mathcal{W}}\left[g\right],	
\end{equation}
for any Schwartz functions $f,g$ on commutative Euclidean space. But we can guarantee this by ensuring it for plane waves, the calculation of which we've essentially already done above in Equation \ref{eqn:planewaves}:
\begin{align}
e^{ikx} \star_\theta e^{ik'x} &= \hat{\mathcal{W}}^{-1}\left[\hat{\mathcal{W}}\left[e^{ikx}\right] \cdot \hat{\mathcal{W}}\left[e^{ik'x}\right]\right] \nonumber\\
&= \hat{\mathcal{W}}^{-1}\left[e^{-\frac{i}{2} \theta^{ij} k_i k'_j}e^{i\left(k+k'\right)\cdot \hat{x}}\right] \nonumber \\
e^{ikx} \star_\theta e^{ik'x} &\equiv e^{-\frac{i}{2} \theta^{\mu\nu} k_\mu k'_\nu}e^{i\left(k+k'\right)\cdot x},
\end{align}

\noindent where the $\theta$ subscript merely tells us the star-product will depend on the noncommutativity tensor, and this will henceforth be dropped. This gives a position-space representation of the star-product,
\begin{equation}\label{eqn:starprod}
f(x) \star g(x) = \left. \exp\left(\frac{i}{2} \theta_{\mu\nu} \partial_y^\mu \partial_z^\nu\right) f(y) g(z) \right|_{y = z = x} = f(x) \exp\left(\frac{i}{2} \overleftarrow{\partial}^\mu \theta_{\mu\nu} \overrightarrow{\partial}^\nu\right) g(x).
\end{equation}
The general procedure to construct a noncommutative field theory from a commutative one is then by application of the Weyl-Wigner map. As an example, for a simple $\phi^n$ theory we find
\begin{equation}
\mathcal{L}^{(NC)}_\text{int} = \frac{\lambda}{n!} \hat{\mathcal{W}}^{-1}\left[\hat{\mathcal{W}}\left(\phi(x)\right)^n\right]\ = \frac{\lambda}{n!} \overbrace{\phi(x)\star\phi(x)\star\cdots\star\phi(x)}^{\text{n copies}}.
\end{equation}


\chapter{Wilsonian Interpretations of NCFTs from Auxiliary Fields}\label{app:auxfield}
In this appendix we discuss various generalizations of the procedure introduced in \cite{Minwalla:1999px,VanRaamsdonk:2000rr} to account for the new structures appearing in the noncommutative quantum effective action via the introduction of additional auxiliary fields. 

\section{Scalar Two-Point Function}

It is simple to generalize the procedure discussed in Section \ref{sec:phi4} to add to the quadratic effective action of $\phi$ any function we wish through judicious choice of the two-point function for an auxiliary field $\sigma$ which linearly mixes with it. In position space, if we wish to add to our effective Lagrangian 
\begin{equation}
\Delta \mathcal{L}_{\text{eff}} = \half c^2 \phi(x) f(-i\partial) \phi(x),
\end{equation} 
where $f(-i\partial)$ is any function of momenta, and $c$ is a coupling we've taken out for convenience, then we simply add to our tree-level Lagrangian 
\begin{equation}
\Delta \mathcal{L} = \half \sigma(x) f^{-1}(-i \partial) \sigma(x) + i c \sigma(x) \phi(x),
\end{equation}
where $f^{-1}$ is the operator inverse of $f$.  It should be obvious that this procedure is entirely general. As applied to the Euclidean $\phi^4$ model, we may use this procedure to add a second auxiliary field to account for the logarithmic term in the quadratic effective action as
\begin{equation}
\Delta \mathcal{L} = \half \sigma(x) \frac{1}{\log\left[1 - \frac{4}{\Lambda^2 \partial \circ \partial }\right]} \sigma(x) - \frac{g M}{\sqrt{96 \pi^2}} \sigma(x) \phi(x),
\end{equation}
where we point out that the argument of the log is just $4/(\Lambda_{\text{eff}}^2 p \circ p)$ in position space. We may then try to interpret $\sigma$ also as a new particle. As discussed in \cite{VanRaamsdonk:2000rr}, its logarithmic propagator may be interpreted as propagation in an additional dimension of spacetime.

Alternatively, we may simply add a single auxiliary field which accounts for both the quadratic and logarithmic IR singularities by formally applying the above procedure. But having assigned them an exotic propagator, it then becomes all the more difficult to interpret such particles as quanta of elementary fields.

\section{Fermion Two-Point Function} \label{app:auxferm}

To account for the IR structure in the fermion two-point function, we must add an auxiliary fermion $\xi$. If we wish to find a contribution to our effective Lagrangian of 
\begin{equation}
\Delta \mathcal{L}_{\text{eff}} = c^2 \bar \psi \mathcal{O} \psi,
\end{equation}
where $\mathcal{O}$ is any operator on Dirac fields, then we should add to our tree-level Lagrangian
\begin{equation}
\Delta \mathcal{L} = - \bar \xi \mathcal{O}^{-1} \xi + c \left( \bar \xi \psi + \bar \psi \xi\right),
\end{equation}
with $\mathcal{O}^{-1}$ the operator inverse of $\mathcal{O}$. In the Lorentzian Yukawa theory of Section \ref{sec:yukawa}, if we add to the Lagrangian
\begin{equation}
\Delta \mathcal{L} =  - \overline{\xi} \frac{M - i \slashed{\partial}/2}{M^2 - \partial^2/4} \left[\log\left(1 - \frac{4}{\Lambda^2 \partial \circ \partial}\right)\right]^{-1} \xi + \frac{g}{2\sqrt{2}\pi} \left( \overline{\xi} \psi + \overline{\psi} \xi\right).
\end{equation}
we again find a one-loop quadratic effective Lagrangian which is equal to the $\Lambda \rightarrow \infty$ value of the original, but now for any value of $\Lambda$.

\section{Three-Point Function} \label{app:aux3pt}

We may further generalize the procedure for introducing auxiliary fields to account for IR poles to the case of poles in the three-point effective action. It's clear from the form of the IR divergences in Equation \ref{eqn:3ptfct} that they `belong' to each leg, and so na\"{i}vely one might think this means that the divergences we've already found in the two point functions already fix them. However those corrections only appear in the internal lines and were already proportional to $g^2$, and so they will be higher order corrections. Instead we must generate a correction to the vertex function itself which only corrects one of the legs. 

To do this we must introduce auxiliary fields connecting each possible partition of the interaction operator. However, while an auxiliary scalar $\chi$ coupled as $\chi \varphi + \chi \overline{\psi}\psi$ would generate a contribution to the vertex which includes the $\chi$ propagator with the $\varphi$ momentum flowing through it, it would also generate a new $(\overline{\psi}\psi)^2$ contact operator, which we don't want. To avoid this we introduce \emph{two} auxiliary fields with off-diagonal two-point functions, a trick used for similar purposes in \cite{VanRaamsdonk:2000rr}. By abandoning minimality, we can essentially use an auxiliary sector to surgically introduce insertions of functions of momenta wherever we want them.

We can first see how this works on the scalar leg. We add to our tree-level Lagrangian 
\begin{equation}
\Delta \mathcal{L} = - \chi_1(x) f^{-1}(-i\partial) \chi_2(x) + \kappa_1 \chi_1(x) \varphi(x) + \kappa_2 \chi_2(x) \star \overline{\psi}(x) \star \psi(x).
\end{equation}
Now to integrate out the auxiliary fields we note that for a three point vertex, one may use momentum conservation to put all the noncommutativity between two of the fields. That is, $\chi_2(x) \star \overline{\psi}(x) \star \psi(x) = \chi_2(x)  (\overline{\psi}(x) \star \psi(x)) =  (\overline{\psi}(x) \star \psi(x)) \chi_2(x)$ as long as this is not being multiplied by any other functions of $x$. So we may use this form of the interaction to simply integrate out the auxiliary fields. We end up with 
\begin{equation}
\Delta \mathcal{L}_{\text{eff}} = \kappa_1 \kappa_2 \overline{\psi} \star \psi \star f(-i\partial) \varphi 
\end{equation}
which is exactly of the right form to account for an IR divergence in the three-point function which only depends on the $\varphi$ momentum.

For the fermionic legs, we need to add fermionic auxiliary fields which split the Yukawa operator in the other possible ways. We introduce Dirac fields $\xi,\xi'$ and a differential operator on such fields $\mathcal{O}^{-1}(-i\partial)$. Then if we add to the Lagrangian
\begin{equation}
\Delta \mathcal{L} = - \overline{\xi} \mathcal{O}^{-1}\xi' - \overline{\xi'} \mathcal{O}^{-1} \xi + c_1 (\overline{\xi} \star \psi \star \varphi + \overline{\psi} \star \xi\star \varphi) + c_2 (\overline{\xi} \star \varphi \star \psi  + \overline{\psi} \star \varphi \star \xi) +  c_3(\overline{\xi'} \psi + \overline{\psi} \xi'),
\end{equation}
we now end up with a contribution to the effective Lagrangian
\begin{equation}
\Delta \mathcal{L}_{\text{eff}} = c_1 c_3 \left( \bar \psi \star \mathcal{O} \left(\psi\right) \star \varphi  + \bar \psi \star \mathcal{O}\left(\psi \star \varphi\right)\right) + c_2 c_3 \left( \bar \psi \star \varphi \star \mathcal{O} \left(\psi\right)   + \bar \psi \star \mathcal{O}\left(\varphi \star \psi \right)\right),
\end{equation}
where we have abused notation and now the argument of $\mathcal{O}$ specifies which fields it acts on. These terms have the right form to correct both vertex orderings.

Now that we've introduced interactions between auxiliary fields and our original fields, the obvious question to ask is whether we can utilize the \textit{same} auxiliary fields to correct both the two-point and three-point actions. In fact, using two auxiliary fields with off-diagonal propagators per particle we may insert any corrections we wish. The new trick is to endow the auxiliary field interactions with extra momentum dependence. 

For a first example with a scalar, consider differential operators $f$, $\Phi$, and add to the Lagrangian
\begin{equation}
\Delta \mathcal{L} = - \chi_1 f^{-1}(-i\partial)\chi_2 + \kappa_1 \chi_1 \varphi + \kappa_2 \chi_2 \overline{\psi} \star \psi + g \varphi\Phi(-i\partial) \chi_2.
\end{equation}
We may now integrate out the auxiliary fields and find 
\begin{equation}
\Delta \mathcal{L}_{\text{eff}} = g \kappa_1 \varphi f(\Phi(\varphi)) + \kappa_1 \kappa_2 \overline{\psi} \star \psi \star f(\varphi) 
\end{equation}
where we've assumed that $f$ and $\Phi$ commute. If we take $\Phi = \mathds{1}$ then we have the interpretation of merely inserting the $\chi$ two-point function in both the two-and three-point functions. But we are also free to use some nontrivial $\Phi$, and thus to make the corrections to the two- and three-point functions have whatever momentum dependence we wish. It should be obvious how to generalize this to insert momentum dependence into the scalar lines of arbitrary $n-$point functions.

The case of a fermion is no more challenging in principle. For differential operators $\mathcal{O}, \mathcal{F}$, we add
\begin{multline}
\Delta \mathcal{L} = - \overline{\xi} \mathcal{O}^{-1}\xi' - \overline{\xi'} \mathcal{O}^{-1} \xi + c_1 (\overline{\xi} \star \psi \star \varphi + \overline{\psi} \star \xi \star \varphi) + c_2 (\overline{\xi} \star \varphi \star \psi  + \overline{\psi} \star \varphi \star \xi ) \\ + c_3(\overline{\xi'} \psi + \overline{\psi} \xi') +  \frac{g}{2} \left( \bar \xi \mathcal{O}^{-1}\mathcal{F}\psi + \bar \psi \mathcal{O}^{-1} \mathcal{F} \xi\right),
\end{multline}
and upon integrating out the auxiliary fields we find
\begin{equation}
\Delta \mathcal{L}_{\text{eff}} = g c_3 \bar \psi \mathcal{F} \psi + c_1 c_3 \left( \bar \psi \star \mathcal{O} \left(\psi\right) \star \varphi  + \bar \psi \star \mathcal{O}\left(\psi \star \varphi\right)\right) + c_2 c_3 \left( \bar \psi \star \varphi \star \mathcal{O} \left(\psi\right)   + \bar \psi \star \mathcal{O}\left(\varphi \star \psi \right)\right),
\end{equation}
where the generalization to $n$-points is again clear. Note that in the fermionic case it's crucial that we be allowed to insert different momentum dependence in the corrections to the two- and three-point functions, as these have different Lorentz structures.

Now we cannot quite implement this for the two- and three-point functions calculated in Section \ref{sec:yukawa}, for the simple reason that we regulated these quantities differently. That is, we have abused notation and the symbol `$\Lambda$' means different things in the results for the two- and three-point functions. In order to carry out this procedure, we could simply regulate the two-point functions in $3d$ Schwinger space, though we run into the technical obstruction that the integration method above only calculates the leading divergence, which is not good enough for the scalar case.
\end{mainmatter}

\ssp
\bibliographystyle{JHEP}
\bibliography{dissertation}

\end{document}